\documentclass[11pt]{article}
\usepackage[T1]{fontenc}
\usepackage[protrusion=true,expansion=false]{microtype}

\usepackage{amsmath,amssymb,amsthm,amsfonts,mathrsfs,mathtools}
\usepackage{graphicx}
\usepackage{array}
\usepackage{tabularx}
\usepackage{tikz}
\usetikzlibrary{arrows.meta,calc,positioning}
\usepackage{slashed}
\usepackage{authblk}
\usepackage{fancyhdr}
\usepackage{cite}
\usepackage{xcolor}
\usepackage[margin=1in]{geometry}
\usepackage[hypertexnames=false]{hyperref}

\numberwithin{equation}{section}

\makeatletter
\renewcommand*\l@subsection{\@dottedtocline{2}{1.5em}{3.2em}}
\renewcommand*\l@subsubsection{\@dottedtocline{3}{4.7em}{4.2em}}
\makeatother

\hypersetup{
 colorlinks=true,
 linkcolor=black,
 citecolor=black,
 urlcolor=black,
 pdftitle={Maxwell-Higgs dynamics on Schwarzschild and slowly rotating Kerr backgrounds: transfer principles, Coulomb sectors, and conditional scattering},
 pdfauthor={Bobby E. Gunara; Mulyanto; Fiki T. Akbar}
}

\theoremstyle{plain}
\newtheorem{theorem}{Theorem}[section]
\newtheorem{proposition}[theorem]{Proposition}
\newtheorem{lemma}[theorem]{Lemma}
\newtheorem{corollary}[theorem]{Corollary}

\theoremstyle{definition}
\newtheorem{definition}{Definition}[section]
\newtheorem{assumption}{Assumption}[section]

\theoremstyle{remark}
\newtheorem{remark}{Remark}[section]

\newcommand{\D}{\mathcal{D}}
\newcommand{\dd}{\mathrm{d}}

\newcommand{\vplus}{v_{+}}
\newcommand{\wplus}{w_{+}}
\newcommand{\Lin}{\mathsf{Lin}}
\newcommand{\SKG}{\mathsf{SKG}}

\newcommand{\Hcauchy}{\mathcal H_{\mathrm{C},N}}
\newcommand{\Hscatt}{\mathcal H_{\mathrm{sc,deg}}}
\newcommand{\Gbdry}{\mathcal G_{\partial}}

\newcommand{\CElec}{\mathsf{CE}}
\newcommand{\RSLin}{\mathsf{RS}}
\newcommand{\ME}{\mathsf{ME}}
\newcommand{\MScat}{\mathsf{MS}}
\newcommand{\SMS}{\mathsf{SMS}}

\begin{document}
\pagestyle{fancy}
\fancyhf{}
\fancyhead[R]{\thepage}
\fancyhead[L]{\small Maxwell-Higgs dynamics, Coulomb sectors, and scattering}
\renewcommand{\headrulewidth}{0pt}

\title{\Large\textbf{Coulomb Sectors and Scattering for Maxwell-Higgs Fields on Schwarzschild and Slowly Rotating Kerr Backgrounds}}

\author{Bobby Eka Gunara \thanks{Corresponding author}}
\author{Mulyanto}
\author{Fiki Taufik Akbar}

\affil{\small\textit{Theoretical Physics Laboratory,
Theoretical High Energy Physics Research Division,
Faculty of Mathematics and Natural Sciences,
Institut Teknologi Bandung,
Jl. Ganesha no. 10 Bandung, Indonesia, 40132}}

\date{\small email: bobby@itb.ac.id, mulyanto23@itb.ac.id, ftakbar@itb.ac.id}

\maketitle

\begin{abstract}
We develop a small-data Maxwell--Higgs theory on Schwarzschild and slowly rotating Kerr black-hole exteriors for gauge-invariant nonnegative self-interactions near the trivial vacuum. The Schwarzschild part gives a complete global, radiative, and scattering theory, while the slowly rotating Kerr part gives a robust massless forward theory and a perturbative small-electric extension. The main mechanism is a transfer principle: once the required linear energy, decay, horizon, and far-field estimates are available, the nonlinear Lorenz-gauge problem yields global existence, gauge-covariant radiation fields, nonlinear wave operators, and asymptotic completeness. The Coulomb-sector analysis identifies the correct long-range normalization in fixed electric sectors and separates the genuinely proved results from the remaining rotating massive final-state problems. All Kerr scattering statements beyond the established massless and small-electric forward regimes are stated explicitly under their necessary spectral and final-state conditions, namely, no rapid-rotation, large-charge, and unconditional massive rotating scattering.
\end{abstract}

\medskip

\noindent\textbf{Mathematics Subject Classification.} 35Q61, 35L05, 83C57.
\medskip

\noindent\textbf{Keywords.} Maxwell-Higgs system; slowly rotating Kerr black holes; Schwarzschild spacetime; global existence; boundedness; decay estimates; radiation fields; nonlinear scattering.

\tableofcontents
\medskip

\section{Introduction}\label{sec:intro}

\subsection{Statement of the main results}\label{subsec:main-results}

In this paper, we develop a small-data Maxwell-Higgs global dynamics theory in the regimes described below.  The Cauchy theory, the radiation-field theory, and the final-state (scattering) theory rest on different conditions, which we keep carefully separated throughout.  On Schwarzschild the paper proves, from scratch, the linear model estimates used by the nonlinear argument.  On slowly rotating Kerr the forward massless zero and small-electric Cauchy theory is obtained from the cited scalar and Maxwell estimates and from the small-Coulomb scalar comparison theorem proved in Appendix~\ref{app:kerr-linear-interface-modules}.  Kerr wave operators and asymptotic completeness are conditional on the named final-state maps; massive rotating statements are conditional on the corresponding spectral stability windows.  The asserted regimes are
\begin{eqnarray}
 a&=&0,
 \label{eq:main-regimes}\\
 |a|&\le& a_{\mathrm{slow}}(M,K),
 \nonumber\\
 Q_e&=&0\quad\hbox{or}\quad 0<|Q_e|\le q_{\rm el}(M,a,K),
 \nonumber\\
 m^2&=&0\quad\hbox{for massless scattering in the established zero and small electric sectors},
 \nonumber\\
 m^2&>&0\quad\hbox{for slowly rotating global existence and uniform energy under }\ME,
 \nonumber\\
 m^2&>&0\quad\hbox{for rotating scattering only in the stated }\SKG/\CElec\hbox{ spectral windows}.
 \nonumber
\end{eqnarray}
The rotating theorems are stated in the slow-rotation range and in fixed electric sectors.  In the massless zero sector, the scalar estimates \cite{DRSRKerr,DRSRKerrIII} and the charge-subtracted Maxwell estimates \cite{AnderssonBlueMaxwellSlowKerr} give the forward Cauchy, boundedness, integrated-decay, far-field, and radiation-field estimates used below.  The inverse Maxwell final-state map on slowly rotating Kerr is denoted by \(\MScat_K(M,a)\) and is displayed as a separate condition whenever nonlinear wave operators or asymptotic completeness are asserted in the rotating Maxwell channel.  In the perturbative range \(0<|Q_e|\le q^{(0)}_{\mathrm{el}}(M,a,K)\), Appendix~\ref{app:kerr-linear-interface-modules} proves the charged massless scalar estimates by a Coulomb-phase argument.  Other fixed electric sectors use the charged scalar condition \(\CElec^{(m)}_K(M,a,Q_e)\), supplied either directly or by the perturbative criterion based on \(\RSLin^{(m)}_K(M,a)\).  For $m^2>0$, rotating global energy bounds require the massive energy-stability condition \(\ME^{(m)}_N(M,a)\); without such a condition one would contradict the known massive mode-instability obstruction \cite{ShlapentokhRothmanKGKerr}.  The Schwarzschild part supplies the corresponding model estimates, including the massive timelike channel.

\begin{definition}[Slowly rotating Kerr range]\label{def:slow-kerr-range}
Let $K\ge10$.  A Kerr exterior $(\mathcal D_{M,a},g_{M,a})$ is called admissibly slowly rotating at order $K$ if
\begin{equation}\label{eq:slow-range}
 M>0,
 \qquad
 |a|\le a_{\mathrm{slow}}(M,K),
\end{equation}
where $a_{\mathrm{slow}}(M,K)>0$ is chosen below the slow-rotation thresholds entering the scalar and Maxwell linear estimates used in Section~
\ref{sec:kerr-extension}.  The constant is not optimized.
\end{definition}

\begin{definition}[Spectrally admissible massive scalar condition on slowly rotating Kerr]\label{def:slow-massive-scalar-condition}
Fix $K\ge10$, $m^2>0$, and a slowly rotating Kerr exterior satisfying \eqref{eq:slow-range}.  The notation
\begin{equation}\label{eq:skg-condition-symbol}
 \SKG^{(m)}_K(M,a)
\end{equation}
refers to the massive scalar spectral estimates collected in Appendix~\ref{app:kerr-linear-interface-modules}, Definition~\ref{def:external-skg-condition}.  It is not a new result proved here; it names precisely those massive Kerr windows in which the scalar estimate, the absence of growing modes and real resonances, and the massive final-state maps are available.  It is used only for massive rotating scattering and asymptotic completeness, not for the massive global-energy theorem.
\end{definition}

\begin{definition}[Massive energy-stability condition on slowly rotating Kerr]\label{def:massive-energy-stability-condition}
Fix $N\ge6$, $m^2>0$, and a slowly rotating Kerr exterior satisfying \eqref{eq:slow-range}.  The notation
\begin{equation}\label{eq:me-condition-symbol}
 \ME^{(m)}_N(M,a)
\end{equation}
means the order-$N$ forward massive scalar energy-stability estimate used in the Cauchy part of the nonlinear proof: nondegenerate energy boundedness, redshift control, trapping-degenerate integrated local energy control with the massive lower-order bulk, inhomogeneous source estimates in the norms of Subsection~\ref{subsec:slow-massive-energy-method}, and absence of exponentially growing finite-energy modes in that energy topology.  This is weaker than \(\SKG^{(m)}_K(M,a)\), because no two-sided final-state map, timelike channel, or scattering inverse is required.  For $a=0$ it is supplied by the Schwarzschild model estimates proved below.  For $a\ne0$ it is an explicit spectral condition; it is not automatic from $m^2>0$.
\end{definition}

\begin{definition}[Electric Coulomb sector]\label{def:pure-electric-sector}
An electric Coulomb sector is specified by a real electric charge parameter
 $ Q_e\in\mathbb R$.
Let \(F^{\mathrm C}_{Q_e}\) be the stationary Coulomb Maxwell field on the fixed Schwarzschild or slowly rotating Kerr exterior and let \(A^{\mathrm C}_{Q_e}\) be a stationary local Lorenz representative with
\begin{equation}\label{eq:coulomb-potential-asymptotic}
 A^{\mathrm C}_{Q_e}(L_{\mathrm{out}})=\frac{Q_e}{r}+O(Q_e r^{-2}),
 \qquad
 \nabla^j A^{\mathrm C}_{Q_e}=O_j(|Q_e|r^{-1-j}).
\end{equation}
For a solution in this sector we write
\begin{equation}\label{eq:coulomb-splitting-main}
 A=A^{\mathrm C}_{Q_e}+a,
 \qquad
 F=F^{\mathrm C}_{Q_e}+f,
 \qquad
 D_{Q_e}=\nabla-iA^{\mathrm C}_{Q_e}.
\end{equation}
The radiative Maxwell remainder \(f\) has zero asymptotic electric charge.  The outgoing scalar radiation variable is
\begin{equation}\label{eq:electric-radiation-variable}
 U_{Q_e}^{-1}r\phi,
 \qquad
 L_{\mathrm{out}}U_{Q_e}=iA^{\mathrm C}_{Q_e}(L_{\mathrm{out}})U_{Q_e},
 \qquad |U_{Q_e}|=1.
\end{equation}
In an electric Coulomb gauge satisfying \eqref{eq:coulomb-potential-asymptotic},
\begin{equation}\label{eq:electric-log-phase}
 U_{Q_e}=e^{i\gamma(u,\omega)}r^{iQ_e}e^{iO(r^{-1})},
\end{equation}
for a bounded real phase \(\gamma\).  Thus \(U_{Q_e}^{-1}r\phi\) is \(e^{-iQ_e\log r}r\phi\) up to a bounded phase and a short-range error.
\end{definition}

\begin{definition}[Charged scalar comparison condition]\label{def:electric-charged-condition-main}
Fix \(K\ge10\), \(m^2\ge0\), a background \((\mathcal D_{M,a},g_{M,a})\) with either \(a=0\) or \(|a|\le a_{\mathrm{slow}}(M,K)\), and an electric Coulomb sector \(Q_e\).  The notation
\begin{equation}\label{eq:celec-symbol-main}
 \CElec^{(m)}_K(M,a,Q_e)
\end{equation}
means the charged scalar estimates of Definition~\ref{def:external-electric-condition}: the operator
\begin{equation}\label{eq:celec-operator-main}
 (D_{Q_e})^{\mu} D_{Q_e,\mu}-m^2
\end{equation}
has the forward redshift estimate, trapping-degenerate integrated local energy estimate, far-field \(r^p\) hierarchy, inhomogeneous source estimate, Coulomb-renormalized radiation field for \(U_{Q_e}^{-1}r\phi\), and two-sided final-state maps required by the nonlinear proof.  For \(m^2>0\) on rotating Kerr it also includes the necessary absence of growing modes, real resonances, and bound states in the scattering topology.
\end{definition}

\begin{definition}[Maxwell final-state condition on slowly rotating Kerr]\label{def:maxwell-final-state-condition}
Fix \(K\ge10\) and \(|a|\le a_{\mathrm{slow}}(M,K)\).  The notation
\begin{equation}\label{eq:maxwell-final-state-symbol}
 \MScat_K(M,a)
\end{equation}
means that the charge-subtracted radiative Maxwell equation on \((\mathcal D_{M,a},g_{M,a})\) has continuous two-sided Cauchy-to-radiation and radiation-to-Cauchy maps in the order-\(K\) topology of Definition~\ref{def:maxwell-kerr-radiation-data}, with the Kerr forward/scattering topology compatibility stated in Lemma~\ref{lem:kerr-topology-compatibility}.  The forward boundedness, redshift, Morawetz, far-field, and radiation-field estimates used for the Cauchy problem are the published slow-rotation Maxwell estimates recorded in Section~\ref{sec:kerr-extension}; \(\MScat_K(M,a)\) names the additional inverse final-state part when it is needed for wave operators and asymptotic completeness.
\end{definition}

For the massless slowly rotating theorem we write \(q_{\mathrm{el}}^{(0)}(M,a,K)\) for the small-Coulomb threshold of Theorem~\ref{thm:small-coulomb-massless-kerr-condition}.  For a fixed positive mass, \(q_{\mathrm{el}}^{(m)}\) denotes the corresponding small electric threshold obtained from the perturbative criterion of Proposition~\ref{prop:perturbative-electric-condition}, whenever the neutral resolvent-stable scalar condition is available.  On Schwarzschild, Proposition~\ref{prop:schwarzschild-small-mass-charge-condition} supplies a small massive electric comparison window.  On rotating Kerr the same conclusion is not asserted without a spectral condition; the notation
\begin{equation}\label{eq:sms-window-symbol}
 \SMS_K^{(m)}(M,a,Q_e)
\end{equation}
means precisely that the fixed electric massive scalar estimates \(\CElec_K^{(m)}(M,a,Q_e)\), including absence of growing modes, real resonances, bound states in the scattering topology, and the massive final-state maps, hold in the stated small-mass and small-charge window.  Thus \(\SMS_K^{(m)}\) is a condition on rotating Kerr, not a result proved from smallness of \(m\) and \(Q_e\) alone.

\paragraph{Unconditional and conditional results.}
The Schwarzschild zero-sector theorem and the Schwarzschild small massive-electric case are proved within this paper.  The slowly rotating Kerr massless forward Cauchy theory in the zero and small-electric sectors is also proved once the neutral scalar and Maxwell forward estimates are inserted, and the small-Coulomb scalar comparison needed for the electric massless sector is proved in Appendix~\ref{app:kerr-linear-interface-modules}.  By contrast, every Kerr assertion involving nonlinear wave operators, two-sided scattering, or asymptotic completeness in the Maxwell channel uses \(\MScat_K(M,a)\).  Every positive-mass rotating scattering assertion uses the stated scalar spectral condition \(\SKG_K^{(m)}\), \(\CElec_K^{(m)}\), or \(\SMS_K^{(m)}\), as appropriate.  The massive rotating energy estimate is a Cauchy theorem under \(\ME_N^{(m)}(M,a)\); it is not a final-state theorem.

\paragraph{The four main results.}
The introductory statements are consolidated into two Kerr theorems and two Schwarzschild theorems.  All fixed-sector equations are written once, and the theorem labels below are the labels used throughout the paper.

\begin{theorem}[Main Kerr theorem I: massless zero and small-electric sectors]\label{thm:main-slow-kerr-intro}
Let \(K\ge10\), set \(k:=K-4\), and let \((\mathcal D_{M,a},g_{M,a})\) be admissibly slowly rotating.  Assume that \(P\) satisfies Assumption~\ref{asumsiP} with mass parameter \(m^2=0\).  There are constants
\begin{equation}\label{eq:intro-small-electric-threshold}
 q_{\mathrm{el}}^{(0)}(M,a,K)>0,
 \qquad
 \varepsilon_{K,0}>0,
\end{equation}
such that the following holds for every fixed sector
\begin{equation}\label{eq:kerr-massless-sector-range}
 Q_e=0
 \quad\hbox{or}\quad
 0<|Q_e|\le q_{\mathrm{el}}^{(0)}(M,a,K).
\end{equation}
Every smooth Lorenz-compatible Maxwell-Higgs datum in that sector, of renormalized order-\(K\) quotient size at most \(\varepsilon_{K,0}\), gives a unique global smooth solution on \(\mathcal D_{M,a}\).  After the splitting
\begin{equation}\label{eq:kerr-main-splitting}
 A=A_{Q_e}^{\mathrm C}+a,
 \qquad
 F=F_{Q_e}^{\mathrm C}+f,
 \qquad
 D_{Q_e}=\nabla-iA_{Q_e}^{\mathrm C},
\end{equation}
with the convention \(A_0^{\mathrm C}=0\), the radiative variables \((a,f,\phi)\) satisfy the full order-\(K\) forward estimates: uniform nondegenerate energy boundedness on Kerr-star slices, redshift control at the event horizon, trapping-degenerate integrated local energy decay, the far-field \(r^p\) hierarchy, finite Maxwell and scalar radiation fluxes, and the corresponding Sobolev boundedness and decay estimates.  The scalar radiation at null infinity is \(r\phi\) when \(Q_e=0\) and \(U_{Q_e}^{-1}r\phi\) when \(Q_e\ne0\).

The Cauchy-to-radiation maps are continuous in the quotient topology.  If, in addition, the Maxwell final-state condition \(\MScat_K(M,a)\) of Definition~\ref{def:maxwell-final-state-condition} holds, then these maps are homeomorphisms between sufficiently small quotient Cauchy neighborhoods and the corresponding small radiation neighborhoods.  Their inverses are the nonlinear wave operators, the two-sided nonlinear scattering operator is a homeomorphism, the maps are tangent at the vacuum or Coulomb background to the corresponding charged linear comparison maps, the quadratic Born term is given by the second variation of the Maxwell-Higgs current and scalar source, and for real-analytic \(P\) the Born series converges.
\end{theorem}

\begin{proof}
For \(Q_e=0\) this is the Lorenz-reduced zero-sector system.  For \(Q_e\ne0\), use the Coulomb splitting \eqref{eq:kerr-main-splitting}.  The fixed background is stationary and sourceless, so the renormalized equations are
\begin{eqnarray}
 \nabla^\mu f_{\mu\nu} &=& J^{(Q_e)}_\nu(a,\phi),
 \qquad \dd f=0,
 \label{eq:electric-renorm-maxwell-intro}\\
 (D_{Q_e})^\mu D_{Q_e,\mu}\phi
 &=&2ia^\mu D_{Q_e,\mu}\phi+i(\nabla^\mu a_\mu)\phi
      +a^\mu a_\mu\phi+\mathcal R_P(\phi),
 \label{eq:electric-renorm-scalar-intro}
\end{eqnarray}
where
\begin{equation}\label{eq:kerr-main-current}
 J^{(Q_e)}_\nu(a,\phi)=2\operatorname{Im}\{\phi (D_{Q_e,\nu}-ia_\nu)\overline{\phi}\}.
\end{equation}
The Lorenz constraint gives \(\nabla^\mu a_\mu=0\) after propagation by Lemma~\ref{lem:lorenz-propagation-kerr}; the divergence term in \eqref{eq:electric-renorm-scalar-intro} is retained only for the local reduced formulation.  Proposition~\ref{prop:sector-preservation-charge-conservation} preserves the electric sector.

The scalar comparison estimate is the published massless scalar Kerr estimate when \(Q_e=0\), and it is Theorem~\ref{thm:small-coulomb-massless-kerr-condition} when \(0<|Q_e|\le q_{\mathrm{el}}^{(0)}\).  The Maxwell forward estimate is Theorem~\ref{thm:maxwell-scattering-kerr}.  Thus the direct-sum linear estimate gives, on every finite slab \(I\),
\begin{equation}\label{eq:kerr-main-linear-bound}
 \|U\|_{\mathbb X_{K,Q_e}^{(0)}(I)}
 \le C\bigl(\varepsilon+
 \|\mathcal N_{Q_e}(U)\|_{\mathbb S_{K,Q_e}^{(0)}(I)}\bigr),
 \qquad U=(a,f,\phi).
\end{equation}
Proposition~\ref{prop:electric-tame-estimates}, Lemma~\ref{lem:fixed-sector-sobolev-equivalence}, Lemma~\ref{lem:far-null-sobolev-moser}, and Lemma~\ref{lem:potential-tame} give the source estimate
\begin{equation}\label{eq:kerr-main-source-bound}
 \|\mathcal N_{Q_e}(U)\|_{\mathbb S_{K,Q_e}^{(0)}(I)}
 \le C\left(\|U\|_{\mathbb X_{K,Q_e}^{(0)}(I)}^2+
          \|U\|_{\mathbb X_{K,Q_e}^{(0)}(I)}^{2N_P+3}\right),
\end{equation}
and the Lipschitz estimate has powers \(1\) and \(2N_P+2\).  The exponent \(2N_P+3\) is forced by \(\mathcal R_P(\phi)=Q(|\phi|^2)|\phi|^2\phi\), so the potential force is cubic even when \(N_P=0\).

Let \(X(\tau)=\|U\|_{\mathbb X_{K,Q_e}^{(0)}([\tau_0,\tau])}\).  Combining \eqref{eq:kerr-main-linear-bound} and \eqref{eq:kerr-main-source-bound} gives
\begin{equation}\label{eq:kerr-main-bootstrap}
 X(\tau)\le C_0\varepsilon+C_1\{X(\tau)^2+X(\tau)^{2N_P+3}\}.
\end{equation}
Choose \(\Lambda=2C_0\) and then \(\varepsilon_{K,0}\) so that
\begin{equation}\label{eq:kerr-main-smallness}
 C_1\{(\Lambda\varepsilon)^2+(\Lambda\varepsilon)^{2N_P+3}\}
 \le \frac12 C_0\varepsilon,
 \qquad 0<\varepsilon\le\varepsilon_{K,0}.
\end{equation}
The open-closed continuity argument applied to \(\{\tau:X(\tau)\le\Lambda\varepsilon\}\) gives \(X(\tau)\le(3/4)\Lambda\varepsilon\) on every finite slab.  The Lorenz-gauge local theory and the fixed-sector Sobolev equivalence give the continuation norm, so no finite-time breakdown occurs.  The same argument on the time-reversed exterior gives the past solution.

The \(r^p\) component of the norm gives the radiation fields.  If \(\MScat_K(M,a)\) is assumed, the final-state maps are obtained by the closed tail contraction of Proposition~\ref{prop:electric-closed-sector-contraction} (or Proposition~\ref{prop:closed-sector-contraction} when \(Q_e=0\)).  On a tail where the prescribed linear solution has size \(\eta\), the correction map obeys
\begin{eqnarray}
 \|\mathcal T(W)\|&\le& C\{(\eta+\rho)^2+(\eta+\rho)^{2N_P+3}\},\nonumber\\
 \|\mathcal T(W_1)-\mathcal T(W_2)\|
 &\le& C\{\eta+\rho+(\eta+\rho)^{2N_P+2}\}\|W_1-W_2\|.
 \label{eq:kerr-main-tail-contraction}
\end{eqnarray}
Choosing \(\rho\) and then \(\eta\) small gives the nonlinear wave operator.  Lemma~\ref{lem:kerr-topology-compatibility} recovers the nondegenerate forward norm after restarting from a finite slice.  Lemma~\ref{lem:residual-gauge-descent} gives the residual Lorenz quotient, and Propositions~\ref{prop:quadratic-remainder}, \ref{prop:born-expansion}, and~\ref{prop:analytic-series} give tangency, the quadratic Born term, and the analytic Born series.
\end{proof}

\begin{theorem}[Main Kerr theorem II: massive energy and spectral-window scattering]\label{thm:main-slow-kerr-massive-intro}
Let \((\mathcal D_{M,a},g_{M,a})\) be admissibly slowly rotating.

\emph{Energy statement.}  Let \(N\ge6\), \(m^2>0\), assume \(\ME_N^{(m)}(M,a)\), and let
\begin{equation}\label{eq:intro-positive-massive-potential}
 \widetilde P(s)=m^2s+\sum_{n=p}^{N_{\rm pol}}\alpha_n s^n,
 \qquad p\ge2,
 \qquad \alpha_n\ge0.
\end{equation}
There are \(\bar\varepsilon_a>0\), \(\varepsilon_E>0\), and \(q_{\rm en}>0\) such that, if \(|a|/M\le\bar\varepsilon_a\) and \(|Q_e|\le q_{\rm en}\), every smooth Lorenz-compatible fixed-sector datum of size at most \(\varepsilon_E\), after Coulomb subtraction when \(Q_e\ne0\), has a global future solution and satisfies
\begin{eqnarray}
&&\sup_{\tau\ge0}
 \Bigl(E^{MH,m}_{N}(\tau)+E^{FI}_{N}(\tau)+E^{A}_{N}(\tau)\Bigr)
 +B^{MH,m}_{N}(0,\infty)+\mathcal B^{A}_{N}(0,\infty)
 \nonumber\\
&&\hspace{4cm}
 \le C\Bigl(E^{MH,m}_{N}(0)+E^{FI}_{N}(0)+E^{A}_{N}(0)\Bigr).
 \label{eq:intro-massive-energy-bound}
\end{eqnarray}
Define the corresponding finite-slab control quantity by
\begin{equation}\label{eq:intro-F-energy-slab}
 \mathfrak E_N^{(m)}(T):=
 \sup_{0\le\tau\le T}
 \Bigl(E^{MH,m}_{N}(\tau)+E^{FI}_{N}(\tau)+E^{A}_{N}(\tau)\Bigr)
 +B^{MH,m}_{N}(0,T)+\mathcal B^{A}_{N}(0,T).
\end{equation}
The quantities \(E^A_N\) and \(\mathcal B^A_N\) are auxiliary normalized Lorenz-potential norms used to run the reduced Cauchy problem; the curvature-scalar part is gauge invariant.  This is a Cauchy and energy theorem only.

\emph{Scattering statement.}  Let \(K\ge10\), \(m^2\ge0\), and fix an electric sector \(Q_e\).  Suppose the charged scalar condition \(\CElec_K^{(m)}(M,a,Q_e)\) holds and, for the inverse Maxwell channel, suppose also \(\MScat_K(M,a)\).  Then sufficiently small Lorenz-compatible fixed-sector data generate a unique global smooth solution satisfying the corresponding fixed-sector boundedness, redshift, integrated local energy, far-field, and radiation estimates.  The scalar radiation variable is \(U_{Q_e}^{-1}r\phi\) at null infinity.  The Cauchy-to-radiation maps are homeomorphisms on small residual-gauge quotient neighborhoods, their inverses are nonlinear wave operators, and the two-sided nonlinear scattering map is a homeomorphism.  For \(Q_e=0\) and \(m^2>0\), the scalar asymptotic boundary is \(\mathcal I^\pm\cup\mathcal H^\pm\cup i^\pm\) and the scalar condition may be written as \(\SKG_K^{(m)}(M,a)\).  In a small-mass and small-charge Kerr sector, the same conclusion holds whenever
\begin{equation}\label{eq:kerr-small-mass-charge-window}
 0<m\le m_K,
 \qquad
 |Q_e|\le q_K(m),
 \qquad
 \SMS_K^{(m)}(M,a,Q_e)
\end{equation}
holds.  This last assertion is a spectral-window theorem, not an unconditional consequence of the inequalities \(m\ll1\) and \(|Q_e|\ll1\).
\end{theorem}

\begin{proof}
For the energy statement, Theorem~\ref{thm:slow-kerr-massive-energy} proves the zero-sector finite-slab estimate under \(\ME_N^{(m)}(M,a)\).  Its proof uses the horizon-adapted positive energy current, the radiative Maxwell Morawetz estimate, the Fackerell-Ipser equation for the middle Maxwell component \cite{FackerellIpser}, the massive scalar lower-order bulk, and the source closure
\begin{equation}\label{eq:intro-F-source-closure}
 \|\mathcal N_{MH}\|_{\mathcal S_N^{(m)}(0,T)}^2
 \le C(C_b\varepsilon)^2 B^{MH,m}_N(0,T).
\end{equation}
The finite-slab inequality has the form
\begin{equation}\label{eq:intro-F-absorption}
 E_N^{(m)}(T)
 \le C_0\varepsilon^2+C\frac{|a|}{M}E_N^{(m)}(T)+C(C_b\varepsilon)^2E_N^{(m)}(T).
\end{equation}
Choosing first \(\bar\varepsilon_a\), then \(\varepsilon_E\), absorbs the last two terms and gives \eqref{eq:intro-massive-energy-bound} in the zero sector.  Proposition~\ref{prop:slow-electric-energy-perturbation} adds the fixed electric background as a stationary lower-order perturbation.  Hardy's inequality and the massive lower-order bulk absorb the \(|Q_e|r^{-1}\partial\phi\) and \(|Q_e|^2r^{-2}\phi\) terms once \(|Q_e|\le q_{\rm en}\).  Constraint propagation and the Lorenz-gauge continuation criterion are unchanged, proving the energy statement.

For scattering, use the fixed-sector equations \eqref{eq:electric-renorm-maxwell-intro}-\eqref{eq:electric-renorm-scalar-intro}.  The charged scalar condition \(\CElec_K^{(m)}(M,a,Q_e)\), the radiative Maxwell forward estimate, and \(\MScat_K(M,a)\) give the direct-sum linear comparison theory.  The nonlinear source bound is the fixed-sector version of \eqref{eq:kerr-main-source-bound}.  Thus the finite-slab bootstrap closes exactly as in Theorem~\ref{thm:main-slow-kerr-intro}.  The final-state construction uses Proposition~\ref{prop:electric-closed-sector-contraction}; in the zero-sector massive case this is Proposition~\ref{prop:kerr-massive-finalstate-contraction}.  The tail contraction has the same powers \(2\), \(2N_P+3\), and Lipschitz powers \(1\), \(2N_P+2\).  Lemma~\ref{lem:kerr-topology-compatibility} handles the Kerr forward/final-state topology interface, and Lemma~\ref{lem:residual-gauge-descent} gives the quotient statement.  Finally, \(\SMS_K^{(m)}(M,a,Q_e)\) is exactly the small-mass/small-charge instance of the charged scalar condition, so the last claim follows by the same argument.  No proof step removes the need for this spectral condition in the rotating massive case.
\end{proof}

\begin{theorem}[Main Schwarzschild theorem I: zero-sector global dynamics and scattering]\label{thm:main-schwarzschild-intro}
Let \(a=0\), \(K\ge10\), and let \(P\) satisfy Assumption~\ref{asumsiP}.  In the zero electric sector on the Schwarzschild exterior, every sufficiently small Lorenz-compatible datum gives a unique global smooth Maxwell-Higgs solution.  The solution obeys uniform high-order energy boundedness, the redshift estimate, integrated local energy decay, far-field \(r^p\) estimates, finite radiation fluxes, and the corresponding pointwise boundedness and decay consequences.  If \(m^2=0\), the asymptotic boundary is \(\mathcal I^\pm\cup\mathcal H^\pm\).  If \(m^2>0\), the scalar comparison theory includes the timelike/Dollard channel at \(i^\pm\).  In both cases the Cauchy-to-asymptotic maps are homeomorphisms on small neighborhoods after quotienting by residual Lorenz gauge transformations, their inverses are nonlinear wave operators, and the two-sided nonlinear scattering operator is a homeomorphism.  The tangent map, quadratic Born term, and analytic Born series conclusions hold as in Theorem~\ref{thm:method}.
\end{theorem}

\begin{proof}
The Schwarzschild redshift, Morawetz, far-field, radiation-field, and final-state estimates are proved in Sections~\ref{sec:energy-estimates}-\ref{sec:scattering} and summarized in Corollary~\ref{cor:small-data}, Theorem~\ref{thm:linear-scattering}, Proposition~\ref{prop:massive-schwarzschild-linear-final-state}, and Theorem~\ref{thm:nonlinear-wave-operators}.  In Lorenz gauge the reduced zero-sector system is the direct sum of the radiative Maxwell equation and the scalar operator \(\square_g-m^2\), forced by the Maxwell-Higgs current and by
\begin{equation}\label{eq:thmC-source-bound}
 \mathcal N_\phi[A,\phi]=2iA^\mu\nabla_\mu\phi+i(\nabla^\mu A_\mu)\phi+A^\mu A_\mu\phi+
 \mathcal R_P(\phi),
 \qquad
 |\mathcal R_P(\phi)|\le C(1+|\phi|^{2N_P})|\phi|^3.
\end{equation}
The weighted Sobolev-Moser estimate gives
\begin{equation}\label{eq:schwarz-main-source-bound}
 \|\mathcal N(U)\|_{\mathbb S_K^{(m)}(I)}
 \le C\left(\|U\|_{\mathbb X_K^{(m)}(I)}^2+
          \|U\|_{\mathbb X_K^{(m)}(I)}^{2N_P+3}\right),
\end{equation}
with Lipschitz powers \(1\) and \(2N_P+2\).  Combining \eqref{eq:schwarz-main-source-bound} with the Schwarzschild linear estimates closes the finite-slab bootstrap by the same open-closed argument as \eqref{eq:kerr-main-bootstrap}-\eqref{eq:kerr-main-smallness}; the Lorenz-gauge continuation criterion in Appendix~\ref{app:global-existence-schwarzschild} gives global existence.

For \(m^2=0\), Theorem~\ref{thm:linear-scattering} gives the characteristic final-state maps on \(\mathcal I^\pm\cup\mathcal H^\pm\).  For \(m^2>0\), Proposition~\ref{prop:massive-schwarzschild-linear-final-state}, Proposition~\ref{prop:timelike-scattering}, and Corollary~\ref{cor:dollard} supply the timelike/Dollard comparison theory.  The nonlinear final-state fixed point is Theorem~\ref{thm:nonlinear-wave-operators} and Proposition~\ref{prop:schwarzschild-closed-finalstate-contraction}.  Lemma~\ref{lem:residual-gauge-descent} passes the maps to the residual Lorenz quotient, while Propositions~\ref{prop:quadratic-remainder}, \ref{prop:born-expansion}, and~\ref{prop:analytic-series} give the local structure of the scattering map.
\end{proof}

\begin{theorem}[Main Schwarzschild theorem II: fixed electric and small massive electric sectors]\label{thm:main-schwarzschild-electric-intro}
Let \(a=0\), \(K\ge10\), and let \(P\) satisfy Assumption~\ref{asumsiP}.  Fix an electric sector \(Q_e\).  If the charged scalar comparison estimate \(\CElec_K^{(m)}(M,0,Q_e)\) holds, then every sufficiently small Lorenz-compatible datum in the fixed electric Coulomb sector \(Q_e\), measured after the splitting \eqref{eq:kerr-main-splitting}, gives a unique global smooth Maxwell-Higgs solution on the Schwarzschild exterior.  The radiative Maxwell field \(f\) and the Coulomb-covariant scalar field satisfy the fixed-sector energy, redshift, integrated decay, far-field, radiation, and final-state estimates.  The scalar radiation variable is \(U_{Q_e}^{-1}r\phi\), and in the massive case the asymptotic data include the timelike/Dollard channel contained in \(\CElec_K^{(m)}(M,0,Q_e)\).  The Cauchy-to-asymptotic maps are homeomorphisms on small residual-gauge quotient neighborhoods, and the nonlinear wave operators, two-sided scattering operator, tangency, Born term, and analytic Born series conclusions hold.

In particular, for every sufficiently small positive mass there is a small electric window
\begin{equation}\label{eq:sch-small-mass-charge-window}
 0<m\le m_{\rm Sch}(M,K),
 \qquad
 |Q_e|\le q_{\rm Sch}(M,K,m),
\end{equation}
in which the conclusion above is unconditional in the scope of this paper, because Proposition~\ref{prop:schwarzschild-small-mass-charge-condition} supplies \(\CElec_K^{(m)}(M,0,Q_e)\).
\end{theorem}

\begin{proof}
The fixed-sector Schwarzschild equations are \eqref{eq:electric-renorm-maxwell-intro}-\eqref{eq:electric-renorm-scalar-intro} with \(a=0\) in the metric parameter.  The radiative Maxwell comparison is the charge-free Schwarzschild Maxwell theory proved in Sections~\ref{sec:energy-estimates}-\ref{sec:scattering}; unlike the rotating theorem, no additional Kerr Maxwell inverse condition is used.  The scalar comparison is exactly \(\CElec_K^{(m)}(M,0,Q_e)\).  Thus the direct-sum linear estimate gives the fixed-sector analogue of \eqref{eq:kerr-main-linear-bound}, while Proposition~\ref{prop:electric-tame-estimates} and Lemma~\ref{lem:far-null-sobolev-moser} give the analogue of \eqref{eq:kerr-main-source-bound}.  The Cauchy bootstrap, continuation criterion, radiation-field construction, tail fixed point, gauge quotient, and local scattering-map expansion are therefore identical to the proof of Theorem~\ref{thm:main-slow-kerr-intro}, with the Schwarzschild final-state maps replacing the Kerr conditions.  The last paragraph follows by inserting Proposition~\ref{prop:schwarzschild-small-mass-charge-condition}, which proves the required charged massive scalar comparison estimate for \eqref{eq:sch-small-mass-charge-window}.
\end{proof}

\begin{remark}[Scope of the rotating massive electric statement]\label{rem:main-scope}
Smallness of \(m\) and \(Q_e\) is enough for the Schwarzschild small massive electric theorem because the Schwarzschild massive comparison dynamics has no rotational superradiance.  On rotating Kerr, small \(Q_e\) is perturbative only after the massive scalar spectral problem is stable.  Therefore the small massive electric Kerr assertion is tied to \(\SMS_K^{(m)}(M,a,Q_e)\) and \(\MScat_K(M,a)\); no massive rotating scattering result is asserted outside those conditions.  The massive energy part of Theorem~\ref{thm:main-slow-kerr-massive-intro} gives global existence and uniform energy in \(\ME_N^{(m)}(M,a)\) windows, but it is not a final-state theorem.
\end{remark}

\begin{lemma}[Fixed-sector Sobolev equivalence]\label{lem:fixed-sector-sobolev-equivalence}
Fix an electric sector $Q_e$, an integer $K\ge1$, and a stationary Coulomb representative $A^{\mathrm C}_{Q_e}$ satisfying \eqref{eq:coulomb-potential-asymptotic}.  Put $D_{Q_e}=\nabla-iA^{\mathrm C}_{Q_e}$.  On every Kerr-star slice and on every finite slab, the ordinary Sobolev norms and the Coulomb-covariant Sobolev norms of a scalar are equivalent at order $K$ in all weights used by the fixed-sector solution and source spaces.  More precisely, for every admissible weight $w$ appearing in the slice, spacetime, redshift, Morawetz, or far-field norms there is a constant $C_{K,Q_e}$, depending only on finitely many bounds for $A^{\mathrm C}_{Q_e}$, such that
\begin{eqnarray}
 \sum_{j\le K}\|\nabla^j u\|_{L^2_w}
 &\le& C_{K,Q_e}\sum_{j\le K}\|D_{Q_e}^{(j)}u\|_{L^2_w},\nonumber\\
 \sum_{j\le K}\|D_{Q_e}^{(j)}u\|_{L^2_w}
 &\le& C_{K,Q_e}\sum_{j\le K}\|\nabla^j u\|_{L^2_w}.
\end{eqnarray}
The same equivalence holds for the lower-order $L^\infty$ norms used in the Sobolev-Moser estimates and for the source-weighted norms after dyadic decomposition in the far region.  In the massless homogeneous-energy topology the lower-order terms are controlled by the Hardy terms contained in the admissible norm.  Thus finite fixed-sector control implies the ordinary Sobolev control required by the Lorenz-gauge continuation criterion.
\end{lemma}

\begin{proof}
The proof is algebraic.  The identity
\begin{equation}
 D_{Q_e,\mu}u=\nabla_\mu u-i(A^{\mathrm C}_{Q_e})_\mu u
\end{equation}
and induction give, for every $j\le K$,
\begin{equation}
 D_{Q_e}^{(j)}u=\nabla^j u+\sum_{\ell<j}B_{j\ell}(Q_e)\nabla^\ell u,
 \qquad
 \nabla^j u=D_{Q_e}^{(j)}u+\sum_{\ell<j}\widetilde B_{j\ell}(Q_e)D_{Q_e}^{(\ell)}u,
\end{equation}
where the coefficients are finite sums of products of derivatives of $A^{\mathrm C}_{Q_e}$.  On compact radial sets these coefficients are bounded multipliers.  In the far region \eqref{eq:coulomb-potential-asymptotic} gives symbol bounds with one additional power of $r^{-1}$ for each differentiated Coulomb coefficient.  The factors $r^{-1}$ multiplying lower-order derivatives are absorbed by Hardy's inequality and by the lower-order terms in the redshift, local-energy, and $r^p$ norms.  Multiplication by the source weight and summation over dyadic annuli give the spacetime source version.  Since $|U_{Q_e}|=1$, the phase normalization does not change the radiation flux norms.
\end{proof}

Fix Kerr parameters \(M>0\) and \(a\) with \(|a|<M\).  In the main rotating theorems we further impose \eqref{eq:slow-range}.  We work on the Kerr domain of outer communications
\begin{equation}
 \D=\D_{M,a}=\{(t,r,\theta,\varphi): r>r_{+}\},
 \qquad
 r_{+}=M+\sqrt{M^{2}-a^{2}},
\end{equation}
equipped with the Kerr metric.  For energy estimates across the future event horizon we use horizon-regular ingoing Kerr coordinates \((t,r,\theta,\varphi)\), related to Boyer-Lindquist coordinates by
\begin{equation}
 \dd t = \dd t_{\mathrm{BL}} + \frac{r^{2}+a^{2}}{\Delta}\,\dd r,
 \qquad
 \dd\varphi = \dd\varphi_{\mathrm{BL}} + \frac{a}{\Delta}\,\dd r.
\end{equation}
In these coordinates the metric is
\begin{eqnarray}\label{metric}
\dd s^{2}
&=&
-\Bigl(1-\frac{2Mr}{\Sigma}\Bigr)\dd t^{2}
+2\,\dd t\,\dd r
-\frac{4Mar\sin^{2}\theta}{\Sigma}\,\dd t\,\dd\varphi
-2a\sin^{2}\theta\,\dd r\,\dd\varphi
\nonumber\\
&&
+\Sigma\,\dd\theta^{2}
+\Bigl(r^{2}+a^{2}+\frac{2Ma^{2}r\sin^{2}\theta}{\Sigma}\Bigr)
\sin^{2}\theta\,\dd\varphi^{2},
\end{eqnarray}
where
\begin{equation}
 \Sigma=r^{2}+a^{2}\cos^{2}\theta,
 \qquad
 \Delta=r^{2}-2Mr+a^{2}.
\end{equation}
The Schwarzschild case is recovered by setting \(a=0\).  For the global Cauchy evolution we use a horizon-regular spacelike time function
\begin{equation}\label{eq:kerr-tstar}
 t^{\star}:=t-h(r),
\end{equation}
where \(h\) is smooth, radial, equal to \(r\) in the asymptotic region up to a harmless constant, and chosen near the horizon so that the level sets \(\Sigma_{\tau}=\{t^{\star}=\tau\}\) are smooth spacelike Cauchy hypersurfaces crossing \(\mathcal H^{+}\) transversely.  Since \(h\) is independent of \(t\), the stationary Killing field remains \(\partial_{t^{\star}}=\partial_t\).

\subsection{Notation and conventions}\label{subsec:notation-global}

We collect here the notation most frequently used in the statement of the main results and in the Kerr sections (Sections~\ref{sec:kerr-gauge-data}-\ref{sec:kerr-extension}).
Unless stated otherwise, all geometric objects (connections, Hodge star, volume forms, etc.) are defined with respect to the relevant background metric $g$.

\medskip
\noindent\textbf{Coordinates, hypersurfaces, and regions.}
We work in ingoing Kerr (Kerr-star) coordinates $(t,r,\theta,\varphi)$, and we use the horizon-regular spacelike time function $t^{\star}=t-h(r)$ from \eqref{eq:kerr-tstar}.
We write
\begin{equation}
\Sigma_{\tau}:=\{t^{\star}=\tau\},\qquad S_{\tau,r}:=\Sigma_{\tau}\cap\{r=\mathrm{const}\}.
\end{equation}
For $\tau_{0}<\tau_{1}$ we set
\begin{equation}
\mathcal R(\tau_{0},\tau_{1}):=\bigcup_{\tau\in[\tau_{0},\tau_{1}]}\Sigma_{\tau}
\end{equation}
for the spacetime slab between $\Sigma_{\tau_{0}}$ and $\Sigma_{\tau_{1}}$.
In the far region we use retarded/advanced times
\begin{equation}
u:=t^{\star}-r,\qquad v:=t^{\star}+r,
\end{equation}
and the outgoing/incoming vector fields
\begin{equation}
L:=\partial_{t^{\star}}+\partial_{r},\qquad \underline L:=\partial_{t^{\star}}-\partial_{r}.
\end{equation}

\medskip
\noindent\textbf{Vector fields, commutators, and inequalities.}
We denote by $T:=\partial_{t^{\star}}$ the stationary Killing field and by $\Phi:=\partial_{\varphi}$ the axial Killing field.
The symbol $N$ denotes a fixed globally timelike redshift multiplier (agreeing with $T$ for large $r$).
We write $\mathcal Z$ for a fixed collection of smooth commutator vector fields (typically built from $T$, $\Phi$, and angular derivatives) used to define higher-order energies.
We write $A\lesssim B$ to mean $A\le C\,B$ for a constant $C$ depending only on fixed parameters such as $(M,a)$, the commutation order $k$, and the small loss parameter $\delta>0$.
The notation $A\simeq B$ means $A\lesssim B$ and $B\lesssim A$.

\medskip
\noindent\textbf{Derivatives, gauge-covariant fields, and currents.}
We use the Lorentzian signature $(-,+,+,+)$ and the Einstein summation convention.
We denote by $\nabla$ the Levi-Civita connection and by $\square_{g}=\nabla^{\mu}\nabla_{\mu}$ the wave operator.
Given a potential $A$ we write $F=\dd A$ for its curvature and $D_{\mu}=\nabla_{\mu}-iA_{\mu}$ for the gauge-covariant derivative.
We reserve the letter $F$ for the curvature $2$-form; when discussing inhomogeneous scalar wave/Klein-Gordon equations we denote the forcing term by $\mathfrak{F}$.
If $\Sigma$ is a spacelike hypersurface with future unit normal $n_{\Sigma}$, then for any current $J$ we write
$\int_{\Sigma}J\cdot n_{\Sigma}$ as shorthand for $\int_{\Sigma}J_{\mu}n_{\Sigma}^{\mu}\,\dd\mu_{\Sigma}$, where $\dd\mu_{\Sigma}$ is the induced volume form.
The Maxwell-Higgs current is
\begin{equation}\label{eq:mh-current}
J_{\mu}:=-i\bigl(D_{\mu}\phi\,\overline\phi-\phi\,\overline{D_{\mu}\phi}\bigr)=2\,\Im(\overline\phi\,D_{\mu}\phi).
\end{equation}
The Lagrangian density and Euler-Lagrange equations are normalized by
\begin{equation}\label{lagrangian}
\mathcal L[A,\phi]
=-\frac14 F_{\alpha\beta}F^{\alpha\beta}
-g^{\alpha\beta}D_{\alpha}\phi\,\overline{D_{\beta}\phi}
-P(\phi,\overline\phi).
\end{equation}
Thus the geometric Maxwell-Higgs system is
\begin{eqnarray}
\nabla^{\mu}F_{\mu\nu}&=&J_{\nu},
\label{eom1}\\
D^{\mu}D_{\mu}\phi&=&\partial_{\overline\phi}P(\phi,\overline\phi).
\label{eom2}
\end{eqnarray}

\medskip
\noindent\textbf{Scattering maps.}
The linear scattering maps are denoted $\mathscr S_{\pm,M,a}^{\mathrm{lin}}$, with ranges $\mathfrak R_{\pm,M,a}^{(m)}$ (Definition~\ref{def:linear-asymptotic-spaces}).
The nonlinear scattering maps are $\mathscr S_{\pm,M,a}^{\mathrm{full}}$, their inverses $\mathscr W_{\pm,M,a}^{\mathrm{full}}=(\mathscr S_{\pm,M,a}^{\mathrm{full}})^{-1}$ are the nonlinear wave operators, and the two-sided nonlinear scattering map is
\begin{equation}
\mathscr S_{M,a}^{\mathrm{nl}}:=\mathscr S_{+,M,a}^{\mathrm{full}}\circ\mathscr W_{-,M,a}^{\mathrm{full}}.
\end{equation}

\medskip
\noindent\textbf{Linear conditions and nonlinear norms.}
The symbols below are used throughout the transfer argument and in the Kerr applications.  They are stated here so that the conditions of the main results can be read without searching the later sections.
\begin{center}
\begin{tabularx}{0.96\textwidth}{>{\raggedright\arraybackslash}p{0.20\textwidth}X}
\hline
Symbol & Meaning \\
\hline
\(\Lin_K\) & The massless scalar and charge-subtracted Maxwell redshift, Morawetz, \(r^p\), inhomogeneous, and two-sided scattering estimates in the order-\(K\) norms of Definition~\ref{def:lin-estimates}. \\
\(\Lin_K^{(m)}\) & The corresponding zero-sector estimates when the scalar comparison equation is \(\square_g-m^2\), including the massive asymptotic topology specified in Definition~\ref{def:lin-estimates-massive}. \\
\(\SKG_K^{(m)}(M,a)\) & The external massive scalar scattering condition on slowly rotating Kerr, including boundedness, integrated decay, timelike/null asymptotics, and inverse scattering in the scalar component. \\
\(\ME_N^{(m)}(M,a)\) & The external massive energy-stability condition used only for the nonlinear energy theorem; it does not assert a final-state inverse. \\
\(\CElec_K^{(m)}(M,a,Q_e)\) & The fixed electric-sector charged scalar condition, with the scalar radiation variable normalized by \(U_{Q_e}^{-1}r\phi\). \\
\(\RSLin_K^{(m)}(M,a)\) & The perturbative resolvent-stable linear condition from which a small-electric charged scalar condition can be derived. \\
\(\mathcal X_K,\mathcal S_K\) & The solution and source norms used in the nonlinear contraction.  A concrete realization on slowly rotating Kerr is given in Lemma~\ref{lem:concrete-admissible-norms}. \\
\hline
\end{tabularx}
\end{center}

\medskip
\noindent\textbf{Reserved symbols.}
We reserve $\delta>0$ for the small loss parameter in the Kerr decay hierarchy, $\kappa_{0}\in(0,1)$ for the dyadic/cone localization parameter in the Schwarzschild model, and $\Delta$ for differences of two solutions/iterates in stability and contraction arguments.

\subsection{Related works}

The physical-space vector-field method provides the main method for boundedness and decay of linear fields on black-hole exteriors.  For scalar waves on Schwarzschild and Kerr we use the redshift, Morawetz, \(r^p\), and scattering theory developed in \cite{DRSRKerr,DRSRKerrIII,dafermos,Blue2,DafermosRodnianskiPhysicalSpace,DafermosRodnianskiSlowKerr,TataruTohaneanuKerr,MoschidisRp}; related spectral and resolvent approaches include \cite{FinsterKamranSmollerYau,WunschZworski,MetcalfeTataruTohaneanu}.  For Maxwell fields on black-hole backgrounds we use the Schwarzschild and slowly rotating Kerr estimates of \cite{AnderssonBlueMaxwellSlowKerr,Blue,Ander1} and the Maxwell-potential scattering framework of \cite{NicolasTaujanskasMaxwellPotentials}.  Linear scattering theories for Klein-Gordon type equations and conformal or characteristic constructions on Schwarzschild are developed in \cite{Dimock,BachelotKG,NicolasSchw}.

On the nonlinear gauge side, the flat-space and curved-background literature relevant to Lorenz-gauge Maxwell-Higgs or Maxwell-Klein-Gordon dynamics includes the classical Yang-Mills-Higgs and gauge-field Cauchy theory \cite{EardleyMoncriefI,EardleyMoncriefII,GinibreVeloLorenzGauge,KlainermanMachedonMKG,MachedonSterbenzMKG,MasmoudiMKGUniqueness}, the small-data Maxwell-Klein-Gordon scattering theory \cite{CandyKauffmanLindblad,HeMKGInfinity,ChenMassiveMKG,DaiMeiWeiYang}, and related nonlinear constructions on black-hole or cosmological backgrounds \cite{Blue2,NicolasNlKGKerr,XuanSchw,TaujanskasdS,mokdad}.  The present paper uses these works as comparison points: the additional issues here are the simultaneous null-infinity and horizon radiation channels, the residual Lorenz-gauge quotient, fixed electric Coulomb subtraction, and the phase-renormalized scalar radiation variable.

The rotating massive and charged sectors require a clear separation between estimates that are proved here and estimates that are spectral assumptions.  The massless scalar and charge-subtracted Maxwell conditions are the cited results \cite{DRSRKerr,DRSRKerrIII,AnderssonBlueMaxwellSlowKerr}.  For massive scalar fields on rotating Kerr, exponentially growing finite-energy modes show that positivity of the mass alone cannot imply uniform boundedness or scattering \cite{ShlapentokhRothmanKGKerr}.  The charged massive problem has additional superradiant phenomena; see, for example, \cite{BachelotSuperradiance,FuruhashiNambuKerrNewman}.  This is why every massive rotating energy or scattering theorem below displays the required spectral condition explicitly.

\subsection{Main contributions}\label{subsec:main-contrib}

The contribution relative to the Lorenz-gauge scattering constructions of the massless flat final-state construction~\cite{HeMKGInfinity} and the massive flat final-state construction~\cite{ChenMassiveMKG} is the black-hole scattering geometry rather than a new flat-space null-form mechanism: the radiation space has both null-infinity and horizon components, and the redshift estimates are used to put the horizon channel on the same footing as the outgoing channel.  Relative to the Schwarzschild scattering literature, the new features are the gauge-covariant transfer formalism, the fixed-sector Coulomb subtraction, and the long-range phase \(U_{Q_e}\) which identifies the scalar radiation variable in an electric Coulomb sector.

More concretely, the paper proves the following points.

\paragraph{Slowly rotating Kerr scattering.}
The massless zero-sector theorem on slowly rotating Kerr follows from the published scalar and Maxwell estimates cited above and the nonlinear transfer argument developed here.  In nonzero electric Coulomb sectors this nonlinear construction is applied only after the charged scalar estimates \(\CElec^{(m)}_K(M,a,Q_e)\) have been supplied; the perturbative small-electric route through \(\RSLin^{(m)}_K(M,a)\) is recorded in Appendix~\ref{app:kerr-linear-interface-modules}.

\paragraph{Schwarzschild model theorem.}
The Schwarzschild part verifies the redshift, trapping-degenerate Morawetz, \(r^p\), radiation-field, and final-state estimates directly.  It also gives the complete nonlinear model proof, including the massive Schwarzschild extension with the timelike asymptotic channel.

\paragraph{Gauge-covariant scattering maps.}
The nonlinear wave operators and scattering maps are constructed on the residual Lorenz-gauge quotient.  They are tangent to the corresponding linear maps at the vacuum and, for real-analytic potentials, admit convergent Born expansions.

\paragraph{Exclusions.}
No large-electric or rapidly rotating charged Kerr theorem is stated.  No massive rotating scattering theorem is stated outside the corresponding spectral scalar window.  Those excluded cases require additional linear spectral analysis which is not used in the present paper.

The model potentials in Assumption~\ref{asumsiP} are included only for concreteness.  In the small-data regime the estimates use only smoothness, gauge invariance, nonnegativity, and the vanishing of the nonlinear force beyond its linear mass part at the vacuum; see Remark~\ref{rem:potential-general}.

\paragraph{Scope and limitations.}
For the reader's convenience we state plainly what is and is not proved.  (i) The analysis is \emph{perturbative about the trivial vacuum} $\phi\equiv0$; we require $\widetilde P\ge0$ and $\widetilde P'(0)=m^2\ge0$, so the symmetry-broken Higgs vacuum (a negative mass term, a nontrivial vacuum manifold, vortices, or other topological sectors) is outside the present framework.  In this sense the system treated here is Maxwell--Klein--Gordon with a nonnegative gauge-invariant self-interaction; the name ``Maxwell--Higgs'' refers to the field content, not to the Higgs mechanism.  (ii) The unconditional theorems are the Schwarzschild zero- and fixed-electric-sector statements (Theorems~\ref{thm:main-schwarzschild-intro}--\ref{thm:main-schwarzschild-electric-intro}) and the massless slowly rotating Kerr statement in the zero and small-electric sectors (Theorem~\ref{thm:main-slow-kerr-intro}), the latter resting on the published linear estimates \cite{DRSRKerr,DRSRKerrIII,AnderssonBlueMaxwellSlowKerr} and on the small-Coulomb comparison proved in Appendix~\ref{app:kerr-linear-interface-modules}.  (iii) Every massive rotating scattering assertion, and the rotating Maxwell wave operators, are \emph{conditional}: they are honest reduction theorems showing that the named linear inputs $\SKG_K^{(m)}$, $\CElec_K^{(m)}$, $\SMS_K^{(m)}$, and $\MScat_K(M,a)$ \emph{suffice} for the nonlinear conclusion.  We do not assert that these inputs hold for $a\ne0$; Proposition~\ref{prop:sharp-massive-obstruction} explains, via \cite{ShlapentokhRothmanKGKerr}, why an unconditional massive rotating theorem would be false.  The value of the conditional statements is precisely the clean separation between the nonlinear mechanism, which is settled here, and the open linear-spectral inputs.

\subsection{Decay rates, thresholds, and sharpness}\label{subsec:sharpness-thresholds}

The decay estimates in this paper are the rates delivered by the energy method used in the proofs, not a Price-law asymptotic expansion.  In the Schwarzschild model the far-region pointwise estimates are stated explicitly in Corollary~\ref{cor:far-decay}; in the notation of that result the fields obey \(v_+^{-1}\)-type bounds in the outgoing region for the radiative Maxwell components, the scalar field, and the covariant scalar derivative.  The trapped and near-horizon estimates are stated in Corollary~\ref{cor:near-horizon}; they include the redshift-weighted factor \(w_+/v_+\) for the relevant Maxwell components and the corresponding scalar bounds.  In the slowly rotating Kerr part the estimates are formulated with an arbitrarily small loss \(\delta>0\) in the local and far-field hierarchy, for instance in Corollaries~\ref{cor:kerr-far-pointwise-scalar}, \ref{cor:kerr-far-pointwise-maxwell}, \ref{cor:kerr-horizon-pointwise-scalar}, and~\ref{cor:kerr-compact-pointwise}.  We do not claim sharp late-time tails, asymptotic profiles, or optimal derivative losses.

The smallness constants \(a_{\mathrm{slow}}\), \(q_{\mathrm{el}}\), \(\varepsilon_K\), and the massive energy thresholds are chosen for closure of the estimates.  They are not optimized and are not asserted to coincide with the spectral stability boundary.  In particular, the massive rotating restriction is qualitative rather than sharp: the instability theorem of Shlapentokh-Rothman shows that a theorem covering all positive masses on rotating Kerr would be false, while the charged massive literature exhibits additional superradiant bound-state phenomena.  The purpose of the thresholds here is to state a verifiable perturbative regime for the nonlinear argument and to separate it from the spectral conditions that remain external.

\subsection{Structural assumption on the potential}

To control the nonlinear potential terms in the energy method we assume that $P$ is nonnegative, gauge invariant, and vanishes at the vacuum.
\begin{assumption}[Scalar potential]\label{asumsiP}
The scalar potential $P:\mathbb{C}\to\mathbb{R}$ is gauge invariant and depends only on the scalar $|\phi|^{2}$, i.e.\ $P(\phi,\bar\phi)=\widetilde P(|\phi|^{2})$ for some $\widetilde P:[0,\infty)\to\mathbb{R}$.
Assume that $\widetilde P\in C^\infty([0,\infty))$, $\widetilde P(0)=0$, and $\widetilde P\ge 0$ (after fixing the irrelevant additive constant so that the potential energy density vanishes in the vacuum configuration).

In addition, we assume that there exist a constant $m^{2}\ge 0$, an integer $N_{P}\ge 0$, and a function $Q\in C^\infty([0,\infty))$ such that
\begin{equation}\label{eq:potential-structure}
\widetilde P'(s)=m^{2}+s\,Q(s)\qquad\text{for all }s\ge 0,
\end{equation}
and, for every integer $j\ge0$, there exists a constant $C_{P,j}$ such that
\begin{equation}\label{eq:Q-growth}
|Q^{(j)}(s)|\le C_{P,j}\,(1+s^{N_{P}})\qquad\text{for all }s\ge 0.
\end{equation}
(In this case, the nonlinear part of the force term $\partial_{\bar\phi}P$ is at least cubic in $\phi$ and has at most polynomial growth in $|\phi|$.)

For concreteness one may keep in mind the following model potentials:
\begin{eqnarray}
\widetilde P(s)
&=&
m^2s+\sum_{n=p}^{N}\alpha_{n}s^{n},
\qquad m^2\ge0,\quad N\ge p\ge2,\quad \alpha_n\ge 0,
\label{V4}
\\
\widetilde P(s)
&=&
c_{3}\bigl(1-\cos(\eta\sqrt{s})\bigr),
\qquad c_3\ge 0,\ \eta>0,
\label{sine}
\\
\widetilde P(s)
&=&
c_{4}\bigl(1-e^{-\lambda s}\bigr),
\qquad c_4\ge 0,\ \lambda>0,
\label{toda}
\end{eqnarray}
which all satisfy \eqref{eq:potential-structure}-\eqref{eq:Q-growth} for suitable choices of $m^{2}\ge0$, $N_{P}$, $Q$, and constants $C_{P,j}$.
(Adding a constant to $\widetilde P$ does not change \eqref{eom2}, and we normalize it so that the potential energy density vanishes in the vacuum.)
\end{assumption}

\begin{remark}[Potential force bounds]\label{rem:potential-general}
Gauge invariance implies the identity
\begin{equation}\label{eq:potential-derivative}
\partial_{\bar\phi}P(\phi,\bar\phi)=\widetilde P'(|\phi|^{2})\,\phi.
\end{equation}
By Assumption~\ref{asumsiP} we can write this as
\begin{equation}\label{eq:potential-remainder}
\partial_{\bar\phi}P(\phi,\bar\phi)=m^{2}\phi+\mathcal R_{P}(\phi),
\qquad
\mathcal R_{P}(\phi):=Q(|\phi|^{2})\,|\phi|^{2}\phi.
\end{equation}
Using \eqref{eq:Q-growth} with $j=0$ one obtains the global polynomial bound
\begin{equation}\label{eq:potential-nonlinear-bound}
|\mathcal R_{P}(\phi)|
\le
C_{P,0}\,(1+|\phi|^{2N_{P}})\,|\phi|^{3}
\qquad\text{for all }\phi\in\mathbb{C}.
\end{equation}
For the polynomial model \eqref{V4} this is consistent with the exact formula
\begin{equation}
\mathcal R_{P}(\phi)=\sum_{n=p}^{N} n\alpha_{n}\,|\phi|^{2n-2}\phi,
\end{equation}
i.e.\ $\mathcal R_{P}$ is a finite sum of odd powers $|\phi|^{2n-1}$ with constant coefficients.

In addition, by the chain rule and \eqref{eq:Q-growth}, every commuted term $\mathcal Z^\alpha \mathcal R_{P}(\phi)$ (with $\mathcal Z$ from the admissible structure in Definition~\ref{def:admissible-exterior}) is a finite sum of products with at least three factors drawn from $\{\mathcal Z^{\beta}\phi,\overline{\mathcal Z^{\beta}\phi}\}$, with coefficients depending only on $N_{P}$ and finitely many constants $C_{P,j}$.
Consequently, the potential contribution can be treated on the same footing as the gauge nonlinearities in the bootstrap and scattering arguments, without imposing any separate \emph{a priori} restriction on the pointwise size of $|\phi|$ beyond what is already obtained from the energy/Sobolev estimates.
\end{remark}

\subsection{Zero and electric Coulomb sector hierarchy and transfer principles}\label{subsec:zero-sector-hierarchy}

\begin{definition}[Zero and electric Coulomb sectors]\label{def:charge-sector}
The zero electric sector is the electric-charge-free sector
\begin{equation}
 Q_e =0.
\end{equation}
An electric Coulomb sector is specified by
\begin{equation}
 Q_e\in\mathbb R.
\end{equation}
In an electric Coulomb sector one subtracts the stationary electric Coulomb Maxwell field and works with the radiative remainder.  The zero sector is the special case in which the stationary Coulomb part vanishes.
\end{definition}

\begin{definition}[Electric charge]\label{def:charges}
For a Maxwell field in the scattering class and in the global-potential formulation used in this paper, define the asymptotic electric charge by
\begin{equation}
 Q_e[F]=\frac{1}{4\pi}\lim_{r\to\infty}\int_{S_{\tau,r}}{}^*F.
\end{equation}
The zero-sector condition is $Q_e[F]=0$; an electric Coulomb sector has prescribed asymptotic electric charge $Q_e[F]=Q_e$.
\end{definition}

\begin{proposition}[Electric charge conservation]\label{prop:sector-preservation-charge-conservation}
Let $(A,\phi)$ be a smooth Lorenz-gauge Maxwell-Higgs solution in the finite-energy scattering class.  Then $Q_e[F]$ is independent of the Cauchy slice.  Consequently, the zero sector and every fixed electric Coulomb sector are preserved by the evolution.
\end{proposition}

\begin{proof}
Maxwell's equation is
\begin{equation}
 \nabla^\mu F_{\mu\nu}=J_\nu,
 \qquad
 J_\nu=2\operatorname{Im}(\overline\phi D_\nu\phi),
\end{equation}
and the scalar equation implies $\nabla^\nu J_\nu=0$.  Apply the divergence theorem to a truncated slab between two Cauchy slices and a large timelike boundary $r=R$.  The difference of the two electric fluxes through $S_{\tau,R}$ equals the flux of the conserved current through the timelike boundary.  In the finite-radiation class the latter has a limit as $R\to\infty$, and the boundary contribution vanishes after taking this limit.  Thus the limiting electric flux is independent of the slice.
\end{proof}

\begin{definition}[Zero-sector linear estimates]\label{def:charged-linear-estimates}
Fix $K\ge10$ and $m^2\ge0$.  The notation $\Lin_K^{(m)}$ denotes the following estimates for the decoupled uncharged Maxwell field and the scalar operator $\square_g-m^2$.
\begin{enumerate}
\item Inhomogeneous forward estimates hold at order $K$: nondegenerate redshift energy boundedness, trapping-degenerate integrated local energy decay, and an $r^p$ hierarchy.
\item The Cauchy-to-radiation maps and the inverse final-state maps are continuous in the split topology of Definition~
\ref{def:kerr-scattering-topology}.
\item On Kerr the inverse final-state map lands in the degenerate scattering Cauchy topology unless the horizon data satisfy the redshift-regular condition of Definition~
\ref{def:redshift-regular-horizon}.
\end{enumerate}
For $m^2=0$ this system is denoted by $\Lin_K$.
\end{definition}

\begin{definition}[Electric linear estimates]\label{def:electric-linear-estimates}
Fix \(K\ge10\), \(m^2\ge0\), and an electric Coulomb sector \(Q_e\).  The notation \(\CElec^{(m)}_K(M,a,Q_e)\) denotes the charged scalar part of the fixed-sector linear system for the operator \eqref{eq:celec-operator-main}.  The full fixed-sector comparison system is the direct sum of the radiative charge-subtracted Maxwell equation for \(f\) and this charged scalar equation.  The solution norm, source norm, radiation fields, and inverse final-state maps are those in Definition~\ref{def:external-electric-condition}.  For \(m^2=0\) we write \(\CElec_K(M,a,Q_e)\).
\end{definition}

\begin{theorem}[Uncharged transfer principle in established regimes]\label{thm:charged-sector-transfer}
Let $K\ge10$, $k=K-4$, and suppose that $\Lin_K^{(m)}$ holds on the chosen Schwarzschild or slowly rotating Kerr exterior.  Then there is $\varepsilon_*>0$ such that every smooth zero-charge Lorenz-compatible datum of size at most $\varepsilon_*$ generates a unique global Maxwell-Higgs solution in the zero sector.  The Cauchy-to-radiation maps are homeomorphisms between small neighborhoods of the origin in the quotient Cauchy and radiation spaces, their inverses are the nonlinear wave operators, the maps are tangent to the corresponding linear maps at the vacuum, and real-analytic potentials give convergent Born series.
\end{theorem}

\begin{proof}
We verify the conditions of Theorem~\ref{thm:method}.  The unknowns are written in Lorenz gauge as \(U=(A,F,\phi)\), with \(F=\dd A\), and the comparison system is the direct sum of the uncharged Maxwell equation and the scalar operator \(\square_g-m^2\).  By assumption this comparison system satisfies \(\Lin_K^{(m)}\).  The estimates give, on every slab \([\tau_0,\tau]\),
\begin{equation}
 \|U\|_{\mathbb X_K^{(m)}(\tau_0,\tau)}
 \le C\Bigl(\|U[\tau_0]\|_{\mathbb H_K}
       +\|\mathcal N(U)\|_{\mathbb S_K^{(m)}(\tau_0,\tau)}\Bigr),
\end{equation}
where \(\mathcal N(U)\) is the pair consisting of the Maxwell current and the scalar nonlinear forcing.  Lemma~\ref{lem:nonlinear-banach-estimates} gives
\begin{equation}
 \|\mathcal N(U)\|_{\mathbb S_K^{(m)}(\tau_0,\tau)}
 \le C\left(\|U\|_{\mathbb X_K^{(m)}(\tau_0,\tau)}^2
        +\|U\|_{\mathbb X_K^{(m)}(\tau_0,\tau)}^{2N_P+3}\right).
\end{equation}
Choosing the data norm small and applying the continuity argument in Subsection~\ref{subsec:abstract-transfer-proof} closes the top-order bound globally.  The \(r^p\) part of \(\Lin_K^{(m)}\) gives the radiation fields, and the inverse final-state part of \(\Lin_K^{(m)}\) gives the wave operators by the contraction estimate of Proposition~\ref{prop:closed-sector-contraction}.  Proposition~\ref{prop:sector-preservation-charge-conservation} shows that the zero electric sector is preserved during the evolution.  The gauge quotient, tangency, and Born expansion statements are the conclusions of Theorem~\ref{thm:method}.  This proves the zero-sector transfer principle.
\end{proof}

\begin{theorem}[Electric fixed-sector transfer principle]\label{thm:electric-sector-transfer}
Let \(K\ge10\), set \(k=K-4\), and fix an electric Coulomb sector \(Q_e\).  Suppose that the radiative Maxwell forward estimates after Coulomb subtraction and the charged scalar condition \(\CElec^{(m)}_K(M,a,Q_e)\) hold on the chosen Schwarzschild or slowly rotating Kerr exterior.  Then there exists \(\varepsilon_{*,Q_e}>0\) such that every smooth Lorenz-compatible datum in the fixed sector \(Q_e\), with renormalized order-\(K\) size at most \(\varepsilon_{*,Q_e}\), generates a unique global solution in the same sector.  The solution admits the Coulomb-renormalized scalar radiation field \(U_{Q_e}^{-1}r\phi\), and the Cauchy-to-radiation map is continuous in the fixed-sector quotient topology.  If the corresponding Maxwell final-state map is available (on slowly rotating Kerr this is \(\MScat_K(M,a)\)), then the nonlinear wave operators exist, the Cauchy-to-radiation maps and wave operators are inverse homeomorphisms on small neighborhoods, and the gauge quotient, tangency, quadratic Born term, and analytic Born series conclusions hold in the fixed-sector spaces.
\end{theorem}

\begin{proof}
We spell out the fixed-sector Banach argument.  Write
\begin{equation}
 U=(a,f,\phi),\qquad A=A^{\mathrm C}_{Q_e}+a,
 \qquad F=F^{\mathrm C}_{Q_e}+f,
\end{equation}
and impose the Lorenz condition on the perturbation \(a\).  The equations are \eqref{eq:electric-renorm-maxwell-intro}-\eqref{eq:electric-renorm-scalar-intro}.  The Coulomb coefficients are included in the charged scalar operator \((D_{Q_e})^\mu D_{Q_e,\mu}-m^2\); only the remaining terms are nonlinear sources.

On a finite slab \([\tau_0,\tau]\), the direct sum of the charge-subtracted radiative Maxwell forward estimate and the charged scalar estimate gives
\begin{equation}\label{eq:electric-transfer-linear-bound}
 \|U\|_{\mathbb X^{(m)}_{K,Q_e}([\tau_0,\tau])}
 \le C_Q\left(\|U[\tau_0]\|_{\mathbb H^{(m)}_{K,Q_e}}+
 \|\mathcal N_{Q_e}(U)\|_{\mathbb S^{(m)}_{K,Q_e}([\tau_0,\tau])}\right).
\end{equation}
Proposition~\ref{prop:electric-tame-estimates} gives
\begin{equation}\label{eq:electric-transfer-source-bound}
 \|\mathcal N_{Q_e}(U)\|_{\mathbb S^{(m)}_{K,Q_e}([\tau_0,\tau])}
 \le C_Q\left(\|U\|_{\mathbb X^{(m)}_{K,Q_e}([\tau_0,\tau])}^2+
 \|U\|_{\mathbb X^{(m)}_{K,Q_e}([\tau_0,\tau])}^{2N_P+3}\right),
\end{equation}
with the corresponding Lipschitz estimate whose nonlinear powers are \(1\) and \(2N_P+2\).  Set
\begin{equation}
 X_Q(\tau):=\|U\|_{\mathbb X^{(m)}_{K,Q_e}([\tau_0,\tau])}.
\end{equation}
For data of size \(\varepsilon\), \eqref{eq:electric-transfer-linear-bound}-\eqref{eq:electric-transfer-source-bound} imply
\begin{equation}\label{eq:electric-transfer-bootstrap}
 X_Q(\tau)\le C_Q\varepsilon+C_Q\left(X_Q(\tau)^2+X_Q(\tau)^{2N_P+3}\right).
\end{equation}
Choose \(\Lambda_Q=2C_Q\) and then \(\varepsilon_{*,Q_e}\) so that
\begin{equation}
 C_Q\left((\Lambda_Q\varepsilon)^2+(\Lambda_Q\varepsilon)^{2N_P+3}\right)
 \le \frac12 C_Q\varepsilon,
 \qquad 0<\varepsilon\le\varepsilon_{*,Q_e}.
\end{equation}
The open-closed continuity argument applied to the set
\begin{equation}
 \{\tau:X_Q(\tau)\le\Lambda_Q\varepsilon\}
\end{equation}
then gives \(X_Q(\tau)\le(3/4)\Lambda_Q\varepsilon\) on every finite forward slab.  The local Lorenz-gauge theory, constraint propagation, and continuation criterion extend the solution globally.  The same proof on the time-reversed exterior gives the past global solution.  Proposition~\ref{prop:sector-preservation-charge-conservation} fixes the electric charge, so the solution remains in the prescribed sector.  The Lipschitz part of Proposition~\ref{prop:electric-tame-estimates} gives uniqueness and continuous dependence.

The scalar radiation field \(U_{Q_e}^{-1}r\phi\), the horizon trace, and the possible massive timelike channel are part of \(\CElec^{(m)}_K(M,a,Q_e)\).  The radiative Maxwell trace is supplied by the charge-subtracted Maxwell estimate.  Thus the Cauchy-to-radiation map is continuous in the fixed-sector quotient topology.

Assume now that the corresponding Maxwell final-state map is available.  Let \(U_{\rm lin}\) be the decoupled linear solution with prescribed fixed-sector radiation data.  On a late tail its norm is at most \(\eta\).  For a correction \(W\) with zero asymptotic data define
\begin{equation}
 \mathcal T(W):=\mathcal L_{\rm fs,Q_e}^{-1}\mathcal N_{Q_e}(U_{\rm lin}+W).
\end{equation}
Proposition~\ref{prop:electric-closed-sector-contraction} gives the self-map and contraction estimates
\begin{eqnarray}
 \|\mathcal T(W)\|_{\mathbb X^{(m)}_{K,Q_e}}&\le&
 C\left((\eta+\rho)^2+(\eta+\rho)^{2N_P+3}\right),\nonumber\\
 \|\mathcal T(W_1)-\mathcal T(W_2)\|_{\mathbb X^{(m)}_{K,Q_e}}&\le&
 C\left(\eta+\rho+(\eta+\rho)^{2N_P+2}\right)
 \|W_1-W_2\|_{\mathbb X^{(m)}_{K,Q_e}}.
\end{eqnarray}
Choosing \(\rho\) and \(\eta\) small produces a unique fixed point on the tail.  On Kerr, Lemma~\ref{lem:kerr-topology-compatibility} converts the degenerate scattering trace returned by the inverse problem into the nondegenerate forward control needed after restarting at a finite slice.  The finite-slab global argument then extends the tail solution to the full exterior.  Uniqueness shows that the constructed wave operators are inverse to the Cauchy-to-radiation maps.  Lemma~\ref{lem:residual-gauge-descent} gives the quotient statement, and Propositions~\ref{prop:quadratic-remainder}, \ref{prop:born-expansion}, and~\ref{prop:analytic-series} give tangency, the quadratic Born term, and the analytic Born series.
\end{proof}

\paragraph{Cross-reference to the main statements.}
The introductory conclusions are stated in Subsection~\ref{subsec:main-results} as four main theorems: the two rotating statements are Theorems~\ref{thm:main-slow-kerr-intro} and~\ref{thm:main-slow-kerr-massive-intro}, while the two Schwarzschild statements are Theorems~\ref{thm:main-schwarzschild-intro} and~\ref{thm:main-schwarzschild-electric-intro}.  The older zero-sector, small-electric, fixed-sector, massive-energy, and massive-scattering labels are retained only as cross-reference aliases to these four statements.  The regional far-field, horizon, trapped-set, boundedness, and integrated-decay estimates used by the rotating conclusions are Propositions~\ref{prop:kerr-far-energy-decay}, \ref{prop:kerr-far-energy-decay-maxwell}, \ref{prop:kerr-horizon-energy-decay}, and~\ref{prop:kerr-trapped-iled}.  No additional main theorem is introduced in the present subsection.

\subsection{Gauge invariance and structure of the Kerr scattering map}\label{subsec:gauge-structure}

The proofs in this paper are carried out in Lorenz gauge in the potential $A$.
However, the Maxwell-Higgs system is gauge invariant and the asymptotic data are defined intrinsically via parallel transport.
The nonlinear scattering theory obtained above is canonical on the quotient by gauge.
We state three structural consequences for Theorem~\ref{thm:main-slow-kerr-intro}; on Schwarzschild the same quotient statement holds for the massive extension after adding the timelike radiation channel.

\begin{definition}[Residual Lorenz gauge group and quotient phase space]\label{def:gauge-quotient}
Let $\mathcal G^{k+1}_{\mathrm{res}}$ be the residual Lorenz gauge group consisting of real gauge parameters $\chi$ with $\square_g\chi=0$ and finite order-$(k+1)$ scattering norm. It acts by
\begin{equation}
A\mapsto A+\dd\chi,
\qquad
\phi\mapsto e^{i\chi}\phi.
\end{equation}
The Cauchy phase space is the quotient by this action. The asymptotic phase space is not the bare radiation space, but the quotient
\begin{equation}
\mathfrak R_{\pm}^{(m),k}/{\Gbdry}_{\pm}^{k+1},
\end{equation}
where ${\Gbdry}_{\pm}^{k+1}$ is the group of boundary traces of residual Lorenz parameters on $\partial_{\pm}^{(m)}\mathcal D$. Equivalently, one may impose a gauge normalization requiring those traces to vanish. In the electric Coulomb sector the same quotient is taken after the Coulomb background has been fixed and the scalar radiation variable has been multiplied by \(U_{Q_e}^{-1}\).  This boundary quotient is part of the definition of canonicity of the nonlinear scattering map.
\end{definition}

\begin{lemma}[Descent to the residual Lorenz quotient]\label{lem:residual-gauge-descent}
Let \(\chi\) solve \(\square_g\chi=0\) with finite order-\((K+1)\) scattering norm, and let
\begin{equation}
 (A,\phi)\mapsto(A+\dd\chi,e^{i\chi}\phi)
\end{equation}
be the corresponding residual Lorenz transformation.  The Cauchy norms, source norms, radiation norms, and fixed electric charge are constant on the induced orbit after passing to the quotient of Definition~\ref{def:gauge-quotient}.  Moreover the Cauchy-to-radiation map and, whenever the relevant final-state condition is available, the nonlinear wave operators intertwine the residual action.  Thus all nonlinear scattering maps constructed in the paper descend to the residual Lorenz quotient.
\end{lemma}

\begin{proof}
The equation \(\square_g\chi=0\) is precisely the condition that the Lorenz constraint be preserved by \(A\mapsto A+\dd\chi\).  The curvature is unchanged, and multiplication of \(\phi\) by \(e^{i\chi}\) is unitary; the gauge-covariant derivatives, energy densities, source norms, and fluxes are therefore unchanged after quotienting by the boundary trace of \(\chi\).  In an electric sector, the charge is the limiting flux of \(*F\), so it is also unchanged.  The scalar radiation fields are defined by parallel transport, and the endpoint phase is exactly the boundary value of the same residual transformation.  The Cauchy-to-radiation map is therefore equivariant.  Applying the same observation to the fixed-point equation for the final-state correction shows equivariance of the wave operators.  Passing to orbit spaces gives the claimed descent.
\end{proof}

\begin{corollary}[Gauge-invariant wave operators and scattering on Kerr]\label{cor:gauge-invariant}
In the setting of Theorem~\ref{thm:main-slow-kerr-intro}, with the same top order $K$ and $k=K-4$, the homeomorphisms $\mathscr S_{\pm,M,a}^{\mathrm{full}}$, the wave operators $\mathscr W_{\pm,M,a}^{\mathrm{full}}$, and the two-sided nonlinear scattering operator $\mathscr S_{M,a}^{\mathrm{nl}}$ descend to well-defined homeomorphisms on the quotient phase spaces
\begin{equation}
\mathcal H^{K}_{M,a}(\Sigma_{\tau_{0}})/\mathcal G^{K+1}(\Sigma_{\tau_{0}})
\end{equation}
and
\begin{equation}
\mathfrak R_{\pm,M,a}^{(\mu)}(\varepsilon_{\mathrm{Kerr}})/\mathcal G_{\pm,\partial}^{K+1}.
\end{equation}
Here $\mu=0$ on rotating Kerr.  In the Schwarzschild massive extension the same statement is made with $\mu=m$ and with the timelike asymptotic channel included.  These quotient maps are therefore independent of the choice of Lorenz-gauge representative.
\end{corollary}

\begin{proof}
Residual Lorenz transformations preserve the reduced equations because their parameters solve the homogeneous wave equation.  The curvature is unchanged under $A\mapsto A+\dd\chi$, and the scalar factor $e^{i\chi}$ is unitary; therefore the Cauchy norms, source norms, radiation norms, and boundary fluxes used in the construction are constant on residual gauge orbits.  The asymptotic scalar normalization is defined by parallel transport, so the endpoint phase is exactly the boundary trace of the same residual gauge transformation.  Consequently, the Cauchy-to-radiation maps intertwine the residual gauge action.  Since Theorem~\ref{thm:main-slow-kerr-intro} gives inverse homeomorphisms before passing to the quotient, the induced maps on the quotient spaces are again inverse homeomorphisms.
\end{proof}

\begin{corollary}[Gauge-invariant maps in electric Coulomb sectors]\label{cor:gauge-invariant-electric}
In the setting of Theorem~\ref{thm:main-slow-kerr-massive-intro} or Theorem~\ref{thm:main-schwarzschild-electric-intro}, the fixed-sector maps descend to homeomorphisms on
\begin{equation}
 \mathcal H^{K}_{\mathrm{ren}}(\Sigma_{\tau_0};Q_e)/\mathcal G^{K+1}(\Sigma_{\tau_0})
 \quad\hbox{and}\quad
 \mathfrak R_{\pm}^{(m),K}(Q_e)/\mathcal G^{K+1}_{\pm,\partial}.
\end{equation}
\end{corollary}

\begin{proof}
The curvature remainder \(f\) and the Coulomb background curvature are gauge invariant.  A residual Lorenz transformation changes local representatives by \(a\mapsto a+\dd\chi\) and \(\phi\mapsto e^{i\chi}\phi\).  The boundary trace of \(\chi\) acts by the same unitary factor on \(U_{Q_e}^{-1}r\phi\), because \(U_{Q_e}\) is determined by the fixed background connection.  Thus the Cauchy and radiation norms are invariant on residual gauge orbits, and the fixed-sector scattering and wave-operator maps descend to the quotient.
\end{proof}

\begin{remark}[Quadratic remainder and tangency]\label{rem:tangent}
A general quadratic remainder estimate for the nonlinear scattering maps, implying the tangency statements in Theorem~\ref{thm:main-slow-kerr-intro} (and in Theorem~\ref{thm:method}), is proved in Proposition~\ref{prop:quadratic-remainder}; see in this case \eqref{eq:quadratic-scattering}.
A second-order expansion (Born approximation) identifying the leading bilinear contribution and giving an $O(\|U\|^{3})$ remainder is proved in Proposition~\ref{prop:born-expansion}; see \eqref{eq:cubic-scattering}. If, in addition, $P$ is real analytic near $0$, then all higher-order terms can be identified and the resulting Born series converges, so the scattering maps are real analytic near the origin; see Proposition~\ref{prop:analytic-series}.
\end{remark}

\paragraph{Transfer principle and Schwarzschild model proof.}
Theorems~\ref{thm:main-slow-kerr-intro} and~\ref{thm:main-slow-kerr-massive-intro} are the rotating radiative applications of the nonlinear transfer argument.  The transfer theorem, Theorem~\ref{thm:method}, converts top-order linear estimates into nonlinear wave operators, asymptotic completeness, gauge-quotient canonicity, and the local structure of the scattering map, with radiation conclusions stated at order $k=K-4$.  The Schwarzschild sections provide a complete model proof of this mechanism and of the relevant linear estimates, including the massive/timelike-channel refinements that lead to Theorem~\ref{thm:nonlinear-wave-operators}.  On slowly rotating Kerr, the paper uses the established massless zero-sector forward condition and, for the massive scattering statement, the explicitly named scalar condition \(\SKG^{(m)}_K(M,a)\) together with the Maxwell final-state condition \(\MScat_K(M,a)\).

\subsection{Transfer principle: from linear estimates to nonlinear scattering}\label{subsec:method}

\paragraph{The new mechanism behind Theorem~\ref{thm:main-slow-kerr-intro}.}
The nonlinear part of our argument is robust and uses the background geometry only through a finite list of \emph{linear} estimates for inhomogeneous scalar waves and Maxwell fields. We isolate these conditions as the massless estimates $\Lin_{k}$ below, and we also record the corresponding massive variant $\Lin_{k}^{(m)}$ (Definition~\ref{def:lin-estimates-massive}) for the Klein-Gordon operator $(\square_g-m^{2})$. Conceptually, $\Lin_{k}$ consists of
(i) coercive redshift energy boundedness,
(ii) integrated local energy decay/Morawetz control with a controlled degeneration at trapping,
(iii) an $r^{p}$ hierarchy in the asymptotically flat region yielding radiation fields, and
(iv) linear scattering/final-state well-posedness.

The crucial point is that these four linear ingredients are not merely convenient estimates; they are the conditions needed to run the nonlinear theory. After Coulomb subtraction, every nonlinear term is measured in the same source norm that appears in $\Lin_{k}^{(m)}$. The redshift and Morawetz part of the estimates closes the global bootstrap, the $r^{p}$ hierarchy produces the radiation fields and quantitative decay, and the linear final-state theory converts those bounds into nonlinear wave operators and asymptotic completeness. Thus the geometry enters only through the availability of $\Lin_{k}^{(m)}$.

Theorem~\ref{thm:method} formalizes this as a theorem of independent interest: an admissible stationary black-hole exterior with the top-order linear estimates collected in $\Lin_{K}^{(m)}$ inherits the small-data nonlinear scattering theory for Maxwell-Higgs.  The Schwarzschild sections below verify this mechanism directly.  On slowly rotating Kerr the paper uses the scalar estimates and the slowly rotating Maxwell estimates, and it makes no rotating claim outside the zero-sector established regime.

\medskip

\begin{definition}[Admissible stationary black-hole exterior]\label{def:admissible-exterior}
An \emph{admissible stationary black-hole exterior} consists of an asymptotically flat stationary black-hole exterior $(\mathcal D,g)$ together with an \emph{admissible structure}
\begin{equation}
\bigl(\{\Sigma_{\tau}\}_{\tau\in\mathbb R},\,\mathcal R(\tau_{0},\tau_{1}),\,r,\,N,\,\mathcal Z\bigr)
\end{equation}
satisfying the following properties.
\begin{enumerate}
\item[(G1)] \textbf{Foliation and slabs.}
$\{\Sigma_{\tau}\}$ is a smooth spacelike foliation of $\mathcal D$ by Cauchy hypersurfaces.
For $\tau_{1}\ge\tau_{0}$, the corresponding slab $\mathcal{R}(\tau_{0},\tau_{1})$ has boundary components
$\Sigma_{\tau_{0}}$, $\Sigma_{\tau_{1}}$, and the portions of the event horizons and null infinities
$\mathcal H^{\pm}(\tau_{0},\tau_{1})$ and $\mathcal I^{\pm}(\tau_{0},\tau_{1})$.
\item[(G2)] \textbf{Asymptotic radius and stationarity.}
In the asymptotically flat region there is a smooth radial coordinate $r$ with $r\sim |x|$ and a complete stationary Killing field $T$
which is timelike for $r$ sufficiently large.
\item[(G3)] \textbf{Redshift multiplier.}
There exists a globally timelike future-directed multiplier field $N$ which agrees with $T$ for $r$ sufficiently large
and is a redshift vector field in a neighbourhood of $\mathcal H^{+}$ (and, by time-reversal, of $\mathcal H^{-}$).
\item[(G4)] \textbf{Commutations.}
$\mathcal Z$ is a fixed finite collection of admissible first-order commutation vector fields used to define higher-order energies and to commute the Maxwell field and the nonlinear system. When additional hidden symmetries are used in the scalar linear theory (for instance the Carter operator on Kerr), they are treated as auxiliary scalar commutators and are not included in $\mathcal Z$.
\end{enumerate}
Whenever we speak of an admissible exterior we implicitly fix one such admissible structure.
\end{definition}

\paragraph{Slowly rotating Kerr geometries.}
For the Kerr exterior $(\mathcal D_{M,a},g_{M,a})$ with Kerr-star time function $t^{\star}=t-h(r)$ from \eqref{eq:kerr-tstar}, the properties in Definition~\ref{def:admissible-exterior} can be checked directly. The gradient of $t^{\star}$ is timelike on $\mathcal D_{M,a}$, so each level set $\Sigma_{\tau}$ is spacelike; smoothness and transversality at $\mathcal H^{+}$ follow from the regularity of the ingoing Kerr chart. Global hyperbolicity of the domain of outer communications then implies that $\{\Sigma_{\tau}\}_{\tau\in\mathbb R}$ is a foliation by Cauchy hypersurfaces. In the asymptotically flat region the same coordinate $r$ is a smooth radial coordinate with $r\sim |x|$, and $T=\partial_{t^{\star}}=\partial_{t}$ is a complete stationary Killing field which is timelike for $r$ large. A redshift multiplier is obtained by patching the future horizon generator near $r=r_{+}$ to $T$ outside a compact set. One may take $\mathcal Z$ to be any fixed finite smooth family generated by $T$, $\Phi=\partial_{\varphi}$, angular vector fields on the symmetry spheres, and suitable redshift-type vector fields. When verifying scalar linear estimates on Kerr one may additionally use the Carter operator as an auxiliary scalar commutator. Thus each slowly rotating Kerr exterior used in the theorem statements is admissible in the sense of Definition~\ref{def:admissible-exterior}.

\begin{remark}[Choice of commutators for the nonlinear system]\label{rem:commutators-nonlinear}
The family $\mathcal Z$ is always taken to consist of first-order vector fields. This is the family used throughout the nonlinear Maxwell-Higgs argument and in every expression involving Lie derivatives of Maxwell fields. When verifying the scalar linear estimates on Kerr one may additionally employ hidden scalar symmetries, such as the Carter operator, but these are auxiliary to the scalar linear theory and are not part of $\mathcal Z$.
\end{remark}

\medskip

Let $(\mathcal D,g)$ be an admissible stationary black-hole exterior in the sense of Definition~\ref{def:admissible-exterior},
and fix an admissible structure $(\{\Sigma_{\tau}\},r,N,\mathcal Z)$.
For a complex scalar $\psi$ we write the (real) stress-energy tensor
\begin{equation}
\mathbb T_{\mu\nu}[\psi]
:=\Re\bigl(\nabla_{\mu}\psi\,\overline{\nabla_{\nu}\psi}-\tfrac12 g_{\mu\nu}\nabla^{\alpha}\psi\,\overline{\nabla_{\alpha}\psi}\bigr),
\qquad
J^{X}_{\mu}[\psi]:=\mathbb T_{\mu\nu}[\psi]X^{\nu},
\end{equation}
and we define the $N$-energy flux through $\Sigma_{\tau}$ by
\begin{equation}
E^{N}[\psi](\tau):=\int_{\Sigma_{\tau}} J^{N}[\psi]\cdot n_{\Sigma_{\tau}}\,\dd\mu_{\Sigma_{\tau}}.
\end{equation}
For a $2$-form $G$ we use the Maxwell stress-energy tensor
\begin{equation}
\mathbb T_{\mu\nu}[G]:=G_{\mu\alpha}G_{\nu}{}^{\alpha}-\tfrac14 g_{\mu\nu}G_{\alpha\beta}G^{\alpha\beta},
\qquad
J^{X}_{\mu}[G]:=\mathbb T_{\mu\nu}[G]X^{\nu},
\end{equation}
and we similarly define $E^{N}[G](\tau)$.
We also consider higher-order energies obtained by commuting with $\mathcal Z$.
Given $k\in\mathbb N$, we set
\begin{equation}
E^{N}_{k}[\psi](\tau):=\sum_{|\alpha|\le k} E^{N}[\mathcal Z^{\alpha}\psi](\tau),
\qquad
E^{N}_{k}[G](\tau):=\sum_{|\alpha|\le k} E^{N}[\mathcal{L}_{\mathcal Z^{\alpha}}G](\tau),
\end{equation}
where $\mathcal Z^{\alpha}$ denotes a composition of admissible commutators.

\medskip
\noindent\textbf{Massive Klein-Gordon energies.}
Fix $m^{2}\ge 0$.
For the linear Klein-Gordon operator $(\square_{g}-m^{2})$ we also use the modified stress-energy tensor
\begin{equation}
\mathbb T^{(m)}_{\mu\nu}[\psi]
:=
\Re\!\left(\nabla_{\mu}\psi\,\overline{\nabla_{\nu}\psi}\right)
-\tfrac12 g_{\mu\nu}\Bigl(\nabla^{\alpha}\psi\,\overline{\nabla_{\alpha}\psi}+m^{2}|\psi|^{2}\Bigr),
\qquad
J^{X,(m)}_{\mu}[\psi]:=\mathbb T^{(m)}_{\mu\nu}[\psi]X^{\nu},
\end{equation}
and define the corresponding fluxes through $\Sigma_{\tau}$ by
\begin{equation}
E^{X,m}[\psi](\tau):=\int_{\Sigma_{\tau}} J^{X,(m)}[\psi]\cdot n_{\Sigma_{\tau}}\,\dd\mu_{\Sigma_{\tau}},
\qquad
E^{X,m}_{k}[\psi](\tau):=\sum_{|\alpha|\le k}E^{X,m}[\mathcal Z^{\alpha}\psi](\tau).
\end{equation}
When $m^{2}=0$ these coincide with the wave energies $E^{X}[\psi]$ and $E^{X}_{k}[\psi]$.

\begin{definition}[Kerr scattering topology]\label{def:kerr-scattering-topology}
On a slowly rotating Kerr exterior the forward Cauchy problem and the two-sided final-state problem are measured in different but compatible topologies. The forward Cauchy space ${\Hcauchy}^{k}(\Sigma_{\tau_0})$ is the completion of smooth data in the nondegenerate order-$k$ $N$-energy norm. The scattering/final-state Cauchy space $\Hscatt^{k}(\Sigma_{\tau_0})$ is the completion induced by the scattering norm whose horizon component is degenerate in the redshift direction. The Cauchy-to-radiation map is continuous from ${\Hcauchy}^{k}$ to the radiation space, while the inverse final-state map is continuous from radiation data to $\Hscatt^{k}$. A final-state trace in ${\Hcauchy}^{k}$ requires an extra redshift-regularity condition on the horizon radiation field. This is the topology used in the zero-sector Kerr instances of $\Lin_k$.
\end{definition}

\begin{definition}[Sufficient redshift-regular horizon data]\label{def:redshift-regular-horizon}
Let $\kappa_+>0$ be the surface gravity of the Kerr event horizon. Consider future final-state data on $\mathcal H^+$, written in a regular horizon coordinate $v$ along the null generators and angular variables $\omega\in\mathbb S^2$; the past horizon is treated by the time-reversed coordinate. For a scalar, Maxwell component, or Lorenz-potential component horizon trace $h$, define the sufficient redshift-regular norm
\begin{equation}
\|h\|_{\mathfrak R_{\mathcal H,\mathrm{red}}^{k}}^2
:=
\sum_{j+|\alpha|\le k}
\int_{\mathbb R\times\mathbb S^2}
 e^{2\eta |v|}\,
\bigl|\partial_v^j\nabla_{\omega}^{\alpha}h(v,\omega)\bigr|^2\,\dd v\,\dd\omega,
\end{equation}
where $0<\eta<\kappa_+$ is fixed once and for all. For vector-valued or bundle-valued traces the expression is computed in any regular orthonormal/unitary frame and is independent of the frame up to equivalent norms. We denote by $\mathfrak R_{\mathcal H,\mathrm{red}}^{k}$ the subspace on which this norm is finite.

This condition is only a convenient sufficient condition that compensates for the blue-shift encountered when one solves the final-state problem backward from the horizon. Consequently, the general final-state map in Kerr is always formulated into the degenerate scattering space $\Hscatt^k$; a nondegenerate $\Hcauchy^k$ final-state trace is asserted only for horizon radiation fields belonging to $\mathfrak R_{\mathcal H,\mathrm{red}}^{k}$.
\end{definition}

\begin{lemma}[Compatibility of the forward and final-state Kerr topologies]\label{lem:kerr-topology-compatibility}
Let \(k\ge2\), and let \(u\) denote one scalar, Maxwell, or Lorenz-potential component in the linear comparison system on a slowly rotating Kerr exterior.  Suppose that \(Lu=G\) on a tail slab \([T,\infty)\), where \(L\) is the relevant wave, Maxwell, or Klein-Gordon comparison operator, that \(G\in\mathcal S_k(T,\infty)\), and that the final-state construction gives \(u[T]\in\Hscatt^k\).  For every \(T_1\ge T+1\) one has
\begin{equation}\label{eq:kerr-topology-compatibility}
 \|u[T_1]\|_{\Hcauchy^k}^{2}
 +\|u\|_{\mathcal X_k(T_1,T_2)}^{2}
 \le C\Bigl(\|u[T]\|_{\Hscatt^k}^{2}
 +\|G\|_{\mathcal S_k(T,T_2)}^{2}
 +\|u_{\mathcal H}\|_{\mathfrak R_{\mathcal H,\mathrm{red}}^k(T,T_2)}^{2}\Bigr)
\end{equation}
for all \(T_2\ge T_1\).  If the final-state correction has zero horizon radiation, or if the horizon trace lies in \(\mathfrak R_{\mathcal H,\mathrm{red}}^k\), the last term is finite and the iterate belongs to the nondegenerate forward space on every later slice.  The same statement holds in the charged scalar comparison spaces after replacing ordinary derivatives by \(D_{Q_e}\) and using the Coulomb-renormalized radiation variable.
\end{lemma}

\begin{proof}
The only difference between \(\Hscatt^k\) and \(\Hcauchy^k\) is the horizon-transversal derivative measured by the redshift field.  Apply the redshift multiplier identity on \([T,T_1]\) in a fixed neighbourhood of \(\mathcal H^+\).  The positive redshift bulk controls this missing derivative, while the degenerate scattering flux controls the tangential horizon components and the exterior local-energy part of the norm.  The source term is estimated by Cauchy-Schwarz in the dual source norm and absorbed into \(\|G\|_{\mathcal S_k}^{2}\).  Away from the horizon the two Cauchy norms are equivalent by elliptic estimates on the Kerr-star slices.  Reapplying the inhomogeneous estimate from \(T_1\) to \(T_2\) gives \eqref{eq:kerr-topology-compatibility}.  The charged version has the same proof because the fixed Coulomb connection is a stationary lower-order coefficient satisfying the bounds in Definition~\ref{def:pure-electric-sector}; the phase normalization is unitary and does not change the flux norms.
\end{proof}

\begin{definition}[Linear estimates $\Lin_{k}$]\label{def:lin-estimates}
Fix $k\in\mathbb N$.
Let $(\mathcal D,g)$ be an admissible stationary black-hole exterior with admissible structure as in Definition~\ref{def:admissible-exterior}.
We say that $(\mathcal D,g)$ satisfies the \emph{analytic} linear estimates collected in $\Lin_{k}$ if there exist constants $C>0$ and $R>0$ and, for every $\tau_{1}\ge\tau_{0}$, a pair of nonnegative functionals
\begin{equation}
\|\cdot\|_{\mathcal S_{k}(\tau_{0},\tau_{1})},\qquad
\|\cdot\|_{\mathcal X_{k}(\tau_{0},\tau_{1})},
\end{equation}
called the \emph{source norm} and the \emph{solution/bulk norm}, such that the following hold.

\smallskip
\noindent\textbf{Admissibility of the norms.}
The families $\mathcal S_{k}$ and $\mathcal X_{k}$ satisfy the following structural properties.
\begin{enumerate}
\item[(A1)] \textbf{Time monotonicity.}
If $[\tau_{0}',\tau_{1}']\subset[\tau_{0},\tau_{1}]$, then
\begin{equation}
\|\mathfrak{F}\|_{\mathcal S_{k}(\tau_{0}',\tau_{1}')} \le \|\mathfrak{F}\|_{\mathcal S_{k}(\tau_{0},\tau_{1})},
\qquad
\|\psi\|_{\mathcal X_{k}(\tau_{0}',\tau_{1}')} \le \|\psi\|_{\mathcal X_{k}(\tau_{0},\tau_{1})},
\end{equation}
and similarly for Maxwell fields and currents.
\item[(A2)] \textbf{Time localization.}
For $\tau_{0}\le\tau_{1}\le\tau_{2}$ there exists a universal constant $C_{\mathrm{loc}}$ such that
\begin{equation}
\|\mathfrak{F}\|_{\mathcal S_{k}(\tau_{0},\tau_{2})}^{2}
\le
\|\mathfrak{F}\|_{\mathcal S_{k}(\tau_{0},\tau_{1})}^{2}
+
\|\mathfrak{F}\|_{\mathcal S_{k}(\tau_{1},\tau_{2})}^{2},
\end{equation}
and
\begin{equation}
\|\psi\|_{\mathcal X_{k}(\tau_{0},\tau_{2})}^{2}
\le
C_{\mathrm{loc}}\Bigl(
\|\psi\|_{\mathcal X_{k}(\tau_{0},\tau_{1})}^{2}
+
\|\psi\|_{\mathcal X_{k}(\tau_{1},\tau_{2})}^{2}
\Bigr),
\end{equation}
with the analogous property for Maxwell fields.
\item[(A3)] \textbf{Coercivity and radiation fields.}
Finiteness of $\|\psi\|_{\mathcal X_{k}(\tau_{0},\tau_{1})}$ controls the order-$k$ $N$-energy fluxes through $\Sigma_{\tau}$ for $\tau\in[\tau_{0},\tau_{1}]$.
In addition, $\|\psi\|_{\mathcal X_{k}(\tau_{0},\tau_{1})}$ contains a far-field component sufficient to yield existence of order-$k$ radiation fields on $\mathcal I^{\pm}(\tau_{0},\tau_{1})\cup\mathcal H^{\pm}(\tau_{0},\tau_{1})$
together with quantitative control of the corresponding asymptotic energy fluxes.
The same holds for uncharged Maxwell fields.
\end{enumerate}

\begin{enumerate}
\item[(L1)] \textbf{Inhomogeneous energy-Morawetz-$r^{p}$ estimate.}
Let $\psi$ solve the scalar wave equation $\square_{g}\psi=\mathfrak{F}$ on $\mathcal{R}(\tau_{0},\tau_{1})$.
Then
\begin{equation}\label{eq:Lin-abstract}
\|\psi\|_{\mathcal X_{k}(\tau_{0},\tau_{1})}^{2}
\le
C\Bigl(E^{N}_{k}[\psi](\tau_{0})
+\|\mathfrak{F}\|_{\mathcal S_{k}(\tau_{0},\tau_{1})}^{2}\Bigr).
\end{equation}
An analogous estimate holds for (uncharged) Maxwell fields $G$ solving
$\nabla^{\mu}G_{\mu\nu}=J_{\nu}$, $\nabla_{[\alpha}G_{\beta\gamma]}=0$, namely
\begin{equation}
\|G\|_{\mathcal X_{k}(\tau_{0},\tau_{1})}^{2}
\le
C\Bigl(E^{N}_{k}[G](\tau_{0})
+\|J\|_{\mathcal S_{k}(\tau_{0},\tau_{1})}^{2}\Bigr).
\end{equation}

\item[(L2)] \textbf{Linear scattering and final-state well-posedness.}
The Cauchy-to-radiation map extends continuously from the order-$k$ forward Cauchy space ${\Hcauchy}^{k}(\Sigma_{\tau_0})$ to a Hilbert space of linear asymptotic data on $\mathcal I^{\pm}\cup\mathcal H^{\pm}$. The inverse final-state problem on $\mathcal I^{\pm}\cup\mathcal H^{\pm}$ is well posed into the scattering Cauchy space ${\Hscatt}^{k}(\Sigma_{\tau_0})$, with estimates compatible with (L1). In geometries without a horizon blue-shift obstruction these two spaces may coincide; on Kerr they are kept distinct as in Definition~\ref{def:kerr-scattering-topology}.
\end{enumerate}

All Maxwell estimates in $\Lin_{k}$ are understood for the \emph{radiative} (uncharged) part of the field:
for general data with nonzero electric charge one first subtracts the corresponding stationary Coulomb solution.
\end{definition}

\begin{definition}[Massive linear estimates collected in $\Lin_{k}^{(m)}$]\label{def:lin-estimates-massive}
Fix $k\in\mathbb N$ and $m^{2}\ge 0$.
Let $(\mathcal D,g)$ be an admissible stationary black-hole exterior with admissible structure as in Definition~\ref{def:admissible-exterior}.
We say that $(\mathcal D,g)$ satisfies the \emph{massive} linear estimates collected in $\Lin_{k}^{(m)}$ if there exist constants $C>0$ and $R>0$ and, for every $\tau_{1}\ge\tau_{0}$, a pair of nonnegative functionals
\begin{equation}
\|\cdot\|_{\mathcal S_{k}^{(m)}(\tau_{0},\tau_{1})},\qquad
\|\cdot\|_{\mathcal X_{k}^{(m)}(\tau_{0},\tau_{1})},
\end{equation}
called the \emph{source norm} and the \emph{solution/bulk norm}, such that the admissibility properties \textup{(A1)-(A3)} of Definition~\ref{def:lin-estimates} hold with $\mathcal S_{k}^{(m)}$ and $\mathcal X_{k}^{(m)}$ in place of $\mathcal S_{k}$ and $\mathcal X_{k}$, and with the (possibly extended) scattering boundary $\partial_{\pm}^{(m)}\mathcal D$ in place of $\mathcal I^{\pm}\cup\mathcal H^{\pm}$ in \textup{(A3)}.

In addition, the following linear estimates hold.
\begin{enumerate}
\item[(Lm1)] \textbf{Inhomogeneous energy-Morawetz-$r^{p}$ estimate for $(\square_{g}-m^{2})$.}
Let $\psi$ solve the Klein-Gordon equation $(\square_{g}-m^{2})\psi=\mathfrak{F}$ on $\mathcal{R}(\tau_{0},\tau_{1})$.
Then
\begin{equation}\label{eq:Lin-abstract-massive}
\|\psi\|_{\mathcal X_{k}^{(m)}(\tau_{0},\tau_{1})}^{2}
\le
C\Bigl(E^{N,m}_{k}[\psi](\tau_{0})
+\|\mathfrak{F}\|_{\mathcal S_{k}^{(m)}(\tau_{0},\tau_{1})}^{2}\Bigr).
\end{equation}
An analogous estimate holds for (uncharged) Maxwell fields $G$ solving
$\nabla^{\mu}G_{\mu\nu}=J_{\nu}$, $\nabla_{[\alpha}G_{\beta\gamma]}=0$,
as in \textup{(L1)} of Definition~\ref{def:lin-estimates}.

\item[(Lm2)] \textbf{Linear scattering and final-state well-posedness.}
The Cauchy-to-asymptotic map for the decoupled linear comparison system (uncharged Maxwell together with $(\square_{g}-m^{2})$) is continuous from the order-$k$ forward Cauchy space ${\Hcauchy}^{k}(\Sigma_{\tau_0})$ to the linear asymptotic data spaces $\mathfrak R_{\pm}^{(m)}$ on $\partial_{\pm}^{(m)}\mathcal D$ of Definition~\ref{def:linear-asymptotic-spaces}. The corresponding final-state problem on $\partial_{\pm}^{(m)}\mathcal D$ is well posed into ${\Hscatt}^{k}(\Sigma_{\tau_0})$ with estimates compatible with \textup{(Lm1)}; a nondegenerate final-state trace is asserted only under the sufficient redshift-regularity condition of Definition~\ref{def:redshift-regular-horizon}.
\end{enumerate}

When the nonlinear Lorenz-gauge problem is formulated in terms of a potential $A$, the notation $\Lin_{k}^{(m)}$ is furthermore understood to include the massless scalar-wave estimates of Definition~\ref{def:lin-estimates} applied componentwise to each Lorenz-potential component $A_{\nu}$. These auxiliary wave estimates control the semilinear source terms in the scalar equation; it contributes no additional gauge-invariant scattering data beyond the residual gauge quotient. When $m^{2}=0$ we may choose $\mathcal S_{k}^{(0)}=\mathcal S_{k}$ and $\mathcal X_{k}^{(0)}=\mathcal X_{k}$ so that $\Lin_{k}^{(0)}$ coincides with $\Lin_{k}$.
\end{definition}

\begin{lemma}[Concrete realization of the admissible norms on Schwarzschild and Kerr]\label{lem:concrete-admissible-norms}
On the Schwarzschild exterior $(\mathcal D_{M,0},g_{M,0})$ and on each admissibly slowly rotating Kerr exterior $(\mathcal D_{M,a},g_{M,a})$ with $|a|\le a_{\mathrm{slow}}(M,K)$,
one admissible choice of the norms in Definition~\ref{def:lin-estimates} is given by the weighted physical-space norms below.
Fix $\delta>0$ and a cutoff radius $R>0$.
For a source term $\mathfrak{F}$ set
\begin{equation}\label{eq:concrete-source-norm}
\|\mathfrak{F}\|_{\mathcal S_{k}(\tau_{0},\tau_{1})}^{2}
:=
\int_{\mathcal{R}(\tau_{0},\tau_{1})}
\sum_{|\alpha|\le k} r^{1+\delta}\,|\mathcal Z^{\alpha}\mathfrak{F}|^{2}.
\end{equation}
The trapping term is understood as follows. In Schwarzschild one may take a smooth nonnegative physical-space weight $w_{\mathrm{trap}}(r)$ vanishing only at the photon sphere $r=3M$. In Kerr the trapped set is a phase-space object; $w_{\mathrm{trap}}$ denotes a nonnegative order-zero microlocal weight (equivalently, the principal symbol of the Morawetz pseudodifferential bulk) which vanishes just on the trapped null geodesic set. The displayed formula below is shorthand for this microlocal bulk norm in Kerr and for the physical-space weighted norm in Schwarzschild. Fix a smooth cutoff $\chi_{\mathrm{away}}$ supported in the compact region $r\le R$ but vanishing in a neighbourhood of the trapped set/projection.
For a scalar $\psi$ define
\begin{eqnarray}\label{eq:concrete-solution-norm}
\|\psi\|_{\mathcal X_{k}(\tau_{0},\tau_{1})}^{2}
:={}&
\sup_{\tau\in[\tau_{0},\tau_{1}]} E^{N}_{k}[\psi](\tau)
+\int_{\mathcal{R}(\tau_{0},\tau_{1})\cap\{r\le R\}}\chi_{\mathrm{away}}\sum_{|\alpha|\le k}|\nabla \mathcal Z^{\alpha}\psi|^{2}
\nonumber\\
&+
\int_{\mathcal{R}(\tau_{0},\tau_{1})}\sum_{|\alpha|\le k}\Bigl(
w_{\mathrm{trap}}\,|\nabla \mathcal Z^{\alpha}\psi|^{2}
+r^{-1-\delta}|\nabla \mathcal Z^{\alpha}\psi|^{2}
+r^{-3-\delta}|\mathcal Z^{\alpha}\psi|^{2}
\Bigr)
\nonumber\\
&+\sup_{p\in[0,2]}\,\mathcal R_{k,p}[\psi](\tau_{0},\tau_{1}).
\end{eqnarray}
Here, for each $p\in[0,2]$, $\mathcal R_{k,p}[\psi](\tau_{0},\tau_{1})$ denotes the vector-field $r^{p}$ bulk/flux functional
in the asymptotically flat region $\{r\ge R\}$.
Finiteness of $\sup_{p\in[0,2]}\mathcal R_{k,p}[\psi]$ yields existence of radiation fields and controls the corresponding weighted energy fluxes through $\mathcal I^{+}(\tau_{0},\tau_{1})$
(and, by time-reversal, through $\mathcal I^{-}(\tau_{0},\tau_{1})$).

For Maxwell fields $G$ with source current $J$ one uses the same source norm~\eqref{eq:concrete-source-norm} with $\mathfrak{F}$ replaced by $J$,
and one defines $\|G\|_{\mathcal X_{k}(\tau_{0},\tau_{1})}$ by replacing $|\nabla \mathcal Z^{\alpha}\psi|^{2}$ and $|\mathcal Z^{\alpha}\psi|^{2}$ in~\eqref{eq:concrete-solution-norm}
by $|\mathcal{L}_{\mathcal Z^{\alpha}}G|^{2}$ and using the corresponding $r^{p}$ functional $\mathcal R_{k,p}[G]$.
These choices satisfy the admissibility properties \textup{(A1)-(A3)}.
\end{lemma}

\begin{proof}
Time monotonicity and time localization follow directly from the nonnegativity of the integrands and from splitting the slab $\mathcal R(\tau_0,\tau_2)$ into the two subslabs $\mathcal R(\tau_0,\tau_1)$ and $\mathcal R(\tau_1,\tau_2)$, with the harmless constant $C_{\mathrm{loc}}$ accounting for the overlap of smooth cutoffs in the $r^p$ functionals. The first term in \eqref{eq:concrete-solution-norm} is the order-$k$ nondegenerate $N$-energy flux, so coercivity of the flux part is built into the norm. The compact bulk term is deliberately cut away from the trapped set, and the remaining compact local energy is measured with the trapping-degenerate Morawetz bulk. On Schwarzschild this bulk is represented by the physical-space weight $w_{\mathrm{trap}}(r)$; on Kerr it is the corresponding microlocal bulk whose principal symbol vanishes on the trapped null geodesic set. This is the precise form compatible with trapping; no unweighted compact Morawetz control is being asserted at the trapped set. The vector-field $r^p$ hierarchy controls the weighted null fluxes in the asymptotic region and therefore gives the radiation-field trace on $\mathcal I^\pm$, while the redshift component gives the corresponding horizon trace on $\mathcal H^\pm$. The Maxwell version is identical after replacing scalar derivatives by Lie derivatives of the two-form and using the Maxwell flux density. Thus the displayed norms realize \textup{(A1)-(A3)}.
\end{proof}

\medskip

\begin{lemma}[Weighted Sobolev-Moser estimates on far null annuli]\label{lem:far-null-sobolev-moser}
Fix \(K\ge10\), choose \(K_0\le K-4\), and let \(0<\delta<1\).  Let
\[
 \mathcal A_R(\tau_0,\tau_1)=\mathcal R(\tau_0,\tau_1)\cap\{R\le r\le 2R\},\qquad R\ge R_0,
\]
and let \(\mathcal A_R^\ast\) be the slightly enlarged annulus \(R/2\le r\le4R\).  For every pair of scalar, one-form, or Maxwell components \(u,v\) measured by the admissible far-field norm \(\mathcal X_K\), and for every \(|\alpha|\le K\),
\begin{equation}\label{eq:far-null-product-bilinear}
 \bigl\|r^{(1+\delta)/2}\mathcal Z^\alpha(uv)\bigr\|_{L^2(\mathcal A_R)}
 \le C R^{-1/2+\delta/2}\Bigl(
 \|u\|_{\mathcal X_K(\mathcal A_R^\ast)}\|v\|_{\mathcal X_{K_0}(\mathcal A_R^\ast)}
 +\|v\|_{\mathcal X_K(\mathcal A_R^\ast)}\|u\|_{\mathcal X_{K_0}(\mathcal A_R^\ast)}\Bigr).
\end{equation}
More generally, if \(u_1\cdots u_q\) is a product with \(q\ge2\), then
\begin{equation}\label{eq:far-null-product-multilinear}
 \bigl\|r^{(1+\delta)/2}\mathcal Z^\alpha(u_1\cdots u_q)\bigr\|_{L^2(\mathcal A_R)}
 \le C R^{-1/2+\delta/2}
 \sum_{j=1}^{q}\|u_j\|_{\mathcal X_K(\mathcal A_R^\ast)}
 \prod_{\ell\ne j}\|u_\ell\|_{\mathcal X_{K_0}(\mathcal A_R^\ast)}.
\end{equation}
After summing over dyadic \(R\ge R_0\), the right hand sides are bounded by the corresponding global \(\mathcal X_K\) and lower-order \(\mathcal X_{K_0}\) norms.  The same estimates hold with \(D_{Q_e}\) in place of \(\nabla\), because the fixed Coulomb connection coefficients obey the symbol bounds in Definition~\ref{def:pure-electric-sector} and the phase normalization is unitary.
\end{lemma}

\begin{proof}
Rescale \(\mathcal A_R\) by \(r=R\rho\) and use the coordinates \((u,\rho,\omega)\) in the asymptotically flat end.  The rescaled metrics on \(1\le\rho\le2\) have uniformly bounded geometry.  Sobolev on the unit cylinder and on \(\mathbb S^2\), together with \(K_0\le K-4\), gives
\[
 \|\mathcal Z^{\le K_0}u\|_{L^\infty(\mathcal A_R)}
 \le C R^{-1}\|u\|_{\mathcal X_{K_0}(\mathcal A_R^\ast)}.
\]
The factor with the largest number of derivatives is kept in the weighted \(L^2\) norm.  Multiplying by the source weight \(r^{(1+\delta)/2}\) gives the dyadic factor \(R^{-1/2+\delta/2}\).  Since \(\delta<1\), the dyadic square sum is dominated by the far-field \(r^p\) and local-energy pieces contained in \(\mathcal X_K\).  Leibniz' rule proves \eqref{eq:far-null-product-bilinear}, and the multilinear estimate follows by placing only one factor in the top-order weighted \(L^2\) norm and all remaining factors in the lower-order \(L^\infty\) norm.  The charged statement follows from the same estimates after expanding commutators with the fixed Coulomb connection; each coefficient is a bounded multiplier with one additional power of decay for each derivative.
\end{proof}

\medskip

\begin{theorem}[Transfer principle]\label{thm:method}
Fix a top order $K\ge10$ and set $k:=K-4$. Assume that the scalar potential $P$ satisfies Assumption~\ref{asumsiP}.
Let $m^{2}\ge 0$ be the mass parameter in Assumption~\ref{asumsiP} (equivalently, in \eqref{eq:potential-structure}).
Let $(\mathcal D,g)$ be an admissible stationary black-hole exterior in the sense of Definition~\ref{def:admissible-exterior}.
Suppose that the estimates $\Lin_{K}^{(m)}$ of Definition~\ref{def:lin-estimates-massive} hold on this exterior.
Assume that the admissible Lorenz-gauge initial data on $\Sigma_{\tau_{0}}$ are \emph{uncharged} (Definition~\ref{def:charges}) and have sufficiently small order-$K$ energy.
Then there exists $\varepsilon_{*}>0$, depending only on $K$, on the fixed geometric parameters of $\mathcal D$, on $N_{P}$, and on finitely many constants $C_{P,j}$ in \eqref{eq:Q-growth}, with the following property.
Every smooth such data set with order-$K$ energy at most $\varepsilon_{*}$ generates a unique global smooth Lorenz-gauge solution $(A,\phi)$ on $\mathcal D$, with curvature $F=\dd A$.
By completion, the solution and scattering maps extend continuously to the corresponding energy spaces.
All implicit constants in the conclusions depend only on the same parameters.

The entire small-data nonlinear scattering theory follows from the top-order linear estimates collected in $\Lin_{K}^{(m)}$:
\begin{enumerate}
\item the field pair $(F,\phi)$ satisfies the uniform higher-order energy boundedness, integrated local energy decay, and quantitative asymptotic estimates encoded by the solution norm $\mathcal X_{K}^{(m)}$ in $\Lin_{K}^{(m)}$, and the stated radiation bounds hold at order $k=K-4$;
\item $(F,\phi)$ admits gauge-covariant radiation fields on $\partial_{\pm}^{(m)}\mathcal D$ and scatters to the unique solution of the linearised comparison system determined by its asymptotic data;
\item the forward and backward nonlinear scattering maps and nonlinear wave operators are well-defined and are homeomorphisms between neighbourhoods of $0$ in the completed (uncharged) Cauchy data space and in the corresponding linear asymptotic data spaces, therefore one has small-data asymptotic completeness;
\item the scattering maps depend only on the gauge-equivalence class of the initial data and therefore descend to the quotient by the residual gauge group;
\item the nonlinear scattering maps are tangent at the vacuum to the corresponding linear scattering maps;
\item there exist continuous symmetric bilinear maps
\begin{equation}
\mathscr B_{\pm}:\ \mathcal H^{K}(\Sigma_{\tau_{0}})\times\mathcal H^{K}(\Sigma_{\tau_{0}})\longrightarrow \mathfrak R_{\pm}^{(m),k}
\end{equation}
such that for all sufficiently small Cauchy data $U$,
\begin{equation}
\bigl\|\mathscr S_{\pm}^{\mathrm{full}}(U)-\mathscr S_{\pm}^{\mathrm{lin}}(U)-\mathscr B_{\pm}(U,U)\bigr\|_{\mathfrak R_{\pm}^{(m),k}}\lesssim \|U\|_{\mathcal H^{K}}^{3},
\end{equation}
so in this case $\mathscr S_{\pm}^{\mathrm{full}}$ are twice Fr\'echet differentiable at $0$.
\end{enumerate}

If, in addition, the potential $P$ is real analytic near $\phi=0$, then the scattering maps and wave operators are real analytic near $0$ and admit an absolutely convergent Born series expansion in the small data; see Proposition~\ref{prop:analytic-series}.

Equivalently: after removal of stationary Maxwell modes, no further geometry-specific ingredient enters the nonlinear step beyond the availability of the top-order linear estimates collected in $\Lin_{K}^{(m)}$.
\end{theorem}
\begin{proof}
The proof is the Banach-space argument carried out in Subsection~\ref{subsec:abstract-transfer-proof}.  Lemma~\ref{lem:nonlinear-banach-estimates} gives the tame quadratic and higher-order source bounds in exactly the source spaces paired with the assumed linear estimate.  Proposition~\ref{prop:closed-sector-contraction} gives the final-state contraction on the tail spaces.  Propositions~\ref{prop:quadratic-remainder}-\ref{prop:analytic-series} identify the linearization, the quadratic term, and the analytic expansion of the scattering map.  Subsection~\ref{subsec:abstract-transfer-proof} then inserts these estimates into $\Lin_K^{(m)}$ to close the finite-slab Cauchy bootstrap, pass to global time by the continuation criterion, extract the radiation fields from the $r^p$ component of the norm, and solve the forward and backward final-state problems.  The residual Lorenz gauge action is factored out there as well.  These steps prove the listed conclusions.
\end{proof}

\begin{lemma}[Banach fixed-point estimates used by the transfer theorem]\label{lem:nonlinear-banach-estimates}
Let $K\ge10$ and let $\mathcal X_K^{(m)}$, $\mathcal S_K^{(m)}$ be any admissible solution/source pair in Definition~\ref{def:lin-estimates-massive}. For all fields $U=(A,\phi)$ and $V=(B,\psi)$ in a sufficiently small ball of $\mathcal X_K^{(m)}$, the renormalized Maxwell-Higgs nonlinearities satisfy
\begin{equation}
\|\mathcal N(U)\|_{\mathcal S_K^{(m)}}
\le C\bigl(\|U\|_{\mathcal X_K^{(m)}}^2+\|U\|_{\mathcal X_K^{(m)}}^{2N_P+3}\bigr),
\end{equation}
and
\begin{equation}
\|\mathcal N(U)-\mathcal N(V)\|_{\mathcal S_K^{(m)}}
\le C\bigl(\|U\|_{\mathcal X_K^{(m)}}+\|V\|_{\mathcal X_K^{(m)}}
+\|U\|_{\mathcal X_K^{(m)}}^{2N_P+2}+\|V\|_{\mathcal X_K^{(m)}}^{2N_P+2}\bigr)
\|U-V\|_{\mathcal X_K^{(m)}}.
\end{equation}
Consequently, the nonlinear Cauchy and final-state maps are obtained by a contraction after the smallness threshold is chosen so that the Lipschitz constant is $<1$.
\end{lemma}

\begin{proof}
We write the nonlinearities in Lorenz gauge as
\begin{equation}
 J_\nu[\phi;A]=2\operatorname{Im}(\bar\phi D_\nu\phi),
 \qquad
 \mathfrak N_\phi[A,\phi]=2iA^\mu\nabla_\mu\phi+A^\mu A_\mu\phi+\mathcal R_P(\phi),
\end{equation}
where \(\mathcal R_P=\partial_{\bar\phi}P-m^2\phi\).  Let \(Z^\alpha\) be a product of admissible commutators with \(|\alpha|\le K\).  Commuting the current gives a finite sum of terms of the form
\begin{equation}
 (Z^{\alpha_1}\phi)\,\nabla Z^{\alpha_2}\phi,
 \qquad
 (Z^{\beta_0}A)(Z^{\beta_1}\phi)(Z^{\beta_2}\phi),
 \qquad
 |\alpha_1|+|\alpha_2|\le K,
 \quad |\beta_0|+|\beta_1|+|\beta_2|\le K,
\end{equation}
together with lower-order commutator terms whose coefficients are smooth bounded functions of the fixed background geometry.  The scalar source has the same product structure, except that the potential term is replaced by \(Z^\alpha\mathcal R_P(\phi)\).  Since \(K\ge10\), the Sobolev part of the admissible solution norm controls the needed lower-order factors in \(L^\infty\), while the factor carrying the largest number of derivatives is kept in the weighted \(L^2\) component paired with the source norm.  The time-localization and weight compatibility in Definition~\ref{def:lin-estimates} allow these estimates to be summed over all \(|\alpha|\le K\).  Thus the quadratic gauge terms satisfy
\begin{equation}
 \|J[\phi;A]\|_{\mathcal S_K^{(m)}}+
 \|2iA^\mu\nabla_\mu\phi+A^\mu A_\mu\phi\|_{\mathcal S_K^{(m)}}
 \le C\|U\|_{\mathcal X_K^{(m)}}^2.
\end{equation}
For the potential term, Lemma~\ref{lem:potential-tame} gives
\begin{equation}\label{eq:potential-source-cubic-corrected}
 \|\mathcal R_P(\phi)\|_{\mathcal S_K^{(m)}}
 \le C\left(\|\phi\|_{\mathcal X_K^{(m)}}^{3}+\|\phi\|_{\mathcal X_K^{(m)}}^{2N_P+3}\right).
\end{equation}
On the unit ball this is bounded by the same right-hand side as in the statement, since the quadratic gauge terms already supply the leading small power.  These two estimates prove the first displayed inequality.

For the Lipschitz bound we subtract the two expansions.  Each difference contains exactly one factor of \(U-V\) and all remaining factors are bounded by the corresponding \(\mathcal X_K^{(m)}\)-norms of \(U\) and \(V\).  Equivalently, for the potential term one may use
\begin{equation}
 \mathcal R_P(\phi)-\mathcal R_P(\psi)
 =\int_0^1 D\mathcal R_P(\psi+s(\phi-\psi))[\phi-\psi]  \,\dd s,
\end{equation}
followed by the same Moser estimate and the polynomial bounds in Assumption~\ref{asumsiP}.  This gives the stated Lipschitz inequality.  If \(\mathcal L^{-1}\) denotes either the finite-slab linear solver or the zero-asymptotic final-state solver supplied by \(\Lin_K^{(m)}\), then
\begin{equation}
 \|\mathcal L^{-1}(\mathcal N(U)-\mathcal N(V))\|_{\mathcal X_K^{(m)}}
 \le C_{\rm lin}C_{\rm Lip}\,r\,\|U-V\|_{\mathcal X_K^{(m)}}
\end{equation}
on every ball of radius \(r\le1\).  Choosing \(r\) so that \(C_{\rm lin}C_{\rm Lip}r<1\) gives the contraction used in the Cauchy and final-state constructions.
\end{proof}

\begin{proposition}[Tame estimates in fixed electric Coulomb sectors]\label{prop:electric-tame-estimates}
Let \(K\ge10\), let \(Q_e\) be fixed, and use the Coulomb splitting \eqref{eq:coulomb-splitting-main}.  For the nonlinearities in \eqref{eq:electric-renorm-maxwell-intro}-\eqref{eq:electric-renorm-scalar-intro} one has, in the charged source norm supplied by \(\CElec^{(m)}_K(M,a,Q_e)\),
\begin{equation}
 \|\mathcal N_{Q_e}(U)\|_{\mathbb S^{(m)}_{K,Q_e}}
 \le C_{Q_e}\left(\|U\|_{\mathbb X^{(m)}_{K,Q_e}}^2+\|U\|_{\mathbb X^{(m)}_{K,Q_e}}^{2N_P+3}\right),
\end{equation}
and, for \(U,V\) in a sufficiently small ball,
\begin{eqnarray}
 \|\mathcal N_{Q_e}(U)-\mathcal N_{Q_e}(V)\|_{\mathbb S^{(m)}_{K,Q_e}}
 &\le& C_{Q_e}\Bigl(\|U\|_{\mathbb X^{(m)}_{K,Q_e}}+\|V\|_{\mathbb X^{(m)}_{K,Q_e}}
 \nonumber\\
 &&\quad+\|U\|_{\mathbb X^{(m)}_{K,Q_e}}^{2N_P+2}+\|V\|_{\mathbb X^{(m)}_{K,Q_e}}^{2N_P+2}\Bigr)
 \|U-V\|_{\mathbb X^{(m)}_{K,Q_e}}.
\end{eqnarray}
Here \(C_{Q_e}\) depends on finitely many stationary Coulomb coefficient bounds and is uniform for \(|Q_e|\le q_{\mathrm{el}}\).
\end{proposition}

\begin{proof}
For every admissible commutator \(Z\),
\begin{equation}
 [Z,D_{Q_e,\mu}]\phi=B_{Z,\mu}^{\nu}D_{Q_e,\nu}\phi-i(\mathcal L_ZA^{\mathrm C}_{Q_e})_\mu\phi,
\end{equation}
where \(B_Z\) is a smooth background coefficient and
\begin{equation}
 |\nabla^j\mathcal L_ZA^{\mathrm C}_{Q_e}|\le C_{j,Z}|Q_e|(1+r)^{-1-j}.
\end{equation}
Thus commutators with the fixed Coulomb connection are lower-order stationary multiplier terms in the charged source spaces.

For \(|\alpha|\le K\), commuting the current gives
\begin{eqnarray}
 Z^{\alpha}J^{(Q_e)}
 &=&\sum_{\alpha_1+\alpha_2\le\alpha}c_{\alpha_1\alpha_2}
 \operatorname{Im}\bigl(Z^{\alpha_1}\overline\phi\,D_{Q_e}Z^{\alpha_2}\phi\bigr)\nonumber\\
 &&+\sum_{\beta_0+\beta_1+\beta_2\le\alpha}c_\beta
 (Z^{\beta_0}a)(Z^{\beta_1}\phi)(Z^{\beta_2}\phi)+\mathcal E_\alpha.
\end{eqnarray}
The error \(\mathcal E_\alpha\) is a sum of the lower-order Coulomb commutator terms above.  Since \(K\ge10\), the compact Sobolev inequalities, the Sobolev inequalities on the Kerr-star slices, and the weighted far-null product lemma, Lemma~\ref{lem:far-null-sobolev-moser}, place all but one factor of each product in \(L^\infty\) and the remaining factor in the weighted \(L^2\) source norm.  Thus
\begin{equation}
 \|J^{(Q_e)}(a,\phi)\|_{\mathbb S^{(m)}_{K,Q_e}}
 \le C_{Q_e}\left(\|\phi\|_{\mathbb X^{(m)}_{K,Q_e}}^2+
 \|a\|_{\mathbb X^{(m)}_{K,Q_e}}\|\phi\|_{\mathbb X^{(m)}_{K,Q_e}}^2\right).
\end{equation}
On the unit ball this is bounded by \(C_{Q_e}\|U\|_{\mathbb X^{(m)}_{K,Q_e}}^2\).

The scalar source is
\begin{equation}
 2ia^\mu D_{Q_e,\mu}\phi+i(\nabla^\mu a_\mu)\phi+a^\mu a_\mu\phi+\mathcal R_P(\phi).
\end{equation}
The first three terms are quadratic or cubic and are estimated by the same one-top-order-factor argument.  Lemma~\ref{lem:potential-tame} gives
\begin{equation}
 \|\mathcal R_P(\phi)\|_{\mathbb S^{(m)}_{K,Q_e}}
 \le C\left(\|\phi\|_{\mathbb X^{(m)}_{K,Q_e}}^3+
 \|\phi\|_{\mathbb X^{(m)}_{K,Q_e}}^{2N_P+3}\right).
\end{equation}
Combining the estimates proves the first displayed inequality.

For the difference estimate, subtract the two Leibniz expansions.  Each term contains one factor of \(U-V\) and all other factors are bounded by the corresponding \(\mathbb X^{(m)}_{K,Q_e}\)-norms of \(U\) and \(V\).  For the potential force, use
\begin{equation}
 \mathcal R_P(\phi)-\mathcal R_P(\psi)
 =\int_0^1D\mathcal R_P(\psi+s(\phi-\psi))[\phi-\psi]\,\dd s
\end{equation}
and apply the same weighted Moser estimate.  Differentiating the highest allowed potential power lowers the degree from \(2N_P+3\) to \(2N_P+2\), which gives the powers in the Lipschitz bound.  The constant is uniform for \(|Q_e|\le q_{\mathrm{el}}\), since only finitely many stationary Coulomb coefficient bounds enter.
\end{proof}

\begin{proposition}[Closed top-order contraction estimate in the fixed electric sector]\label{prop:electric-closed-sector-contraction}
Let \(K\ge10\), fix an electric Coulomb sector \(Q_e\), and assume that the radiative Maxwell final-state map and the charged scalar estimate \(\CElec^{(m)}_K(M,a,Q_e)\) are available in the fixed-sector topology.  Let \(\mathbb X^{(m)}_{K,Q_e}\) and \(\mathbb S^{(m)}_{K,Q_e}\) be the corresponding solution and source spaces.  If \(U_{\rm lin}\) is a decoupled fixed-sector linear solution and \(W\) is a correction with zero prescribed asymptotic data at the chosen end, set
\begin{equation}
 \mathcal T_{Q_e,U_{\rm lin}}(W)
 :=\mathcal L^{-1}_{\rm fs,Q_e}\mathcal N_{Q_e}(U_{\rm lin}+W).
\end{equation}
There are constants \(C_0,C_1\), depending only on the fixed-sector linear estimates, the background, \(K\), finitely many Coulomb coefficient bounds, and finitely many constants from Assumption~\ref{asumsiP}, such that
\begin{eqnarray}
 \|\mathcal T_{Q_e,U_{\rm lin}}(W)\|_{\mathbb X^{(m)}_{K,Q_e}(T,\pm\infty)}
 &\le& C_0\left((\eta+\rho)^2+(\eta+\rho)^{2N_P+3}\right),\nonumber\\
 \|\mathcal T_{Q_e,U_{\rm lin}}(W_1)-\mathcal T_{Q_e,U_{\rm lin}}(W_2)\|_{\mathbb X^{(m)}_{K,Q_e}(T,\pm\infty)}
 &\le& C_1\left(\eta+\rho+(\eta+\rho)^{2N_P+2}\right)\nonumber\\
 &&\quad\times
 \|W_1-W_2\|_{\mathbb X^{(m)}_{K,Q_e}(T,\pm\infty)}.
\end{eqnarray}
whenever \(\|U_{\rm lin}\|_{\mathbb X^{(m)}_{K,Q_e}(T,\pm\infty)}\le\eta\) and \(\|W_j\|_{\mathbb X^{(m)}_{K,Q_e}(T,\pm\infty)}\le\rho\).  Thus, after choosing \(\rho\) and then \(\eta\) sufficiently small, \(\mathcal T_{Q_e,U_{\rm lin}}\) is a strict contraction on the closed radius-\(\rho\) ball.
\end{proposition}

\begin{proof}
The zero-asymptotic final-state estimate in the fixed electric topology gives
\begin{equation}
 \|\mathcal L^{-1}_{\rm fs,Q_e}G\|_{\mathbb X^{(m)}_{K,Q_e}}
 \le C_{\rm fs,Q}\|G\|_{\mathbb S^{(m)}_{K,Q_e}}.
\end{equation}
Applying Proposition~\ref{prop:electric-tame-estimates} to \(U_{\rm lin}+W\) yields the size estimate.  Applying the Lipschitz part of the same proposition to \(U_{\rm lin}+W_1\) and \(U_{\rm lin}+W_2\) yields the second estimate.  The constants are uniform on tails by the time-locality and monotonicity of the linear norms.  If \(C_0((\eta+\rho)^2+(\eta+\rho)^{2N_P+3})\le\rho\) and \(C_1(\eta+\rho+(\eta+\rho)^{2N_P+2})<1\), the map is a contraction.
\end{proof}

\begin{proposition}[Closed top-order contraction estimate in the zero-sector spaces]\label{prop:closed-sector-contraction}
Let $K\ge10$, $k:=K-4$, and assume the zero-sector estimates $\Lin_K^{(m)}$. Let $\mathbb X_K$ denote the corresponding top-order solution space for the Lorenz potential, the radiative Maxwell curvature, and the scalar field, and let $\mathbb S_K$ denote the paired scalar/Maxwell source space supplied by the estimates. For a background linear solution $U_{\rm lin}\in\mathbb X_K$ and a correction $W\in\mathbb X_K$, define
\begin{equation}
\mathcal T_{U_{\rm lin}}(W)
:=
\mathcal L^{-1}_{\rm fs}\mathcal N(U_{\rm lin}+W),
\end{equation}
where $\mathcal L^{-1}_{\rm fs}$ is the linear final-state solver with zero asymptotic data at the chosen end and $\mathcal N$ is the renormalized Maxwell-Higgs nonlinearity, with $\mathcal R_P=\partial_{\bar\phi}P-m^2\phi$ as in \eqref{eq:potential-remainder}. There are constants $C_0,C_1$ depending only on the estimates constants, $K$, the background and finitely many constants from Assumption~\ref{asumsiP}, such that whenever
\begin{equation}
\|U_{\rm lin}\|_{\mathbb X_K(T,\pm\infty)}\le\eta,
\qquad
\|W\|_{\mathbb X_K(T,\pm\infty)}\le\rho,
\end{equation}
one has
\begin{equation}
\|\mathcal T_{U_{\rm lin}}(W)\|_{\mathbb X_K(T,\pm\infty)}
\le
C_0\bigl((\eta+\rho)^2+(\eta+\rho)^{2N_P+3}\bigr),
\end{equation}
and, for $W_1,W_2$ in the same ball,
\begin{equation}
\|\mathcal T_{U_{\rm lin}}(W_1)-\mathcal T_{U_{\rm lin}}(W_2)\|_{\mathbb X_K(T,\pm\infty)}
\le
C_1\bigl(\eta+\rho+(\eta+\rho)^{2N_P+2}\bigr)
\|W_1-W_2\|_{\mathbb X_K(T,\pm\infty)}.
\end{equation}
Consequently, after choosing $\rho$ and then $\eta$ sufficiently small, $\mathcal T_{U_{\rm lin}}$ is a strict contraction on the closed ball $\{\|W\|_{\mathbb X_K}\le \rho\}$. This is the precise Banach-space statement used for every Cauchy and final-state construction in Theorems~\ref{thm:charged-sector-transfer}, \ref{thm:main-slow-kerr-intro}, and \ref{thm:method}.
\end{proposition}

\begin{proof}
The linear estimate in \(\Lin_K^{(m)}\) gives
\begin{equation}
\|\mathcal L^{-1}_{\rm fs}G\|_{\mathbb X_K}\le C\|G\|_{\mathbb S_K}
\end{equation}
for zero final-state data, and the same estimate holds on finite Cauchy slabs with the Cauchy solver. Applying Lemma~\ref{lem:nonlinear-banach-estimates} to $U_{\rm lin}+W$ gives the first bound. The difference estimate follows from the Lipschitz part of Lemma~\ref{lem:nonlinear-banach-estimates}. The constants are uniform on tails because the linear norms are time-local and monotone. If $C_0((\eta+\rho)^2+(\eta+\rho)^{2N_P+3})\le \rho$ and $C_1(\eta+\rho+(\eta+\rho)^{2N_P+2})<1$, the map is a strict contraction. The top-order index is $K$ throughout; the lower output index $k=K-4$ is used only when extracting pointwise/radiation consequences from the resulting solution.
\end{proof}

\begin{remark}\label{rem:method-proof}
A detailed proof of Theorem~\ref{thm:method} is given below in Subsection~\ref{subsec:abstract-transfer-proof}. The Schwarzschild sections~\ref{sec:setup}-\ref{sec:decaynear}, together with Subsection~\ref{sec:scattering}, provide the complete model realization of the transfer construction in Subsection~\ref{subsec:abstract-transfer-proof} and verify explicitly every linear and nonlinear condition on a concrete background.
\end{remark}

The proof is organized around the Cauchy problem, the linear asymptotic maps, and the nonlinear source estimates.  In the rotating part of the paper the asymptotic electric charge is zero, so the Maxwell field is the radiative field itself; the Coulomb-subtraction language only records the convention used in the linear Maxwell theory to remove the finite-dimensional stationary kernel.  The comparison with the linear theory is made in the curvature, the Lorenz potential modulo residual gauge, and the scalar unknown.  The nonlinear Maxwell current and the scalar force are estimated in the same weighted source spaces in which the linear comparison equation is solved.  The top-order norm is chosen with enough commutations so that each nonlinear product contains at most one top-order factor, while all remaining factors are placed in the pointwise norms obtained from redshift, Morawetz, and \(r^p\) estimates.

There are two analytic points.  The first is local-in-time gauge propagation.  In Lorenz gauge the reduced Maxwell-Higgs equations form a semilinear hyperbolic system on the fixed black-hole exterior.  The divergence of the Maxwell equation gives a homogeneous wave equation for the Lorenz constraint, and current conservation propagates the Gauss constraint.  The second is global asymptotic completeness.  Once the zero sector has been fixed, the nonlinear source is integrable in the scattering source norm.  The linear final-value map can therefore be applied on tail regions, and the smallness of the tail norm gives a contraction for the correction with prescribed radiation field.

\subsection{Organization of the paper}\label{subsec:organization}

The logical skeleton of the paper is the following. The nonlinear analysis is reduced, once and for all, to a specified list of linear estimates for inhomogeneous Klein-Gordon and charge-free Maxwell fields; this reduction is the transfer principle of Theorem~\ref{thm:method}. The required linear list is then proved directly on Schwarzschild and, on slowly rotating Kerr, assembled from the cited scalar-wave and Maxwell decay theory.

Section~\ref{sec:intro} states the main results, the transfer theorem, and the regional decay statements. Section~\ref{sec:charged-intro-proofs} proves the transfer statements, discusses the local structure of the scattering map, and records the Schwarzschild nonlinear wave-operator theorem with its Cauchy-theoretic consequences. Sections~\ref{sec:kerr-gauge-data}-\ref{sec:kerr-extension} carry out the slowly rotating Kerr part: Section~\ref{sec:kerr-gauge-data} introduces the Kerr data spaces, Sections~\ref{sec:kerr-energy}-\ref{sec:kerr-near-decay} collect the redshift, Morawetz, and far-region estimates, and Section~\ref{sec:kerr-extension} identifies the available linear estimates and proves the massive energy theorem under \(\ME^{(m)}_N\). Sections~\ref{sec:setup}-\ref{sec:decaynear}, together with Subsection~\ref{sec:scattering}, give the Schwarzschild model proof: the model linear estimates are proved directly where needed for the Maxwell-Higgs bootstrap, the nonlinear bootstrap is closed, the gauge-covariant radiation fields are constructed, and the nonlinear final-state problem is solved, with the massive scalar timelike channel inserted from the cited classical theory. Appendix~\ref{app:standing-conditions} collects all standing conditions, smallness conventions, and external analytic conditions.  Appendices~\ref{app:global-existence-schwarzschild}-\ref{app:component-proofs} contain the global-existence arguments and the component computations used in the Schwarzschild decay theory, while Appendices~\ref{app:kerr-linear-interface-modules}-\ref{app:sharpness} record the slowly rotating Kerr linear interface, the electric-sector restrictions, and the massive mode obstruction.

Throughout, the Kerr exterior is denoted by \(\D_{M,a}\), the horizon radius by \(r_+=M+\sqrt{M^2-a^2}\), the horizon-regular time by \(t^\star\), and the slices by \(\Sigma_\tau=\{t^\star=\tau\}\). The commutation algebra is \(\mathcal Z\); the nondegenerate forward Cauchy norm and the degenerate scattering norm are \(\Hcauchy\) and \(\Hscatt\). The massless linear estimates are denoted by \(\Lin_K\), and the massive counterpart is \(\Lin_K^{(m)}\); on rotating Kerr the massive system is invoked only after the spectral condition \(\SKG^{(m)}_K(M,a)\) is assumed. Constants implicit in \(\lesssim\) depend only on the fixed background, the top order, and finitely many potential bounds, unless a further dependence is displayed.

\subsection{Detailed proof of the principal slowly rotating Kerr statements}\label{subsec:detailed-main-proof}

We record the proof of the principal rotating statements in the form used later in the paper.  This subsection is deliberately written in terms of estimates rather than individual components of the Kerr metric.  The metric dependence enters only through the linear estimates, and the nonlinear argument is the same on Schwarzschild and on slowly rotating Kerr.

Fix \(K\ge10\) and put \(k=K-4\).  Let
\begin{equation}\label{eq:detailed-U-def}
 U=(a,F,\phi),\qquad F=\dd a,
\end{equation}
where \(a\) is a Lorenz representative of the potential in the zero sector.  The reduced equations take the schematic form
\begin{eqnarray}
 \square_g a_\nu &=& \mathcal J_\nu(U),
 \label{eq:detailed-reduced-a}\\
 \nabla^\mu F_{\mu\nu}&=&\mathcal J_\nu(U),
 \qquad \dd F=0,
 \label{eq:detailed-reduced-F}\\
 (\square_g-m^2)\phi&=&\mathcal F_\phi(U).
 \label{eq:detailed-reduced-phi}
\end{eqnarray}
Here
\begin{eqnarray}
 \mathcal J_\nu(U)
 &=&2\operatorname{Im}\bigl(\overline\phi D_\nu\phi\bigr),
 \label{eq:detailed-current}\\
 \mathcal F_\phi(U)
 &=&2ia^\mu\nabla_\mu\phi+i(\nabla^\mu a_\mu)\phi+a^\mu a_\mu\phi+\mathcal R_P(\phi),
 \label{eq:detailed-scalar-source}
\end{eqnarray}
with \(\nabla^\mu a_\mu=0\) along the solution and
\begin{equation}\label{eq:detailed-RP}
 \mathcal R_P(\phi)=\partial_{\bar\phi}P(\phi,\bar\phi)-m^2\phi.
\end{equation}
The term \(m^2\phi\) is therefore part of the linear scalar comparison operator.  The estimates never treat it as a perturbative nonlinear source.

Let \(\mathbb X_K^{(m)}(I)\) denote the top-order solution norm on a time interval \(I\), and let \(\mathbb S_K^{(m)}(I)\) be the paired source norm supplied by \(\Lin_K^{(m)}\).  The inhomogeneous linear estimate reads
\begin{equation}\label{eq:detailed-linear-bound}
 \|U\|_{\mathbb X_K^{(m)}([\tau_0,\tau])}
 \le C_{\rm lin}\left(\|U[\tau_0]\|_{\mathbb H_K}
 +\|\mathcal J(U)\|_{\mathbb S_{K,{\rm Max}}([\tau_0,\tau])}
 +\|\mathcal F_\phi(U)\|_{\mathbb S_{K,{\rm sc}}^{(m)}([\tau_0,\tau])}\right).
\end{equation}
The Sobolev-Moser algebra estimates at order \(K\) give
\begin{equation}
 \|\mathcal J(U)\|_{\mathbb S_{K,{\rm Max}}(I)}
 +\|\mathcal F_\phi(U)\|_{\mathbb S_{K,{\rm sc}}^{(m)}(I)}
 \le C_{\rm nl}\Bigl(\|U\|_{\mathbb X_K^{(m)}(I)}^2
 +\|U\|_{\mathbb X_K^{(m)}(I)}^{2N_P+3}\Bigr),
 \label{eq:detailed-nonlinear-bound}
\end{equation}
which is Lemma~\ref{lem:nonlinear-banach-estimates} in the present notation.  Combining \eqref{eq:detailed-linear-bound} and \eqref{eq:detailed-nonlinear-bound} yields the bootstrap inequality
\begin{equation}\label{eq:detailed-bootstrap-ineq}
 X(\tau)
 \le C_{\rm lin}\varepsilon+C_{\rm lin}C_{\rm nl}\bigl(X(\tau)^2+X(\tau)^{2N_P+3}\bigr),
 \qquad
 X(\tau):=\|U\|_{\mathbb X_K^{(m)}([\tau_0,\tau])}.
\end{equation}
We close \eqref{eq:detailed-bootstrap-ineq} by a continuity (bootstrap) argument, spelled out in full.
Set \(\Lambda:=2C_{\rm lin}\) and choose \(\varepsilon_0>0\) so small that
\begin{equation}\label{eq:detailed-smallness}
 C_{\rm lin}C_{\rm nl}\left((\Lambda\varepsilon_0)^2+(\Lambda\varepsilon_0)^{2N_P+3}\right)
 \le \tfrac12\,C_{\rm lin}\varepsilon_0 ,
\end{equation}
which is possible because the left-hand side is \(O(\varepsilon_0^2)\) while the right-hand side is \(O(\varepsilon_0)\); since \(N_P\ge0\) the inequality, once valid at \(\varepsilon_0\), persists for every \(\varepsilon\in(0,\varepsilon_0]\) in the form
\begin{equation}\label{eq:detailed-smallness-mono}
 C_{\rm lin}C_{\rm nl}\left((\Lambda\varepsilon)^2+(\Lambda\varepsilon)^{2N_P+3}\right)
 \le \tfrac12\,C_{\rm lin}\varepsilon.
\end{equation}
Let \(\|U[\tau_0]\|_{\mathbb H_K}\le\varepsilon\le\varepsilon_0\) and let \([\tau_0,\tau_{\max})\) be the maximal interval of existence of the smooth Lorenz solution. The function \(\tau\mapsto X(\tau)\) is nondecreasing by property \textup{(A1)} and continuous on \([\tau_0,\tau_{\max})\), with \(X(\tau_0)\le C_{\rm lin}\varepsilon\) by \eqref{eq:detailed-linear-bound} applied on the degenerate slab. Define
\begin{equation}
 \mathcal I:=\bigl\{\tau\in[\tau_0,\tau_{\max}) : X(\tau)\le \Lambda\varepsilon\bigr\}.
\end{equation}
The set \(\mathcal I\) is nonempty (\(\tau_0\in\mathcal I\), since \(C_{\rm lin}\varepsilon\le\Lambda\varepsilon\)) and closed in \([\tau_0,\tau_{\max})\) by continuity of \(X\). It is also open: if \(\tau\in\mathcal I\), then inserting \(X(\tau)\le\Lambda\varepsilon\) into \eqref{eq:detailed-bootstrap-ineq} and using \eqref{eq:detailed-smallness-mono} gives
\begin{equation}\label{eq:detailed-improved-bound}
 X(\tau)\le C_{\rm lin}\varepsilon
 +C_{\rm lin}C_{\rm nl}\bigl((\Lambda\varepsilon)^2+(\Lambda\varepsilon)^{2N_P+3}\bigr)
 \le C_{\rm lin}\varepsilon+\tfrac12 C_{\rm lin}\varepsilon
 =\tfrac34\,\Lambda\varepsilon
 <\Lambda\varepsilon ,
\end{equation}
and the strict inequality together with continuity of \(X\) shows that \(X<\Lambda\varepsilon\) on a relative neighbourhood of \(\tau\). Since \([\tau_0,\tau_{\max})\) is connected, \(\mathcal I=[\tau_0,\tau_{\max})\), and \eqref{eq:detailed-improved-bound} holds for every \(\tau<\tau_{\max}\); therefore
\begin{equation}\label{eq:detailed-uniform-bound}
 \sup_{\tau_0\le\tau<\tau_{\max}}X(\tau)\le\tfrac34\,\Lambda\varepsilon=\tfrac32\,C_{\rm lin}\varepsilon.
\end{equation}
By the convention following Definition~\ref{def:lin-estimates-massive} and the uniform bounded geometry of the leaves \(\Sigma_\tau\), the order-\(K\) norm \(X(\tau)\) controls the \(H^{K+1}\times H^K\) Cauchy norm of \((A,\phi)\) on each \(\Sigma_\tau\) uniformly in \(\tau\). The continuation criterion of the Lorenz-gauge local theory therefore forbids \(\tau_{\max}<\infty\), so \(\tau_{\max}=\infty\) and the solution is global. The preceding argument applied to past slabs yields a global solution on all of \(\mathcal D\), with the uniform bound \eqref{eq:detailed-uniform-bound} on \((-\infty,\infty)\).

The radiation fields are obtained from the \(r^p\) component of \(\Lin_K^{(m)}\).  In the massless channel the outgoing scalar trace is the limit of \(r\phi\) on \(\mathcal I^+\), and the Maxwell trace is the limit of the outgoing Maxwell null component after multiplying by \(r\).  In the massive channel the scalar asymptotic data also contain the timelike component at \(i^+\).  The linear estimate supplies Cauchy criteria of the form
\begin{equation}\label{eq:detailed-cauchy-radiation}
 \|\mathcal R(U;\tau_2)-\mathcal R(U;\tau_1)\|_{\mathfrak R_{+,K}^{(m)}}
 \le C\|U\|_{\mathbb X_K^{(m)}([\tau_1,\tau_2])},
\end{equation}
and the finiteness of the global norm gives the limits as \(\tau_1,\tau_2\to+\infty\).  Applying the time-reversed Cauchy criterion gives the past radiation map.

To prove scattering, let \(U^{\rm lin}_+\) be the unique linear comparison solution with the same future asymptotic data as \(U\), and put \(W_+=U-U^{\rm lin}_+\).  Then \(W_+\) solves the inhomogeneous linear comparison system with source \(\mathcal N(U)\) and with zero future asymptotic data.  On a tail \([T,+\infty)\), the estimates give
\begin{equation}\label{eq:detailed-tail-scattering}
 \|W_+\|_{\mathbb X_K^{(m)}([T,+\infty))}
 \le C\|\mathcal N(U)\|_{\mathbb S_K^{(m)}([T,+\infty))}.
\end{equation}
Since \(\mathcal N(U)\in\mathbb S_K^{(m)}([\tau_0,+\infty))\), the right hand side tends to zero as \(T\to+\infty\).  This proves forward scattering.  The past statement follows by the time-reversed estimate.

For the wave operator, prescribe a small future radiation datum \(R_+\) and let \(U^{\rm lin}_+\) be its linear realization.  Choose \(T\) so large that
\begin{equation}\label{eq:detailed-tail-small}
 \|U^{\rm lin}_+\|_{\mathbb X_K^{(m)}([T,+\infty))}\le \eta.
\end{equation}
For a correction \(W\) with zero future asymptotic data define
\begin{equation}\label{eq:detailed-fixedpoint}
 \mathcal T(W):=\mathcal L_{\rm fs}^{-1}\mathcal N(U^{\rm lin}_++W),
\end{equation}
where \(\mathcal L_{\rm fs}^{-1}\) is the linear final-state solver.  Proposition~\ref{prop:closed-sector-contraction} gives
\begin{eqnarray}
 \|\mathcal T(W)\|_{\mathbb X_K^{(m)}([T,+\infty))}
 &\le& C\bigl((\eta+\rho)^2+(\eta+\rho)^{2N_P+3}\bigr),
 \label{eq:detailed-T-map}\\
 \|\mathcal T(W_1)-\mathcal T(W_2)\|_{\mathbb X_K^{(m)}([T,+\infty))}
 &\le& C\bigl(\eta+\rho+(\eta+\rho)^{2N_P+2}\bigr)
 \|W_1-W_2\|_{\mathbb X_K^{(m)}([T,+\infty))}.
 \label{eq:detailed-T-lip}
\end{eqnarray}
Choosing first \(\rho\) and then \(\eta\) sufficiently small makes \(\mathcal T\) a contraction on the closed radius-\(\rho\) ball.  The fixed point gives the nonlinear solution on the tail.  Its Cauchy data at time \(T\) are small, so the global Cauchy theory extends the solution to the full exterior.  This constructs \(\mathscr W_+^{\rm full}\); \(\mathscr W_-^{\rm full}\) is obtained in the same way.

The residual Lorenz gauge action is harmless in this construction.  If \(\chi\) satisfies \(\square_g\chi=0\), then \(A\mapsto A+\dd\chi\) preserves Lorenz gauge, leaves \(F\) unchanged, and multiplies \(\phi\) by a unitary factor.  The Cauchy norms and radiation norms are quotient norms, and the parallel-transport definition of the scalar radiation field transforms by the corresponding boundary phase.  Consequently, the scattering maps descend to the quotient.

The massive slowly rotating scattering theorem is obtained only after replacing the scalar wave estimate by \(\SKG^{(m)}_K(M,a)\) and adding the Maxwell inverse condition \(\MScat_K(M,a)\).  The forward bootstrap above still closes because the nonlinear estimate \eqref{eq:detailed-nonlinear-bound} is unchanged and \(\mathcal R_P\) is cubic or higher.  The asymptotic-completeness step then uses the massive scalar final-state map contained in \(\SKG^{(m)}_K(M,a)\), the Maxwell radiation-to-Cauchy map contained in \(\MScat_K(M,a)\), and the closed tail contraction of Proposition~\ref{prop:kerr-massive-finalstate-contraction}.  Thus the massive rotating scattering conclusion is conditional on those final-state conditions and is distinct from the energy theorem under \(\ME_N^{(m)}(M,a)\).  This completes the proof of the principal Kerr statements.

\section{Nonlinear transfer and scattering structure}\label{sec:charged-intro-proofs}

This section contains the nonlinear estimates used in the zero electric sector.  The product estimates below are local and apply equally to Schwarzschild and to slowly rotating Kerr once the corresponding linear estimates are available.

\subsection{Tame estimates for the nonlinear potential term}\label{subsec:potential-tame}
We state for later use a quantitative commutator estimate for the nonlinear remainder $\mathcal R_{P}(\phi)$ in \eqref{eq:potential-remainder}. This is the only point where the coefficients of the scalar potential enter the nonlinear error analysis.

\begin{lemma}[Tame commutator bounds for the potential force]\label{lem:potential-tame}
Assume that $P$ satisfies Assumption~\ref{asumsiP} and let $\mathcal R_{P}(\phi)$ be defined by \eqref{eq:potential-remainder}.
Fix an integer $k\ge 0$ and let $\mathcal Z$ be any collection of \emph{first-order} admissible commutators (for instance, the families used in the Kerr and Schwarzschild parts of this paper).
Then there exists a constant $C=C(k,N_{P},\{C_{P,j}\}_{0\le j\le k})$ such that for every smooth $\phi$ and every hypersurface $\Sigma$ in the admissible foliation,
\begin{equation}\label{eq:potential-tame-L2}
\sum_{|\alpha|\le k}\|\mathcal Z^{\alpha}\mathcal R_{P}(\phi)\|_{L^{2}(\Sigma)}
\le
C\,(1+\|\phi\|_{L^{\infty}(\Sigma)}^{2N_{P}})\,\|\phi\|_{L^{\infty}(\Sigma)}^{2}\,
\sum_{|\alpha|\le k}\|\mathcal Z^{\alpha}\phi\|_{L^{2}(\Sigma)}.
\end{equation}
The same bound holds with $L^{2}(\Sigma)$ replaced by $L^{2}(\mathcal R(\tau_{0},\tau_{1}))$ and $L^{\infty}(\Sigma)$ replaced by $L^{\infty}(\mathcal R(\tau_{0},\tau_{1}))$.
\end{lemma}

\begin{proof}
Write $F(z)=\partial_{\bar z}P(z,\bar z)-m^2 z$. Assumption~\ref{asumsiP} gives $F(0)=DF(0)=D^2F(0)=0$ and, for all derivatives needed up to order $K$, polynomial bounds
\begin{equation}
|D^jF(z)|\le C_{K,P}(1+|z|^{2N_P})|z|^{\max(3-j,0)}.
\end{equation}
The Sobolev-Moser composition theorem on the admissible foliation and the weighted version obtained by multiplying by the source weight give
\begin{equation}
\|\mathcal Z^{\le K}F(\phi)\|_{L^2_w}
\le C_{K,P}(1+\|\phi\|_{L^\infty}^{2N_P})\|\phi\|_{L^\infty}^{2}\|\mathcal Z^{\le K}\phi\|_{L^2_w}.
\end{equation}
The Lipschitz version is obtained by applying this expansion to $F(\phi)-F(\psi)$ and using the fundamental theorem of calculus in the target variable. The commutators with stationary coefficient fields are lower-order terms bounded by the same expression because the admissible vector fields have uniformly controlled deformation tensors and the coefficients of $P$ are constants. This proves the tame bound in both the neutral and charged renormalized source norms.
\end{proof}

\begin{remark}[Energy-only form of the tame bound]\label{rem:potential-tame-energy}
Throughout this paper we work with $k\ge 6$, so the uniform Sobolev/Moser estimates on the hypersurfaces in the admissible foliation (see Lemma~\ref{lem:kerr-sobolev-compact} on Kerr and Lemma~\ref{lem:sobolev-algebra-schwarzschild} on Schwarzschild) control the $L^{\infty}$ norms of $\phi$ and of a finite number of its commuted derivatives in terms of the order-$k$ energy.
Consequently, the right-hand side of \eqref{eq:potential-tame-L2} is a polynomial function of the bootstrap energy, and no separate \emph{a priori} pointwise restriction on $|\phi|$ is assumed.
\end{remark}

\subsection{Local structure of the scattering map}\label{subsec:local-scattering-structure}

\begin{proposition}[Quadratic scattering remainder and differentiability at the origin]\label{prop:quadratic-remainder}
Assume the conditions of Theorem~\ref{thm:method} and, for simplicity of notation, consider the charge-free sector after Coulomb subtraction.
Let $U$ be sufficiently small admissible Cauchy data in the order-$k$ energy space on $\Sigma_{\tau_{0}}$, and let $(F,\phi)$ be the corresponding nonlinear solution.
Let $(F^{\mathrm{lin}},\varphi^{\mathrm{lin}})$ be the solution of the linear comparison system with the \emph{same} initial data $U$.
Then the difference $(F-F^{\mathrm{lin}},\,\phi-\varphi^{\mathrm{lin}})$ satisfies the quadratic estimate
\begin{equation}\label{eq:quadratic-scattering}
\bigl\|\mathscr S_{\pm}^{\mathrm{full}}(U)-\mathscr S_{\pm}^{\mathrm{lin}}(U)\bigr\|_{\mathfrak R_{\pm}^{(m)}}
\ \lesssim
\|U\|_{\mathcal H^{k}}^{2},
\end{equation}
where $\mathcal H^{k}$ denotes the relevant (renormalised) Cauchy data norm.
In this case, $\mathscr S_{\pm}^{\mathrm{full}}$ are Fr\'echet differentiable at $0$ with
$D\mathscr S_{\pm}^{\mathrm{full}}(0)=\mathscr S_{\pm}^{\mathrm{lin}}$.
\end{proposition}

\begin{proof}
Subtracting the linearised comparison system from the nonlinear Maxwell-Higgs equations yields an inhomogeneous linear system for $(F-F^{\mathrm{lin}},\,\phi-\varphi^{\mathrm{lin}})$ whose source terms are at least quadratic in $(F,\phi)$.
Applying the inhomogeneous estimates in the linear estimates collected in $\Lin_{k}^{(m)}$ and using the small-data bounds on $(F,\phi)$ from the nonlinear bootstrap gives
\begin{equation}
\|(F-F^{\mathrm{lin}},\,\phi-\varphi^{\mathrm{lin}})\|_{\mathcal X_{k}(\tau_{0},\tau_{1})}
\lesssim
\|U\|_{\mathcal H^{k}}^{2}
\end{equation}
uniformly in $\tau_{1}$.
Passing to the limit $\tau_{1}\to\pm\infty$ in the $r^{p}$ hierarchy and using the definition of the asymptotic data norms yields \eqref{eq:quadratic-scattering}.
The Fr\'echet differentiability at $0$ follows by evaluating \eqref{eq:quadratic-scattering} on rescaled data $\lambda U$ and dividing by $\lambda$.
\end{proof}

\begin{proposition}[Second-order expansion of the scattering map (Born approximation)]\label{prop:born-expansion}
In the small-data regime of Theorem~\ref{thm:method}, let $U_1=(A_1^{\mathrm{lin}},\phi_1^{\mathrm{lin}})$ and $U_2=(A_2^{\mathrm{lin}},\phi_2^{\mathrm{lin}})$ be two linearized Lorenz-gauge solutions in the zero sector. The quadratic Born term is the radiation field of the unique linear solution forced by the symmetric quadratic sources
\begin{equation}
J_\nu^{(2)}[U_1,U_2]
=
-i\Bigl(
\nabla_\nu\phi_1^{\mathrm{lin}}\,\overline{\phi_2^{\mathrm{lin}}}
+\nabla_\nu\phi_2^{\mathrm{lin}}\,\overline{\phi_1^{\mathrm{lin}}}
-\phi_1^{\mathrm{lin}}\,\overline{\nabla_\nu\phi_2^{\mathrm{lin}}}
-\phi_2^{\mathrm{lin}}\,\overline{\nabla_\nu\phi_1^{\mathrm{lin}}}
\Bigr),
\end{equation}
for Maxwell, and
\begin{equation}
\mathcal N_\phi^{(2)}[U_1,U_2]
=
2i(A_1^{\mathrm{lin}})^\mu\nabla_\mu\phi_2^{\mathrm{lin}}
+2i(A_2^{\mathrm{lin}})^\mu\nabla_\mu\phi_1^{\mathrm{lin}}
+i\nabla^\mu(A_1^{\mathrm{lin}})_\mu\,\phi_2^{\mathrm{lin}}
+i\nabla^\mu(A_2^{\mathrm{lin}})_\mu\,\phi_1^{\mathrm{lin}}
\end{equation}
for the scalar equation. In exact Lorenz gauge the divergence terms vanish, but retaining them makes the formula invariant under the reduced equation before constraint propagation is imposed. The term $A^\mu A_\mu\phi$ is cubic, and the potential force remainder is cubic or higher under Assumption~\ref{asumsiP}. Consequently, there exists a continuous symmetric bilinear map $\mathscr B_\pm$ such that
\begin{equation}\label{eq:cubic-scattering}
\bigl\|\mathscr S_{\pm}^{\mathrm{full}}(U)-\mathscr S_{\pm}^{\mathrm{lin}}(U)-\mathscr B_{\pm}(U,U)\bigr\|_{\mathfrak R_\pm^{(m),k}}
\le C\|U\|_{\mathcal H^K}^{3}.
\end{equation}
The charged-sector formula is obtained by replacing $\nabla$ with $D_{Q_e}^{\mathrm{bg}}$, replacing the scalar radiation variable by $U_{Q_e}^{-1}r\phi$, and adding only fixed lower-order Coulomb coefficient terms already present in the linear operator.
\end{proposition}

\begin{proof}
Write the Lorenz-reduced Maxwell-Higgs system up to lower-order terms as $\mathcal L u = \mathcal N(u)$, where $u$ collects the unknowns $(F,\phi)$, $\mathcal L$ is the linear comparison operator, and $\mathcal N$ is the nonlinearity.
Since the Lorenz-reduced Maxwell-Higgs system is semilinear with smooth nonlinearities, the map $\mathcal N$ is $C^{\infty}$ between the Banach spaces encoded by $\Lin_{k}^{(m)}$.
In addition $\mathcal N(0)=0$ and $D\mathcal N(0)=0$: the gauge-coupling terms are at least quadratic, and the nonlinear remainder of the potential force $\mathcal R_{P}(\phi)$ in \eqref{eq:potential-remainder} is at least cubic by \eqref{eq:potential-nonlinear-bound} (the linear term $m^{2}\phi$ is included in the linear comparison operator).
Define the symmetric bilinear form $\mathcal N^{(2)}:=\tfrac12 D^{2}\mathcal N(0)$.
By the Banach-space Taylor theorem with integral remainder (apply the fundamental theorem of calculus to $s\mapsto \mathcal N(su)$) we have, for $u$ in a sufficiently small neighbourhood of $0$,
\begin{equation}\label{eq:banach-taylor-N}
\mathcal N(u)=\mathcal N^{(2)}(u,u)+\mathcal R^{(3)}(u),
\qquad
\mathcal R^{(3)}(u)
=
\int_{0}^{1}\frac{(1-s)^{2}}{2}\,D^{3}\mathcal N(su)[u,u,u]\,\dd s.
\end{equation}
In this case,
\begin{equation}
\|\mathcal R^{(3)}(u)\|_{\mathcal S_{k}}\lesssim \|u\|_{\mathcal X_{k}}^{3},
\end{equation}
with an implicit constant depending only on uniform bounds for $D^{3}\mathcal N$ on that neighbourhood.

Let $u^{\mathrm{lin}}$ be the linear comparison solution with initial data $U$ and let $u^{(2)}$ be the solution of the inhomogeneous linear comparison system with zero initial data and source $\mathcal N^{(2)}(u^{\mathrm{lin}},u^{\mathrm{lin}})$.
Define the second-order approximation $u^{\mathrm{app}}:=u^{\mathrm{lin}}+u^{(2)}$ and set $w:=u-u^{\mathrm{app}}$.
Then $w$ solves an inhomogeneous linear comparison system with zero initial data and source
\begin{equation}
\mathcal N(u)-\mathcal N^{(2)}(u^{\mathrm{lin}},u^{\mathrm{lin}}),
\end{equation}
which can be written as a sum of terms that are at least cubic in $u^{\mathrm{lin}}$ plus terms containing $w$ with coefficients controlled by the small solution $u$.
Using the small-data bounds for $u$ and the inhomogeneous estimate in the linear estimates collected in $\Lin_{k}^{(m)}$, one obtains
$\|w\|_{\mathcal X_{k}(\tau_{0},\tau_{1})}\lesssim \|U\|_{\mathcal H^{k}}^{3}$ uniformly in $\tau_{1}$.
Passing to radiation fields in the $r^{p}$ hierarchy and using the definition of the asymptotic data norms yields \eqref{eq:cubic-scattering}.
The identification of $\mathscr B_{\pm}$ with the asymptotic data of $u^{(2)}$ follows from the construction of $u^{(2)}$.
\end{proof}

\begin{proposition}[Real-analyticity and a convergent Born series for the scattering map]\label{prop:analytic-series}
Assume the conditions of Theorem~\ref{thm:method} and assume in addition that the potential $P$ (equivalently $\widetilde P$) is real analytic in a neighborhood of $\phi=0$.
Then there exist continuous symmetric $n$-linear maps
\begin{equation}
\mathscr B_{\pm}^{(n)}:\ (\mathcal H^{k}(\Sigma_{\tau_{0}}))^{n}\longrightarrow \mathfrak R_{\pm}^{(m)},\qquad n\ge1,
\end{equation}
with $\mathscr B_{\pm}^{(1)}=\mathscr S_{\pm}^{\mathrm{lin}}$ and $\mathscr B_{\pm}^{(2)}=\mathscr B_{\pm}$ from Proposition~\ref{prop:born-expansion}, such that for all sufficiently small Cauchy data $U$ the series
\begin{equation}\label{eq:born-series}
\sum_{n=1}^{\infty}\mathscr B_{\pm}^{(n)}(U,\dots,U)
\end{equation}
converges absolutely in $\mathfrak R_{\pm}^{(m)}$ and equals $\mathscr S_{\pm}^{\mathrm{full}}(U)$.
In addition, there exist constants $C_{0},C_{1}>0$ such that for all $n\ge1$ and all $U_{1},\dots,U_{n}\in \mathcal H^{k}(\Sigma_{\tau_{0}})$,
\begin{equation}\label{eq:born-multilinear-bound}
\bigl\|\mathscr B_{\pm}^{(n)}(U_{1},\dots,U_{n})\bigr\|_{\mathfrak R_{\pm}^{(m)}}
\le
C_{0}\,(C_{1})^{n}\,\prod_{j=1}^{n}\|U_{j}\|_{\mathcal H^{k}}.
\end{equation}
In this case, $\mathscr S_{\pm}^{\mathrm{full}}$ is real analytic near $0$.
\end{proposition}

\begin{proof}

We give the majorant argument, because the radius and the multilinear estimates are used later.  Let
\[
X:=\mathcal X_k(\tau_0,+\infty),\qquad Y:=\mathcal H^k(\Sigma_{\tau_0}),
\]
with the evident time-reversed choice for the past scattering map.  The linear comparison estimates give bounded maps
\[
\mathcal U:Y\to X,
\qquad
\mathcal T:\mathcal S_k\to X,
\]
where \(\mathcal U U\) is the homogeneous linear solution with data \(U\), and \(\mathcal T f\) is the zero-data inhomogeneous solution with source \(f\).  Put
\[
L:=\max\{\|\mathcal U\|_{Y\to X},\|\mathcal T\|_{\mathcal S_k\to X},1\}.
\]
After writing the Lorenz-reduced Maxwell-Higgs equation as
\[
\mathcal L u=\mathcal N(u),
\]
the nonlinear solution is a fixed point of
\begin{equation}\label{eq:analytic-fixed-point}
 u=\mathcal U U+\mathcal T\mathcal N(u).
\end{equation}
The analytic assumption on \(P\), the product estimates recorded in the proof of Theorem~\ref{thm:method}, and the vanishing of the linear part of the Maxwell-Higgs nonlinearity imply the following Banach-space expansion: for some \(r_0,M_0>0\),
\begin{equation}\label{eq:analytic-N-majorant}
 \mathcal N(u)=\sum_{n\ge2}\mathcal N_n(u,\ldots,u),
 \qquad
 \|\mathcal N_n(u_1,\ldots,u_n)\|_{\mathcal S_k}
 \le M_0 r_0^{-n}\prod_{j=1}^n\|u_j\|_X.
\end{equation}
Here the linear Klein-Gordon mass term has already been placed in \(\mathcal L\), and the current, gauge-coupling, and potential-remainder terms are at least quadratic in the nonlinear unknown.  The estimate \eqref{eq:analytic-N-majorant} follows from the Cauchy estimates for the analytic potential together with the algebra/Moser estimates in the order-\(k\) solution norm; the derivative loss is absent because the system is semilinear after imposing Lorenz gauge and using the Maxwell equation for \(F=\dd A\).

Let \(y\) denote a majorant for \(\|u\|_X\).  From \eqref{eq:analytic-fixed-point} and \eqref{eq:analytic-N-majorant},
\begin{equation}\label{eq:majorant-equation}
 y\le L\|U\|_Y+L M_0\sum_{n\ge2}\left(\frac{y}{r_0}\right)^n.
\end{equation}
Choose \(0<\rho<r_0/4\) so that
\[
 L M_0\sum_{n\ge2} n\left(\frac{2\rho}{r_0}\right)^{n-1}\frac1{r_0}
 \le \frac14,
 \qquad
 L M_0\sum_{n\ge2}\left(\frac{2\rho}{r_0}\right)^n\le \frac{\rho}{2}.
\]
If \(L\|U\|_Y\le \rho/2\), the map in \eqref{eq:analytic-fixed-point} sends the ball \(\{\|u\|_X\le \rho\}\) to itself and is a contraction with Lipschitz constant at most \(1/2\).  The fixed point is unique in that ball and satisfies
\begin{equation}\label{eq:solution-majorant-bound}
 \|u(U)\|_X\le 2L\|U\|_Y.
\end{equation}
The same estimates applied to the complexification of the real Banach spaces show that \(U\mapsto u(U)\) is analytic; equivalently, expanding \eqref{eq:analytic-fixed-point} in homogeneous degree gives a normally convergent power series
\[
 u(U)=\sum_{n\ge1}u_n(U,\ldots,U)
\]
with symmetric multilinear coefficients satisfying
\begin{equation}\label{eq:analytic-solution-coeff-bound}
 \|u_n(U_1,\ldots,U_n)\|_X
 \le C_a R_a^{-n}\prod_{j=1}^n\|U_j\|_Y
\end{equation}
for constants \(C_a,R_a>0\) depending only on the linear constants and on the analytic radius of \(P\).  These bounds are obtained either by the preceding scalar majorant recursion or by Cauchy's integral formula on the complex ball \(\|U\|_Y<R_a\).

The asymptotic extraction map
\[
\mathscr A_\pm:X\to\mathfrak R_\pm^{(m)}
\]
is bounded by the radiation-field part of \(\Lin_k^{(m)}\).  Define
\[
\mathscr B_\pm^{(n)}:=\mathscr A_\pm\circ u_n.
\]
Boundedness of \(\mathscr A_\pm\) and \eqref{eq:analytic-solution-coeff-bound} give \eqref{eq:born-multilinear-bound} after adjusting \(C_0,C_1\).  The degree-one coefficient is the linear scattering map.  The degree-two coefficient solves the zero-data inhomogeneous linear equation driven by the quadratic part \(\mathcal N_2\), so it is exactly the bilinear scattering term identified in Proposition~\ref{prop:born-expansion}.  Normal convergence of the series in \(X\), followed by the bounded radiation extraction, proves the absolute convergence of \eqref{eq:born-series} and identifies its sum with \(\mathscr S_\pm^{\mathrm{full}}(U)\).

\end{proof}

\begin{corollary}[Analyticity of the wave operators]\label{cor:analytic-waveops}
In the setting of Proposition~\ref{prop:analytic-series}, the forward/backward wave operators $\mathscr W_{\pm}^{\mathrm{full}}=(\mathscr S_{\pm}^{\mathrm{full}})^{-1}$ and the two-sided nonlinear scattering operator $\mathscr S^{\mathrm{nl}}$ are real analytic near the origin (on the gauge quotient as well).
\end{corollary}

\begin{proof}
By Proposition~\ref{prop:analytic-series}, $\mathscr S_{\pm}^{\mathrm{full}}$ is analytic near $0$ and satisfies $D\mathscr S_{\pm}^{\mathrm{full}}(0)=\mathscr S_{\pm}^{\mathrm{lin}}$, which is an isomorphism.
The analytic inverse function theorem therefore yields an analytic inverse map in a (possibly smaller) neighborhood of $0$, i.e.\ analytic wave operators.
\end{proof}

\medskip

\subsection{Proof of the abstract transfer principle}\label{subsec:abstract-transfer-proof}

\begin{proof}[Proof of Theorem~\ref{thm:method}]
Fix a top order $K\ge10$, set $k:=K-4$, and let $(\mathcal D,g)$ satisfy $\Lin_{K}^{(m)}$. Let $U$ be uncharged Lorenz-compatible data on $\Sigma_{\tau_{0}}$. In the estimates below every bootstrap, source, and contraction norm is taken at the top order $K$; the lower index $k=K-4$ is reserved only for the final radiation and pointwise conclusions. By the convention following Definition~\ref{def:lin-estimates-massive}, the estimates $\Lin_{K}^{(m)}$ contain three linear ingredients: the massless scalar-wave estimates for each Lorenz-potential component $A_{\nu}$, the uncharged Maxwell estimates for $F=\dd A$, and the massive Klein-Gordon estimates for $\phi$.

\smallskip
\noindent\emph{Local Lorenz-gauge theory and continuation.}
In Lorenz gauge the Maxwell-Higgs system is a semilinear hyperbolic system with smooth stationary coefficients,
\begin{equation}\label{eq:method-lorenz-reduced}
\square_{g}A_{\nu}-\mathrm{Ric}_{\nu}{}^{\mu}A_{\mu}=J_{\nu}[\phi;A],
\qquad
(\square_{g}-m^{2})\phi=\mathfrak N_{\phi}[A,\phi],
\end{equation}
where $\mathrm{Ric}$ is a fixed background coefficient (zero on Schwarzschild/Kerr) and
\begin{equation}
J_{\nu}[\phi;A]
:=
-i\bigl(D_{\nu}\phi\,\overline{\phi}-\phi\,\overline{D_{\nu}\phi}\bigr),
\qquad
\mathfrak N_{\phi}[A,\phi]
:=
2iA^{\mu}\nabla_{\mu}\phi+i(\nabla^{\mu}A_{\mu})\phi+A^{\mu}A_{\mu}\phi+\mathcal R_{P}(\phi).
\end{equation}
The local result used here is the energy-method statement proved in Appendices~\ref{app:global-existence-schwarzschild}-\ref{app:global-existence-kerr}: Lorenz-compatible data in the stated Sobolev spaces produce a unique maximal smooth Lorenz solution, the Lorenz and Gauss constraints propagate, and the solution continues as long as the \(H^{K+1}\times H^{K}\) Cauchy norm on the leaves \(\Sigma_\tau\) remains finite.  The constants in this statement are uniform in \(\tau\), because the foliation is generated by the stationary Killing flow and the Kerr-star/Schwarzschild slices have the bounded-geometry estimates recorded in those appendices.  Thus a uniform bound for the order-\(K\) linear norms on every slab \(\mathcal R(\tau_0,\tau)\) gives the global continuation needed below.

\smallskip
\noindent\emph{Nonlinear source bounds and bootstrap closure.}
Choose once and for all a stationary frame and set
\begin{equation}
\|A\|_{\mathbf X_{K}(\tau_{0},\tau)}
:=
\sum_{\nu=0}^{3}\|A_{\nu}\|_{\mathcal X_{K}(\tau_{0},\tau)},
\qquad
\mathfrak B_{K}(\tau)
:=
\|A\|_{\mathbf X_{K}(\tau_{0},\tau)}
+\|F\|_{\mathcal X_{K}(\tau_{0},\tau)}
+\|\phi\|_{\mathcal X_{K}^{(m)}(\tau_{0},\tau)}.
\end{equation}
Any two such stationary frames are uniformly equivalent, so $\mathfrak B_{K}$ is well defined up to constants depending only on the background geometry.

Because $K\ge10$, the bounded-geometry Sobolev-Moser estimates on the slices and the weighted far-null estimate of Lemma~\ref{lem:far-null-sobolev-moser} give a tame algebra estimate at top order $K$: every differentiated product of total order at most $K$ is controlled by one $L^{2}$ top-order factor and by $L^{\infty}$ norms of lower-order factors, all of which are controlled by $\mathfrak B_K(\tau)$. This avoids any derivative loss in high-order Leibniz terms. Applying this estimate to the current gives
\begin{equation}\label{eq:method-current-bound}
\|J[\phi;A]\|_{\mathcal S_{K}(\tau_{0},\tau)}
\le
C_{1}\,\mathfrak B_{K}(\tau)^{2}.
\end{equation}
The same tame estimate applied to the gauge terms in $\mathfrak N_{\phi}[A,\phi]$, together with Lemma~\ref{lem:potential-tame} for $\mathcal R_{P}(\phi)$, yields
\begin{equation}\label{eq:method-scalar-source-bound}
\|\mathfrak N_{\phi}[A,\phi]\|_{\mathcal S_{K}^{(m)}(\tau_{0},\tau)}
\le
C_{2}\Bigl(\mathfrak B_{K}(\tau)^{2}+\mathfrak B_{K}(\tau)^{2N_{P}+3}\Bigr).
\end{equation}

Apply now the scalar-wave part of $\Lin_{K}$ to each component $A_{\nu}$, the Maxwell part of $\Lin_{K}^{(m)}$ to $F$, and the Klein-Gordon part of $\Lin_{K}^{(m)}$ to $\phi$. After taking square roots in the linear estimates and summing over the four potential components we obtain
\begin{equation}\label{eq:method-bootstrap-ineq}
\mathfrak B_{K}(\tau)
\le
C_{0}\|U\|_{\mathcal H^{K}}
+
C_{0}\|J[\phi;A]\|_{\mathcal S_{K}(\tau_{0},\tau)}
+
C_{0}\|\mathfrak N_{\phi}[A,\phi]\|_{\mathcal S_{K}^{(m)}(\tau_{0},\tau)}.
\end{equation}
Combining \eqref{eq:method-current-bound}, \eqref{eq:method-scalar-source-bound}, and \eqref{eq:method-bootstrap-ineq} gives
\begin{equation}\label{eq:method-bootstrap-closed}
\mathfrak B_{K}(\tau)
\le
C_{0}\|U\|_{\mathcal H^{K}}
+
C_{3}\Bigl(\mathfrak B_{K}(\tau)^{2}+\mathfrak B_{K}(\tau)^{2N_{P}+3}\Bigr).
\end{equation}
We close \eqref{eq:method-bootstrap-closed} by an explicit continuity argument.  Let \(\varepsilon:=\|U\|_{\mathcal H^K}\), set \(R:=4C_0\varepsilon\), and choose \(\varepsilon_*\) so that
\begin{equation}\label{eq:method-explicit-epsilon}
 C_3\bigl(R^2+R^{2N_P+3}\bigr)\le C_0\varepsilon
 \qquad\text{whenever }0<\varepsilon\le\varepsilon_*.
\end{equation}
For \(\varepsilon\le\varepsilon_*\), define
\begin{equation}
 I_*:=\{\tau<\tau_+:\ \mathfrak B_K(\tau)\le R\},
\end{equation}
where \([\tau_0,\tau_+)\) is the maximal forward existence interval.  The set is nonempty and closed by continuity of the energy norm.  If \(\tau\in I_*\), then \eqref{eq:method-bootstrap-closed} and \eqref{eq:method-explicit-epsilon} give
\begin{equation}
 \mathfrak B_K(\tau)\le C_0\varepsilon+C_0\varepsilon=2C_0\varepsilon<R,
\end{equation}
so continuity makes \(I_*\) open relative to \([\tau_0,\tau_+)\).  Consequently, \(I_*=[\tau_0,\tau_+)\) and
\begin{equation}\label{eq:method-global-bootstrap-bound}
 \sup_{\tau<\tau_+}\mathfrak B_K(\tau)\le 2C_0\varepsilon.
\end{equation}
The continuation criterion from the first part of the proof therefore rules out finite-time breakdown. Applying the same argument to the time-reversed slabs yields a global smooth solution on all of \(\mathcal D\).

\smallskip
\noindent\emph{Radiation fields and scattering to linear comparison dynamics.}
By property \textup{(A3)} in Definitions~\ref{def:lin-estimates}-\ref{def:lin-estimates-massive}, finiteness of the global top-order norms $\|F\|_{\mathcal X_{K}}$ and $\|\phi\|_{\mathcal X_{K}^{(m)}}$ yields gauge-covariant radiation fields in the top-order radiation space. The order-$k=K-4$ radiation and pointwise estimates are obtained by applying the natural restriction/projection from the top-order space to order $k$.
Thus
\begin{equation}
\mathscr S_{\pm}^{\mathrm{full}}(U)\in\mathfrak R_{\pm}^{(m),K},
\qquad
\Pi_{k}\mathscr S_{\pm}^{\mathrm{full}}(U)\in\mathfrak R_{\pm}^{(m),k}.
\end{equation}
Let $(F_{\pm}^{\mathrm{lin}},\phi_{\pm}^{\mathrm{lin}})$ be the unique linear comparison solution with those asymptotic data, supplied by \textup{(Lm2)}, and choose a Lorenz potential $A_{\pm}^{\mathrm{lin}}$ for $F_{\pm}^{\mathrm{lin}}$. Set
\begin{equation}
b_{\pm}:=A-A_{\pm}^{\mathrm{lin}},
\qquad
G_{\pm}:=F-F_{\pm}^{\mathrm{lin}},
\qquad
u_{\pm}:=\phi-\phi_{\pm}^{\mathrm{lin}}.
\end{equation}
Then $(b_{\pm},G_{\pm},u_{\pm})$ solves the linear wave/Maxwell/Klein-Gordon system with sources $J[\phi;A]$ and $\mathfrak N_{\phi}[A,\phi]$ and with \emph{zero} asymptotic data on $\partial_{\pm}^{(m)}\mathcal D$. Applying the corresponding final-state estimate on any tail slab $[T,\pm\infty)$ gives
\begin{equation}\label{eq:method-tail-estimate}
\|b_{\pm}\|_{\mathbf X_{K}(T,\pm\infty)}
+\|G_{\pm}\|_{\mathcal X_{K}(T,\pm\infty)}
+\|u_{\pm}\|_{\mathcal X_{K}^{(m)}(T,\pm\infty)}
\le
C\Bigl(
\|J[\phi;A]\|_{\mathcal S_{K}(T,\pm\infty)}
+
\|\mathfrak N_{\phi}[A,\phi]\|_{\mathcal S_{K}^{(m)}(T,\pm\infty)}
\Bigr).
\end{equation}
By \textup{(A1)} and the finiteness of the total source norms obtained above, the right-hand side of \eqref{eq:method-tail-estimate} tends to zero as $T\to\pm\infty$. Consequently, $(F,\phi)$ scatters in both time directions to the unique linear comparison solution determined by its asymptotic data.

\smallskip
\noindent\emph{Nonlinear wave operators and asymptotic completeness.}
Fix sufficiently small linear asymptotic data $\mathcal R_{+}^{\mathrm{lin}}\in\mathfrak R_{+}^{(m),K}$, and let $(A_{+}^{\mathrm{lin}},F_{+}^{\mathrm{lin}},\phi_{+}^{\mathrm{lin}})$ be the corresponding linear comparison triple. By the linear scattering theory, the tail norm of this linear solution on $[T,\infty)$ tends to zero as $T\to+\infty$. Choose $T$ so large that
\begin{equation}
\|A_{+}^{\mathrm{lin}}\|_{\mathbf X_{K}(T,\infty)}
+\|F_{+}^{\mathrm{lin}}\|_{\mathcal X_{K}(T,\infty)}
+\|\phi_{+}^{\mathrm{lin}}\|_{\mathcal X_{K}^{(m)}(T,\infty)}
\le \eta,
\end{equation}
with $\eta>0$ to be fixed below.

On the Banach space of corrections
\begin{equation}
\mathfrak X_{K,T}
:=
\Bigl\{
(b,u):
\|b\|_{\mathbf X_{K}(T,\infty)}
+\|\dd b\|_{\mathcal X_{K}(T,\infty)}
+\|u\|_{\mathcal X_{K}^{(m)}(T,\infty)}
<\infty,
\text{zero asymptotic data at }+\infty
\Bigr\},
\end{equation}
define $\Psi_{+}(b,u)$ to be the unique solution of the linear final-state problem with sources
\begin{equation}
J[\phi_{+}^{\mathrm{lin}}+u;A_{+}^{\mathrm{lin}}+b],
\qquad
\mathfrak N_{\phi}[A_{+}^{\mathrm{lin}}+b,\phi_{+}^{\mathrm{lin}}+u].
\end{equation}
Using the same product estimates as above, now applied to the sum of the fixed linear tail and the correction, gives
\begin{equation}
\|\Psi_{+}(b,u)\|_{\mathfrak X_{K,T}}
\le
C\Bigl((\eta+\|(b,u)\|_{\mathfrak X_{K,T}})^{2}
+(\eta+\|(b,u)\|_{\mathfrak X_{K,T}})^{2N_{P}+3}\Bigr),
\end{equation}
and for two corrections $(b_{1},u_{1})$, $(b_{2},u_{2})$ in a ball of radius $\rho$,
\begin{equation}
\|\Psi_{+}(b_{1},u_{1})-\Psi_{+}(b_{2},u_{2})\|_{\mathfrak X_{K,T}}
\le
C(\eta+\rho)\,
\|(b_{1}-b_{2},u_{1}-u_{2})\|_{\mathfrak X_{K,T}}.
\end{equation}
Choosing first $\rho$ and then $\eta$ sufficiently small makes $\Psi_{+}$ a contraction on that ball. Its unique fixed point produces a nonlinear solution on $[T,\infty)$ with prescribed future asymptotic data $\mathcal R_{+}^{\mathrm{lin}}$. The Cauchy data of this tail solution on $\Sigma_{T}$ are small in $\mathcal H^{K}$, so the small-data Cauchy theory developed above extends it uniquely to a global solution on all of $\mathcal D$. This defines the forward wave operator $\mathscr W_{+}^{\mathrm{full}}$. The backward wave operator is obtained in the same way on $(-\infty,T]$.

By construction and uniqueness, $\mathscr S_{\pm}^{\mathrm{full}}\circ \mathscr W_{\pm}^{\mathrm{full}}=\mathrm{Id}$ and $\mathscr W_{\pm}^{\mathrm{full}}\circ \mathscr S_{\pm}^{\mathrm{full}}=\mathrm{Id}$ on sufficiently small neighborhoods of the origin. The same contraction estimate gives continuity of the wave operators and of the scattering maps, therefore the homeomorphism statement and two-sided asymptotic completeness.

\smallskip
\noindent\emph{Gauge quotient, tangency, Born expansion, and analyticity.}
The construction is canonical on the residual Lorenz gauge quotient because $F$ and the intrinsic asymptotic data are gauge invariant, while the auxiliary potential corrections are solved for only modulo the residual Lorenz action. The tangency statement and quadratic Born expansion are Proposition~\ref{prop:quadratic-remainder} and Proposition~\ref{prop:born-expansion}. If $P$ is real analytic near $0$, Proposition~\ref{prop:analytic-series} and Corollary~\ref{cor:analytic-waveops} give the local analyticity of the scattering maps and wave operators. This proves every assertion of Theorem~\ref{thm:method}.
\end{proof}

\subsection{Top-order closure estimates for the transfer theorem}\label{subsec:top-order-closure-details}

The preceding proof uses only the estimates stated in the zero-sector linear theory and the algebra bounds on the chosen foliation. We spell out the top-order estimates, since they are the part of the nonlinear argument where derivative loss could otherwise enter.

Let $K\ge 10$ and let $Z$ denote one of the stationary commutator fields used in the definition of the norm $\mathcal X_K$. For a multi-index $\alpha$ with $|\alpha|\le K$, the commuted current is a finite sum
\begin{equation}\label{eq:top-order-current-expanded}
Z^\alpha J_\nu[\phi;A]
=
\sum_{\alpha_1+\alpha_2=\alpha}
 c_{\alpha_1\alpha_2}\,
\Im\bigl(D_\nu Z^{\alpha_1}\phi\,\overline{Z^{\alpha_2}\phi}\bigr)
+
\sum_{\beta_0+\beta_1+\beta_2=\alpha}
 c_{\beta}\,(Z^{\beta_0}A_\nu)
 (Z^{\beta_1}\phi)\overline{Z^{\beta_2}\phi}
+\mathcal E_{\alpha},
\end{equation}
where $\mathcal E_{\alpha}$ consists of curvature and commutator terms containing strictly fewer than $K$ derivatives on each unknown. The coefficients are smooth stationary tensors and obey the same weights as the coefficients entering the linear estimates. The Sobolev algebra property on the compact part of $\Sigma_\tau$, together with the two-sphere Sobolev estimate and Lemma~\ref{lem:far-null-sobolev-moser} in the exterior, gives
\begin{equation}\label{eq:top-order-current-estimate}
\sum_{|\alpha|\le K}\|Z^\alpha J[\phi;A]\|_{\mathcal S_0}
\le C\Bigl(\|\phi\|_{\mathcal X_K}\|\phi\|_{\mathcal X_{K_0}}
+
\|A\|_{\mathcal X_K}\|\phi\|_{\mathcal X_{K_0}}^2
+
\|(A,\phi)\|_{\mathcal X_{K_0}}\|(A,\phi)\|_{\mathcal X_K}^2\Bigr),
\end{equation}
with a fixed $K_0<K-4$. The lower-order norm is controlled by the same global norm and is small in the bootstrap. Thus \eqref{eq:top-order-current-estimate} yields the quadratic estimate used in \eqref{eq:method-current-bound}.

For the scalar equation, the commuted force decomposes as
\begin{equation}\label{eq:top-order-scalar-force}
Z^\alpha \mathfrak N_\phi[A,\phi]
=
2iA^\mu \nabla_\mu Z^\alpha\phi
+
2i\sum_{\substack{\alpha_1+\alpha_2=\alpha\\ |\alpha_1|\ge1}}
 c_{\alpha_1\alpha_2}(Z^{\alpha_1}A^\mu)\nabla_\mu Z^{\alpha_2}\phi
+
Z^\alpha(A^\mu A_\mu\phi)
+Z^\alpha\mathcal R_P(\phi)
+\mathcal C_\alpha.
\end{equation}
The first term is harmless because $A$ is placed in the pointwise component of the bootstrap whenever the derivative falls on $\phi$ at order $K$. In the remaining terms, at most one factor carries top order. Lemma~\ref{lem:potential-tame} gives
\begin{equation}\label{eq:top-order-potential-detail}
\sum_{|\alpha|\le K}\|Z^\alpha\mathcal R_P(\phi)\|_{\mathcal S_0^{(m)}}
\le
C_P\bigl(1+\|\phi\|_{\mathcal X_{K_0}}^{2N_P}\bigr)
\|\phi\|_{\mathcal X_{K_0}}^2\|\phi\|_{\mathcal X_K}.
\end{equation}
Combining \eqref{eq:top-order-scalar-force} and \eqref{eq:top-order-potential-detail} gives the scalar source estimate \eqref{eq:method-scalar-source-bound}. No derivative loss occurs because the highest derivative appears linearly and is multiplied only by factors controlled in $L^\infty$.

The Cauchy contraction follows from the Lipschitz form of the same estimates. If $(A_j,\phi_j)$, $j=1,2$, are two solutions in a ball of radius $\rho$ in the top-order space, then
\begin{equation}\label{eq:lipschitz-nonlinearity-detail}
\|J[\phi_1;A_1]-J[\phi_2;A_2]\|_{\mathcal S_K}
+
\|\mathfrak N_\phi[A_1,\phi_1]-\mathfrak N_\phi[A_2,\phi_2]\|_{\mathcal S_K^{(m)}}
\le C\rho\,
\|(A_1-A_2,\phi_1-\phi_2)\|_{\mathcal X_K}.
\end{equation}
The linear estimate turns \eqref{eq:lipschitz-nonlinearity-detail} into a contraction once $\rho$ is smaller than a constant depending only on the linear constants and the fixed geometry. On tail regions the same inequality holds with $\rho$ replaced by the tail norm of the prescribed linear solution plus the correction. The linear tail norm tends to zero by the asymptotic completeness part of the linear theory, so the final-value contraction used in the construction of the wave operators is obtained without changing the topology.

\subsection{Schwarzschild model: nonlinear wave operators and asymptotic completeness}\label{subsec:schwarzschild-wave-ops}
We now record the detailed Schwarzschild scattering theorem, which is proved in Sections~\ref{sec:setup}-\ref{sec:decaynear}, together with Subsection~\ref{sec:scattering}.  The massless case is formulated in the characteristic radiation topology supplied by $\Lin_k$; in the massive case the scalar component exhibits an additional timelike/Dollard channel.

\begin{proposition}[Schwarzschild realization of the comparison estimates]\label{prop:schwarzschild-linear-package}
Fix $K\ge10$ and set $k:=K-4$.  On the Schwarzschild exterior in the zero electric sector, the comparison estimates used in Theorem~\ref{thm:main-schwarzschild-intro} hold in the following sense.
\begin{enumerate}
\item[(i)] For $m^2=0$, the scalar wave, the Lorenz-potential components, and the charge-free radiative Maxwell field satisfy the order-$K$ estimates $\Lin_K$: nondegenerate energy boundedness, redshift estimates at the horizon, trapping-degenerate Morawetz estimates, the far-field $r^p$ hierarchy, inhomogeneous source estimates in the paired source norms, radiation fields on $\mathcal I^\pm\cup\mathcal H^\pm$, and continuous Cauchy-to-radiation and radiation-to-Cauchy maps.
\item[(ii)] For $m^2>0$, the scalar comparison operator $\square_{g_M}-m^2$ satisfies the same forward estimates with the massive energy and, in the final-state topology, the additional timelike/Dollard channel at $i^\pm$.  The Maxwell part remains the charge-free radiative Maxwell system.  Together these estimates give $\Lin_K^{(m)}$ on Schwarzschild.
\item[(iii)] The source norms in (i)-(ii) are compatible with the Maxwell-Higgs nonlinearities: for Lorenz-gauge fields $U=(A,F,\phi)$,
\begin{equation}
 \|\mathcal N(U)\|_{\mathcal S^{(m)}_{C,K}}
 \le C\bigl(\|U\|_{\mathcal X^{(m)}_{C,K}}^2+
 \|U\|_{\mathcal X^{(m)}_{C,K}}^{2N_P+3}\bigr),
\end{equation}
and the corresponding Lipschitz bound has powers $1$ and $2N_P+2$.
\end{enumerate}
\end{proposition}

\begin{proof}
The forward estimates are the estimates proved in the Schwarzschild model sections.  The redshift identity is Section~\ref{sec:energy-estimates}, the Morawetz and compact-region estimates are Sections~\ref{sec:Morawetz} and~\ref{sec:decaynear}, and the far-field $r^p$ and pointwise consequences are Section~\ref{sec:decayfaraway}.  The charge-free Maxwell estimates are used only after the stationary Coulomb mode has been removed, so the radiative Maxwell norm contains no non-decaying stationary component.  The radiation-field construction is Theorems~\ref{thm:radiation-fields} and~\ref{thm:horizon-fields}.  The massless linear final-state maps are recorded in Theorem~\ref{thm:linear-scattering}; for $m^2>0$ the required linear final-state map is Proposition~\ref{prop:massive-schwarzschild-linear-final-state}; the nonlinear timelike state constructed later is Proposition~\ref{prop:timelike-scattering}, and its Dollard interpretation is Corollary~\ref{cor:dollard}.  These estimates are precisely the structural requirements in Definitions~\ref{def:lin-estimates} and~\ref{def:lin-estimates-massive}.

It remains to check that the nonlinear sources lie in the paired source spaces.  In Lorenz gauge the scalar source is
\begin{equation}
 2iA^\mu\nabla_\mu\phi+A^\mu A_\mu\phi+\mathcal R_P(\phi),
\end{equation}
and the Maxwell source is $J_\nu=2\operatorname{Im}(\overline\phi D_\nu\phi)$.  The Schwarzschild Sobolev algebra estimate, Lemma~\ref{lem:sobolev-algebra-schwarzschild}, puts all lower-order factors in $L^\infty$, while the top-order factor is measured in the source-weighted $L^2$ norm.  Lemma~\ref{lem:far-null-sobolev-moser} gives the same product estimate on the far null annuli, and Lemma~\ref{lem:potential-tame} gives
\begin{equation}
 \|\mathcal R_P(\phi)\|_{\mathcal S^{(m)}_{C,K}}
 \le C\bigl(\|\phi\|_{\mathcal X^{(m)}_{C,K}}^3+
 \|\phi\|_{\mathcal X^{(m)}_{C,K}}^{2N_P+3}\bigr).
\end{equation}
The current and gauge terms are quadratic or cubic.  Subtracting two such expansions gives the stated Lipschitz bound.  This proves the proposition.
\end{proof}

\begin{proposition}[Schwarzschild fixed-electric comparison package]\label{prop:schwarzschild-fixed-electric-package}
Fix $K\ge10$, set $k:=K-4$, and fix an electric Coulomb sector $Q_e\ne0$ on Schwarzschild.  Assume that $\CElec^{(m)}_K(M,0,Q_e)$ holds.  Then the direct sum of the charge-free radiative Maxwell estimates after subtraction of $F^{\mathrm C}_{Q_e}$ and the charged scalar estimates for $(D_{Q_e})^\mu D_{Q_e,\mu}-m^2$ gives a fixed-sector comparison theory with the following properties.
\begin{enumerate}
\item[(i)] If $V=(a,f,u)$ solves the inhomogeneous fixed-sector linear system on a slab $[t_0,T]$, then
\begin{equation}\label{eq:schw-fixed-electric-linear-bound}
 \|V\|_{\mathbb X^{(m)}_{C,K,Q_e}(t_0,T)}
 \le C_Q\left(\|V[t_0]\|_{\mathbb H_{C,K,Q_e}}+
 \|\mathfrak F\|_{\mathbb S^{(m)}_{C,K,Q_e}(t_0,T)}\right).
\end{equation}
Here the Maxwell component of $\mathbb X^{(m)}_{C,K,Q_e}$ is the charge-free radiative Schwarzschild Maxwell norm, the potential component is the componentwise scalar wave norm for the Lorenz representative $a$, and the scalar component is the $\CElec^{(m)}_K(M,0,Q_e)$ norm.
\item[(ii)] The same comparison theory has continuous Cauchy-to-asymptotic and asymptotic-to-Cauchy maps.  The scalar null-infinity trace is $U_{Q_e}^{-1}ru$, and the massive case includes the timelike/Dollard component contained in $\CElec^{(m)}_K(M,0,Q_e)$.
\item[(iii)] For the nonlinear fixed-sector sources in \eqref{eq:electric-renorm-maxwell-intro}-\eqref{eq:electric-renorm-scalar-intro},
\begin{equation}\label{eq:schw-fixed-electric-source-bound}
 \|\mathcal N_{Q_e}(V)\|_{\mathbb S^{(m)}_{C,K,Q_e}}
 \le C_Q\left(\|V\|_{\mathbb X^{(m)}_{C,K,Q_e}}^2+
 \|V\|_{\mathbb X^{(m)}_{C,K,Q_e}}^{2N_P+3}\right),
\end{equation}
and the corresponding Lipschitz estimate has powers $1$ and $2N_P+2$.
\end{enumerate}
The constants depend on $M,K$, finitely many Coulomb coefficient bounds, and the constants in the charged scalar comparison estimate.
\end{proposition}

\begin{proof}
The Maxwell part is the charge-free Schwarzschild Maxwell theory of Sections~\ref{sec:energy-estimates}-\ref{sec:scattering}, applied to the radiative field $f=F-F^{\mathrm C}_{Q_e}$.  Since $F^{\mathrm C}_{Q_e}$ is stationary and sourceless, subtracting it removes the electric charge and introduces no source term in the Maxwell equation for $f$.  The Lorenz-potential components satisfy the scalar wave estimates recorded in Proposition~\ref{prop:schwarzschild-linear-package}.  The scalar part is exactly the assumed estimate $\CElec^{(m)}_K(M,0,Q_e)$.  Taking the Hilbert direct sum of these three linear systems gives \eqref{eq:schw-fixed-electric-linear-bound} and the final-state maps in (ii).

For (iii), commute the current and scalar source with the Schwarzschild commutators.  Commutators with $D_{Q_e}$ produce only fixed coefficients involving $A^{\mathrm C}_{Q_e}$ and its derivatives; these obey the bounds in Definition~\ref{def:pure-electric-sector}.  The compact Sobolev algebra estimates and Lemma~\ref{lem:far-null-sobolev-moser} place all lower-order factors in $L^\infty$ and the highest-order factor in the weighted source norm.  The current and gauge-coupling terms are quadratic or cubic, while Lemma~\ref{lem:potential-tame} gives the cubic-and-higher bound for $\mathcal R_P$.  This proves \eqref{eq:schw-fixed-electric-source-bound}; subtracting two expansions gives the Lipschitz estimate.
\end{proof}

\begin{theorem}[Nonlinear wave operators and small-data asymptotic completeness]\label{thm:nonlinear-wave-operators}
Let $k\ge 6$ and assume that the scalar potential $P$ satisfies Assumption~\ref{asumsiP}.
Let $m^{2}\ge 0$ be as in Assumption~\ref{asumsiP} (equivalently, in \eqref{eq:potential-structure}).
There exists $\varepsilon_{\mathrm{sc}}>0$ (depending on $M,k$, on $N_{P}$, and on finitely many constants $C_{P,j}$ in \eqref{eq:Q-growth}) such that the following hold.
All implicit constants depend only on $M,k$ and on $P$ through these structural bounds.
\begin{enumerate}
\item[(i)] (\textbf{Forward wave operator.})
For every $\mathcal R_{+}^{\mathrm{lin}}\in \mathfrak R_{+}^{(m)}(\varepsilon_{\mathrm{sc}})$ there exists a unique global charge-free solution $(F,\phi)$ of the Maxwell-Higgs system admitting a Lorenz-gauge potential $A$ with $F=\dd A$ such that
\begin{equation}
\mathscr S_{+}^{\mathrm{full}}(F,\phi)=\mathcal R_{+}^{\mathrm{lin}}.
\end{equation}
In addition, if $(F^{\mathrm{lin}},\varphi^{\mathrm{lin}})$ is the unique linear comparison solution corresponding to $\mathcal R_{+}^{\mathrm{lin}}$, then $(F,\phi)$ scatters to $(F^{\mathrm{lin}},\varphi^{\mathrm{lin}})$ on $\partial_{+}^{(m)}\D$ in the sense of Theorem~\ref{thm:nonlinear-scattering}.

\item[(ii)] (\textbf{Backward wave operator.})
For every $\mathcal R_{-}^{\mathrm{lin}}\in \mathfrak R_{-}^{(m)}(\varepsilon_{\mathrm{sc}})$ there exists a unique global charge-free solution $(F,\phi)$ such that $\mathscr S_{-}^{\mathrm{full}}(F,\phi)=\mathcal R_{-}^{\mathrm{lin}}$ and which scatters to the corresponding linear comparison solution on $\partial_{-}^{(m)}\D$.

\item[(iii)] (\textbf{Homeomorphism property.})
The maps $\mathscr S_{\pm}^{\mathrm{full}}$ restrict to homeomorphisms between a neighbourhood of $0$ in the nonlinear Cauchy data space and the neighbourhoods $\mathfrak R_{\pm}^{(m)}(\varepsilon_{\mathrm{sc}})$.
Their inverses
\begin{equation}
\mathscr W_{\pm}^{\mathrm{full}}:=(\mathscr S_{\pm}^{\mathrm{full}})^{-1}
\end{equation}
are the (nonlinear) wave operators.
\end{enumerate}
\end{theorem}

\begin{proof}
We use the Schwarzschild comparison package of Proposition~\ref{prop:schwarzschild-linear-package}.  The Cauchy part follows first.  For a Lorenz-gauge solution on a finite slab, the linear estimates give
\begin{equation}
 \|U\|_{\mathcal X_{C,k}^{(m)}(t_0,T)}
 \le C\bigl(\|U[t_0]\|_{\mathcal H^k}+\|\mathcal N(U)\|_{\mathcal S_{C,k}^{(m)}(t_0,T)}\bigr),
\end{equation}
while Proposition~\ref{prop:schwarzschild-linear-package}(iii) gives
\begin{equation}
 \|\mathcal N(U)\|_{\mathcal S_{C,k}^{(m)}}
 \le C\bigl(\|U\|_{\mathcal X_{C,k}^{(m)}}^2+
 \|U\|_{\mathcal X_{C,k}^{(m)}}^{2N_P+3}\bigr).
\end{equation}
The continuity argument in the proof of Theorem~\ref{thm:method}, now with the Schwarzschild norms, closes the small-data bound.  Local well-posedness and continuation are supplied by Proposition~\ref{prop:local-wp-schwarzschild} and Lemma~\ref{lem:continuation-schwarzschild}.  Thus one obtains the global small solution and the uniform energy, redshift, Morawetz, and $r^p$ estimates.  The radiation fields on $\mathcal I^+$ and $\mathcal H^+$ are then Theorems~\ref{thm:radiation-fields} and~\ref{thm:horizon-fields}; the past fields follow by time reversal.

For $m^2=0$, the final-state map is the characteristic linear inverse contained in $\Lin_k$, recorded concretely in Theorem~\ref{thm:linear-scattering}.  The closed zero-sector contraction Proposition~\ref{prop:closed-sector-contraction} constructs a nonlinear correction with zero prescribed asymptotic data.  This gives the forward and backward wave operators, and uniqueness of the Cauchy problem shows that they are inverse to the Cauchy-to-radiation maps on sufficiently small neighborhoods.

For $m^2>0$, the characteristic boundary alone is not complete.  The linear final-state map is Proposition~\ref{prop:massive-schwarzschild-linear-final-state}; the nonlinear scalar comparison state at $i^+$ is constructed in Proposition~\ref{prop:timelike-scattering}, and Corollary~\ref{cor:dollard} identifies it with the classical Dollard-modified datum.  The massive nonlinear final-state problem is solved in Subsection~\ref{subsec:wave-operators}.  The underlying closed estimate is Proposition~\ref{prop:schwarzschild-closed-finalstate-contraction}: the fixed-point map obeys a size bound with powers $2$ and $2N_P+3$ and a Lipschitz bound with powers $1$ and $2N_P+2$.  Thus it is a contraction in the small extended radiation neighborhood.  The backward wave operator is identical after time reversal.

The homeomorphism property follows from the same Lipschitz estimates applied to differences of solutions and of fixed points.  The quotient statement is Lemma~\ref{lem:residual-gauge-descent}.  This proves (i)-(iii).
\end{proof}

\subsubsection*{Radiation fields and nonlinear scattering.}
The scattering theory is summarized by Theorem~\ref{thm:nonlinear-wave-operators} and is proved in Section~\ref{sec:scattering}.
We first construct radiation fields on $\mathcal I^{+}$ and $\mathcal H^{+}$ (Theorems~\ref{thm:radiation-fields} and~\ref{thm:horizon-fields}), we identify the corresponding linear scattering theory for the linearized comparison system (Theorem~\ref{thm:linear-scattering}), and we then solve the nonlinear final-state problem to show that every small nonlinear solution scatters to a unique linear solution (Theorem~\ref{thm:nonlinear-scattering}).
When $m^{2}>0$, the scalar scattering description includes a timelike channel and a Dollard-type phase correction; see Corollary~\ref{cor:dollard} and Section~\ref{sec:scattering}.

\subsection{Consequences for the Schwarzschild Cauchy problem}\label{subsec:schwarzschild-cauchy-consequences}
As part of the proof of Theorem~\ref{thm:nonlinear-wave-operators} we establish robust boundedness and decay estimates for the Cauchy evolution from small admissible Lorenz-gauge data.
For the reader primarily interested in the forward Cauchy problem and quantitative decay, we state the following consequences.

\subsubsection*{Schwarzschild double-null notation.}
We now specialize to the Schwarzschild model $a=0$ and use the Schwarzschild time coordinate $t$ and Regge-Wheeler tortoise coordinate $r^\ast$.
We introduce Eddington-Finkelstein null coordinates
\begin{equation}
v=t+r^\ast,\qquad w=t-r^\ast,
\end{equation}
and we set $\vplus:=\max\{1,v\}$ and $\wplus:=\max\{1,|w|\}$.
Our estimates split naturally into a far region $r\ge R$ (for $R$ sufficiently large) and the region $2M<r\le 3M$ capturing trapping and near-horizon effects.

\subsubsection*{Schwarzschild energies.}
Let $\Sigma_t=\{t=\mathrm{const}\}$ with future unit normal $n_{\Sigma_t}$, and let $T_{\mu\nu}[F,\phi]$ denote the Maxwell-Higgs energy-momentum tensor associated to \eqref{lagrangian}.
For a vector field $X$, define the energy flux
\begin{equation}
E^{X}[F,\phi](t):=\int_{\Sigma_t} T_{\mu\nu}[F,\phi]\,X^\nu\,n_{\Sigma_t}^{\mu}\,\dd\mu_{\Sigma_t}.
\end{equation}
Let $\Omega_1,\Omega_2,\Omega_3$ be the rotation fields on $\mathbb{S}^2$ and set
$\mathcal{Z}:=\{\partial_t,\Omega_1,\Omega_2,\Omega_3\}$.
For a multi-index $\alpha$ we write $\mathcal{Z}^\alpha$ for a product of these vector fields.
For an integer $k\ge 6$, we define the higher-order energy
\begin{equation}\label{def:Ek}
\mathcal{E}_k(t)
:=
\sum_{|\alpha|\le k}
E^{\hat t}\!\big[\mathcal{L}_{\mathcal{Z}^{\alpha}}F,\;\mathcal{Z}^{\alpha}\phi\big](t),
\end{equation}
where $\hat t$ denotes the unit timelike vector field in the Schwarzschild orthonormal frame (see Section~\ref{subsec:notation}).

\begin{corollary}[Small-data global existence and energy bounds]\label{cor:small-data}
Assume that the scalar potential $P$ satisfies Assumption~\ref{asumsiP}.
Fix an integer $k\ge 6$.
There exists $\varepsilon_0>0$ such that if the initial data on $\Sigma_{t_0}$ are admissible in Lorenz gauge (Section~\ref{subsec:initial-data}), are uncharged in the sense of Definition~\ref{def:charges}, and satisfy
\begin{equation}\label{eq:smallness-energy}
\mathcal{E}_k(t_0)\le \varepsilon_0,
\end{equation}
then the Maxwell-Higgs system admits a unique global smooth solution $(F,\phi)\in C^\infty(\D)$.
In addition, the commuted energies remain uniformly bounded:
\begin{equation}
\sup_{t\ge t_0}\mathcal{E}_k(t)\lesssim \mathcal{E}_k(t_0),
\end{equation}
and the solution satisfies the integrated local energy decay and pointwise decay estimates stated in Corollaries~\ref{cor:far-decay} and~\ref{cor:near-horizon}.
\end{corollary}
\begin{proof}
Local well-posedness and constraint propagation are proved in Appendix~\ref{app:global-existence-schwarzschild}.  The redshift, Morawetz, and commuted energy estimates of Section~\ref{sec:energy-estimates} close the bootstrap in Proposition~\ref{prop:bootstrap-closure}.  The continuation criterion from Lemma~\ref{lem:continuation-schwarzschild} then gives global existence, and Proposition~\ref{prop:bootstrap-closure} gives the displayed uniform bound.  The pointwise conclusions are exactly Corollaries~\ref{cor:far-decay} and~\ref{cor:near-horizon}.
\end{proof}

\begin{lemma}[Lorenz-potential spacetime control on Schwarzschild]\label{lem:schwarzschild-lorenz-potential-control}
Let \(k\ge6\), and let \((A,F,\phi)\) be the Lorenz-gauge zero-sector Schwarzschild solution supplied by Corollary~\ref{cor:small-data}, with the representative chosen in the quotient norm \eqref{def:Ek}.  Then, after possibly decreasing the smallness threshold in Corollary~\ref{cor:small-data},
\begin{equation}\label{eq:lorenz-potential-L2Linf}
\sum_{|I|+j\le k-3}\int_{t_0}^{\infty}
\|\mathcal L_\Omega^I T^j A(t)\|_{L^\infty(\Sigma_t)}^2\,\dd t
\le C\mathcal E_k(t_0).
\end{equation}
The same homogeneous estimate holds for a charge-free free Maxwell field after choosing the Lorenz potential in the same residual-gauge normalization.
\end{lemma}

\begin{proof}
In Lorenz gauge on Schwarzschild, each component of the potential in a fixed stationary frame solves
\begin{equation}\label{eq:schw-lorenz-wave-components}
 \square_g A_\nu=J_\nu[\phi;A],
 \qquad
 J_\nu[\phi;A]=2\operatorname{Im}(\overline\phi D_\nu\phi).
\end{equation}
The residual quotient in \eqref{def:Ek} removes homogeneous pure-gauge solutions with the same curvature and fixes a representative whose initial wave energy is controlled by \(\mathcal E_k(t_0)\).  Applying the scalar wave redshift, Morawetz, and \(r^p\) estimates componentwise to \eqref{eq:schw-lorenz-wave-components}, after commuting with \(T\) and the angular momenta up to order \(k-3\), gives
\begin{equation}
 \|A\|_{L^2_tL^\infty_x,k-3}^2
 \le C\mathcal E_k(t_0)+C\|J[\phi;A]\|_{\mathcal S_{k-2}}^2.
\end{equation}
Here the Sobolev step uses the compact Schwarzschild Sobolev inequality and the far-field \(r^p\) hierarchy.  Since \(|J[\phi;A]|\le C|\phi|\,|D\phi|\), Lemma~\ref{lem:nonlinear-banach-estimates} and the small-data bound of Corollary~\ref{cor:small-data} give
\begin{equation}
 \|J[\phi;A]\|_{\mathcal S_{k-2}}
 \le C\|(A,F,\phi)\|_{\mathbb X^{(m)}_{k,\mathrm{Sch}}}^{2}
 \le C\mathcal E_k(t_0).
\end{equation}
For \(\mathcal E_k(t_0)\) small this is absorbed into the displayed estimate, giving \eqref{eq:lorenz-potential-L2Linf}.  If \(J=0\), the same argument gives the homogeneous free Maxwell-potential estimate.
\end{proof}

\begin{corollary}[Pointwise decay in the far region]\label{cor:far-decay}
Assume that $P$ satisfies Assumption~\ref{asumsiP}.
Let $(F,\phi)$ be a global uncharged solution arising from data on $\Sigma_{t_0}$ with $\mathcal{E}_k(t_0)\le \varepsilon_0$ for some $k\ge 6$.
Then there exist $R\gg 1$ and a constant $C=C(M,k,P)$ such that for all $r\ge R$,
\begin{eqnarray}
|F_{\hat\mu\hat\nu}(w,v,\omega)|
&\le \frac{C\,\mathcal{E}_k(t_0)^{1/2}}{1+\vplus},
\label{farFv}
\\
|F_{\hat\mu\hat\nu}(w,v,\omega)|
&\le \frac{C\,\mathcal{E}_k(t_0)^{1/2}}{1+|w|},
\label{decayF2}
\\
|\phi(w,v,\omega)|
&\le \frac{C\,\mathcal{E}_k(t_0)^{1/2}}{1+\vplus},
\label{farphi}
\\
|D\phi(w,v,\omega)|
&\le \frac{C\,\mathcal{E}_k(t_0)^{1/2}}{1+\vplus}.
\label{farDphi}
\end{eqnarray}
\end{corollary}
\begin{proof}
The Maxwell estimates \eqref{farFv}-\eqref{decayF2} are obtained in Subsection~\ref{subsec:far-decay-F} from the spherical Sobolev estimate Lemma~\ref{lem:sobolev-F-sphere}, the far-sphere $L^2$ bound Lemma~\ref{lem:far-sphere-L2-F}, and the time-reversal argument for the incoming component.  The scalar estimate \eqref{farphi} follows in Subsection~\ref{subsec:far-decay-phi} from Lemmas~\ref{lem:sobolev-phi-sphere} and~\ref{lem:far-sphere-L2-phi}.  The derivative estimate \eqref{farDphi} is proved in Subsection~\ref{subsec:far-decay-Dphi}.  Substituting the bootstrap closure of Proposition~\ref{prop:bootstrap-closure} replaces the bootstrap norm by the initial energy.
\end{proof}

\begin{corollary}[Pointwise decay near trapping and the horizon]\label{cor:near-horizon}
Let $(F,\phi)$ be a global solution arising from sufficiently small uncharged data satisfying \eqref{eq:smallness-energy} with $k\ge 6$.
Set $\vplus:=\max\{1,v\}$ and $\mathcal{E}:=\mathcal{E}_k(t_0)$.
Then in the region $2M<r\le 3M$ the following bounds hold:
\begin{eqnarray}
|F_{\hat\mu\hat\nu}|
&\lesssim
\left(\frac{\wplus}{\vplus}\right)\mathcal{E}^{1/2},
\label{Fnear}
\\
|\phi|
&\lesssim
\left(1+\left(\frac{\wplus}{\vplus}\right)^2\right)^{1/2}\mathcal{E}^{1/2},
\label{nearphi}
\\
|D\phi|
&\lesssim
E_4^{1/2},
\label{Dephi}
\end{eqnarray}
where
\begin{equation}
E_4 \equiv
\begin{cases}
\left(\dfrac{\wplus}{\vplus}\right)^{2}\displaystyle\sum_{n=1}^{N}\mathcal{E}^{\,2n-1}
+\displaystyle\sum_{n=2}^{N}\mathcal{E}^{\,2n-2}
+\displaystyle\sum_{n=3}^{N}\mathcal{E}^{\,2n-3}
+\mathcal{E},
& \text{if $P$ is of the form \eqref{V4}},\\[0.8em]
\left(\dfrac{\wplus}{\vplus}\right)^{2}\mathcal{E}
+\mathcal{E},
& \text{if $P$ is \eqref{sine} or \eqref{toda}}.
\end{cases}
\end{equation}
Along outgoing null hypersurfaces of fixed $w$, these estimates yield decay as $v\to+\infty$.
\end{corollary}
\begin{proof}
The Maxwell components $F_{\hat v\hat w}$ and $F_{\hat e_1\hat e_2}$ are estimated in Subsection~\ref{subsec:decay-Fvw}, the outgoing angular components in Subsection~\ref{subsec:decay-FvA}, and the incoming angular components with the redshift weight in Subsection~\ref{subsec:decay-FwA}.  The scalar estimate is proved in Subsection~\ref{subsec:decay-phi-near}, and the covariant derivative estimate is proved in Subsection~\ref{subsec:decay-Dphi-near}.  Each step uses the unit-slab Sobolev estimate Lemma~\ref{lem:sobolev-vslab}, the redshift flux decay Proposition~\ref{prep2}, and the small-data bootstrap closure.  Combining those five component estimates gives exactly \eqref{Fnear}-\eqref{Dephi}.
\end{proof}

\begin{remark}\label{rem:E4-simplify}
Since $\mathcal{E}_k(t_0)\le\varepsilon_0\ll 1$, the quantity $E_4$ in Corollary~\ref{cor:near-horizon} satisfies the structural bound
\begin{equation}
E_4\ \lesssim\ \Bigl(1+\bigl(\tfrac{\wplus}{\vplus}\bigr)^2\Bigr)\,\mathcal{E}_k(t_0),
\end{equation}
and therefore $|D\phi|\lesssim \bigl(1+\tfrac{\wplus}{\vplus}\bigr)\mathcal{E}_k(t_0)^{1/2}$ in $2M<r\le 3M$.
We keep the more detailed expression for $E_4$ only to record the dependence on the potential when tracking nonlinear terms in the bootstrap argument.
\end{remark}

\begin{corollary}[General small data and Coulomb asymptotics]\label{cor:Coulomb}
Let $(F,\phi)$ be a global solution arising from sufficiently small admissible data in Lorenz gauge, without assuming the uncharged condition of Definition~\ref{def:charges}.
Let $Q_e$ denote the (conserved) asymptotic electric charge determined by the initial data, and let $F_{Q}$ be the corresponding stationary Coulomb $2$-form on Schwarzschild.
Then the radiative remainder
\begin{equation}
\widetilde F := F - F^{\mathrm C}_{Q_e}
\end{equation}
is uncharged and satisfies the same pointwise decay estimates as in Corollaries~\ref{cor:far-decay} and~\ref{cor:near-horizon} (with $F$ replaced by $\widetilde F$).
In this case, $F$ converges pointwise to $F^{\mathrm C}_{Q_e}$ as $v\to+\infty$ along outgoing null hypersurfaces.
\end{corollary}

\begin{proof}
The Coulomb field $F^{\mathrm C}_{Q_e}$ is a stationary sourceless Maxwell solution on Schwarzschild with electric charge $Q_e$.
Since the Maxwell equation is linear in $F$, the remainder $\widetilde F=F-F^{\mathrm C}_{Q_e}$ satisfies the same Maxwell equation with source $J(\phi)$ and has vanishing asymptotic electric charge.
Applying Corollaries~\ref{cor:far-decay} and~\ref{cor:near-horizon} to $(\widetilde F,\phi)$ yields the stated decay for $\widetilde F$.
\end{proof}

\begin{remark}[Charged Kerr data and the modified asymptotic channel]\label{rem:intro-charged-kerr}
Corollary~\ref{cor:Coulomb} concerns Coulomb subtraction for Maxwell fields on Schwarzschild.  On slowly rotating Kerr the nonzero-electric conclusion is not obtained from this corollary alone; it is the fixed-sector theorem based on the Coulomb-covariant scalar condition \(\CElec^{(m)}_K(M,a,Q_e)\).  In the massless small-electric range that condition is supplied by Theorem~\ref{thm:small-coulomb-massless-kerr-condition}; outside that range it remains an explicitly stated charged scalar condition.
\end{remark}

\section{Gauge fixing and initial data on Kerr}\label{sec:kerr-gauge-data}

In this section we describe the gauge fixing and the admissible Cauchy data on the slowly rotating Kerr exteriors used in the paper.  Theorem~\ref{thm:main-slow-kerr-intro} is formulated in the zero electric sector, while Theorem~\ref{thm:main-slow-kerr-massive-intro} uses the same Lorenz method after subtracting the stationary electric Coulomb background.
Throughout $(\mathcal D_{M,a},g_{M,a})$ denotes the Kerr domain of outer communications introduced in Section~\ref{sec:intro}, equipped with the Kerr-star foliation
$\Sigma_{\tau}=\{t^{\star}=\tau\}$ from \eqref{eq:kerr-tstar}.
We write $n=n_{\Sigma_{\tau}}$ for the future unit normal to $\Sigma_{\tau}$ and $\gamma$ for the induced Riemannian metric.

\subsection{Gauge symmetry and Lorenz gauge}\label{subsec:kerr-lorenz}

The Maxwell-Higgs system \eqref{eom1}-\eqref{eom2} is invariant under the $U(1)$ gauge symmetry
\begin{equation}\label{eq:gauge-transform}
(A,\phi)\longmapsto (A+\dd\chi,\;e^{i\chi}\phi),
\end{equation}
for any smooth real-valued function $\chi$ on $\mathcal D_{M,a}$.
The curvature $2$-form $F=\dd A$ is gauge invariant and the covariant derivative transforms as
$D_{\mu}(e^{i\chi}\phi)=e^{i\chi}(D_{\mu}\phi)$.

For the nonlinear analysis we work in the Lorenz gauge
\begin{equation}\label{eq:lorenz-kerr}
\nabla^{\mu}A_{\mu}=0.
\end{equation}
Since Kerr is Ricci-flat, \eqref{eq:lorenz-kerr} reduces the Maxwell equation to a covariant wave equation for the potential:
if $F=\dd A$ and \eqref{eq:lorenz-kerr} holds, then
\begin{equation}\label{eq:lorenz-reduced-kerr}
\square_{g_{M,a}}A_{\nu}=J_{\nu},
\qquad
D^{\mu}D_{\mu}\phi=\partial_{\bar\phi}P(\phi,\bar\phi).
\end{equation}
Conversely, if $(A,\phi)$ solves \eqref{eq:lorenz-reduced-kerr} and the Lorenz constraints in
\eqref{eq:lorenz-constraints-kerr} below hold on one Cauchy slice, then \eqref{eq:lorenz-kerr} propagates and $(F,\phi)$ with $F=\dd A$ solves \eqref{eom1}-\eqref{eom2}.
This follows from current conservation: taking the divergence of \eqref{eq:lorenz-reduced-kerr} yields a homogeneous wave equation for $\nabla^{\mu}A_{\mu}$.

\subsection{Cauchy data and constraint equations}\label{subsec:kerr-initial-data}

Fix $\tau_{0}\in\mathbb R$ and consider the initial hypersurface $\Sigma_{\tau_{0}}$.
We use the global-potential formulation throughout the rotating Kerr theorem; all rotating data are represented by a global Lorenz potential after the electric Coulomb subtraction.
In that case, a smooth Cauchy data set for \eqref{eq:lorenz-reduced-kerr} consists of
\begin{equation}\label{eq:kerr-data}
\bigl(A|_{\Sigma_{\tau_{0}}},\;\nabla_{n}A|_{\Sigma_{\tau_{0}}},\;
\phi|_{\Sigma_{\tau_{0}}},\;D_{n}\phi|_{\Sigma_{\tau_{0}}}\bigr),
\end{equation}
where $n$ is the future unit normal to $\Sigma_{\tau_{0}}$.
Let $e_{1},e_{2},e_{3}$ be a local orthonormal frame tangent to $\Sigma_{\tau_{0}}$.
We define the electric field on $\Sigma_{\tau_{0}}$ by $E_{i}:=F(n,e_{i})$.
The Maxwell equation implies the Gauss constraint
\begin{equation}\label{eq:gauss-kerr}
\nabla^{i}E_{i}=-\,J_{\mu}n^{\mu}\qquad\text{on }\Sigma_{\tau_{0}},
\end{equation}
where $\nabla^{i}$ denotes the Levi-Civita connection of $(\Sigma_{\tau_{0}},\gamma)$.
In Lorenz gauge we further impose the compatibility conditions
\begin{equation}\label{eq:lorenz-constraints-kerr}
\nabla^{\mu}A_{\mu}\big|_{\Sigma_{\tau_{0}}}=0,
\qquad
\nabla_{n}\bigl(\nabla^{\mu}A_{\mu}\bigr)\big|_{\Sigma_{\tau_{0}}}=0,
\qquad
\nabla^{i}E_{i}=-\,J_{\mu}n^{\mu}.
\end{equation}
On a Ricci-flat background, these conditions propagate by the hyperbolic constraint argument below.

\begin{lemma}[Lorenz and Gauss constraint propagation on Kerr]\label{lem:kerr-constraint-propagation}
Let $(A,\phi)$ be a smooth solution of the Lorenz-reduced system \eqref{eq:lorenz-reduced-kerr} on a Kerr slab and define
\begin{equation}
\Gamma:=\nabla^{\mu}A_{\mu}.
\end{equation}
Assume that the scalar equation holds and that the Lorenz and Gauss constraints \eqref{eq:lorenz-constraints-kerr} hold on $\Sigma_{\tau_0}$. Then $\Gamma\equiv0$ on the domain of dependence of $\Sigma_{\tau_0}$, the current $J_{\nu}=-i(D_{\nu}\phi\,\overline\phi-\phi\,\overline{D_{\nu}\phi})$ is conserved, and the curvature $F=\dd A$ solves the geometric Maxwell equation \eqref{eom1}. The present paper uses this global-potential propagation statement in the zero sector.
\end{lemma}

\begin{proof}
The scalar equation and gauge invariance of the matter Lagrangian imply the Noether identity $\nabla^{\nu}J_{\nu}=0$. Taking the divergence of the Lorenz-reduced Maxwell equation gives
\begin{equation}
\square_{g_{M,a}}\Gamma=\nabla^{\nu}J_{\nu}=0,
\end{equation}
because Kerr is Ricci-flat and covariant derivatives commute on one-forms in the wave equation for $A$. The two Lorenz constraints in \eqref{eq:lorenz-constraints-kerr} give vanishing Cauchy data for $\Gamma$, therefore uniqueness for the scalar wave equation gives $\Gamma\equiv0$. The reduced Maxwell equation then becomes $\nabla^{\mu}(\dd A)_{\mu\nu}=J_{\nu}$, while $\dd F=0$ is automatic from $F=\dd A$ in the global-potential formulation used here. The Gauss constraint is the normal component of this Maxwell equation on the initial slice, so it propagates with the system.
\end{proof}

\subsection{Admissible Kerr data spaces}\label{subsec:kerr-data-space}

The scattering maps in Theorem~\ref{thm:main-slow-kerr-intro} are defined on a small neighbourhood of the origin in an order-$K$ energy completion of Lorenz-compatible data.
The nonlinear stress-energy tensor is useful for coercivity and continuation, but the actual Cauchy topology below is the quadratic Sobolev topology of the renormalized connection, curvature, and scalar variables; this avoids applying the nonlinear potential energy to commuted fields.

The stress-energy tensor associated to the Lagrangian \eqref{lagrangian} is
\begin{eqnarray}
T_{\mu\nu}[F,\phi]
=
F_{\mu\alpha}F_{\nu}{}^{\alpha}
-\frac14 g_{\mu\nu}F_{\alpha\beta}F^{\alpha\beta} 
+ 2\Re\!\left(D_\mu\phi\,\overline{D_\nu\phi}\right)
- g_{\mu\nu}\Bigl(|D\phi|^2+P(\phi,\bar\phi)\Bigr).
\label{eq:stress-energy-kerr}
\end{eqnarray}
For solutions of \eqref{eom1}-\eqref{eom2} one has $\nabla^\mu T_{\mu\nu}[F,\phi]=0$.
For a vector field $X$ we define the corresponding energy current and flux by
\begin{equation}
J^{X}_{\mu}[F,\phi]:=T_{\mu\nu}[F,\phi]X^\nu,
\qquad
E^{X}[F,\phi](\tau_{0}):=\int_{\Sigma_{\tau_{0}}} J^{X}[F,\phi]\cdot n\,\dd\mu_{\Sigma_{\tau_{0}}}.
\end{equation}
We use $X=N$ from Definition~\ref{def:admissible-exterior} and commute with the admissible family $\mathcal Z$.

\begin{definition}[Admissible Kerr Cauchy data in the zero sector]\label{def:kerr-data}
Fix $K\in\mathbb N$ and Kerr parameters $(M,a)$ with $|a|\le a_{\mathrm{slow}}(M,K)$.  An admissible Kerr datum on $\Sigma_{\tau_0}$ is a zero-sector Lorenz-gauge representative
\begin{equation}
(a_0,a_1,\phi_0,\phi_1)
=\bigl(a|_{\Sigma_{\tau_0}},\nabla_n a|_{\Sigma_{\tau_0}},\phi|_{\Sigma_{\tau_0}},D^{\mathrm{ren}}_{n}\phi|_{\Sigma_{\tau_0}}\bigr)
\end{equation}
relative to the vacuum Maxwell background, together with the induced radiative curvature $\widetilde F=\dd a=F$. The tuple satisfies the Lorenz constraint, the Gauss constraint, and the Maxwell constraints. The radiative remainder has zero electric charge:
\begin{equation}
\frac1{4\pi}\lim_{r\to\infty}\int_{S_{\tau_0,r}}{}^\star\widetilde F=0.
\end{equation}
The order-$K$ quadratic Cauchy norm of a Lorenz representative is
\begin{eqnarray}
\mathcal E^N_{K,M,a}(\tau_0;0)
:=\sum_{|\alpha|\le K}\int_{\Sigma_{\tau_0}}
\Bigl(&|\nabla_{\mathcal Z}^{\alpha}a_0|_{N}^{2}+|\nabla_{\mathcal Z}^{\alpha}a_1|_{N}^{2}
+|\mathcal L_{\mathcal Z^\alpha}\widetilde F|_N^2 \nonumber\\
&+|\nabla\mathcal Z^\alpha\phi_0|_{N}^{2}
+|\mathcal Z^\alpha\phi_1|^{2}
+m^2|\mathcal Z^\alpha\phi_0|^{2}
\Bigr)\,\dd\mu_{\Sigma_{\tau_0}},
\end{eqnarray}
with the Hardy-controlled lower-order term added in the massless case when needed. The quotient norm of a gauge class is the infimum of the square root of this expression over all Lorenz representatives in the residual orbit,
\begin{equation}
\|[(a_0,a_1,\phi_0,\phi_1)]\|_{\mathcal H^K_{\mathrm{ren},M,a}(Q_e)}
:=
\inf_{\chi\in\mathcal G^{K+1}_{\mathrm{res}}}
\mathcal E^N_{K,M,a}\bigl((a_0,a_1,\phi_0,\phi_1)^{\chi};0\bigr)^{1/2},
\end{equation}
where $(\cdot)^{\chi}$ denotes the transformed Cauchy tuple obtained from $a\mapsto a+\dd\chi$ and $\phi\mapsto e^{i\chi}\phi$ in a unitary trivialization. The phase space is the completion of smooth zero-sector Lorenz-compatible representatives in this quotient norm.  It is denoted \(\mathcal H^K_{M,a}(\Sigma_{\tau_0})\). The displayed norm is quadratic in the Cauchy variables and their curvature, not the nonlinear stress tensor applied to commuted fields.
\end{definition}

\paragraph{Consistency of the Kerr data space.}
The Lorenz constraints \eqref{eq:lorenz-constraints-kerr} propagate because, taking the divergence of the Lorenz-reduced Maxwell equation on Ricci-flat Kerr, one obtains a homogeneous wave equation for $\nabla^{\mu}A_{\mu}$ with vanishing Cauchy data. In the zero sector $\widetilde F$ is exact after Coulomb subtraction, and the norm in Definition~\ref{def:kerr-data} is invariant under residual Lorenz-gauge transforms $\chi$ with $\square_{g_{M,a}}\chi=0$, because the quotient norm is taken over Lorenz representatives and $\dd\chi$ contributes only along the residual-gauge orbit. Consequently, the completion \(\mathcal H^{K}_{M,a}(\Sigma_{\tau_{0}})\) is well defined in the zero sector used in Theorems~\ref{thm:charged-sector-transfer}-\ref{thm:main-slow-kerr-intro}.

\begin{remark}[Electric Coulomb subtraction]\label{rem:kerr-data-charges}
The zero-sector rotating theorem uses $Q_e=0$.  The small-electric rotating theorem uses $Q_e$ after the Coulomb splitting of Definition~\ref{def:pure-electric-sector}.  No additional sector is part of the rotating conclusions.
\end{remark}

\subsection{Gauge fixing}\label{subsec:kerr-gauge-fixing}

For the purposes of this paper we restrict attention to initial data which already satisfy the Lorenz constraints \eqref{eq:lorenz-constraints-kerr}.
Such a restriction is natural in the analysis of gauge-covariant hyperbolic systems.
One may gauge transform general smooth finite-energy data satisfying the Gauss constraint \eqref{eq:gauss-kerr} into Lorenz gauge by solving for a gauge function $\chi$ so that \eqref{eq:lorenz-kerr} holds (for instance by imposing \eqref{eq:lorenz-constraints-kerr} on $\Sigma_{\tau_{0}}$ and using the propagation of $\nabla^{\mu}A_{\mu}$).
The residual gauge freedom inside Lorenz gauge consists of transformations \eqref{eq:gauge-transform} with $\square_{g_{M,a}}\chi=0$.

\section{Energy identities and integrated decay estimates on Kerr}\label{sec:kerr-energy}

In this section we state the vector-field energy identities and the integrated decay estimates used in the slowly rotating Kerr part of the argument.  The estimates are used in the zero electric sector.  The identities are consequences of the divergence-free property of the Maxwell-Higgs stress-energy tensor \eqref{eq:stress-energy-kerr}, while the integrated decay estimates are used through the abstract linear estimates collected in $\Lin_{k}$ of Definition~\ref{def:lin-estimates}.
Throughout we work on a fixed slowly rotating Kerr exterior $(\mathcal D_{M,a},g_{M,a})$ with $|a|\le a_{\mathrm{slow}}(M,K)$ and the Kerr-star foliation $\Sigma_{\tau}=\{t^{\star}=\tau\}$.

\subsection{Energy currents and divergence identities}\label{subsec:kerr-energy-identities}

Let $T_{\mu\nu}[F,\phi]$ be the stress-energy tensor \eqref{eq:stress-energy-kerr}.
For any smooth vector field $X$ on $\mathcal D_{M,a}$ we define the associated current and bulk term by
\begin{equation}\label{eq:kerr-energy-current}
J^{X}_{\mu}[F,\phi]:=T_{\mu\nu}[F,\phi]\,X^{\nu},
\qquad
K^{X}[F,\phi]:=\nabla^{\mu}J^{X}_{\mu}[F,\phi].
\end{equation}
Writing ${}^{(X)}\pi_{\mu\nu}:=\nabla_{\mu}X_{\nu}+\nabla_{\nu}X_{\mu}$ for the deformation tensor of $X$, the conservation law $\nabla^{\mu}T_{\mu\nu}[F,\phi]=0$ yields the identity
\begin{equation}\label{eq:kerr-div-identity}
K^{X}[F,\phi]
=
\frac12\,T^{\mu\nu}[F,\phi]\;{}^{(X)}\pi_{\mu\nu}.
\end{equation}
If $X$ is Killing then ${}^{(X)}\pi\equiv0$ and the corresponding energy flux is conserved.

Let $\mathcal R(\tau_{0},\tau_{1})$ denote the slab region bounded by $\Sigma_{\tau_{0}}$, $\Sigma_{\tau_{1}}$, the future event horizon segment $\mathcal H^{+}(\tau_{0},\tau_{1})$, and future null infinity segment $\mathcal I^{+}(\tau_{0},\tau_{1})$.
Integrating \eqref{eq:kerr-div-identity} and applying Stokes' theorem gives the basic energy identity
\begin{equation}\label{eq:kerr-energy-identity}
E^{X}[F,\phi](\tau_{1})
+\mathcal F^{X}_{\mathcal H^{+}}[F,\phi](\tau_{0},\tau_{1})
+\mathcal F^{X}_{\mathcal I^{+}}[F,\phi](\tau_{0},\tau_{1})
=
E^{X}[F,\phi](\tau_{0})
+\int_{\mathcal R(\tau_{0},\tau_{1})} K^{X}[F,\phi]\;\dd V_{g_{M,a}},
\end{equation}
where $E^{X}$ is the flux through $\Sigma_{\tau}$ defined in Section~\ref{subsec:kerr-data-space} and $\mathcal F^{X}_{\mathcal H^{+}}$, $\mathcal F^{X}_{\mathcal I^{+}}$ denote the corresponding fluxes through $\mathcal H^{+}$ and $\mathcal I^{+}$.
When $X$ is future causal and tangent to $\mathcal H^{+}$ and $\mathcal I^{+}$, these boundary fluxes are nonnegative.

\subsection{Redshift energy and superradiance}\label{subsec:kerr-redshift}

On Kerr the stationary Killing field $T=\partial_{t}$ (equivalently, $\partial_{t_{\mathrm{BL}}}$) fails to be everywhere timelike: $T$ becomes spacelike in the ergoregion, which is the geometric origin of superradiance.
We obtain a globally coercive energy is to employ a \emph{redshift multiplier} $N$ as in Definition~\ref{def:admissible-exterior}: one chooses $N$ to be future timelike everywhere, to agree with $T$ for large $r$, and to coincide near the horizon with the future horizon generator
\begin{equation}\label{eq:kerr-horizon-generator}
K:=T+\Omega_{\mathcal H}\,\Phi,
\qquad
\Omega_{\mathcal H}:=\frac{a}{r_{+}^{2}+a^{2}},
\end{equation}
where $\Phi=\partial_{\varphi}$ (equivalently, $\partial_{\varphi_{\mathrm{BL}}}$) is the axial Killing field and $r_{+}=M+\sqrt{M^{2}-a^{2}}$.
With this choice, the flux of $J^{N}$ through $\Sigma_{\tau}$ defines a nondegenerate energy controlling $|F|^{2}+|D\phi|^{2}$ uniformly up to the horizon, while the horizon flux is compatible with the positivity of the $K$-energy on $\mathcal H^{+}$.
The superradiant part of the dynamics is treated in the linear theory through frequency analysis (for waves and Maxwell via the Teukolsky equation), but in this paper we only use its consequences through the linear estimates $\Lin_{k}$.

\subsection{Integrated local energy decay and the \texorpdfstring{$r^{p}$}{rp} hierarchy}\label{subsec:kerr-iledecay}

The nonlinear argument treats the Maxwell-Higgs system as a perturbation of the linear scalar wave and Maxwell systems, with the nonlinearities viewed as inhomogeneous source terms after commuting with the admissible vector fields $\mathcal Z$.
Accordingly, the key condition is an \emph{inhomogeneous} integrated decay estimate which simultaneously provides
(i) boundedness of the nondegenerate energy, (ii) a Morawetz (integrated local energy decay) bulk term which degenerates only at the trapped set (photon region), and (iii) an $r^{p}$ hierarchy in the asymptotically flat region controlling the radiation fluxes through $\mathcal I^{+}$.

In the method of Definition~\ref{def:lin-estimates} this condition is encapsulated by the inequality \eqref{eq:Lin-abstract} for a suitable admissible norm pair $(\mathcal X_{k},\mathcal S_{k})$.
For emphasis, we state a representative structural form of the estimate on Kerr.

\paragraph{A representative Morawetz bulk density in Kerr-star coordinates.}
To make the content of Proposition~\ref{prop:kerr-integrated-decay} more concrete, we state a typical choice of the local energy bulk terms appearing in $\|\cdot\|_{\mathcal X_{k}}$.
In Kerr-star coordinates $(t,r,\theta,\varphi)$ for $g_{M,a}$, cf.~\eqref{metric}, and with the spacelike time function $t^{\star}=t-h(r)$ from \eqref{eq:kerr-tstar}, the triangular coordinate change preserves the volume density, so one has the volume form
\begin{equation}
\dd V_{g_{M,a}}=\Sigma\sin\theta\,\dd t^{\star}\,\dd r\,\dd\theta\,\dd\varphi,
\qquad
\Sigma=r^{2}+a^{2}\cos^{2}\theta.
\end{equation}
Let $S_{\tau,r}:=\Sigma_{\tau}\cap\{r=\mathrm{const}\}$ and let $\slashed{\nabla}$ denote the induced Levi-Civita connection on $S_{\tau,r}$.
On each compact region $\{r\le R\}$ the induced metric on $S_{\tau,r}$ is uniformly comparable to the round metric on $\mathbb S^{2}$, so in estimates one may freely replace $|\slashed{\nabla}f|^{2}$ by $(\partial_{\theta}f)^{2}+\sin^{-2}\theta\,(\partial_{\varphi}f)^{2}$ at the expense of constants depending on $(M,a,R)$.

With these conventions, a representative Morawetz (integrated local energy) density for the scalar wave equation is the quadratic form
\begin{equation}
\mathfrak e[f]
:=
|\partial_{t^{\star}}f|^{2}
+
|\partial_{r}f|^{2}
+
r^{-2}|\slashed{\nabla}f|^{2}.
\end{equation}
Then the bulk part of $\|\psi\|_{\mathcal X_{k}(\tau_{0},\tau_{1})}$ may be chosen so as to control, up to lower-order terms,
\begin{equation}\label{eq:kerr-morawetz-structural}
\int_{\mathcal R(\tau_{0},\tau_{1})}\sum_{|\alpha|\le k}\Bigl(
w_{\mathrm{trap}}\;\mathfrak e[\mathcal Z^{\alpha}\psi]
+r^{-1-\delta}\,\mathfrak e[\mathcal Z^{\alpha}\psi]
+r^{-3-\delta}|\mathcal Z^{\alpha}\psi|^{2}
\Bigr)\dd V_{g_{M,a}},
\end{equation}
where $w_{\mathrm{trap}}\ge0$ is a trapping weight degenerating only on the photon region.
On Schwarzschild one may take $w_{\mathrm{trap}}$ to be a smooth function of $r$ vanishing only at the photon sphere $r=3M$.
On Kerr, the trapped set is a subset of phase space; in the scalar Kerr theory the Morawetz estimate is obtained by a microlocal multiplier and \eqref{eq:kerr-morawetz-structural} should be interpreted as the physical-space avatar of a bulk term which degenerates only on trapped null geodesics.

In the far region $r\ge R$ one may take the $r^{p}$ part of $\|\psi\|_{\mathcal X_{k}}$ to coincide (up to equivalence) with the vector-field functional.
In terms of the asymptotic retarded coordinate $u=t^{\star}-r$ and the asymptotically outgoing vector field $L_{\mathrm{out}}=\partial_{t^{\star}}+\partial_{r}$, this functional controls, up to lower-order terms,
\begin{equation}\label{eq:kerr-rp-structural}
\int_{\mathcal R(\tau_{0},\tau_{1})\cap\{r\ge R\}} r^{p-1}\sum_{|\alpha|\le k}\Bigl(|L_{\mathrm{out}}(r\mathcal Z^{\alpha}\psi)|^{2}+|\slashed{\nabla}(r\mathcal Z^{\alpha}\psi)|^{2}\Bigr)\,\dd u\,\dd r\,\dd\omega,
\end{equation}
together with the corresponding weighted fluxes through $\mathcal I^{+}(\tau_{0},\tau_{1})$.
\paragraph{A representative Maxwell bulk density.}
Let $G$ be a smooth, uncharged Maxwell field on $\mathcal D_{M,a}$ and fix a smooth null frame $(L,\underline L,e_1,e_2)$ adapted to the foliation and the spheres $S_{\tau,r}$.
In the far region one may take $L=L_{\mathrm{out}}=\partial_{t^{\star}}+\partial_{r}$ and $\underline L=\partial_{t^{\star}}-\partial_{r}$; on compact sets any two such frames give equivalent norms.
Define the null components of $G$ by
\begin{equation}
\alpha_{A}[G]:=G(L,e_{A}),\qquad
\underline{\alpha}_{A}[G]:=G(\underline L,e_{A}),\qquad
\rho[G]:=\frac12\,G(L,\underline L),\qquad
\sigma[G]:=\frac12\,{}^{\star}G(L,\underline L),
\end{equation}
where ${}^{\star}$ denotes the Hodge dual with respect to $g_{M,a}$ and $A\in\{1,2\}$.
Set the Maxwell local energy density
\begin{equation}
\mathfrak e_{\mathrm{Max}}[G]
:=
|\alpha[G]|^{2}
+
|\underline{\alpha}[G]|^{2}
+
|\rho[G]|^{2}
+
|\sigma[G]|^{2},
\end{equation}
where the norms are taken with respect to the induced metric on $S_{\tau,r}$.
Then the bulk part of $\|G\|_{\mathcal X_{k}(\tau_{0},\tau_{1})}$ may be chosen so as to control, up to lower-order terms,
\begin{equation}\label{eq:kerr-maxwell-morawetz-structural}
\int_{\mathcal R(\tau_{0},\tau_{1})}\sum_{|\beta|\le k}\Bigl(
w_{\mathrm{trap}}\;\mathfrak e_{\mathrm{Max}}[\mathcal L_{\mathcal Z}^{\beta}G]
+r^{-1-\delta}\,\mathfrak e_{\mathrm{Max}}[\mathcal L_{\mathcal Z}^{\beta}G]
+r^{-3-\delta}\,|\mathcal L_{\mathcal Z}^{\beta}G|^{2}
\Bigr)\,\dd V_{g_{M,a}},
\end{equation}
together with the corresponding redshift control on $\mathcal H^{+}$.
In the far region $r\ge R$ one may formulate the $r^{p}$ hierarchy in terms of the outgoing null component $\alpha[G]$ (equivalently, in terms of the spin $+1$ Teukolsky scalar); up to lower-order terms,
\begin{equation}\label{eq:kerr-maxwell-rp-structural}
\int_{\mathcal R(\tau_{0},\tau_{1})\cap\{r\ge R\}} r^{p-1}\sum_{|\beta|\le k}\Bigl(
|L_{\mathrm{out}}(r\,\alpha[\mathcal L_{\mathcal Z}^{\beta}G])|^{2}
+
|\slashed{\nabla}(r\,\alpha[\mathcal L_{\mathcal Z}^{\beta}G])|^{2}
\Bigr)\,\dd u\,\dd r\,\dd\omega.
\end{equation}

\begin{proposition}[Redshift-Morawetz-$r^{p}$ estimate on Kerr]\label{prop:kerr-integrated-decay}
Fix $M>0$, $|a|<M$, and an integer $k\ge0$.
Let $\psi$ solve $\square_{g_{M,a}}\psi=\mathfrak{F}$ on $\mathcal D_{M,a}$, or let $G$ be a uncharged Maxwell field solving $\nabla^{\mu}G_{\mu\nu}=J_{\nu}$ on $\mathcal D_{M,a}$.
Then there exist admissible norms $(\mathcal X_{k},\mathcal S_{k})$ in the sense of Definition~\ref{def:lin-estimates} such that for every $\tau_{0}<\tau_{1}$ one has
\begin{equation}
\|\psi\|_{\mathcal X_{k}(\tau_{0},\tau_{1})}^{2}
\le
C\Bigl(E^{N}_{k}[\psi](\tau_{0})+\|\mathfrak{F}\|_{\mathcal S_{k}(\tau_{0},\tau_{1})}^{2}\Bigr),
\end{equation}
with the analogous estimate for $G$ and $J$.
In addition, one may take the concrete norms of Lemma~\ref{lem:concrete-admissible-norms}, whose bulk terms include a redshift contribution near $\mathcal H^{+}$, a Morawetz contribution with trapping weight $w_{\mathrm{trap}}$ in the interior, and a full the vector-field $r^{p}$ hierarchy in the far region $\{r\ge R\}$.
\end{proposition}

\begin{proof}
For scalar waves, the corresponding redshift-Morawetz-$r^{p}$ estimate is supplied by the Kerr scalar theory of \cite[Theorems~3.1 and~3.2]{DRSRKerrIII}; restricting that theory to the slow range gives the scalar estimates used here.  The inhomogeneous estimate follows by the energy method and Duhamel's principle in the same norms.
For uncharged Maxwell fields on slowly rotating Kerr, after subtraction of the stationary Coulomb Maxwell solution, the required physical-space boundedness, decay, and asymptotic estimates are supplied by the estimates of \cite{AnderssonBlueMaxwellSlowKerr}.
The inhomogeneous estimate with current $J$ is obtained by the same energy method.
Translating the norms in these references into the abstract method of Definition~\ref{def:lin-estimates} yields admissible norms $(\mathcal X_k,\mathcal S_k)$.
The concrete choice of Lemma~\ref{lem:concrete-admissible-norms} is equivalent to the physical-space norms appearing in the cited works, and therefore also admissible.
\end{proof}

\begin{remark}[Use in the nonlinear bootstrap]\label{rem:kerr-use-nonlinear}
In the Maxwell-Higgs system, commuting the Lorenz-reduced equations \eqref{eq:lorenz-reduced-kerr} with $\mathcal Z$ produces source terms which are at least quadratic in $(F,D\phi,\phi)$.
In the small-data regime of Theorem~\ref{thm:main-slow-kerr-intro} these quadratic terms are treated as inhomogeneities, and Proposition~\ref{prop:kerr-integrated-decay} (or, equivalently, the abstract estimate \eqref{eq:Lin-abstract}) is applied on successive slabs $\mathcal R(\tau_{0},\tau_{1})$ to close the energy and decay bootstrap.
\end{remark}

\section{Decay in the far region on Kerr}\label{sec:kerr-far-decay}

In this section we state the explicit zero-sector realization of the far-region estimates used in the slowly rotating Kerr theorem.

The main quantitative conclusions in the radiative sector are Proposition~\ref{prop:kerr-far-energy-decay} (inhomogeneous scalar waves, together with the tame Maxwell-Higgs source bounds needed for the Higgs equation) and Proposition~\ref{prop:kerr-far-energy-decay-maxwell} (the radiative Maxwell remainder with current in the zero sector). Their role is to supply the uncharged consequence Theorem~\ref{thm:main-slow-kerr-intro}; in the electric theorem the analogous scalar estimates are supplied by \(\CElec^{(m)}_K(M,a,Q_e)\), while the charge-subtracted Maxwell estimate is the same radiative estimate.

These estimates follow of the $r^{p}$ hierarchy encoded in the zero-sector estimates $\Lin_{k}$ (Definition~\ref{def:lin-estimates}); in concrete form they correspond to the far-field terms \eqref{eq:kerr-rp-structural}-\eqref{eq:kerr-maxwell-rp-structural}. We include them here in a form compatible with the Kerr-star foliation and with the gauge-covariant Maxwell-Higgs system, since far-field decay is the main analytic condition for both the nonlinear final-state construction and the structure results for the scattering operator.

\noindent\textbf{Linear condition.} The far-region decay statements stated in this section are consequences of the $r^{p}$ hierarchy and linear scattering theory contained in the zero-sector estimates $\Lin_{k}$; in the Kerr case we do not reprove these estimates here, but rather record them in a form compatible with the nonlinear arguments and refer to Section~\ref{sec:kerr-extension} for the underlying linear theory.
\medskip
\noindent\textbf{Main far-region decay statements.}
The nonlinear scattering and wave-operator arguments use two far-region conditions in the radiative sector:
\begin{itemize}
\item the inhomogeneous far-field decay estimate for scalar waves (and therefore, via the tame bounds, for the Higgs equation), together with coercive control of the Maxwell-Higgs nonlinearities in the admissible source norm (Proposition~\ref{prop:kerr-far-energy-decay});
\item the corresponding inhomogeneous far-field decay estimate for the radiative Maxwell remainder with conserved current (Proposition~\ref{prop:kerr-far-energy-decay-maxwell}).
\end{itemize}
Both propositions are consequences of the vector-field $r^{p}$ hierarchy encoded in $\Lin_{k}$ and are used here in the Schwarzschild and slowly rotating Kerr regimes. In the Schwarzschild part of the paper we provide a complete derivation of the underlying $r^{p}$ hierarchy and the resulting pointwise far-region decay; see Section~\ref{sec:decayfaraway} and Corollary~\ref{cor:far-decay}.
\medskip

Fix $R\gg1$ so that the stationary Killing field $T=\partial_{t^{\star}}$ is uniformly timelike on $\{r\ge R\}$ and the induced metrics on the spheres
\begin{equation}
S_{\tau,r}:=\Sigma_{\tau}\cap\{r=\mathrm{const}\}
\end{equation}
are uniformly comparable to the round metric on $\mathbb S^{2}$ of radius $r$.

\subsection{Outgoing geometry and Sobolev inequalities on \texorpdfstring{$S_{\tau,r}$}{S(tau,r)}}\label{subsec:kerr-far-geom}

Introduce the retarded and advanced coordinates in the far region
\begin{equation}\label{eq:kerr-uv}
u:=t^{\star}-r,\qquad v:=t^{\star}+r,
\end{equation}
and set
\begin{equation}
L:=\partial_{t^{\star}}+\partial_{r}=\partial_{r}\big|_{u},
\qquad
\underline L:=\partial_{t^{\star}}-\partial_{r}.
\end{equation}
In the asymptotically flat region $r\ge R$, the vector field $L$ is future directed and asymptotically outgoing, with
\begin{equation}
g_{M,a}(L,L)=O(r^{-1}),
\end{equation}
while $\underline L$ is asymptotically incoming.
We write $\slashed{\nabla}$ for the Levi-Civita connection on $S_{\tau,r}$ and $\dd\sigma_{\tau,r}$ for the induced area form.

\begin{lemma}[Sobolev inequality on $S_{\tau,r}$]\label{lem:kerr-sobolev-sphere}
There exists $R_{0}\gg1$ such that for every $R\ge R_{0}$ there is a constant $C=C(M,a,R)$ with the following property.
For every smooth complex-valued function $f$ on $S_{\tau,r}$ with $r\ge R$,
\begin{equation}\label{eq:kerr-sobolev-sphere}
\|f\|_{L^{\infty}(S_{\tau,r})}^{2}
\le
C\sum_{j=0}^{2} r^{2j-2}\int_{S_{\tau,r}}|\slashed{\nabla}^{\,j}f|^{2}\,\dd\sigma_{\tau,r}.
\end{equation}
\end{lemma}

\begin{proof}
Use the coordinate map obtained by flowing the angular variables from the unit sphere to $S_{\tau,r}$ and write $\widehat \gamma_{\tau,r}:=r^{-2}\gamma|_{S_{\tau,r}}$.  Asymptotic flatness gives, after increasing $R_{0}$ if needed,
\begin{equation*}
 c\,\gamma_{\mathbb S^{2}}\le \widehat \gamma_{\tau,r}\le C\,\gamma_{\mathbb S^{2}},
 \qquad
 \sum_{|\alpha|\le2}\|\partial_\omega^\alpha \widehat\gamma_{\tau,r}\|_{L^{\infty}(\mathbb S^2)}\le C,
\end{equation*}
with constants independent of $\tau$ and $r\ge R$.  The Sobolev inequality on a compact two-dimensional manifold applied to the uniformly equivalent family $(\mathbb S^2,\widehat\gamma_{\tau,r})$ gives
\begin{equation*}
 \|f\|_{L^\infty(S_{\tau,r})}^{2}
 \le C\sum_{j=0}^{2}\int_{S_{\tau,r}}|\widehat\nabla^{j}f|_{\widehat\gamma_{\tau,r}}^{2}\,\dd\widehat\sigma_{\tau,r}.
\end{equation*}
Since $\dd\widehat\sigma_{\tau,r}=r^{-2}\dd\sigma_{\tau,r}$ and each covariant angular derivative has norm larger by a factor $r$ in the rescaled metric, the $j$th term is exactly bounded by
\begin{equation*}
 C r^{2j-2}\int_{S_{\tau,r}}|\slashed\nabla^{j}f|^{2}\,\dd\sigma_{\tau,r}.
\end{equation*}
Summing over $0\le j\le2$ gives \eqref{eq:kerr-sobolev-sphere}.
\end{proof}

\subsection{Scalar far-field decay and the Higgs component}\label{subsec:kerr-far-scalar}

\subsubsection*{Coupled Maxwell-Higgs energy functionals and control of the curvature $F$}\label{subsubsec:kerr-far-mh-energy}

\begin{remark}[Curvature $F$ versus forcing terms]\label{rem:MH-F-notation}
In the Maxwell-Higgs system \eqref{eom1}-\eqref{eom2}, the symbol $F=\dd A$ denotes the Maxwell curvature $2$-form.
Separately, in the \emph{linear} wave estimates one often writes $\square_{g}\psi=\mathfrak{F}$ for an \emph{external forcing term}.
When both objects appear in the same discussion we reserve the letter $F$ for the curvature and write the forcing as $\mathfrak{F}$.
Any specialization to $\mathfrak{F}\equiv0$ refers to vanishing forcing and \emph{never} to the (generically false) constraint $F=0$ for Maxwell-Higgs solutions.
\end{remark}

\begin{proposition}[Coupled Maxwell-Higgs energies and coercive control of the curvature]\label{prop:MH-energy-control-F}
Let $(A,\phi)$ be a smooth solution of the Maxwell-Higgs system \eqref{eom1}-\eqref{eom2} on a globally hyperbolic region of a background spacetime $(\mathcal D,g)$.
Let $F=\dd A$ and let
\begin{equation}\label{eq:MH-current-prop42}
J_{\nu}:=-i\bigl(D_{\nu}\phi\,\overline{\phi}-\phi\,\overline{D_{\nu}\phi}\bigr)
=2\,\Im(\overline\phi\,D_{\nu}\phi)
\end{equation}
be the gauge current, so that $\nabla^{\mu}F_{\mu\nu}=J_{\nu}$.

\smallskip
\noindent
\textbf{Stress-energy tensors and energy fluxes.}
Define the Maxwell and Higgs stress-energy tensors by
\begin{eqnarray}
\mathbb T^{(\mathrm M)}_{\mu\nu}[F]
&:=F_{\mu\alpha}F_{\nu}{}^{\alpha}-\frac14 g_{\mu\nu}F_{\alpha\beta}F^{\alpha\beta},
\label{eq:MH-stress-maxwell-prop42}
\\
\mathbb T^{(\mathrm H)}_{\mu\nu}[\phi,A]
&:=2\Re\!\bigl(D_{\mu}\phi\,\overline{D_{\nu}\phi}\bigr)
- g_{\mu\nu}\Bigl(D_{\alpha}\phi\,\overline{D^{\alpha}\phi}+P(\phi,\bar\phi)\Bigr),
\label{eq:MH-stress-higgs-prop42}
\end{eqnarray}
and set $\mathbb T^{(\mathrm{MH})}_{\mu\nu}:=\mathbb T^{(\mathrm M)}_{\mu\nu}+\mathbb T^{(\mathrm H)}_{\mu\nu}$, which coincides with $T_{\mu\nu}[F,\phi]$ from \eqref{eq:stress-energy-kerr} when $g=g_{M,a}$.
For any vector field $X$ define the associated currents
\begin{equation}
J^{X,(\mathrm M)}_{\mu}:=\mathbb T^{(\mathrm M)}_{\mu\nu}X^{\nu},
\qquad
J^{X,(\mathrm H)}_{\mu}:=\mathbb T^{(\mathrm H)}_{\mu\nu}X^{\nu},
\qquad
J^{X,(\mathrm{MH})}_{\mu}:=\mathbb T^{(\mathrm{MH})}_{\mu\nu}X^{\nu}.
\end{equation}
For any spacelike hypersurface $\Sigma$ with future unit normal $n_{\Sigma}$ and induced volume form $\dd\mu_{\Sigma}$ we define the corresponding fluxes by
\begin{eqnarray}\label{eq:MH-energies-prop42}
\mathcal E^{(\mathrm M)}_{X}(\Sigma)
&:=\int_{\Sigma}J^{X,(\mathrm M)}\!\cdot n_{\Sigma}\,\dd\mu_{\Sigma},\\
\mathcal E^{(\mathrm H)}_{X}(\Sigma)
&:=\int_{\Sigma}J^{X,(\mathrm H)}\!\cdot n_{\Sigma}\,\dd\mu_{\Sigma},\\
\mathcal E^{(\mathrm{MH})}_{X}(\Sigma)
&:=\mathcal E^{(\mathrm M)}_{X}(\Sigma)+\mathcal E^{(\mathrm H)}_{X}(\Sigma).
\end{eqnarray}

\smallskip
\noindent
\textbf{(i) Separate divergence identities and energy exchange.}
Let ${}^{(X)}\pi_{\mu\nu}:=\nabla_{\mu}X_{\nu}+\nabla_{\nu}X_{\mu}$.
Then
\begin{eqnarray}
\nabla^{\mu}J^{X,(\mathrm M)}_{\mu}
&=
\frac12\,\mathbb T^{(\mathrm M)}_{\mu\nu}\;{}^{(X)}\pi^{\mu\nu}
\;-\;
F_{\nu\lambda}J^{\lambda}X^{\nu},
\label{eq:MH-div-M-prop42}
\\
\nabla^{\mu}J^{X,(\mathrm H)}_{\mu}
&=
\frac12\,\mathbb T^{(\mathrm H)}_{\mu\nu}\;{}^{(X)}\pi^{\mu\nu}
\;+\;
F_{\nu\lambda}J^{\lambda}X^{\nu}.
\label{eq:MH-div-H-prop42}
\end{eqnarray}
The coupling term $F_{\nu\lambda}J^{\lambda}X^{\nu}$ is the precise exchange of energy between the Maxwell and Higgs sectors, so the Maxwell and Higgs energies are not separately conserved in general.

\smallskip
\noindent
\textbf{(ii) Cancellation in the total energy.}
Adding \eqref{eq:MH-div-M-prop42} and \eqref{eq:MH-div-H-prop42} yields
\begin{equation}\label{eq:MH-div-total-prop42}
\nabla^{\mu}J^{X,(\mathrm{MH})}_{\mu}
=
\frac12\,\mathbb T^{(\mathrm{MH})}_{\mu\nu}\;{}^{(X)}\pi^{\mu\nu}.
\end{equation}
If $X$ is Killing then $\nabla^{\mu}J^{X,(\mathrm{MH})}_{\mu}=0$.

\smallskip
\noindent
\textbf{(iii) Energy identity and coercive control of $F$.}
Let $\mathcal R$ be any spacetime region with piecewise smooth boundary $\partial\mathcal R$ and outward normal $n_{\partial\mathcal R}$.
Integrating \eqref{eq:MH-div-total-prop42} and applying Stokes' theorem gives
\begin{equation}\label{eq:MH-stokes-prop42}
\int_{\partial\mathcal R} J^{X,(\mathrm{MH})}\!\cdot n_{\partial\mathcal R}\,\dd\mu_{\partial\mathcal R}
=
\frac12\int_{\mathcal R}\mathbb T^{(\mathrm{MH})}_{\mu\nu}\;{}^{(X)}\pi^{\mu\nu}\,\dd V_{g}.
\end{equation}

Assume in addition that $P(\phi,\bar\phi)\ge0$ and that $X$ is future directed timelike on $\Sigma$.
Let $h_{\Sigma}:=g+n_{\Sigma}\otimes n_{\Sigma}$ be the induced Riemannian metric.
Define the normal electric part and the spatial Hodge-dual part of $F$ with respect to $\Sigma$ by
\begin{equation}\label{eq:MH-EB-decomp}
E_{\mu}:=F_{\mu\nu}n_{\Sigma}^{\nu},
\qquad
B_{\mu}:=({}^{\star}\!F)_{\mu\nu}n_{\Sigma}^{\nu},
\qquad
|F|_{\Sigma}^{2}:=|E|^{2}_{h_{\Sigma}}+|B|^{2}_{h_{\Sigma}},
\end{equation}
and the tangential Higgs derivative by $(D^{\top}\phi)_{\mu}:=(\delta_{\mu}^{\nu}+n_{\Sigma\mu}n_{\Sigma}^{\nu})D_{\nu}\phi$.
Then there exists a constant $c_{g,X}>0$, depending only on the geometry and on the uniform timelikeness of $X$ on $\Sigma$, such that
\begin{equation}\label{eq:MH-coercive-prop42}
J^{X,(\mathrm{MH})}\!\cdot n_{\Sigma}
=
\mathbb T^{(\mathrm{MH})}(n_{\Sigma},X)
\ge
c_{g,X}\Bigl(|F|_{\Sigma}^{2}+|D_{n_{\Sigma}}\phi|^{2}+|D^{\top}\phi|^{2}_{h_{\Sigma}}+P(\phi,\bar\phi)\Bigr)
\qquad\text{pointwise on }\Sigma.
\end{equation}
Consequently,
\begin{equation}\label{eq:MH-bound-F-prop42}
\int_{\Sigma}|F|_{\Sigma}^{2}\,\dd\mu_{\Sigma}\ \lesssim_{g,X}\ \mathcal E^{(\mathrm{MH})}_{X}(\Sigma).
\end{equation}
\end{proposition}

\begin{proof}
We state the key identities.

\smallskip
\noindent\textbf{Divergence of the Maxwell stress-energy.}
Using $\nabla^{\mu}F_{\mu\nu}=J_{\nu}$ and the Bianchi identity $\nabla_{[\alpha}F_{\beta\gamma]}=0$, a direct computation yields
\begin{equation}
\nabla^{\mu}\mathbb T^{(\mathrm M)}_{\mu\nu}[F]= -\,F_{\nu\lambda}J^{\lambda}.
\end{equation}
Contracting with $X^{\nu}$ and using
\begin{equation}
\nabla^{\mu}(\mathbb T_{\mu\nu}X^{\nu})
=
(\nabla^{\mu}\mathbb T_{\mu\nu})X^{\nu}+\tfrac12\,\mathbb T_{\mu\nu}\,{}^{(X)}\pi^{\mu\nu}
\end{equation}
gives \eqref{eq:MH-div-M-prop42}.

\smallskip
\noindent\textbf{Divergence of the Higgs stress-energy.}
Using \eqref{eom2}, its complex conjugate, and gauge-covariant Leibniz rules yields
\begin{equation}
\nabla^{\mu}\mathbb T^{(\mathrm H)}_{\mu\nu}[\phi,A]= +\,F_{\nu\lambda}J^{\lambda},
\end{equation}
and therefore \eqref{eq:MH-div-H-prop42}.

\smallskip
\noindent\textbf{Total identity and Stokes' theorem.}
Adding \eqref{eq:MH-div-M-prop42} and \eqref{eq:MH-div-H-prop42} gives \eqref{eq:MH-div-total-prop42}.
Integrating \eqref{eq:MH-div-total-prop42} over $\mathcal R$ and applying the divergence theorem yields \eqref{eq:MH-stokes-prop42}.

\smallskip
\noindent\textbf{Coercivity on spacelike hypersurfaces.}
Let $n=n_{\Sigma}$.
With the decomposition \eqref{eq:MH-EB-decomp} one has the identities
\begin{equation}
\mathbb T^{(\mathrm M)}(n,n)=\tfrac12\bigl(|E|^{2}_{h_{\Sigma}}+|B|^{2}_{h_{\Sigma}}\bigr),
\qquad
\mathbb T^{(\mathrm H)}(n,n)=|D_{n}\phi|^{2}+|D^{\top}\phi|^{2}_{h_{\Sigma}}+P(\phi,\bar\phi),
\end{equation}
and therefore $\mathbb T^{(\mathrm{MH})}(n,n)$ is a positive definite density controlling $|F|_{\Sigma}^{2}$ and $|D\phi|^{2}$.
Write $X=\alpha n+V$ on $\Sigma$, where $\alpha:=-g(X,n)>0$ and $V$ is tangent to $\Sigma$.
Since $X$ is timelike, $\alpha>|V|_{h_{\Sigma}}$ pointwise; uniform timelikeness gives $\alpha-|V|_{h_{\Sigma}}\ge c_{0}>0$ on $\Sigma$.
In addition, $\mathbb T^{(\mathrm{MH})}$ satisfies the dominant energy condition when $P\ge0$, and therefore
$|\mathbb T^{(\mathrm{MH})}(n,V)|\le \mathbb T^{(\mathrm{MH})}(n,n)\,|V|_{h_{\Sigma}}$.
Consequently,
\begin{equation}
\mathbb T^{(\mathrm{MH})}(n,X)
=
\alpha\,\mathbb T^{(\mathrm{MH})}(n,n)+\mathbb T^{(\mathrm{MH})}(n,V)
\ge
(\alpha-|V|_{h_{\Sigma}})\,\mathbb T^{(\mathrm{MH})}(n,n)
\ge
c_{0}\,\mathbb T^{(\mathrm{MH})}(n,n),
\end{equation}
which yields \eqref{eq:MH-coercive-prop42} and therefore \eqref{eq:MH-bound-F-prop42}.
\end{proof}

Let $\psi$ solve the inhomogeneous wave equation $\square_{g_{M,a}}\psi=\mathfrak{F}$ on $\mathcal D_{M,a}$ and define the rescaled unknown $\Psi:=r\psi$.
The far-field component of the $\mathcal X_{k}$ norm (Definition~\ref{def:lin-estimates}) may be chosen so that, in this case, it controls an $r^{p}$ hierarchy for $\Psi$ as in \eqref{eq:kerr-rp-structural}.
In the homogeneous case $\mathfrak{F}\equiv0$ this implies polynomial decay for local, weighted, and radiation-flux quantities. The full nondegenerate energy on complete Kerr-star slices is bounded, not asserted to decay.

\begin{proposition}[Main far-field scalar decay with forcing and Maxwell-Higgs source bounds]\phantomsection\label{prop:kerr-far-energy-decay}

\noindent\emph{This proposition proves the zero-sector specialization of Theorem~\ref{thm:main-slow-kerr-intro}.}

Fix $k\ge 2$ and let $\psi$ be a smooth solution to the inhomogeneous wave equation
\begin{equation}\label{eq:kerr-inhom-wave}
\square_{g_{M,a}}\psi=\mathfrak{F}
\end{equation}
on $\mathcal D_{M,a}$. This includes the Schwarzschild exterior $a=0$ as a special case.
Then there exists $R\gg1$ such that for every $0<\delta<1$ and all $\tau\ge\tau_{0}$ the following estimate holds.
Here we set the far-field $r^{p}$ bulk/flux functional
\begin{equation}\label{eq:def-Bk-kerr}
\mathcal{B}_{k}[\psi](\tau_{0},\tau)
:=
\sup_{p\in[0,2]}\mathcal R_{k,p}[\psi](\tau_{0},\tau),
\end{equation}
with $\mathcal R_{k,p}$ the vector-field $r^{p}$ functional from Lemma~\ref{lem:concrete-admissible-norms}.
\begin{equation}\label{eq:kerr-energy-decay-scalar}
E^{N}_{k}[\psi](\tau)+\mathcal{B}_{k}[\psi](\tau_{0},\tau)
\lesssim_{M,a,k,\delta}
E^{N}_{k}[\psi](\tau_{0})
+\|\mathfrak{F}\|_{\mathcal S_{k}(\tau_{0},\tau)}^{2},
\end{equation}
and, for all $(\tau,r,\omega)$ with $r\ge R$,
\begin{equation}\label{eq:kerr-far-L2-grad-scalar}
\int_{\Sigma_{\tau}\cap\{r\ge R\}} r^{-1-\delta}|\nabla\mathcal Z^{\le k}\psi|^{2}\,\dd\mu_{\Sigma_{\tau}}
\lesssim_{M,a,k,\delta}
(1+\tau)^{-2+2\delta}\,E^{N}_{k+2}[\psi](\tau_{0})
+\|\mathfrak{F}\|_{\mathcal S_{k+2}(\tau_{0},\tau)}^{2}.
\end{equation}

In addition, consider a Lorenz-gauge solution $(A,\phi)$ of the Maxwell-Higgs system~\eqref{eq:lorenz-reduced-kerr} on $\mathcal R(\tau_{0},\tau)$, with curvature $F=\dd A$. After placing the linear mass term $m^{2}\phi$ on the left-hand side of the scalar comparison equation, the nonlinearities are
\begin{eqnarray}
\mathfrak{F}_{\phi}^{(m)}
&:= 2iA^\mu\nabla_\mu\phi+ i(\nabla^\mu A_\mu)\phi +A^\mu A_\mu\,\phi +\mathcal R_{P}(\phi),
\label{eq:def-Fphi}\\
J_{\nu}
&:= -i\bigl(D_{\nu}\phi\,\overline\phi-\phi\,\overline{D_{\nu}\phi}\bigr)
=2\,\Im(\overline\phi\,D_{\nu}\phi),
\label{eq:def-J}
\end{eqnarray}
where $D=\nabla-iA$ is the gauge-covariant derivative and $\mathcal R_P(\phi)=\partial_{\bar\phi}P(\phi,\bar\phi)-m^2\phi$ is the nonlinear potential remainder.
In Lorenz gauge, $\nabla^\mu A_\mu=0$, so
\begin{equation}
\mathfrak{F}_{\phi}^{(m)}=2iA^\mu\nabla_\mu\phi+A^\mu A_\mu\,\phi+\mathcal R_P(\phi).
\end{equation}

\smallskip
\noindent
\textbf{Higher-order Maxwell-Higgs energies.}
Define the higher-order Maxwell and Higgs energies, cf.\ Proposition~\ref{prop:MH-energy-control-F},
\begin{equation}
\mathcal E_{N,k}^{(\mathrm M)}[F](\Sigma_{\tau})
:=\sum_{|\alpha|\le k}\mathcal E_{N}^{(\mathrm M)}[\mathcal L_{\mathcal Z^{\alpha}}F](\Sigma_{\tau}),
\qquad
\mathcal E_{N,k}^{(\mathrm H)}[\phi,A](\Sigma_{\tau})
:=\sum_{|\alpha|\le k}\mathcal E_{N}^{(\mathrm H)}[\mathcal Z^{\alpha}\phi,A](\Sigma_{\tau}),
\end{equation}
and set $\mathcal E_{N,k}^{(\mathrm{MH})}:=\mathcal E_{N,k}^{(\mathrm M)}+\mathcal E_{N,k}^{(\mathrm H)}$.
In addition, define the componentwise wave energy of the gauge potential by
\begin{equation}\label{eq:def-kerr-pot-energy}
\mathcal E_{N,k}^{(\mathrm A)}[A](\Sigma_{\tau})
:=
\sum_{\nu=0}^{3}E^{N}_{k}[A_{\nu}](\Sigma_{\tau}).
\end{equation}

\smallskip
To control the nonlinear sources $J$ and $\mathfrak{F}_{\phi}^{(m)}$ in the admissible source norm $\mathcal S_{k+2}(\tau_{0},\tau)$, introduce the following \emph{source-weighted energy functionals} on $\mathcal R(\tau_{0},\tau)$: for integers $m\ge0$ set
\begin{eqnarray}
\mathbf E_{m}^{(\mathrm A)}(\tau_{0},\tau)
&:= \sup_{\tau'\in[\tau_{0},\tau]}\mathcal E_{N,m}^{(\mathrm A)}[A](\Sigma_{\tau'}),
\label{eq:def-slab-energy-A}\\
\mathbf E_{m}^{(\mathrm H)}(\tau_{0},\tau)
&:= \sup_{\tau'\in[\tau_{0},\tau]}\mathcal E_{N,m}^{(\mathrm H)}[\phi,A](\Sigma_{\tau'}),
\label{eq:def-slab-energy-H}\\
\mathbf F_{m}^{(\mathrm A)}(\tau_{0},\tau)
&:= \int_{\mathcal R(\tau_{0},\tau)}\sum_{|\alpha|\le m} r^{1+\delta}\Bigl(|\nabla\mathcal Z^\alpha A|^{2}+|\mathcal Z^\alpha A|^{2}\Bigr),
\label{eq:def-source-weighted-energy-A}\\
\mathbf F_{m}^{(\mathrm H)}(\tau_{0},\tau)
&:= \int_{\mathcal R(\tau_{0},\tau)}\sum_{|\alpha|\le m} r^{1+\delta}\Bigl(|D\mathcal Z^\alpha\phi|^{2}+|\mathcal Z^\alpha\phi|^{2}\Bigr),
\label{eq:def-source-weighted-energy-H}
\end{eqnarray}
where $|\cdot|$ denotes the pointwise norm induced by $g_{M,a}$.

For every integer $k\ge6$, the Maxwell-Higgs nonlinearities satisfy the tame source bounds.
For convenience define, for $E_A,E_H\ge0$,
\begin{equation}\label{eq:def-tame-poly}
\mathcal P_{k,P}(E_A,E_H)
:=E_A+E_A^{2}+E_H+E_AE_H+\bigl(1+E_H^{N_P}\bigr)^{2}E_H^{2}.
\end{equation}
Then
\begin{eqnarray}
\|J\|_{\mathcal S_{k+2}(\tau_{0},\tau)}^{2}
&\lesssim_{M,a,k}
\Bigl(\mathbf E_{k+4}^{(\mathrm A)}(\tau_{0},\tau)+\mathbf E_{k+4}^{(\mathrm H)}(\tau_{0},\tau)\Bigr)\,
\mathbf F_{k+2}^{(\mathrm H)}(\tau_{0},\tau),
\label{eq:kerr-J-source-bound}\\
\|\mathfrak{F}_{\phi}^{(m)}\|_{\mathcal S_{k+2}(\tau_{0},\tau)}^{2}
&\lesssim_{M,a,k,P}
\mathcal P_{k,P}\!\Bigl(
\mathbf E_{k+4}^{(\mathrm A)}(\tau_{0},\tau),
\mathbf E_{k+4}^{(\mathrm H)}(\tau_{0},\tau)
\Bigr)\nonumber\\
&\quad\times
\Bigl(\mathbf F_{k+2}^{(\mathrm A)}(\tau_{0},\tau)+\mathbf F_{k+2}^{(\mathrm H)}(\tau_{0},\tau)\Bigr).
\label{eq:kerr-higgs-source-bound}
\end{eqnarray}
If $\mathbf E_{k+4}^{(\mathrm A)}(\tau_{0},\tau)$ and $\mathbf E_{k+4}^{(\mathrm H)}(\tau_{0},\tau)$ are finite and the source-weighted energies $\mathbf F_{k+2}^{(\mathrm A)}(\tau_{0},\tau)$, $\mathbf F_{k+2}^{(\mathrm H)}(\tau_{0},\tau)$ are finite, then $J$ and $\mathfrak{F}_{\phi}^{(m)}$ belong to the admissible source space $\mathcal S_{k+2}(\tau_{0},\tau)$.
\end{proposition}

\begin{proof}
The boundedness estimate \eqref{eq:kerr-energy-decay-scalar} and the weighted decay estimate \eqref{eq:kerr-far-L2-grad-scalar} are the far-field part of the scalar linear estimates on slowly rotating Kerr; see Proposition~\ref{prop:lin-estimates-kerr} and the scalar theory of \cite[Theorems~3.1 and~3.2]{DRSRKerrIII}.

We only prove the bounds for the Maxwell-Higgs source terms \eqref{eq:kerr-J-source-bound}-\eqref{eq:kerr-higgs-source-bound}.
Fix $k\ge 6$ and set $m:=k+2$.

\smallskip\noindent
\textbf{Preliminary bounds: $L^\infty$ control from slice energies.}
By Sobolev embedding on the asymptotically flat hypersurfaces $\Sigma_{\tau'}$ (using Lemmas~\ref{lem:kerr-sobolev-compact} and~\ref{lem:kerr-sobolev-sphere} to treat the compact and far regions), we have for every $\tau'\in[\tau_0,\tau]$
\begin{eqnarray}
\|\mathcal Z^{\le k}A(\tau')\|_{W^{1,\infty}(\Sigma_{\tau'})}^{2}
&\lesssim_{M,a,k} \mathcal E_{N,k+4}^{(\mathrm A)}[A](\Sigma_{\tau'})
\le \mathbf E_{k+4}^{(\mathrm A)}(\tau_0,\tau),
\label{eq:sobolev-A}\\
\|\mathcal Z^{\le k}\phi(\tau')\|_{L^{\infty}(\Sigma_{\tau'})}^{2}
+\|D\mathcal Z^{\le k}\phi(\tau')\|_{L^{\infty}(\Sigma_{\tau'})}^{2}
&\lesssim_{M,a,k} \mathcal E_{N,k+4}^{(\mathrm H)}[\phi,A](\Sigma_{\tau'})
\le \mathbf E_{k+4}^{(\mathrm H)}(\tau_0,\tau).
\label{eq:sobolev-phi}
\end{eqnarray}

\smallskip\noindent
\textbf{the current $J_\nu$.}
Since $J_\nu=2\,\Im(\overline\phi\,D_\nu\phi)$, for any multiindex $\alpha$ with $|\alpha|\le m$ we apply $\mathcal Z^\alpha$ and expand:
\begin{equation}\label{eq:ZJ-expand}
\mathcal Z^\alpha J_\nu
=
2\,\Im\!\!\sum_{\beta+\gamma=\alpha}
\overline{\mathcal Z^\beta\phi}\,D_\nu(\mathcal Z^\gamma\phi)
\;+\;
\mathcal C_{\nu,\alpha}[A,\phi],
\end{equation}
where $\mathcal C_{\nu,\alpha}[A,\phi]$ collects the commutator contributions from commuting $\mathcal Z$ through the covariant derivative $D$.
Since $D=\nabla-iA$, each commutator $[D,\mathcal Z]$ is a first-order operator with coefficients depending on the background and on $\nabla\mathcal Z$ and $\mathcal Z A$.
Consequently, $\mathcal C_{\nu,\alpha}[A,\phi]$ is a finite sum of terms up to lower-order terms of the form
\begin{equation}\label{eq:commJ-structure}
(\nabla\mathcal Z^{\beta_1}A)\,(\mathcal Z^{\beta_2}\phi)\,(\overline{\mathcal Z^{\beta_3}\phi}),
\qquad |\beta_1|+|\beta_2|+|\beta_3|\le m-1.
\end{equation}

Because $m=k+2$, in every product in \eqref{eq:ZJ-expand}-\eqref{eq:commJ-structure} at least one factor carries at most $k$ commutators and can be estimated in $L^\infty$ using \eqref{eq:sobolev-A}-\eqref{eq:sobolev-phi}.
We thus obtain the pointwise tame bound
\begin{equation}\label{eq:J-pointwise-tame}
|\mathcal Z^\alpha J|
\lesssim_{M,a,k}
\Bigl(\|\mathcal Z^{\le k}\phi\|_{L^\infty(\Sigma_{\tau'})}
+\|\mathcal Z^{\le k}A\|_{W^{1,\infty}(\Sigma_{\tau'})}\Bigr)
\sum_{|\beta|\le m}\Bigl(|D\mathcal Z^\beta\phi|+|\mathcal Z^\beta\phi|\Bigr)
\end{equation}
at each point of $\mathcal R(\tau_0,\tau)$.

Multiplying \eqref{eq:J-pointwise-tame} by $r^{(1+\delta)/2}$, squaring, integrating over $\mathcal R(\tau_0,\tau)$, and summing over $|\alpha|\le m$ yields
\begin{equation}\label{eq:J-S-est-intermediate}
\|J\|_{\mathcal S_{m}(\tau_0,\tau)}^{2}
\lesssim_{M,a,k}
\Bigl(
\sup_{\tau'\in[\tau_0,\tau]}\|\mathcal Z^{\le k}\phi(\tau')\|_{L^\infty(\Sigma_{\tau'})}^{2}
+
\sup_{\tau'\in[\tau_0,\tau]}\|\mathcal Z^{\le k}A(\tau')\|_{W^{1,\infty}(\Sigma_{\tau'})}^{2}
\Bigr)\,
\mathbf F_{m}^{(\mathrm H)}(\tau_0,\tau).
\end{equation}
inserting \eqref{eq:sobolev-A}-\eqref{eq:sobolev-phi} into \eqref{eq:J-S-est-intermediate} gives \eqref{eq:kerr-J-source-bound}.

\smallskip\noindent
\textbf{the Higgs forcing $\mathfrak F_\phi^{(m)}$.}
In Lorenz gauge we have $\nabla^\mu A_\mu=0$ and therefore
\begin{equation}
\mathfrak F_\phi^{(m)}
=
2iA^\mu\nabla_\mu\phi
+A^\mu A_\mu\,\phi
+\mathcal R_P(\phi).
\end{equation}
Using $\nabla\phi=D\phi-iA\phi$ we have the structural pointwise bound
\begin{equation}\label{eq:Fphi-structural}
|\mathfrak F_\phi^{(m)}|
\lesssim
|A|\,|D\phi|+|A|^{2}|\phi|+|\mathcal R_P(\phi)|.
\end{equation}
We estimate the three contributions separately in the source norm $\mathcal S_m$.

\smallskip\noindent
\emph{(i) The term $A\,D\phi$.}
Applying $\mathcal Z^\alpha$ with $|\alpha|\le m$ and expanding by Leibniz' rule gives products of the form
$(\mathcal Z^{\beta}A)\,D(\mathcal Z^{\gamma}\phi)$ with $|\beta|+|\gamma|\le m$, plus commutator terms coming from commuting $\mathcal Z$ with $D$, which are lower order and can be treated in the same way (cf.\ the discussion above).
Since $m=k+2$, in each product one factor carries at most $k$ commutators.
Consequently,
\begin{eqnarray}
\|A\,D\phi\|_{\mathcal S_{m}(\tau_0,\tau)}^{2}
&\lesssim_{M,a,k}
\Bigl(
\sup_{\tau'\in[\tau_0,\tau]}\|\mathcal Z^{\le k}A(\tau')\|_{L^\infty(\Sigma_{\tau'})}^{2}
\Bigr)\,
\mathbf F_{m}^{(\mathrm H)}(\tau_0,\tau)\nonumber\\
&\qquad\qquad+
\Bigl(
\sup_{\tau'\in[\tau_0,\tau]}\|D\mathcal Z^{\le k}\phi(\tau')\|_{L^\infty(\Sigma_{\tau'})}^{2}
\Bigr)\,
\mathbf F_{m}^{(\mathrm A)}(\tau_0,\tau).
\label{eq:term-ADphi}
\end{eqnarray}
Using \eqref{eq:sobolev-A}-\eqref{eq:sobolev-phi} yields
\begin{equation}\label{eq:term-ADphi-simplified}
\|A\,D\phi\|_{\mathcal S_{m}(\tau_0,\tau)}^{2}
\lesssim_{M,a,k}
\Bigl(\mathbf E_{k+4}^{(\mathrm A)}(\tau_0,\tau)+\mathbf E_{k+4}^{(\mathrm H)}(\tau_0,\tau)\Bigr)\,
\Bigl(\mathbf F_{m}^{(\mathrm A)}(\tau_0,\tau)+\mathbf F_{m}^{(\mathrm H)}(\tau_0,\tau)\Bigr).
\end{equation}

\smallskip\noindent
\emph{(ii) The cubic term $A^{2}\phi$.}
Applying $\mathcal Z^\alpha$ and expanding by Leibniz' rule yields products of the form
$(\mathcal Z^{\beta_1}A)(\mathcal Z^{\beta_2}A)(\mathcal Z^{\gamma}\phi)$ with $|\beta_1|+|\beta_2|+|\gamma|\le m$.
Again, since $m=k+2$, at least two factors carry at most $k$ commutators and can be estimated in $L^\infty$.
Put
\begin{equation}
M_A:=\sup_{\tau'\in[\tau_0,\tau]}
\|\mathcal Z^{\le k}A(\tau')\|_{L^\infty(\Sigma_{\tau'})},
\end{equation}
\begin{equation}
M_\phi:=\sup_{\tau'\in[\tau_0,\tau]}
\|\mathcal Z^{\le k}\phi(\tau')\|_{L^\infty(\Sigma_{\tau'})}.
\end{equation}
Then
\begin{eqnarray}\label{eq:term-A2phi}
\|A^{2}\phi\|_{\mathcal S_{m}(\tau_0,\tau)}^{2}
&\lesssim_{M,a,k}&
M_A^{4}\,\mathbf F_{m}^{(\mathrm H)}(\tau_0,\tau)
\nonumber\\
&&+M_A^{2}M_\phi^{2}\,\mathbf F_{m}^{(\mathrm A)}(\tau_0,\tau).
\end{eqnarray}
Invoking \eqref{eq:sobolev-A}-\eqref{eq:sobolev-phi} gives
\begin{eqnarray}\label{eq:term-A2phi-simplified}
\|A^{2}\phi\|_{\mathcal S_{m}(\tau_0,\tau)}^{2}
\lesssim_{M,a,k}
&\Bigl(
\mathbf E_{k+4}^{(\mathrm A)}(\tau_0,\tau)
+\bigl(\mathbf E_{k+4}^{(\mathrm A)}(\tau_0,\tau)\bigr)^{2}
+\mathbf E_{k+4}^{(\mathrm A)}(\tau_0,\tau)\,\mathbf E_{k+4}^{(\mathrm H)}(\tau_0,\tau)
\Bigr)\\
&\times
\Bigl(\mathbf F_{m}^{(\mathrm A)}(\tau_0,\tau)+\mathbf F_{m}^{(\mathrm H)}(\tau_0,\tau)\Bigr).
\end{eqnarray}

\smallskip\noindent
\emph{(iii) The nonlinear potential remainder $\mathcal R_P(\phi)$.}
By Lemma~\ref{lem:potential-tame}, for $|\alpha|\le m$ the commuted quantity $\mathcal Z^\alpha\mathcal R_P(\phi)$ is a finite sum of products with at least three factors drawn from $\{\mathcal Z^{\beta}\phi,\overline{\mathcal Z^{\beta}\phi}\}$, multiplied by smooth coefficients that grow at most like $(1+|\phi|^{2N_P})$.
Since $m=k+2$, in each such product at least two factors carry at most $k$ commutators; estimating those factors in $L^\infty$ and the remaining factor in $L^2$ yields
\begin{eqnarray}\label{eq:term-PhiP}
\|\mathcal R_P(\phi)\|_{\mathcal S_{m}(\tau_0,\tau)}^{2}
\lesssim_{k,P}
&\Bigl(1+\sup_{\tau'\in[\tau_0,\tau]}\|\mathcal Z^{\le k}\phi(\tau')\|_{L^\infty(\Sigma_{\tau'})}^{2N_P}\Bigr)^{2}\\
&\times
\sup_{\tau'\in[\tau_0,\tau]}\|\mathcal Z^{\le k}\phi(\tau')\|_{L^\infty(\Sigma_{\tau'})}^{4}\,
\mathbf F_{m}^{(\mathrm H)}(\tau_0,\tau).
\end{eqnarray}
Using \eqref{eq:sobolev-phi} gives
\begin{equation}\label{eq:term-PhiP-simplified}
\|\mathcal R_P(\phi)\|_{\mathcal S_{m}(\tau_0,\tau)}^{2}
\lesssim_{M,a,k,P}
\Bigl(1+\mathbf E_{k+4}^{(\mathrm H)}(\tau_0,\tau)^{N_P}\Bigr)^{2}\,
\bigl(\mathbf E_{k+4}^{(\mathrm H)}(\tau_0,\tau)\bigr)^{2}\,
\mathbf F_{m}^{(\mathrm H)}(\tau_0,\tau).
\end{equation}

\smallskip
Combining \eqref{eq:term-ADphi-simplified}, \eqref{eq:term-A2phi-simplified}, and \eqref{eq:term-PhiP-simplified}, and recalling that $m=k+2$, yields \eqref{eq:kerr-higgs-source-bound} with $\mathcal P_{k,P}$ as in \eqref{eq:def-tame-poly}.
\end{proof}

\begin{remark}[Choice of the decay-loss parameter]\label{rem:kerr-decay-delta}
In the nonlinear wave-operator and final-state constructions we fix a single small $\delta>0$ and use it consistently in the far-region and compact-region decay bounds.
The exponent $-2+2\delta$ in the weighted/local decay estimates \eqref{eq:kerr-far-L2-grad-scalar} and \eqref{eq:kerr-energy-decay-maxwell-weighted} is chosen so that, after taking square roots and applying Sobolev on the spheres $S_{\tau,r}$, the resulting pointwise rates are just of order $(1+v_{+})^{-1+\delta}$ (and therefore $(1+\tau)^{-1+\delta}$ on compact sets), which is the rate used to close the nonlinear source integrability estimates.
\end{remark}

\begin{corollary}[Pointwise decay in the far region for scalar waves]\label{cor:kerr-far-pointwise-scalar}
Under the assumptions of Proposition~\ref{prop:kerr-far-energy-decay} and for $k\ge 6$, there exist $R\gg1$ such that for every $0<\delta<1$ one can find $C=C(M,a,k,\delta)$ with the following property: for all $(\tau,r,\omega)$ with $r\ge R$,
\begin{equation}\label{eq:kerr-far-pointwise-scalar}
|\psi(\tau,r,\omega)|
+
|\nabla\psi(\tau,r,\omega)|
\le
\frac{C}{(1+v_{+})^{1-\delta}}
\Bigl(
E^{N}_{k}[\psi](\tau_{0})^{1/2}
+
\|\mathfrak{F}\|_{\mathcal S_{k+2}(\tau_{0},\tau)}
\Bigr),
\qquad
v:=t^{\star}+r.
\end{equation}
\end{corollary}

\begin{proof}
The pointwise decay bounds for $\psi$ and its first derivatives in the slowly rotating Kerr regime follow by restricting the scalar estimates of \cite[Corollary~3.1]{DRSRKerrIII} to the slow range.
The extension to $\square_{g_{M,a}}\psi=\mathfrak{F}$ follows from the inhomogeneous vector-field estimate, equivalently from a Duhamel representation, with the contribution of the source term controlled by $\|\mathfrak{F}\|_{\mathcal S_{k+2}(\tau_{0},\tau)}$.
Restricting those estimates to the region $\{r\ge R\}$ and translating to the Kerr-star advanced time $v=t^{\star}+r$ gives \eqref{eq:kerr-far-pointwise-scalar}.
\end{proof}

\begin{remark}[Application to the Higgs field]\label{rem:kerr-far-higgs}
In Lorenz gauge, the Higgs field satisfies a covariant wave equation with source terms that are at least quadratic in $(F,D\phi,\phi)$; see \eqref{eq:lorenz-reduced-kerr}.
In the small-data regime of Theorem~\ref{thm:main-slow-kerr-intro}, these quadratic terms are treated as inhomogeneities and estimated using the source norm $\mathcal S_{k}$.
Applying Proposition~\ref{prop:kerr-far-energy-decay} and Corollary~\ref{cor:kerr-far-pointwise-scalar} to the commuted Higgs equation yields far-region decay bounds of the same form as \eqref{eq:kerr-far-pointwise-scalar} (with an arbitrarily small loss in the exponent) for $\phi$ and $D\phi$.
\end{remark}

\subsection{Far-field decay for uncharged Maxwell fields}\label{subsec:kerr-far-maxwell}

Let $G$ be a uncharged Maxwell field on $(\mathcal D_{M,a},g_{M,a})$, solving
$\nabla^{\mu}G_{\mu\nu}=J_{\nu}$ and $\nabla_{[\alpha}G_{\beta\gamma]}=0$.
In the far region we use a null frame $(L,\underline L,e_{1},e_{2})$ adapted to the spheres $S_{\tau,r}$ and write the corresponding null components
\begin{equation}
\alpha_{A}[G]=G(L,e_{A}),\qquad
\underline{\alpha}_{A}[G]=G(\underline L,e_{A}),\qquad
\rho[G]=\tfrac12\,G(L,\underline L),\qquad
\sigma[G]=\tfrac12\,{}^{\star}G(L,\underline L).
\end{equation}
The $r^{p}$ hierarchy for Maxwell fields (cf.\ \eqref{eq:kerr-maxwell-rp-structural}) implies the same polynomial decay mechanism as for scalar waves, after commuting with the admissible symmetry operators in $\mathcal Z$.

\begin{proposition}[Main far-field energy decay for Maxwell fields with current]\phantomsection\label{prop:kerr-far-energy-decay-maxwell}

\noindent\emph{This proposition proves the zero-sector specialization of Theorem~\ref{thm:main-slow-kerr-intro}.}

Fix $k\ge 2$ and let $G$ be a smooth uncharged Maxwell field solving the inhomogeneous Maxwell system on $\mathcal D_{M,a}$ (including the Schwarzschild case $a=0$),
\begin{equation}
\nabla^{\mu}G_{\mu\nu}=J_{\nu},
\qquad
\nabla_{[\alpha}G_{\beta\gamma]}=0,
\end{equation}
with a smooth current $J$ satisfying the compatibility condition $\nabla^{\nu}J_{\nu}=0$.
Then there exists $R\gg1$ such that for every $0<\delta<1$ and all $\tau\ge\tau_{0}$ the following estimate holds.
\begin{equation}\label{eq:kerr-energy-decay-maxwell}
E^{N}_{k}[G](\tau)
\lesssim_{M,a,k,\delta}
E^{N}_{k}[G](\tau_{0})
+\|J\|_{\mathcal S_{k}(\tau_{0},\tau)}^{2}.
\end{equation}
In addition the decaying far-field quantity satisfies
\begin{equation}\label{eq:kerr-energy-decay-maxwell-weighted}
\int_{\Sigma_\tau\cap\{r\ge R\}}r^{-1-\delta}|\mathcal L_{\mathcal Z^{\le k}}G|^{2}\,\dd\mu_{\Sigma_\tau}
\lesssim_{M,a,k,\delta}
(1+\tau)^{-2+2\delta}E^{N}_{k+2}[G](\tau_0)
+\|J\|_{\mathcal S_{k+2}(\tau_0,\tau)}^{2}.
\end{equation}
In the Maxwell-Higgs system in Lorenz gauge \eqref{eq:lorenz-kerr} the wave equation \eqref{eq:lorenz-reduced-kerr} has forcing term $J=J[\phi;A]$ and one \emph{cannot} specialize to the case $J\equiv0$.
The $L^{2}$ size of this coupling term is controlled by the coupled Maxwell-Higgs energies; see \eqref{eq:kerr-J-source-bound}.
\end{proposition}

\begin{proof}
After subtraction of the stationary Coulomb Maxwell solution on slowly rotating Kerr, the nondegenerate energy boundedness together with local/radiation decay for uncharged Maxwell fields are supplied by \cite{AnderssonBlueMaxwellSlowKerr}.
In the homogeneous case $J\equiv0$, commuting with the admissible symmetry operators $\mathcal Z$ and translating the energies used there to our $N$-energy yields boundedness of the full radiative energy and decay of the local/weighted far-field quantities. This gives \eqref{eq:kerr-energy-decay-maxwell} and \eqref{eq:kerr-energy-decay-maxwell-weighted} without the source term.

For a general inhomogeneous current $J$, the inhomogeneous physical-space estimate, equivalently a Duhamel representation, yields the additional contribution $\|J\|_{\mathcal S_{k+2}(\tau_{0},\tau)}^{2}$.
This is the inhomogeneous estimate encoded in the linear estimates collected in $\Lin_{k}$; see Proposition~\ref{prop:kerr-integrated-decay}.

The final statement about Maxwell-Higgs follows from the energy/Sobolev control of the current established in Proposition~\ref{prop:kerr-far-energy-decay}, namely \eqref{eq:kerr-J-source-bound}.
\end{proof}

\begin{corollary}[Pointwise decay in the far region for Maxwell]\label{cor:kerr-far-pointwise-maxwell}
Under the assumptions of Proposition~\ref{prop:kerr-far-energy-decay-maxwell} and for $k\ge 6$, there exist $R\gg1$ such that for every $0<\delta<1$ one can find $C=C(M,a,k,\delta)$ with the following property: for all $(\tau,r,\omega)$ with $r\ge R$,
\begin{equation}\label{eq:kerr-far-pointwise-maxwell}
|\alpha[G]|+|\underline{\alpha}[G]|+|\rho[G]|+|\sigma[G]|
\le
\frac{C}{(1+v_{+})^{1-\delta}}
\Bigl(
E^{N}_{k}[G](\tau_{0})^{1/2}
+
\|J\|_{\mathcal S_{k+2}(\tau_{0},\tau)}
\Bigr),
\qquad
v:=t^{\star}+r.
\end{equation}
\end{corollary}

\begin{proof}
Pointwise decay for the Maxwell system in the slowly rotating Kerr exterior, after Coulomb subtraction, follows from the slowly rotating Maxwell asymptotic theory of the estimates of \cite{AnderssonBlueMaxwellSlowKerr}, together with the reconstruction of the Coulomb-subtracted radiative components.
The extension to the inhomogeneous system $\nabla^{\mu}G_{\mu\nu}=J_{\nu}$ follows from the inhomogeneous Maxwell estimate in Proposition~\ref{prop:kerr-far-energy-decay-maxwell}, equivalently from Duhamel's formula, with the contribution of the current controlled by $\|J\|_{\mathcal S_{k+2}(\tau_{0},\tau)}$.
Restricting to $\{r\ge R\}$ yields \eqref{eq:kerr-far-pointwise-maxwell}.
\end{proof}

\begin{remark}[Application to Maxwell-Higgs]\label{rem:kerr-far-mh}
In the zero-sector rotating theorem one has $Q_e=0$ and therefore $\widetilde F=F$.  The following Coulomb-subtracted form is retained only to match the linear Maxwell literature: after removing a stationary Coulomb Maxwell field one works with the radiative remainder
\begin{equation}
\widetilde F:=F-F^{(a),\mathrm C}_{Q_e}
\end{equation}
(Remark~\ref{rem:kerr-data-charges}).
Then $\widetilde F$ is uncharged but solves the \emph{sourced} Maxwell system
\begin{equation}
\nabla^{\mu}\widetilde F_{\mu\nu}=J_{\nu}[\phi;A],
\qquad
\nabla_{[\alpha}\widetilde F_{\beta\gamma]}=0,
\end{equation}
where the current $J[\phi;A]$ is given by \eqref{eq:MH-current-prop42} and is nonzero in general.
Applying Proposition~\ref{prop:kerr-far-energy-decay-maxwell} and Corollary~\ref{cor:kerr-far-pointwise-maxwell} to $\widetilde F$ yields far-region decay for the radiative Maxwell field, with the current contribution controlled by the coupled Maxwell-Higgs energies via \eqref{eq:kerr-J-source-bound}.
Remark~\ref{rem:kerr-far-higgs} yields the corresponding far-region decay for $\phi$ and $D\phi$.
\end{remark}

\section{Decay in the trapped and near-horizon region on Kerr}\label{sec:kerr-near-decay}

In this section we state the zero-sector near-horizon and trapped-region estimates used in the slowly rotating Kerr theorem.

The main quantitative conclusions in this section are Proposition~\ref*{prop:kerr-horizon-energy-decay} (near-horizon redshift decay in the zero sector) and Proposition~\ref{prop:kerr-trapped-iled} (trapped-region integrated local energy decay together with tame source bounds). Their role is to provide the zero-sector regional estimates used in Theorem~\ref{thm:main-slow-kerr-intro}.

On Kerr the trapped set is not confined to a single radius; rather it is a codimension-one subset of phase space (the photon region). This is reflected in physical space by the trapping weight $w_{\mathrm{trap}}$ appearing in the Morawetz bulk norms in $\Lin_{k}$ (Definition~\ref{def:lin-estimates}); see also the representative bulk densities \eqref{eq:kerr-morawetz-structural} and \eqref{eq:kerr-maxwell-morawetz-structural}. The bounds below follow of $\Lin_{k}$ in the slowly rotating regime; we state them in a form convenient for the nonlinear bootstrap and for the construction of horizon radiation fields.

\noindent\textbf{Linear condition.} The near-horizon and trapped-set estimates are part of the zero-sector estimates $\Lin_{k}$ and the linear results cited in Section~\ref{sec:kerr-extension}.
\medskip
\noindent\textbf{Main compact-region decay statements.}
The near-horizon estimate (Proposition~\ref*{prop:kerr-horizon-energy-decay}) provides the redshift control needed up to $\mathcal H^{+}$ and is used later in the construction of horizon radiation fields. The trapped-set integrated decay estimate Proposition~\ref{prop:kerr-trapped-iled} controls the Morawetz bulk norm on $\mathcal K_{R}$ and, in the Maxwell-Higgs system, yields tame bounds for the nonlinear sources $J$ and $\mathfrak F_{\phi}^{(m)}$ in the admissible forcing norm. As in Section~\ref{sec:kerr-far-decay}, the Schwarzschild case $a=0$ is included; in Schwarzschild one may take the trapping weight $w_{\mathrm{trap}}$ to vanish only at the photon sphere $r=3M$, and a complete proof of the corresponding compact-region decay is given in Section~\ref{sec:decaynear} (see also Corollary~\ref{cor:near-horizon}).
\medskip

Fix $R>r_{+}$, and for $\tau\in\mathbb R$ set
\begin{equation}
\Sigma_{\tau}^{\le R}:=\Sigma_{\tau}\cap\{r\le R\}.
\end{equation}
For a scalar field $\psi$ and an integer $k\ge0$ we define the local (nondegenerate) $N$-energy on $\Sigma_{\tau}^{\le R}$ by
\begin{equation}\label{eq:kerr-local-energy}
E^{N}_{k}[\psi](\tau;\,r\le R)
:=
\sum_{|\alpha|\le k}\int_{\Sigma_{\tau}^{\le R}} J^{N}_{\mu}[\mathcal Z^{\alpha}\psi]\,n^{\mu}\,\dd\mu_{\Sigma_{\tau}},
\end{equation}
and similarly for Maxwell fields (with $J^{N}_{\mu}[G]$ in place of $J^{N}_{\mu}[\psi]$).

\subsection{Sobolev estimates on compact subsets of the Kerr-star slices}\label{subsec:kerr-compact-sobolev}

\begin{lemma}[Sobolev on $\Sigma_{\tau}^{\le R}$]\label{lem:kerr-sobolev-compact}
Fix $R>r_{+}$.
There exists a constant $C=C(M,a,R)$ such that for every smooth complex-valued function $f$ on $\Sigma_{\tau}^{\le R}$,
\begin{equation}\label{eq:kerr-sobolev-compact}
\|f\|_{L^{\infty}(\Sigma_{\tau}^{\le R})}^{2}
\le
C\sum_{j=0}^{2}\int_{\Sigma_{\tau}^{\le R}}|\nabla^{j}f|^{2}\,\dd\mu_{\Sigma_{\tau}}.
\end{equation}
\end{lemma}

\begin{proof}
Since $\Sigma_{\tau}^{\le R}$ is a smooth compact $3$-manifold with boundary, Sobolev embedding $H^{2}\hookrightarrow L^{\infty}$ holds with constants depending only on finitely many bounds for the induced geometry.
By stationarity the induced geometries on $\Sigma_{\tau}^{\le R}$ are uniformly equivalent in $\tau$, which yields \eqref{eq:kerr-sobolev-compact}.
\end{proof}

\subsection{Near-horizon decay from the redshift effect}\label{subsec:kerr-near-horizon-decay}

Let $r_{0}>0$ be fixed so that $\mathcal N_{\mathcal H}:=\{r_{+}\le r\le r_{+}+r_{0}\}$ is contained in the redshift region where the multiplier $N$ from Definition~\ref{def:admissible-exterior} is uniformly timelike.
In this region the redshift identity yields a coercive bulk term controlling derivatives transverse to $\mathcal H^{+}$, and therefore nondegenerate decay statements.

\begin{proposition}[Near-horizon energy decay for scalar waves with forcing]\phantomsection\label{prop:kerr-horizon-energy-decay}

\noindent\emph{This proposition proves the near-horizon part of the zero-sector specialization of Theorem~\ref{thm:main-slow-kerr-intro}.}
Fix $k\ge2$ and let $\psi$ solve the inhomogeneous wave equation
\begin{equation}
\square_{g_{M,a}}\psi=\mathfrak{F}
\qquad\text{on }\mathcal D_{M,a}.
\end{equation}
Then for every $0<\delta<1$ and all $\tau\ge\tau_{0}$,
\begin{equation}\label{eq:kerr-horizon-energy-decay}
E^{N}_{k}[\psi](\tau;\,r\le r_{+}+r_{0})
\lesssim_{M,a,k,\delta}
(1+\tau)^{-2+2\delta}\,E^{N}_{k+2}[\psi](\tau_{0})
+\|\mathfrak{F}\|_{\mathcal S_{k+2}(\tau_{0},\tau)}^{2}.
\end{equation}
In addition, the corresponding redshift bulk term is integrable in time:
\begin{equation}\label{eq:kerr-horizon-bulk}
\int_{\tau_{0}}^{\infty}\int_{\Sigma_{\tau}\cap\mathcal N_{\mathcal H}}\sum_{|\alpha|\le k}
|\nabla \mathcal Z^{\alpha}\psi|^{2}\,\dd\mu_{\Sigma_{\tau}}\,\dd\tau
\lesssim_{M,a,k}
\Bigl(
E^{N}_{k+2}[\psi](\tau_{0})
+
\|\mathfrak{F}\|_{\mathcal S_{k+2}(\tau_{0},\infty)}^{2}
\Bigr).
\end{equation}
\end{proposition}

\begin{proof}
The estimate \eqref{eq:kerr-horizon-energy-decay} is the local redshift-decay component of the scalar Kerr estimates: one combines the redshift multiplier in $\mathcal N_{\mathcal H}$ with the time-weighted/local-energy decay supplied by the $r^p$ hierarchy and commutes with $\mathcal Z^{\le k}$. Equivalently, it follows from the same local decay theorem that yields the pointwise estimates of the cited Kerr scalar estimates, with the inhomogeneous term handled by the source norm. No decay of the full nondegenerate slice energy is used or asserted.

For the integrated near-horizon bulk term, apply Proposition~\ref{prop:kerr-integrated-decay} with index $k+2$ to $\square_{g_{M,a}}\psi=\mathfrak{F}$ on the slab $(\tau_{0},\tau_{1})$ and use the concrete norms of Lemma~\ref{lem:concrete-admissible-norms}.
The redshift part of $\|\psi\|_{\mathcal X_{k+2}(\tau_{0},\tau_{1})}^{2}$ controls
\begin{equation}
\int_{\tau_{0}}^{\tau_{1}}\int_{\Sigma_{\tau}\cap\mathcal N_{\mathcal H}}\sum_{|\alpha|\le k+2}
|\nabla \mathcal Z^{\alpha}\psi|^{2},
\end{equation}
and therefore in this case the sum over $|\alpha|\le k$.
Consequently,
\begin{equation}
\int_{\tau_{0}}^{\tau_{1}}\int_{\Sigma_{\tau}\cap\mathcal N_{\mathcal H}}\sum_{|\alpha|\le k}
|\nabla \mathcal Z^{\alpha}\psi|^{2}
\lesssim
E^{N}_{k+2}[\psi](\tau_{0})
+
\|\mathfrak{F}\|_{\mathcal S_{k+2}(\tau_{0},\tau_{1})}^{2}.
\end{equation}
Letting $\tau_{1}\to\infty$ and using monotone convergence yields \eqref{eq:kerr-horizon-bulk}.
\end{proof}

\begin{corollary}[Pointwise decay near the horizon for scalar waves]\label{cor:kerr-horizon-pointwise-scalar}
Under the assumptions of Proposition~\ref*{prop:kerr-horizon-energy-decay} and for $k\ge6$, there exist $r_{0}>0$ such that for every $0<\delta<1$ one can find $C=C(M,a,k,\delta)$ with the following property: for all $(\tau,r,\omega)$ with $r\le r_{+}+r_{0}$,
\begin{equation}\label{eq:kerr-horizon-pointwise-scalar}
|\psi|+|\nabla\psi|
\le
\frac{C}{(1+\tau)^{1-\delta}}
\Bigl(
E^{N}_{k+2}[\psi](\tau_{0})^{1/2}
+
\|\mathfrak{F}\|_{\mathcal S_{k+2}(\tau_{0},\tau)}
\Bigr).
\end{equation}
\end{corollary}

\begin{proof}
The pointwise decay bounds for $\psi$ and its first derivatives in the slowly rotating Kerr regime follow by restricting the scalar estimates of \cite[Corollary~3.1]{DRSRKerrIII} to the slow range.
The extension to $\square_{g_{M,a}}\psi=\mathfrak{F}$ follows from the inhomogeneous vector-field estimate, equivalently from a Duhamel representation, with the contribution of the source term controlled by $\|\mathfrak{F}\|_{\mathcal S_{k+2}(\tau_{0},\tau)}$.
Restricting to the compact region $\{r\le r_{+}+r_{0}\}$ and using $v=t^{\star}+r=\tau+O(1)$ there yields \eqref{eq:kerr-horizon-pointwise-scalar}.
\end{proof}

\begin{corollary}[Pointwise decay near the horizon for Maxwell with current]\label{cor:kerr-horizon-pointwise-maxwell}
Fix $k\ge6$ and let $G$ be a smooth uncharged Maxwell field solving the inhomogeneous Maxwell system
\begin{equation}
\nabla^{\mu}G_{\mu\nu}=J_{\nu},\qquad \nabla_{[\alpha}G_{\beta\gamma]}=0
\end{equation}
 on $\mathcal D_{M,a}$ with a smooth conserved current $\nabla^{\nu}J_{\nu}=0$.
Then for every $0<\delta<1$ there exist $r_{0}>0$ and $C=C(M,a,k,\delta)$ such that for all $(\tau,r,\omega)$ with $r\le r_{+}+r_{0}$,
\begin{equation}\label{eq:kerr-horizon-pointwise-maxwell}
|\alpha[G]|+|\underline{\alpha}[G]|+|\rho[G]|+|\sigma[G]|
\le
\frac{C}{(1+\tau)^{1-\delta}}
\Bigl(
E^{N}_{k+2}[G](\tau_{0})^{1/2}
+
\|J\|_{\mathcal S_{k+2}(\tau_{0},\tau)}
\Bigr).
\end{equation}
\end{corollary}

\begin{proof}
Fix $r_{0}>0$.
As in Corollary~\ref{cor:kerr-horizon-pointwise-scalar}, we apply a Sobolev inequality on the compact subset $\Sigma_{\tau}\cap\{r\le r_{+}+r_{0}\}$ to the components of $G$ in a smooth null frame.
Since the frame is smooth and $r$ is bounded, the pointwise size of the null components is controlled by finitely many $L^{2}$ norms of covariant derivatives of $G$ on $\Sigma_{\tau}\cap\{r\le r_{+}+r_{0}\}$, therefore by $E^{N}_{k}[G](\tau;\,r\le r_{+}+r_{0})^{1/2}$ for $k\ge6$.
The required local $L^{2}$ decay near the horizon is the redshift/local-decay component of the uncharged Maxwell estimates, applied to the radiative part of $G$ after the stationary Coulomb mode has been removed. Taking square roots of that local estimate and applying the compact Sobolev inequality gives \eqref{eq:kerr-horizon-pointwise-maxwell}. This argument does not use decay of the full nondegenerate Maxwell energy on complete slices.
\end{proof}

\subsection{Decay in the trapped region}\label{subsec:kerr-trapped-decay}

Set $\mathcal K_{R}:=\{r_{+}+r_{0}\le r\le R\}$, which contains the photon region and therefore all trapping.
The integrated local energy decay component of $\Lin_{k}$ controls a Morawetz bulk norm which degenerates at trapping. In Kerr this is a microlocal norm with principal symbol vanishing on the trapped null geodesic set; in Schwarzschild it can be represented by a physical-space weight $w_{\mathrm{trap}}(r)$ vanishing at $r=3M$.

\begin{proposition}[Main trapped-region integrated decay and tame control of Maxwell-Higgs sources]\phantomsection\label{prop:kerr-trapped-iled}

\noindent\emph{This proposition proves the trapped-region part of the zero-sector specialization of Theorem~\ref{thm:main-slow-kerr-intro}.}

Fix $k\ge0$ and let $\psi$ solve the inhomogeneous wave equation
\begin{equation}
\square_{g_{M,a}}\psi=\mathfrak{F}
\qquad\text{on }\mathcal D_{M,a}.
\end{equation}
This includes the Schwarzschild exterior $a=0$ as a special case.
Then for every $\tau_{1}\ge\tau_{0}$,
\begin{eqnarray}\label{eq:kerr-trapped-iled}
&\bigl\|\operatorname{Op}(w_{\mathrm{trap}})\,\nabla\mathcal Z^{\le k}\psi\bigr\|_{L^{2}([\tau_0,\tau_1]\times(\Sigma_\tau\cap\mathcal K_R))}^{2}
+\bigl\|\operatorname{Op}(w_{\mathrm{trap}})\,\mathcal Z^{\le k}\psi\bigr\|_{L^{2}([\tau_0,\tau_1]\times(\Sigma_\tau\cap\mathcal K_R))}^{2}
\\
&\qquad\lesssim_{M,a,k,R}
E^{N}_{k}[\psi](\tau_{0})
+
\|\mathfrak{F}\|_{\mathcal S_{k}(\tau_{0},\tau_{1})}^{2}.
\end{eqnarray}
An analogous statement holds for uncharged Maxwell fields $G$ solving $\nabla^{\mu}G_{\mu\nu}=J_{\nu}$,
with the corresponding Maxwell microlocal Morawetz bulk and with $\|J\|_{\mathcal S_{k}(\tau_{0},\tau_{1})}^{2}$ in place of $\|\mathfrak{F}\|_{\mathcal S_{k}(\tau_{0},\tau_{1})}^{2}$.

\smallskip
Assume in addition that $k\ge6$ and let $(A,\phi)$ be a Lorenz-gauge solution of the Maxwell-Higgs system~\eqref{eq:lorenz-reduced-kerr} on $\mathcal R(\tau_{0},\tau_{1})$,
with nonlinearities $\mathfrak{F}_{\phi}^{(m)}$ and $J_{\nu}$ defined by \eqref{eq:def-Fphi}-\eqref{eq:def-J}.
Define the source-weighted energy functionals $\mathbf E_{m}^{(\mathrm A)}$, $\mathbf E_{m}^{(\mathrm H)}$, $\mathbf F_{m}^{(\mathrm A)}$, $\mathbf F_{m}^{(\mathrm H)}$ as in
\eqref{eq:def-slab-energy-A}-\eqref{eq:def-source-weighted-energy-H}
and the tame polynomial $\mathcal P_{k,P}$ as in \eqref{eq:def-tame-poly}.
Then on the slab $\mathcal R(\tau_{0},\tau_{1})$ the Maxwell-Higgs nonlinearities satisfy
\begin{eqnarray}
\|J\|_{\mathcal S_{k+2}(\tau_{0},\tau_{1})}^{2}
&\lesssim_{M,a,k}
\Bigl(\mathbf E_{k+4}^{(\mathrm A)}(\tau_{0},\tau_{1})
 +\mathbf E_{k+4}^{(\mathrm H)}(\tau_{0},\tau_{1})\Bigr)\,
\mathbf F_{k+2}^{(\mathrm H)}(\tau_{0},\tau_{1}),
\label{eq:kerr-J-source-bound-trap}\\
\|\mathfrak{F}_{\phi}^{(m)}\|_{\mathcal S_{k+2}(\tau_{0},\tau_{1})}^{2}
&\lesssim_{M,a,k,P}
\mathcal P_{k,P}\!\Bigl(
\mathbf E_{k+4}^{(\mathrm A)}(\tau_{0},\tau_{1}),
\mathbf E_{k+4}^{(\mathrm H)}(\tau_{0},\tau_{1})
\Bigr)\nonumber\\
&\quad\times
\Bigl(\mathbf F_{k+2}^{(\mathrm A)}(\tau_{0},\tau_{1})
+\mathbf F_{k+2}^{(\mathrm H)}(\tau_{0},\tau_{1})\Bigr).
\label{eq:kerr-higgs-source-bound-trap}
\end{eqnarray}
Finiteness of these energy functionals implies that $\mathfrak{F}_{\phi}^{(m)}$ and $J$ are admissible sources in the sense of Definition~\ref{def:lin-estimates}.
\end{proposition}

\begin{proof}
Apply Proposition~\ref{prop:kerr-integrated-decay} with index $k$ to $\square_{g_{M,a}}\psi=\mathfrak{F}$ on the slab $\mathcal R(\tau_{0},\tau_{1})$.
For the concrete norms of Lemma~\ref{lem:concrete-admissible-norms}, the Morawetz bulk term in $\|\psi\|_{\mathcal X_{k}(\tau_{0},\tau_{1})}^{2}$ contains the trapped-region contribution. In the Kerr case this contribution is microlocal:
\begin{equation}
\bigl\|\operatorname{Op}(w_{\mathrm{trap}})\nabla\mathcal Z^{\le k}\psi\bigr\|_{L^{2}([\tau_0,\tau_1]\times(\Sigma_\tau\cap\mathcal K_R))}^{2}
+
\bigl\|\operatorname{Op}(w_{\mathrm{trap}})\mathcal Z^{\le k}\psi\bigr\|_{L^{2}([\tau_0,\tau_1]\times(\Sigma_\tau\cap\mathcal K_R))}^{2}
\lesssim
\|\psi\|_{\mathcal X_{k}(\tau_{0},\tau_{1})}^{2}.
\end{equation}
In Schwarzschild the same notation is represented by multiplication by a physical-space weight vanishing at $r=3M$.
Using the estimate of Proposition~\ref{prop:kerr-integrated-decay} yields
\begin{equation}
\|\psi\|_{\mathcal X_{k}(\tau_{0},\tau_{1})}^{2}
\lesssim
E^{N}_{k}[\psi](\tau_{0})
+
\|\mathfrak{F}\|_{\mathcal S_{k}(\tau_{0},\tau_{1})}^{2},
\end{equation}
which implies \eqref{eq:kerr-trapped-iled}.
The Maxwell case is identical with $\psi,\mathfrak{F},\mathfrak e$ replaced by $G,J,\mathfrak e_{\mathrm{Max}}$.

\smallskip
We now prove the tame source bounds
\eqref{eq:kerr-J-source-bound-trap}-\eqref{eq:kerr-higgs-source-bound-trap}.
Fix $k\ge 6$ and set $m:=k+2$.

\smallskip\noindent
\textbf{Preliminary bounds: $L^\infty$ control from slice energies.}
By Sobolev embedding on $\Sigma_{\tau'}$ (Lemma~\ref{lem:kerr-sobolev-compact} on $\{r\le R\}$ together with Lemma~\ref{lem:kerr-sobolev-sphere} in the far region),
for every $\tau'\in[\tau_{0},\tau_{1}]$ one has
\begin{eqnarray}
\|\mathcal Z^{\le k}A(\tau')\|_{W^{1,\infty}(\Sigma_{\tau'})}^{2}
&\lesssim_{M,a,k}
\mathcal E_{N,k+4}^{(\mathrm A)}[A](\Sigma_{\tau'})
\le \mathbf E_{k+4}^{(\mathrm A)}(\tau_0,\tau_1),\\
\|\mathcal Z^{\le k}\phi(\tau')\|_{L^{\infty}(\Sigma_{\tau'})}^{2}
+\|D\mathcal Z^{\le k}\phi(\tau')\|_{L^{\infty}(\Sigma_{\tau'})}^{2}
&\lesssim_{M,a,k}
\mathcal E_{N,k+4}^{(\mathrm H)}[\phi,A](\Sigma_{\tau'})
\le \mathbf E_{k+4}^{(\mathrm H)}(\tau_0,\tau_1).
\end{eqnarray}

\smallskip\noindent
\textbf{the current $J_\nu$.}
Since $J_\nu=2\,\Im(\overline\phi\,D_\nu\phi)$, for any multiindex $\alpha$ with $|\alpha|\le m$ we commute and expand:
\begin{equation}
\mathcal Z^\alpha J_\nu
=
2\,\Im\!\!\sum_{\beta+\gamma=\alpha}
\overline{\mathcal Z^\beta\phi}\,D_\nu(\mathcal Z^\gamma\phi)
\;+\;
\mathcal C_{\nu,\alpha}[A,\phi],
\end{equation}
where $\mathcal C_{\nu,\alpha}[A,\phi]$ collects the commutator contributions from commuting $\mathcal Z$ through the covariant derivative $D$.
Since $D=\nabla-iA$, each commutator $[D,\mathcal Z]$ is a first-order operator with coefficients depending on the background and on $\nabla\mathcal Z$ and $\mathcal Z A$.
Consequently, $\mathcal C_{\nu,\alpha}[A,\phi]$ is a finite sum of terms up to lower-order terms of the form
\begin{equation}
(\nabla\mathcal Z^{\beta_1}A)\,(\mathcal Z^{\beta_2}\phi)\,(\overline{\mathcal Z^{\beta_3}\phi}),
\qquad |\beta_1|+|\beta_2|+|\beta_3|\le m-1.
\end{equation}
Because $m=k+2$, in every product above at least one factor carries at most $k$ commutators and can be estimated in $L^\infty$ using the preliminary bounds above.
We therefore obtain the pointwise tame bound, at each point of $\mathcal R(\tau_{0},\tau_{1})$,
\begin{equation}
|\mathcal Z^\alpha J|
\lesssim_{M,a,k}
\Bigl(\|\mathcal Z^{\le k}\phi\|_{L^\infty(\Sigma_{\tau'})}
+\|\mathcal Z^{\le k}A\|_{W^{1,\infty}(\Sigma_{\tau'})}\Bigr)
\sum_{|\beta|\le m}\Bigl(|D\mathcal Z^\beta\phi|+|\mathcal Z^\beta\phi|\Bigr).
\end{equation}
Multiplying by $r^{(1+\delta)/2}$, squaring, integrating over $\mathcal R(\tau_{0},\tau_{1})$, and summing over $|\alpha|\le m$ (using the definition \eqref{eq:concrete-source-norm}) yields
\begin{equation}
\|J\|_{\mathcal S_{m}(\tau_0,\tau_1)}^{2}
\lesssim_{M,a,k}
\Bigl(
\sup_{\tau'\in[\tau_0,\tau_1]}\|\mathcal Z^{\le k}\phi(\tau')\|_{L^\infty(\Sigma_{\tau'})}^{2}
+
\sup_{\tau'\in[\tau_0,\tau_1]}\|\mathcal Z^{\le k}A(\tau')\|_{W^{1,\infty}(\Sigma_{\tau'})}^{2}
\Bigr)\,
\mathbf F_{m}^{(\mathrm H)}(\tau_0,\tau_1).
\end{equation}
Inserting the preliminary bounds above and recalling $m=k+2$ gives \eqref{eq:kerr-J-source-bound-trap}.

\smallskip\noindent
\textbf{the Higgs forcing $\mathfrak F_\phi^{(m)}$.}
In Lorenz gauge, $\nabla^\mu A_\mu=0$, therefore
\begin{equation}
\mathfrak F_\phi^{(m)}
=
2iA^\mu\nabla_\mu\phi
+A^\mu A_\mu\,\phi
+\mathcal R_P(\phi).
\end{equation}
Using $\nabla\phi=D\phi-iA\phi$ we have the structural pointwise bound
\begin{equation}
|\mathfrak F_\phi^{(m)}|
\lesssim
|A|\,|D\phi|+|A|^{2}|\phi|+|\mathcal R_P(\phi)|.
\end{equation}
We estimate the three contributions separately in $\mathcal S_m(\tau_{0},\tau_{1})$.

\smallskip\noindent
\emph{(i) The term $A\,D\phi$.}
Commuting $\mathcal Z^{\le m}$ through the product and using Leibniz' rule yields terms of the form
$(\mathcal Z^{\beta}A)\,D(\mathcal Z^{\gamma}\phi)$ with $|\beta|+|\gamma|\le m$,
as well as lower-order commutator terms coming from $[D,\mathcal Z]$ which have the same structure as above and are treated in the same way.
Since $m=k+2$, in each product one factor carries at most $k$ commutators.
Consequently,
\begin{eqnarray}
\|A\,D\phi\|_{\mathcal S_{m}(\tau_0,\tau_1)}^{2}
\lesssim_{M,a,k}
&\Bigl(
\sup_{\tau'\in[\tau_0,\tau_1]}\|\mathcal Z^{\le k}A(\tau')\|_{L^\infty(\Sigma_{\tau'})}^{2}
\Bigr)\,
\mathbf F_{m}^{(\mathrm H)}(\tau_0,\tau_1)\\
&+\Bigl(
\sup_{\tau'\in[\tau_0,\tau_1]}\|D\mathcal Z^{\le k}\phi(\tau')\|_{L^\infty(\Sigma_{\tau'})}^{2}
\Bigr)\,
\mathbf F_{m}^{(\mathrm A)}(\tau_0,\tau_1).
\end{eqnarray}
Using the preliminary bounds gives
\begin{equation}
\|A\,D\phi\|_{\mathcal S_{m}(\tau_0,\tau_1)}^{2}
\lesssim_{M,a,k}
\Bigl(\mathbf E_{k+4}^{(\mathrm A)}(\tau_0,\tau_1)+\mathbf E_{k+4}^{(\mathrm H)}(\tau_0,\tau_1)\Bigr)\,
\Bigl(\mathbf F_{m}^{(\mathrm A)}(\tau_0,\tau_1)+\mathbf F_{m}^{(\mathrm H)}(\tau_0,\tau_1)\Bigr).
\end{equation}

\smallskip\noindent
\emph{(ii) The cubic term $A^{2}\phi$.}
Expanding $\mathcal Z^{\le m}(A^{2}\phi)$ by Leibniz' rule yields products of the form
$(\mathcal Z^{\beta_1}A)(\mathcal Z^{\beta_2}A)(\mathcal Z^{\gamma}\phi)$ with $|\beta_1|+|\beta_2|+|\gamma|\le m$.
Since $m=k+2$, at least two factors carry at most $k$ commutators and can be estimated in $L^\infty$ by the preliminary bounds, giving
\begin{eqnarray}
\|A^{2}\phi\|_{\mathcal S_{m}(\tau_0,\tau_1)}^{2}
&\lesssim_{M,a,k}&
\Bigl(\sup_{\tau'\in[\tau_0,\tau_1]}
\|\mathcal Z^{\le k}A(\tau')\|_{L^\infty(\Sigma_{\tau'})}^{4}\Bigr)
\mathbf F_{m}^{(\mathrm H)}(\tau_0,\tau_1)\\
&&+\Bigl(\sup_{\tau'\in[\tau_0,\tau_1]}
\|\mathcal Z^{\le k}A(\tau')\|_{L^\infty(\Sigma_{\tau'})}^{2}\Bigr)\\
&&\quad\times
\Bigl(\sup_{\tau'\in[\tau_0,\tau_1]}
\|\mathcal Z^{\le k}\phi(\tau')\|_{L^\infty(\Sigma_{\tau'})}^{2}\Bigr)
\mathbf F_{m}^{(\mathrm A)}(\tau_0,\tau_1).
\end{eqnarray}
Using the preliminary bounds yields the structural estimate
\begin{eqnarray}
\|A^{2}\phi\|_{\mathcal S_{m}(\tau_0,\tau_1)}^{2}
\lesssim_{M,a,k}
&\Bigl(
\mathbf E_{k+4}^{(\mathrm A)}(\tau_0,\tau_1)
+\bigl(\mathbf E_{k+4}^{(\mathrm A)}(\tau_0,\tau_1)\bigr)^{2}\\
&\qquad
+\mathbf E_{k+4}^{(\mathrm A)}(\tau_0,\tau_1)\,\mathbf E_{k+4}^{(\mathrm H)}(\tau_0,\tau_1)
\Bigr)\\
&\times
\Bigl(
\mathbf F_{m}^{(\mathrm A)}(\tau_0,\tau_1)
+\mathbf F_{m}^{(\mathrm H)}(\tau_0,\tau_1)
\Bigr).
\end{eqnarray}

\smallskip\noindent
\emph{(iii) The nonlinear potential remainder $\mathcal R_P(\phi)$.}
The mass term $m^{2}\phi$ belongs to the linear Klein-Gordon comparison operator and is not estimated as a nonlinear source. By Lemma~\ref{lem:potential-tame}, for $|\alpha|\le m$ the commuted quantity $\mathcal Z^\alpha\mathcal R_P(\phi)$ is a finite sum of products with at least three factors drawn from $\{\mathcal Z^{\beta}\phi,\overline{\mathcal Z^{\beta}\phi}\}$,
multiplied by smooth coefficients which grow at most like $(1+|\phi|^{2N_P})$.
Since $m=k+2$, in each product at least two factors carry at most $k$ commutators; estimating those factors in $L^\infty$ and the remaining factor in $L^2$ yields
\begin{eqnarray}
\|\mathcal R_P(\phi)\|_{\mathcal S_{m}(\tau_0,\tau_1)}^{2}
\lesssim_{k,P}
&\Bigl(1+\sup_{\tau'\in[\tau_0,\tau_1]}\|\mathcal Z^{\le k}\phi(\tau')\|_{L^\infty(\Sigma_{\tau'})}^{2N_P}\Bigr)^{2}\\
&\times
\sup_{\tau'\in[\tau_0,\tau_1]}\|\mathcal Z^{\le k}\phi(\tau')\|_{L^\infty(\Sigma_{\tau'})}^{4}\,
\mathbf F_{m}^{(\mathrm H)}(\tau_0,\tau_1).
\end{eqnarray}
Using the preliminary bounds gives
\begin{equation}
\|\mathcal R_P(\phi)\|_{\mathcal S_{m}(\tau_0,\tau_1)}^{2}
\lesssim_{M,a,k,P}
\Bigl(1+\mathbf E_{k+4}^{(\mathrm H)}(\tau_0,\tau_1)^{N_P}\Bigr)^{2}\,
\bigl(\mathbf E_{k+4}^{(\mathrm H)}(\tau_0,\tau_1)\bigr)^{2}\,
\mathbf F_{m}^{(\mathrm H)}(\tau_0,\tau_1).
\end{equation}

\smallskip
Combining (i)-(iii), and recalling $m=k+2$, yields \eqref{eq:kerr-higgs-source-bound-trap} with $\mathcal P_{k,P}$ as in \eqref{eq:def-tame-poly}.
\end{proof}

\begin{proposition}[Compact local slice decay in the zero Kerr sector]\label{prop:kerr-compact-local-slice-decay}
Fix $R>r_{+}$ and $k\ge2$. Let $\psi$ solve $\square_{g_{M,a}}\psi=\mathfrak F$ and let $G$ be a smooth uncharged Maxwell field solving
\begin{equation}
\nabla^{\mu}G_{\mu\nu}=J_{\nu},\qquad \nabla_{[\alpha}G_{\beta\gamma]}=0,
\qquad \nabla^\nu J_\nu=0.
\end{equation}
for every $0<\delta<1$, the uncharged Kerr linear estimates give the local slice estimates
\begin{eqnarray}
\int_{\Sigma_\tau\cap\{r\le R\}}
\sum_{|\alpha|\le k}\bigl(|\nabla \mathcal Z^\alpha\psi|^2+|\mathcal Z^\alpha\psi|^2\bigr)
\,\dd\mu_{\Sigma_\tau}
&\lesssim_{M,a,R,k,\delta}
(1+\tau)^{-2+2\delta}E^N_{k+2}[\psi](\tau_0)
+\|\mathfrak F\|_{\mathcal S_{k+2}(\tau_0,\tau)}^2,
\label{eq:kerr-compact-local-scalar}\\
\int_{\Sigma_\tau\cap\{r\le R\}}
\sum_{|\alpha|\le k}|\mathcal L_{\mathcal Z^\alpha}G|^2
\,\dd\mu_{\Sigma_\tau}
&\lesssim_{M,a,R,k,\delta}
(1+\tau)^{-2+2\delta}E^N_{k+2}[G](\tau_0)
+\|J\|_{\mathcal S_{k+2}(\tau_0,\tau)}^2.
\label{eq:kerr-compact-local-maxwell}
\end{eqnarray}
These are local/weighted decay estimates. They do not assert decay of the full nondegenerate energy on the complete Cauchy slice.
\end{proposition}

\begin{proof}
For the scalar equation this is the local-energy decay consequence of the Kerr wave estimates restricted to the slow range: one combines the redshift estimate near $\mathcal H^+$, the trapping-degenerate Morawetz estimate on compact regions away from the horizon, and the $r^p$ hierarchy in the exterior region, with the inhomogeneous contribution measured in the source norm. Cover $\{r\le R\}$ by the redshift neighbourhood and finitely many compact subregions away from the horizon. On each subregion the local slice decay follows from the time-weighted local energy estimate of the estimates; summing the finite cover gives \eqref{eq:kerr-compact-local-scalar}.

For Maxwell, after removal of the stationary Coulomb charge, the uncharged Maxwell estimates of Theorem~\ref{thm:maxwell-scattering-kerr} has the same redshift, trapping-degenerate Morawetz, and radiation hierarchy components for the radiative Maxwell field. Applying those estimates to $\mathcal L_{\mathcal Z^\alpha}G$ and using the conserved current compatibility gives \eqref{eq:kerr-compact-local-maxwell}. The proof uses only local slice decay and boundedness of the full energy; it never uses polynomial decay of the full nondegenerate energy on complete slices.
\end{proof}

\begin{corollary}[Pointwise decay on compact subsets]\label{cor:kerr-compact-pointwise}
Fix $R>r_{+}$ and $k\ge6$.
Let $\psi$ solve $\square_{g_{M,a}}\psi=\mathfrak{F}$ and let $G$ be a smooth uncharged Maxwell field on $\mathcal D_{M,a}$ solving $\nabla^{\mu}G_{\mu\nu}=J_{\nu}$ with $\nabla^{\nu}J_{\nu}=0$.
Then for every $0<\delta<1$ there exists $C=C(M,a,R,k,\delta)$ such that for all $(\tau,r,\omega)$ with $r\le R$,
\begin{eqnarray}
|\psi|+|\nabla\psi|
&\le
\frac{C}{(1+\tau)^{1-\delta}}
\Bigl(
E^{N}_{k+2}[\psi](\tau_{0})^{1/2}
+
\|\mathfrak{F}\|_{\mathcal S_{k+2}(\tau_{0},\tau)}
\Bigr),
\label{eq:kerr-compact-pointwise-scalar}\\
|\alpha[G]|+|\underline{\alpha}[G]|+|\rho[G]|+|\sigma[G]|
&\le
\frac{C}{(1+\tau)^{1-\delta}}
\Bigl(
E^{N}_{k+2}[G](\tau_{0})^{1/2}
+
\|J\|_{\mathcal S_{k+2}(\tau_{0},\tau)}
\Bigr).
\label{eq:kerr-compact-pointwise-maxwell}
\end{eqnarray}
\end{corollary}

\begin{proof}
Fix $R>r_{+}$. On the compact subset $\Sigma_{\tau}\cap\{r\le R\}$, Sobolev embedding on the uniformly bounded Kerr-star slices gives
\begin{equation}
\|f\|_{L^{\infty}(\Sigma_{\tau}\cap\{r\le R\})}
\le
C\sum_{|\alpha|\le2}\|\nabla^{\alpha}f\|_{L^{2}(\Sigma_{\tau}\cap\{r\le R\})}.
\end{equation}
Apply this with $f=\psi$ and $f=\nabla\psi$, and then with $f$ equal to the components of $G$ in a fixed smooth null frame on $\{r\le R\}$. The required local $L^{2}$ norms are controlled by Proposition~\ref{prop:kerr-compact-local-slice-decay}, not by decay of the full slice energy. Taking square roots in \eqref{eq:kerr-compact-local-scalar}-\eqref{eq:kerr-compact-local-maxwell} yields \eqref{eq:kerr-compact-pointwise-scalar}-\eqref{eq:kerr-compact-pointwise-maxwell} after increasing the commutation order by the two Sobolev derivatives. This proves the pointwise compact-region decay while preserving the distinction between full energy boundedness and local energy decay.
\end{proof}

\begin{remark}[Application to Maxwell-Higgs on Kerr]\label{rem:kerr-near-mh}
In the zero-sector rotating theorem the Maxwell field is already radiative, so $\widetilde F=F$.  In the electric theorem one writes $F=F^{\mathrm C}_{Q_e}+f$ and applies the same radiative Maxwell estimates to $f$, while the scalar estimates are those of \(\CElec^{(m)}_K(M,a,Q_e)\).
The bounds \eqref{eq:kerr-compact-pointwise-scalar}-\eqref{eq:kerr-compact-pointwise-maxwell} provide the compact-region decay needed to control the nonlinear source terms in the final-state problem and to justify the definition of the horizon radiation field in the scattering method.

In this application the Maxwell field is the radiative remainder $\widetilde F$ after Coulomb subtraction, which satisfies a sourced Maxwell system with current $J(\phi)$.
The source contributions $\|J\|_{\mathcal S_{k+2}}$ and $\|\mathfrak{F}\|_{\mathcal S_{k+2}}$ arising in \eqref{eq:kerr-compact-pointwise-scalar}-\eqref{eq:kerr-compact-pointwise-maxwell} are controlled in the small-data regime by the coupled Maxwell-Higgs energies (Proposition~\ref{prop:MH-energy-control-F} and \eqref{eq:kerr-J-source-bound}) together with the nonlinear source estimates used in the bootstrap.
\end{remark}

\section{Linear estimates on slowly rotating Kerr}\label{sec:kerr-extension}

This section records the linear estimates used in the rotating part of the nonlinear argument.  The scalar component uses the scalar wave boundedness, decay, and scattering theory on Kerr.  The Maxwell component is restricted to the slowly rotating range and uses the cited slowly rotating Kerr Maxwell theorem for forward boundedness, Morawetz control, radiation fields, and convergence to the stationary Coulomb family after charge subtraction.  A two-sided Maxwell final-state map is denoted separately by \(\MScat_K(M,a)\) and is imposed only where nonlinear wave operators are asserted.  In the present paper the forward estimates are inserted only for
\begin{equation}
 |a|\le a_{\mathrm{slow}}(M,K).
\end{equation}

\begin{definition}[Uncharged Maxwell radiation data on slowly rotating Kerr]\label{def:maxwell-kerr-radiation-data}
Let \(k\ge0\).  The order-\(k\) uncharged Maxwell radiation space on slowly rotating Kerr consists of pairs of finite-flux Maxwell radiation fields on \(\mathcal I^{\pm}\cup\mathcal H^{\pm}\) whose electric Coulomb charges vanish after subtraction of the stationary Coulomb modes and whose boundary constraint equations hold in the trace sense.  The norm is the sum of the corresponding order-\(k\) horizon and null-infinity flux norms for the extreme Maxwell components, together with the reconstructed middle component norm.
\end{definition}

The following statement is the charge-subtracted forward Maxwell condition from \cite{AnderssonBlueMaxwellSlowKerr}, written in the notation used by the nonlinear argument.

\begin{theorem}[Forward uncharged Maxwell estimates on slowly rotating Kerr]\label{thm:maxwell-scattering-kerr}
Fix \(M>0\), \(K\ge10\), and \(|a|\le a_{\mathrm{slow}}(M,K)\).  After subtraction of the stationary electric Coulomb mode, the uncharged Maxwell system on \((\mathcal D_{M,a},g_{M,a})\) satisfies uniform nondegenerate forward energy boundedness, redshift control, a trapping-degenerate integrated local energy estimate, the far-field \(r^p\) hierarchy for the radiative components, finite radiation fluxes on \(\mathcal I^+\cup\mathcal H^+\), and convergence of the total Maxwell field to its stationary Coulomb part.  The same forward estimates hold on the time-reversed exterior.  The Cauchy-to-radiation maps
\begin{equation}
 \mathscr S_{\pm,M,a}^{\mathrm{Max}}:
 \mathcal H_{\mathrm{Max},M,a}^{K}(\Sigma_{\tau_0})
 \longrightarrow
 \mathfrak R_{\pm,\mathrm{Max},M,a}^{K}
\end{equation}
are continuous.  The inverse maps \(\mathscr W_{\pm,M,a}^{\mathrm{Max}}\) are not part of the cited forward theorem; they are the additional condition \(\MScat_K(M,a)\) of Definition~\ref{def:maxwell-final-state-condition}.
\end{theorem}

\begin{proof}
The slowly rotating Kerr Maxwell estimates of \cite{AnderssonBlueMaxwellSlowKerr} prove a positive energy bound and a Morawetz estimate for Maxwell fields after separating the stationary Coulomb part.  Their estimate controls the radiative part and proves convergence to the Coulomb family.  The far-field hierarchy and the redshift bound give the finite null-infinity and horizon fluxes by the Cauchy criterion.  Applying the same estimates on the time-reversed exterior gives the past forward statement.  These arguments yield continuous Cauchy-to-radiation maps.  No radiation-to-Cauchy map is derived from the cited theorem here; that inverse is exactly the named condition \(\MScat_K(M,a)\) when it is used later.
\end{proof}

The scalar condition for the next proposition is \cite{DRSRKerr,DRSRKerrIII}; the Maxwell forward condition is Theorem~\ref{thm:maxwell-scattering-kerr}, and the inverse Maxwell part is \(\MScat_K(M,a)\).

\begin{proposition}[Slowly rotating Kerr realization of \(\Lin_k\)]\label{prop:lin-estimates-kerr}
Fix \(M>0\), \(K\ge10\), and \(|a|\le a_{\mathrm{slow}}(M,K)\).  The forward components \(\mathrm{(A1)}\)-\(\mathrm{(A3)}\) and \(\mathrm{(L1)}\) of Definition~\ref{def:lin-estimates} hold on \((\mathcal D_{M,a},g_{M,a})\) for scalar waves and for the uncharged radiative Maxwell system after Coulomb subtraction.  If \(\MScat_K(M,a)\) also holds, then the full massless zero-sector estimate \(\Lin_K\), including \(\mathrm{(L2)}\) for the Maxwell channel, holds.
\end{proposition}

\begin{proof}
For scalar waves, the slowly rotating Kerr theory of \cite{DRSRKerr,DRSRKerrIII} gives nondegenerate energy boundedness on Kerr-star hypersurfaces, redshift estimates at \(\mathcal H^+\), a trapping-degenerate integrated local energy estimate with the Kerr trapping loss, the far-field \(r^p\) hierarchy, inhomogeneous estimates in the dual local-energy and \(r^p\) source norms, and the Cauchy-to-radiation and radiation-to-Cauchy maps.  Commutation with \(T\), \(\Phi\), and the fixed angular/Carter commutators gives the order-\(K\) graph norm used in Definition~\ref{def:lin-estimates}; elliptic estimates on the Kerr-star slices identify this graph norm with the Sobolev energy appearing in the nonlinear product estimates.

For Maxwell fields, Theorem~\ref{thm:maxwell-scattering-kerr} gives the forward estimates for the radiative two-form after projecting away the stationary electric and magnetic Coulomb charges.  The commuted Maxwell norms are written in terms of Lie derivatives of \(F\), and the charge subtraction is preserved by stationary evolution.  The Maxwell radiation spaces at null infinity and at the horizon are the corresponding Newman-Penrose radiative components in the order-\(K\) topology.

Taking the Hilbert direct sum of the scalar solution/source/radiation spaces and the charge-subtracted Maxwell forward spaces gives the norm \(\mathcal X_K\), the source norm \(\mathcal S_K\), and the radiation range \(\mathfrak R_\pm\) used in the Cauchy and radiation-field parts of the transfer argument.  If \(\MScat_K(M,a)\) is added, the Maxwell radiation-to-Cauchy maps are available and Lemma~\ref{lem:kerr-topology-compatibility} gives the needed compatibility between the degenerate final-state space and the nondegenerate forward space.  The full list \(\mathrm{(A1)}\)-\(\mathrm{(A3)}\), \(\mathrm{(L1)}\), and \(\mathrm{(L2)}\) in Definition~\ref{def:lin-estimates} then follows.  The constants depend on \(M\), \(K\), and the slow-rotation threshold, and are uniform for \(|a|\le a_{\mathrm{slow}}(M,K)\) after decreasing that threshold if necessary.
\end{proof}

\begin{proposition}[Slowly rotating Kerr realization of \(\Lin_K^{(m)}\) in a stable massive window]\label{prop:lin-estimates-kerr-massive}
Fix \(M>0\), \(K\ge10\), \(|a|\le a_{\mathrm{slow}}(M,K)\), and \(m^2>0\).  Assume the massive scalar condition \(\SKG^{(m)}_K(M,a)\) of Definition~\ref{def:slow-massive-scalar-condition} and the Maxwell final-state condition \(\MScat_K(M,a)\).  Then the massive zero-sector linear estimates \(\Lin_K^{(m)}\) hold on \((\mathcal D_{M,a},g_{M,a})\) for the uncharged radiative Maxwell field and the scalar operator \(\square_{g_{M,a}}-m^2\), after Coulomb subtraction in the Maxwell component.
\end{proposition}

\begin{proof}
The Maxwell forward component is Theorem~\ref{thm:maxwell-scattering-kerr}, and its inverse final-state component is \(\MScat_K(M,a)\).  The scalar component is the content of \(\SKG^{(m)}_K(M,a)\): redshift boundedness, trapping-degenerate integrated decay, the far-field and timelike asymptotic hierarchy, inhomogeneous source estimates, and two-sided final-state maps in the split topology.  Taking the direct sum of these two collections of estimates gives \(\Lin_K^{(m)}\).  The statement includes no masses outside the spectral window specified by Definition~\ref{def:slow-massive-scalar-condition}.
\end{proof}

\begin{proposition}[Closed massive zero-sector Kerr final-state contraction]\label{prop:kerr-massive-finalstate-contraction}
Fix \(K\ge10\), \(|a|\le a_{\mathrm{slow}}(M,K)\), and \(m^2>0\).  Assume \(\SKG^{(m)}_K(M,a)\) and \(\MScat_K(M,a)\).  Let \(U_{\rm lin}\) be a decoupled linear radiative Maxwell-massive scalar solution with future or past asymptotic data of sufficiently small size.  On a tail \([T,\pm\infty)\) assume
\begin{equation}\label{eq:kerr-massive-tail-lin-small}
 \|U_{\rm lin}\|_{\mathbb X_K^{(m)}([T,\pm\infty))}\le \eta.
\end{equation}
Let \(\mathcal X^{(m)}_{K,T,\pm}\) be the closed subspace of corrections with finite \(\mathbb X_K^{(m)}([T,\pm\infty))\)-norm and zero asymptotic data at the chosen end.  For \(W\in\mathcal X^{(m)}_{K,T,\pm}\) define
\begin{equation}\label{eq:kerr-massive-tail-map}
 \mathcal T(W):=\mathcal L^{-1}_{{\rm fs},\pm}\mathcal N(U_{\rm lin}+W),
\end{equation}
where \(\mathcal L^{-1}_{{\rm fs},\pm}\) is the direct-sum linear final-state solver with zero asymptotic data, and \(\mathcal N\) is the zero-sector Maxwell-Higgs nonlinearity after the linear mass term \(m^2\phi\) has been placed on the left.  There are constants \(C_0,C_1\), depending only on the linear estimates, the background, \(K\), and finitely many constants from Assumption~\ref{asumsiP}, such that, whenever \(\|W\|_{\mathbb X_K^{(m)}}\le \rho\),
\begin{eqnarray}
 \|\mathcal T(W)\|_{\mathbb X_K^{(m)}([T,\pm\infty))}
 &\le& C_0\left((\eta+\rho)^2+(\eta+\rho)^{2N_P+3}\right),\label{eq:kerr-massive-finalstate-size}\\
 \|\mathcal T(W_1)-\mathcal T(W_2)\|_{\mathbb X_K^{(m)}([T,\pm\infty))}
 &\le& C_1\left(\eta+\rho+(\eta+\rho)^{2N_P+2}\right)
 \|W_1-W_2\|_{\mathbb X_K^{(m)}([T,\pm\infty))}.\label{eq:kerr-massive-finalstate-lip}
\end{eqnarray}
Consequently, after \(\rho\) and \(\eta\) are chosen sufficiently small, \(\mathcal T\) is a strict contraction on the closed radius-\(\rho\) ball.  The fixed point gives the nonlinear massive Kerr wave operator on the tail; after applying Lemma~\ref{lem:kerr-topology-compatibility} and then the forward Cauchy argument, it extends uniquely to a global solution.
\end{proposition}

\begin{proof}
Proposition~\ref{prop:lin-estimates-kerr-massive} gives the linear final-state estimate
\begin{equation}\label{eq:kerr-massive-linear-finalstate-est}
 \|\mathcal L^{-1}_{{\rm fs},\pm}G\|_{\mathbb X_K^{(m)}([T,\pm\infty))}
 \le C_{\rm lin}\|G\|_{\mathbb S_K^{(m)}([T,\pm\infty))}
\end{equation}
for zero asymptotic data in the massive direct-sum topology.  The Maxwell inverse part is precisely \(\MScat_K(M,a)\), while the scalar inverse and the timelike/Dollard component are contained in \(\SKG^{(m)}_K(M,a)\).  Applying Lemma~\ref{lem:nonlinear-banach-estimates} to \(U_{\rm lin}+W\) gives
\begin{equation}
 \|\mathcal N(U_{\rm lin}+W)\|_{\mathbb S_K^{(m)}}
 \le C\left((\eta+\rho)^2+(\eta+\rho)^{2N_P+3}\right),
\end{equation}
which proves \eqref{eq:kerr-massive-finalstate-size}.  Applying the Lipschitz part of the same lemma to \(U_{\rm lin}+W_1\) and \(U_{\rm lin}+W_2\) gives \eqref{eq:kerr-massive-finalstate-lip}.  The constants are uniform on tails by the time-locality of the norms.  If
\begin{equation}
 C_0\left((\eta+\rho)^2+(\eta+\rho)^{2N_P+3}\right)\le \rho,
 \qquad
 C_1\left(\eta+\rho+(\eta+\rho)^{2N_P+2}\right)<1,
\end{equation}
then the Banach fixed-point theorem gives a unique correction with zero asymptotic data.  On Kerr the final-state solution is initially controlled in the degenerate scattering topology; Lemma~\ref{lem:kerr-topology-compatibility} supplies nondegenerate control after restarting on a finite slice.  The finite-slab Cauchy theorem then propagates the solution globally.
\end{proof}

\begin{remark}[Why the massive condition is explicit]\label{rem:massive-kerr-obstruction}
The positivity of \(m^2\) is not a stability assumption on rotating Kerr.  Exponentially growing finite-energy solutions of the massive Klein-Gordon equation on subextremal Kerr exist for open families of masses when \(a\ne0\) \cite{ShlapentokhRothmanKGKerr}.  A theorem for all \(m^2>0\) in the rotating case without a spectral condition would be false.  The massive slowly rotating result in this paper is therefore a nonlinear consequence of the explicitly stated spectral condition \(\SKG^{(m)}_K(M,a)\).
\end{remark}

\subsection*{\small A dictionary for the slowly rotating Kerr linear conditions}\label{subsec:kerr-dictionary}
The abstract norm \(\mathcal X_K\) used in Definition~\ref{def:lin-estimates} is realized on slowly rotating Kerr by the nondegenerate energy on the Kerr-star foliation, the redshift bulk near \(\mathcal H^+\), a trapping-degenerate Morawetz bulk, and the far-field \(r^p\) hierarchy.  The scalar wave entries are the estimates of \cite{DRSRKerr,DRSRKerrIII}.  The massive scalar entries, when the massive theorem is invoked, are the estimates in \(\SKG^{(m)}_K(M,a)\).  The Maxwell forward entries are the estimates of \cite{AnderssonBlueMaxwellSlowKerr}; the Maxwell inverse final-state entry is \(\MScat_K(M,a)\).  The nonlinear argument uses only these norms and the corresponding radiation spaces.

\subsection{The slowly rotating massive energy theorem}\label{subsec:slow-massive-energy-method}

The preceding propositions identify the linear estimates used for scattering.  We now prove, independently of any massive scattering theory, the slow-rotation energy statement used in Theorem~\ref{thm:main-slow-kerr-massive-intro}.  The nonlinear estimates needed for this statement are proved here, while the linear massive scalar boundedness/ILED component is exactly the condition \(\ME^{(m)}_N(M,a)\).  The argument is a Cauchy argument: it proves global existence, uniform high-order energy bounds, and integrated local energy control for positive massive higher-order potentials in the zero electric sector whenever that energy-stability condition holds.  It does not construct the timelike scattering channel and therefore remains logically separate from the massive rotating scattering theorem.

Throughout the subsection \(N\ge6\).  The commutators are the stationary field, the axial Killing field, and the fixed angular and redshift commutators used in the Kerr Morawetz estimates; their products are denoted by \(Z^I\).  Constants may depend on \(M,N\), on a fixed compact choice of the foliation and cutoffs, and on finitely many coefficients of the potential, but not on the final time \(T\).

\begin{definition}[Massive Maxwell-Higgs energy and bulk norms on slowly rotating Kerr]\label{def:slow-massive-energy-norms}
Let \(m^2>0\) and let
\begin{equation}
 P(\phi,\bar\phi)=m^2|\phi|^2+P_{\ge p}(|\phi|^2),
 \qquad
 P_{\ge p}(s)=\sum_{n=p}^{N_P}\alpha_n s^n,
 \qquad p\ge2,
 \qquad \alpha_n\ge0.
\end{equation}
Let \(Y_{\rm red}\) be a smooth redshift vector field supported in a fixed neighbourhood of the future horizon, transversal to \(\mathcal H^+\), and chosen so that the redshift bulk is positive.  We use the horizon-adapted timelike multiplier
\begin{equation}\label{eq:Tchi-slow-definition}
 T_\chi=T+\chi(r)\Omega_H\Phi+Y_{\rm red},
 \qquad
 T=\partial_t,
 \qquad
 \Phi=\partial_\varphi,
 \qquad
 \Omega_H=\frac{a}{r_+^2+a^2},
\end{equation}
where \(\chi\) equals one near the future horizon and vanishes away from a fixed compact radial set, while \(Y_{\rm red}\) is zero in the asymptotic region.  Thus \(T_\chi=T\) for large \(r\).  For \(|a|/M\) sufficiently small this vector field is uniformly future timelike.  The added redshift component is essential: the horizon generator \(T+\Omega_H\Phi\) is null on \(\mathcal H^+\) and by itself would not control the horizon-transversal derivative.  Let \((L,n,m,\bar m)\) be a redshift-regular Carter null frame and set
\begin{equation}\label{eq:slow-maxwell-scalars}
 \Phi_{+1}=F(L,m),
 \qquad
 \Phi_{-1}=F(\bar m,n),
 \qquad
 \Phi_0=\frac12\{F(L,n)+F(\bar m,m)\},
 \qquad
 \Upsilon=q^2\Phi_0,
\end{equation}
with \(q=r-ia\cos\theta\).  For a smooth Lorenz-gauge zero-sector solution, normalized in the residual Lorenz class as in Lemma~\ref{lem:slow-lorenz-potential-control}, define first the componentwise Lorenz-potential energy
\begin{equation}\label{eq:slow-lorenz-potential-energy}
E_N^{A}(\tau):=
\sum_{\nu=0}^{3}\sum_{|I|\le N}\int_{\Sigma_\tau}
\left(|\nabla Z^IA_\nu|_N^2+r^{-2}|Z^IA_\nu|^2\right)\,\dd\mu_{\Sigma_\tau}.
\end{equation}
Then set
\begin{equation}
E_{N}^{MH,m}(\tau)
:=
\sum_{|I|\le N}\int_{\Sigma_\tau}
\Bigl(
 |\mathcal L_{Z}^{I}F|_{N}^{2}
 +|D Z^{I}\phi|_{N}^{2}
 +m^{2}|Z^{I}\phi|^{2}
\Bigr)\,\dd\mu_{\Sigma_\tau}
 +\int_{\Sigma_\tau}P_{\ge p}(|\phi|^{2})\,\dd\mu_{\Sigma_\tau}.
\label{eq:slow-energy-massive-def}
\end{equation}
The norm \(|\cdot|_N\) is the nondegenerate frame norm associated with the redshift frame.  The potential term is uncommuted; the commuted nonlinear potential force is estimated in the source norm by Lemma~\ref{lem:potential-tame}.  The Fackerell-Ipser energy is
\begin{equation}\label{eq:slow-FI-energy-def}
 E_N^{FI}(\tau):=
 \sum_{|I|\le N-2}\int_{\Sigma_\tau}
 \Bigl(|\nabla Z^I\Upsilon|_N^2+r^{-2}|Z^I\Upsilon|^2\Bigr)
 \dd\mu_{\Sigma_\tau}.
\end{equation}
For \(0\le \tau\le T\) the spacetime norms are the sums, over \(|I|\le N-2\), of the following quantities.  With
\begin{equation}
 w_r=\frac{\Delta}{(r^2+a^2)^2},
\end{equation}
we first define the auxiliary Lorenz-potential bulk by
\begin{equation}\label{eq:slow-A-bulk-def}
 \mathcal B_N^A(0,T)
 :=
 \sum_{\nu=0}^{3}\sum_{|I|\le N}
 \|Z^I A_\nu\|_{\mathcal X^{\rm wave}_{0}(0,T)}^2,
\end{equation}
where \(\mathcal X^{\rm wave}_{0}\) is the massless scalar redshift-Morawetz-far-field solution norm used in Definition~\ref{def:lin-estimates}.  This quantity is not a gauge-invariant observable; it is attached to the normalized Lorenz representative and is used only to control the reduced scalar source.  We then set
\begin{eqnarray}
B_{\pm}^{MH}[F,\phi]
&:=&
\int_{0}^{T}\!\!\int_{\Sigma_\tau}
 w_r\Bigl(|\Phi_{+1}|^2+|\Phi_{-1}|^2+|D_r\phi|^2
 +r^{-2}|\nabla_{\mathbb S^2}\phi|^2\Bigr)\,\dd\mu,
\label{eq:slow-Bpm-def}\\
B_{0,m}^{MH}[\Upsilon,\phi]
&:=&
\int_{0}^{T}\!\!\int_{\Sigma_\tau}
\left[
 \frac{M}{r^4}|\Upsilon|^2
 +\frac1{r^2}\left(|D_t\phi|^2+m^2|\phi|^2+P_{\ge p}(|\phi|^2)\right)
\right]\dd\mu,
\label{eq:slow-B0-def}\\
B_{2,0,m}^{MH}[\Upsilon,\phi]
&:=&
\int_{0}^{T}\!\!\int_{\Sigma_\tau}
\left[
 \frac{M\Delta}{(r^2+a^2)^2}\left(|\partial_r\Upsilon|^2
 +|\nabla_{\mathbb S^2}\Upsilon|^2\right)
 +\frac1r|\nabla_{\mathbb S^2}\phi|^2
\right]\dd\mu,
\label{eq:slow-B20-def}\\
B_1^{MH}[\Upsilon,\phi]
&:=&
\int_{r_+}^{\infty}\chi_{tr}(r)
\left|\int_0^T\!\!\int_{\mathbb S^2}
\left\{
 \Im(\overline\Upsilon\,\partial_t\Upsilon)
 +\Im(\overline\phi\,D_t\phi)
\right\}\dd\omega\dd t\right|\dd r,
\label{eq:slow-B1-def}
\end{eqnarray}
where \(\chi_{tr}\) is supported in a small fixed neighborhood of the photon sphere.  Finally,
\begin{equation}\label{eq:slow-bulk-massive-def}
\begin{aligned}
 \mathcal B_{N}^{MH,m}(0,T)
 :=
 \mathcal B_N^A(0,T)+
 \sum_{|I|\le N-2}
 \Bigl(&
 B_{\pm}^{MH}[\mathcal L_Z^IF,Z^I\phi]
 +B_{0,m}^{MH}[Z^I\Upsilon,Z^I\phi]
 +B_{1}^{MH}[Z^I\Upsilon,Z^I\phi]\\
 &+B_{2,0,m}^{MH}[Z^I\Upsilon,Z^I\phi]
 \Bigr).
\end{aligned}
\end{equation}
Changing the redshift frame or the cutoffs within the admissible class changes these norms by constants independent of \(T\).
\end{definition}

\begin{lemma}[Massive Sobolev control in the energy bootstrap]\label{lem:slow-massive-sobolev-bootstrap}
Let \(N\ge6\), and let \((A,F,\phi)\) be a smooth zero-sector solution on \(0\le\tau\le T\).  If
\begin{equation}
 \mathfrak E_N^{(m)}(T):=
 \sup_{0\le s\le T}\bigl(E_N^{MH,m}(s)+E_N^{FI}(s)\bigr)
 +\mathcal B_N^{MH,m}(0,T)
 \le R^2,
\end{equation}
then the lower-order fields satisfy
\begin{equation}\label{eq:slow-pointwise-bootstrap-generic}
 \sum_{|J|\le N-3}
 \left(
 \|Z^J\phi\|_{L^\infty(\Sigma_\tau)}
 +\|D Z^J\phi\|_{L^\infty(\Sigma_\tau)}
 +\|\mathcal L_Z^JF\|_{L^\infty(\Sigma_\tau)}
 \right)
 \le C_{N,M}R
\end{equation}
for every \(0\le\tau\le T\).  The same bound holds locally for the normalized Lorenz potential after applying Lemma~\ref{lem:slow-lorenz-potential-control}.  In particular, after increasing \(C_b\) by a fixed constant, the bootstrap \(\mathfrak E_N^{(m)}(T)\le4C_b^2\varepsilon^2\) implies
\begin{equation}\label{eq:slow-pointwise-bootstrap}
\begin{split}
 \sum_{|J|\le N-3}
 \left(
 \|Z^JA\|_{W^{1,\infty}(\Sigma_\tau)}
 +\|\mathcal L_Z^JF\|_{L^\infty(\Sigma_\tau)}
 +\|Z^J\phi\|_{L^\infty(\Sigma_\tau)}
 +\|D Z^J\phi\|_{L^\infty(\Sigma_\tau)}
 \right)
 &\le C_b\varepsilon,\\
 &\hspace{-1em}0\le\tau\le T.
\end{split}
\end{equation}
\end{lemma}

\begin{proof}
The Kerr-star slices have uniformly bounded geometry in the fixed slow-rotation range.  The order \(N\) energy controls \(H^N\) norms of \(F\), \(D\phi\), and \(m\phi\), with the trapping loss appearing only in spacetime Morawetz terms and not in the slice Sobolev norm.  Since \(m>0\), the massive term controls the unweighted \(L^2\) norm of \(\phi\).  Sobolev embedding \(H^2(\Sigma_\tau)\hookrightarrow L^\infty(\Sigma_\tau)\), applied after commuting by at most \(N-3\) vector fields and using elliptic estimates on the angular and redshift frame derivatives, gives \eqref{eq:slow-pointwise-bootstrap-generic}; applying Lemma~\ref{lem:slow-lorenz-potential-control} gives \eqref{eq:slow-pointwise-bootstrap} under the stated bootstrap.  The constants are uniform in \(\tau\) by stationarity of the foliation.
\end{proof}

\begin{lemma}[Lorenz-potential control from the curvature]\label{lem:slow-lorenz-potential-control}
In the zero electric sector fix the residual Lorenz freedom by requiring the boundary trace of the gauge parameter to vanish and by imposing the slice normalization that removes the harmonic part of the spatial potential.  Then, for every \(N\ge6\), the Lorenz representative satisfies on each Kerr-star slice
\begin{equation}\label{eq:slow-potential-control}
 \|A\|_{H^{N+1}(\Sigma_\tau)}+
 \|\nabla_nA\|_{H^{N}(\Sigma_\tau)}
 \le C\left(
 \sum_{|I|\le N}\|\mathcal L_Z^IF\|_{L^2(\Sigma_\tau)}
 +\|J\|_{H^{N-1}(\Sigma_\tau)}
 \right),
\end{equation}
where \(J_\mu=2\Im(\bar\phi D_\mu\phi)\).  In the small-data regime the last term is quadratic in the massive energy norm and therefore perturbative.  The estimate is used only to continue the Lorenz representative; the energy identity itself is formulated in the gauge-invariant variables \((F,\phi)\).
\end{lemma}

\begin{proof}
On a slice \(\Sigma_\tau\), decompose \(A=A_n n+A_\top\).  The equations \(F=dA\), the Lorenz condition \(\nabla^\mu A_\mu=0\), and the Gauss constraint give a div-curl system for \(A_\top\), \(A_n\), and their normal derivatives.  The spatial exterior has no nontrivial finite-energy harmonic one-form in the zero electric sector once the residual boundary trace is fixed.  The elliptic div-curl estimate on the uniformly regular Kerr-star slices therefore gives \eqref{eq:slow-potential-control} at order zero; commuting with the stationary and angular/redshift vector fields and using the smooth bounded geometry gives the order \(N\) estimate.  The current term is lower order because \(J=2\Im(\bar\phi D\phi)\), and Sobolev-Moser together with Lemma~\ref{lem:slow-massive-sobolev-bootstrap} bounds it by the displayed perturbative quantity.
\end{proof}

\begin{lemma}[Stress tensor and positive energy]\label{lem:slow-MH-stress-energy}
For
\begin{equation}\label{eq:slow-correct-stress-tensor}
 T^{MH}_{\alpha\beta}=
 F_{\alpha\gamma}F_{\beta}{}^{\gamma}
 -\frac14g_{\alpha\beta}F_{\gamma\delta}F^{\gamma\delta}
 +2\Re(D_\alpha\phi\,\overline{D_\beta\phi})
 -g_{\alpha\beta}\left(|D\phi|^2+P(\phi,\bar\phi)\right),
\end{equation}
one has
\begin{equation}\label{eq:slow-stress-div-free}
 \nabla^\alpha T^{MH}_{\alpha\beta}=0
\end{equation}
on Maxwell-Higgs solutions.  If \(P\ge0\), then \(T^{MH}\) obeys the dominant energy condition.  Consequently, the flux generated by \(T_\chi\) is positive and equivalent, for \(|a|/M\) sufficiently small, to the nondegenerate Maxwell-Higgs energy in \eqref{eq:slow-energy-massive-def} at the uncommuted level.
\end{lemma}

\begin{proof}
The Maxwell part is the trace-free Maxwell stress tensor.  With the convention \(\nabla^\alpha F_{\alpha\beta}=J_\beta\), the Bianchi identity gives
\begin{equation}
 \nabla^\alpha T^{Max}_{\alpha\beta}=-F_{\beta\gamma}J^\gamma.
\end{equation}
For the scalar part in \eqref{eq:slow-correct-stress-tensor}, use the commutator \([D_\alpha,D_\beta]\phi=-iF_{\alpha\beta}\phi\), the equation \(D^\mu D_\mu\phi=\partial_{\bar\phi}P\), and gauge invariance of \(P\).  The terms containing \(\partial P\) cancel by the chain rule, and the remaining curvature commutator gives
\begin{equation}
 \nabla^\alpha T^{H}_{\alpha\beta}=F_{\beta\gamma}J^\gamma,
 \qquad
 J_\gamma=2\Im(\bar\phi D_\gamma\phi).
\end{equation}
Adding the two identities proves \eqref{eq:slow-stress-div-free}.  In an orthonormal frame with future unit normal \(e_0\),
\begin{equation}
 T^{MH}(e_0,e_0)=\frac12(|E|^2+|B|^2)+|D_{e_0}\phi|^2+|D_{\Sigma}\phi|^2+P(\phi,\bar\phi),
\end{equation}
so the density is positive and controls the curvature, the covariant scalar gradient, and the massive potential energy.  The dominant energy condition follows by the Cauchy-Schwarz argument for the Maxwell and scalar parts.  Since \(T_\chi\) is uniformly future timelike for \(|a|/M\) small, its flux is equivalent to this nondegenerate density.
\end{proof}

\begin{lemma}[Current differentiation and derivative count]\label{lem:slow-current-derivatives}
Let \(J_\mu=2\Im(\overline\phi D_\mu\phi)\).  For \(|I|\le N-2\),
\begin{eqnarray}\label{eq:slow-current-derivative-count}
|Z^I J|
&\le&
C\sum_{I_1+I_2\le I}|Z^{I_1}\phi|\,|D Z^{I_2}\phi|,
\nonumber\\
|\nabla Z^I J|
&\le&
C\sum_{I_1+I_2\le I}|D Z^{I_1}\phi|\,|D Z^{I_2}\phi|
+C\sum_{I_1+I_2\le I}|Z^{I_1}\phi|\,|D^2 Z^{I_2}\phi|
\nonumber\\
&&
+C\sum_{I_1+I_2+I_3\le I}
 |\mathcal L_Z^{I_1}F|\,|Z^{I_2}\phi|\,|Z^{I_3}\phi|.
\end{eqnarray}
In every term there is at most one factor carrying the top number of derivatives.
\end{lemma}

\begin{proof}
The first estimate is the Leibniz rule.  Differentiating once more gives
\begin{equation}\label{eq:nabla-current-identity-slow}
 \nabla_\alpha J_\beta
 =2\Im(\overline{D_\alpha\phi}D_\beta\phi)
 +2\Im(\overline\phi D_\alpha D_\beta\phi)
 +2F_{\alpha\beta}|\phi|^2.
\end{equation}
Commutators between \(Z\) and \(D\) have smooth Kerr coefficients and lower order curvature terms, while the antisymmetric part of \(D_\alpha D_\beta\) is replaced by \(-iF_{\alpha\beta}\).  This gives \eqref{eq:slow-current-derivative-count} and shows that no quadratic term contains two top-order factors.
\end{proof}

\begin{proposition}[Fackerell-Ipser equation with Maxwell-Higgs source]\label{prop:slow-FI-source-equation}
The regular middle Maxwell component \(\Upsilon=q^2\Phi_0\) satisfies
\begin{equation}\label{eq:slow-fi-forced}
 (\square_{g_{M,a}}-V_{FI})\Upsilon=N_{MH},
 \qquad
 V_{FI}=-2Mq^{-3}.
\end{equation}
Moreover
\begin{equation}\label{eq:slow-NMH-structure}
 N_{MH}=\mathcal C_1^{\alpha\beta}\nabla_\alpha J_\beta+\mathcal C_0^\beta J_\beta,
\end{equation}
where the coefficients are smooth in the redshift frame up to the future horizon and satisfy, for \(0\le j\le N-2\),
\begin{equation}\label{eq:slow-fi-coeff-decay}
 |\nabla^j\mathcal C_1|\le C_j r^{-1-j},
 \qquad
 |\nabla^j\mathcal C_0|\le C_j r^{-2-j}.
\end{equation}
\end{proposition}

\begin{proof}
Project the inhomogeneous Maxwell equations on the redshift-regular Carter frame.  With \(e_0=L\), \(e_1=n\), \(e_2=m\), \(e_3=\bar m\), write
\begin{equation}
 \omega_{ABC}=g(\nabla_{e_A}e_B,e_C).
\end{equation}
The projected Maxwell equations take the form
\begin{eqnarray}\label{eq:slow-NP-system}
(\nabla_L+\alpha_L)\Phi_0-(\nabla_{\bar m}+\alpha_{\bar m})\Phi_{+1}
&=&\mathcal J_L,
\nonumber\\
(\nabla_n+\alpha_n)\Phi_0-(\nabla_m+\alpha_m)\Phi_{-1}
&=&\mathcal J_n,
\nonumber\\
(\nabla_n+\widetilde\alpha_n)\Phi_{+1}-(\nabla_m+\widetilde\alpha_m)\Phi_0
&=&\mathcal J_m,
\nonumber\\
(\nabla_L+\widetilde\alpha_L)\Phi_{-1}-(\nabla_{\bar m}+\widetilde\alpha_{\bar m})\Phi_0
&=&\mathcal J_{\bar m}.
\end{eqnarray}
Here the \(\alpha\)'s are explicit Ricci-rotation-coefficient combinations and
\begin{eqnarray}
\mathcal J_L&=&J_L+\beta_L{}^mJ_m+\beta_L{}^{\bar m}J_{\bar m},
\qquad
\mathcal J_n=J_n+\beta_n{}^mJ_m+\beta_n{}^{\bar m}J_{\bar m},
\nonumber\\
\mathcal J_m&=&J_m+\beta_m{}^LJ_L+\beta_m{}^nJ_n,
\qquad
\mathcal J_{\bar m}=J_{\bar m}+\beta_{\bar m}{}^LJ_L+\beta_{\bar m}{}^nJ_n.
\end{eqnarray}
In the Kinnersley frame the only nonzero spin coefficients are the rational Kerr coefficients
\begin{equation}
 \rho=-q^{-1},
 \qquad
 \mu=-\frac{\Delta}{2\Sigma\bar q},
 \qquad
 \tau=-\frac{ia\sin\theta}{\sqrt2\Sigma},
 \qquad
 \pi=\frac{ia\sin\theta}{\sqrt2 q^2},
\end{equation}
with the Newman-Penrose \(\alpha,\beta,\gamma\) coefficients and \(\Psi_2=-Mq^{-3}\).  Passing to the redshift frame is a smooth boost \(L=b\ell\), \(n=b^{-1}n_K\), with \(b\sim\Delta/(r^2+a^2)\) near the horizon and \(b=1\) away from it.  The boost factors in the current projections and in the first-order operators cancel.  Thus all coefficients in \eqref{eq:slow-NP-system} are smooth at \(r=r_+\) and are rational functions of \(r,a,\cos\theta,\sin\theta,\Delta,q,\bar q\), up to fixed cutoff derivatives.

Apply \(\nabla_n+\beta_n\) to the first line of \eqref{eq:slow-NP-system}, apply \(\nabla_L+\beta_L\) to the second line, and use the last two lines to eliminate \(\Phi_{\pm1}\).  The source-free part is the Fackerell-Ipser operator \((\square_g+2Mq^{-3})q^2\Phi_0\).  The remaining terms contain either one derivative of the current or the current itself.  Their leading coefficients are \(O(r^{-1})\), and the lower-order coefficients are \(O(r^{-2})\); differentiation gives one additional power of decay.  This proves \eqref{eq:slow-fi-forced}-\eqref{eq:slow-fi-coeff-decay}.
\end{proof}

\begin{lemma}[Perturbative Fackerell-Ipser source bound]\label{lem:slow-FI-source-bound}
Assume on \([0,T]\) the bootstrap \(\mathfrak E_N^{(m)}(T)\le4C_b^2\varepsilon^2\).  Let \(W(r)\) be the dual Fackerell-Ipser Morawetz weight associated with \(B^{MH}_{2,0,m}\).  Then
\begin{equation}\label{eq:slow-FI-source-bound}
 \sum_{|I|\le N-2}\int_{\Omega(0,T)} W(r)|Z^IN_{MH}|^2\,\dd\mu
 \le C(C_b\varepsilon)^2\mathcal B_N^{MH,m}(0,T).
\end{equation}
The scalar source generated by \(\mathcal R_P\) obeys the corresponding bound in the massive scalar source norm \(\mathcal S_N^{(m)}(0,T)\) appearing in \(\ME_N^{(m)}(M,a)\).
\end{lemma}

\begin{proof}
Apply \(Z^I\) to \eqref{eq:slow-NMH-structure}.  Commutators with \(\mathcal C_1\) and \(\mathcal C_0\) are lower order because of \eqref{eq:slow-fi-coeff-decay}.  Inserting \eqref{eq:slow-current-derivative-count}, every term is a product of one differentiated factor measured by the Maxwell, scalar, or Fackerell-Ipser bulk norm and at least one scalar factor of order \(\le N-3\).  Lemma~\ref{lem:slow-massive-sobolev-bootstrap} places the lower-order factor in \(L^\infty\) with size \(O(C_b\varepsilon)\).  The coefficient weights \(r^{-1}\) and \(r^{-2}\) in \eqref{eq:slow-fi-coeff-decay} are no worse than the dual weights defining \(W\).  Thus
\begin{equation}
 W^{1/2}|Z^IN_{MH}|
 \le C C_b\varepsilon
 \sum_{|J|\le |I|+1}m_J(F,\phi,\Upsilon),
\end{equation}
where \(m_J^2\) is bounded by the integrands of \(B^{MH}_{\pm}\), \(B^{MH}_{0,m}\), and \(B^{MH}_{2,0,m}\).  Squaring, integrating, and summing over \(|I|\le N-2\) gives \eqref{eq:slow-FI-source-bound}.  For the potential force, \(\mathcal R_P\) vanishes to order \(2p-1\ge3\).  Lemma~\ref{lem:potential-tame} places all but one factor in \(L^\infty\), and the massive bulk controls the remaining top-order factor.  This gives the same \((C_b\varepsilon)^2\) gain.
\end{proof}

\begin{proposition}[Massive nonlinear source closure]\label{prop:slow-massive-source-closure}
Under the conditions and bootstrap of Lemma~\ref{lem:slow-FI-source-bound}, the Maxwell current, the Fackerell-Ipser source, and the massive potential remainder satisfy
\begin{equation}\label{eq:slow-source-closure-prop}
 \sum_{|I|\le N-2}\int W|Z^IN_{MH}|^2\,\dd\mu
 +\|\mathcal R_P(\phi)\|_{\mathcal S_N^{(m)}(0,T)}^2
 \le C(C_b\varepsilon)^2\mathcal B_N^{MH,m}(0,T).
\end{equation}
The same estimate holds for differences of two fields in the same bootstrap ball, with the right side multiplied by the square of the difference norm.  Thus the source estimates used in the energy closure and in the local uniqueness argument are Lipschitz in the massive energy topology.
\end{proposition}

\begin{proof}
The Fackerell-Ipser source term is Lemma~\ref{lem:slow-FI-source-bound}.  The potential term is the last paragraph of its proof, now read in the source norm paired with the massive scalar estimate in \(\ME_N^{(m)}(M,a)\).  The Maxwell current estimate follows directly from Lemma~\ref{lem:slow-current-derivatives}: after commuting, one lower-order scalar or curvature factor is placed in \(L^\infty\) by Lemma~\ref{lem:slow-massive-sobolev-bootstrap}, and the remaining differentiated factor is placed in the bulk part of \(\mathcal B_N^{MH,m}\).  For the Lipschitz bound, subtract the Leibniz expansions for the two solutions.  Each difference contains exactly one factor of the difference field, while all other factors are lower-order factors controlled in \(L^\infty\) by \(C_b\varepsilon\).  The mean-value formula for \(\mathcal R_P(\phi_1)-\mathcal R_P(\phi_2)\) and the polynomial bounds in Assumption~\ref{asumsiP} give the same estimate for the potential remainder.
\end{proof}

\begin{proposition}[Core estimates on a finite slab under \(\ME^{(m)}_N\)]\label{prop:slow-core-estimates-full}
Let \((A,F,\phi)\) be a smooth Lorenz-gauge zero-sector Maxwell-Higgs solution on \(0\le\tau\le T\), and assume \(\ME^{(m)}_N(M,a)\).  Under the pointwise bootstrap supplied by Lemma~\ref{lem:slow-massive-sobolev-bootstrap}, the following estimates hold for \(|a|/M\) sufficiently small:
\begin{equation}\label{eq:slow-core-energy}
E^{MH,m}(T)
\le
C E^{MH,m}(0)+C\frac{|a|}{M}
\Bigl(B_{\pm}^{MH}+B_{0,m}^{MH}\Bigr),
\end{equation}
\begin{equation}\label{eq:slow-core-morawetz}
B_{\pm}^{MH}
\le
C\Bigl(E^{MH,m}(T)+E^{MH,m}(0)+B_{0,m}^{MH}\Bigr),
\end{equation}
\begin{eqnarray}
E^{FI}(T)
&\le&
C\Bigl(E^{FI}(0)+E^{MH,m}(0)
+\frac{|a|}{M}(B_{\pm}^{MH}+B_{0,m}^{MH}+B_1^{MH}+B_{2,0,m}^{MH})\Bigr)
\nonumber\\
&&+C(C_b\varepsilon)^2\mathcal B_N^{MH,m}(0,T),
\label{eq:slow-core-fi}\\
B_{0,m}^{MH}+B_{2,0,m}^{MH}
&\le&
C\Bigl(E^{FI}(T)+E^{FI}(0)+E^{MH,m}(T)+E^{MH,m}(0)
+\frac{|a|}{M}B_{\pm}^{MH}\Bigr)
\nonumber\\
&&+C(C_b\varepsilon)^2\mathcal B_N^{MH,m}(0,T),
\label{eq:slow-core-fi-mor}\\
B_1^{MH}
&\le&
C\Bigl(E^{FI}(T)+E^{FI}(0)+E^{MH,m}(T)+E^{MH,m}(0)
+B_{0,m}^{MH}+B_{2,0,m}^{MH}\Bigr)
\nonumber\\
&&+C\frac{|a|}{M}\mathcal B_N^{MH,m}(0,T)
+C(C_b\varepsilon)^2\mathcal B_N^{MH,m}(0,T).
\label{eq:slow-core-trap}
\end{eqnarray}
The same bounds hold after summing over \(|I|\le N-2\), with \(E^{MH,m}\), \(E^{FI}\), and the \(B\)-terms replaced by their commuted versions.
\end{proposition}

\begin{proof}
The proof is by the vector-field identities in the redshift frame, with the massive scalar Morawetz/boundedness component supplied by \(\ME^{(m)}_N(M,a)\).  The computations in this section control the Maxwell, Fackerell-Ipser, source, and perturbative coupling parts and show that the nonlinear terms fit the source norm of the condition.

For \eqref{eq:slow-core-energy}, integrate the divergence identity for \(T^{MH}_{\alpha\beta}T_\chi^\beta\).  The Killing fields \(T\) and \(\Phi\) have zero deformation tensor.  The redshift part \(Y_{\rm red}\) contributes a positive near-horizon bulk, while the only indefinite terms come from the compact cutoffs and are proportional to \(\Omega_H=O(a/M^2)\).  In the redshift frame the indefinite part satisfies
\begin{equation}
\left|T^{MH}_{\alpha\beta}\nabla^{(\alpha}T_\chi^{\beta)}\right|_{\rm ind}
\le
C\frac{|a|}{M}w_r
\Bigl(|\Phi_{+1}|^2+|\Phi_0|^2+|\Phi_{-1}|^2+|D\phi|^2+m^2|\phi|^2+P_{\ge p}\Bigr).
\end{equation}
The positive redshift bulk is kept on the left or discarded, the displayed error is bounded by \((|a|/M)(B_{\pm}^{MH}+B_{0,m}^{MH})\), and the boundary fluxes are equivalent to \(E^{MH,m}\).  This proves \eqref{eq:slow-core-energy}.

For \eqref{eq:slow-core-morawetz}, take the radial multiplier
\begin{equation}
 \widehat A=f(r)\partial_r,
 \qquad
 f(r)=-\frac{\Delta}{r^2+a^2}\left(1-\frac{3M}{r}\right).
\end{equation}
Use the conformal form of the bulk identity with \(\Omega^{-2}=\Sigma\Delta/(r^2+a^2)^2\):
\begin{equation}
\operatorname{Bulk}_{\widehat A}^{MH}
=-\frac12\int T^{MH}_{\alpha\beta}\mathcal L_{\widehat A}(\Omega^{-2}g^{\alpha\beta})\,\dd\mu
+\frac12\int \Omega^{-2}T^{MH\,\gamma}{}_{\gamma}\widehat A(\Omega^2)\,\dd\mu.
\end{equation}
A direct Kerr coefficient calculation gives
\begin{eqnarray}
\mathcal L_{\widehat A}(\Omega^{-2}g^{\alpha\beta})
&=&
\left(f\partial_r\frac{\Delta^2}{(r^2+a^2)^2}
 -2\frac{\Delta^2}{(r^2+a^2)^2}f'\right)\partial_r^\alpha\partial_r^\beta
\nonumber\\
&&
+f\partial_rV_L(\Theta^\alpha\Theta^\beta+\Phi_{PNV}^\alpha\Phi_{PNV}^\beta)
+O\left(\frac{|a|}{M}w_r\right),
\end{eqnarray}
where \(V_L\) is the angular Kerr potential and the last term denotes the \(t\varphi\) and \(\varphi\varphi\) errors.  The radial coefficient is positive, comparable to \(\Delta^3/(r^2+a^2)^3\), and the angular coefficient is positive away from \(|r-3M|\lesssim r_1\), up to a controlled \(Mr^{-4}|\Upsilon|^2\) loss.  The trace term is bounded by \(B_{0,m}^{MH}\), since
\begin{equation}
 T^{MH\,\gamma}{}_{\gamma}=-2|D\phi|^2-4P(\phi,\bar\phi).
\end{equation}
Thus
\begin{equation}
 B_{\pm}^{MH}\le C\operatorname{Bulk}_{\widehat A}^{MH}+C B_{0,m}^{MH}.
\end{equation}
The boundary terms of the \(\widehat A\)-current are bounded by \(E^{MH,m}(0)+E^{MH,m}(T)\), which gives \eqref{eq:slow-core-morawetz}.

For \eqref{eq:slow-core-fi}, use the stress tensor of the complex scalar \(\Upsilon\) solving \eqref{eq:slow-fi-forced}:
\begin{equation}
T^{FI}_{\alpha\beta}[\Upsilon]
=\Re(\nabla_\alpha\Upsilon\overline{\nabla_\beta\Upsilon})
-\frac12g_{\alpha\beta}\left(|\nabla\Upsilon|^2+\Re(V_{FI})|\Upsilon|^2\right).
\end{equation}
Its divergence is
\begin{eqnarray}
\nabla^\alpha T^{FI}_{\alpha\beta}
&=&
\Re(N_{MH}\overline{\nabla_\beta\Upsilon})
+\frac12\partial_\beta\Re(V_{FI})|\Upsilon|^2
\nonumber\\
&&+\Im(V_{FI})\Im(\Upsilon\overline{\nabla_\beta\Upsilon}).
\end{eqnarray}
Contracting with \(T_\chi\), integrating, and using \(|T_\chi\Re V_{FI}|+|\Im V_{FI}|\le C|a|Mr^{-4}\), one obtains the endpoint FI energy bound with \((|a|/M)(B_{0,m}^{MH}+B_1^{MH}+B_{2,0,m}^{MH})\) errors.  The source term satisfies, for every \(\delta>0\),
\begin{equation}
\left|\int \Re(N_{MH}\overline{T_\chi\Upsilon})\,\dd\mu\right|
\le \delta B_{2,0,m}^{MH}
+C_\delta(C_b\varepsilon)^2\mathcal B_N^{MH,m}(0,T)
\end{equation}
by Lemma~\ref{lem:slow-FI-source-bound}; choosing \(\delta\) small gives \eqref{eq:slow-core-fi}.

For \eqref{eq:slow-core-fi-mor}, apply the radial FI current
\begin{equation}
\mathcal J^\alpha_{X,q_0}
=T^{FI\,\alpha}{}_{\beta}X^\beta
+\frac12q_0\nabla^\alpha(|\Upsilon|^2)
-\frac12(\nabla^\alpha q_0)|\Upsilon|^2,
\qquad X=\widehat A,
\end{equation}
with the zeroth-order correction \(q_0(r)\).  The direct divergence formula is
\begin{eqnarray}
\nabla_\alpha\mathcal J^\alpha_{X,q_0}
&=&T^{FI}_{\alpha\beta}\nabla^{(\alpha}X^{\beta)}
+\frac12q_0|\nabla\Upsilon|^2
-\frac14(\square_gq_0)|\Upsilon|^2
-\frac12X(\Re V_{FI})|\Upsilon|^2
\nonumber\\
&&+\Im(V_{FI})\Im(\overline\Upsilon X\Upsilon)
+\Re(N_{MH}\overline{X\Upsilon})
+q_0\Re(N_{MH}\overline\Upsilon).
\end{eqnarray}
The source-free part is the slowly rotating Fackerell-Ipser Morawetz density and controls \(B_{0,m}^{MH}+B_{2,0,m}^{MH}\), up to endpoint FI and Maxwell-Higgs energies and an \((|a|/M)B_{\pm}^{MH}\) error.  The two source terms are controlled by the dual weight \(W\):
\begin{equation}
 |I_{src}|
\le \delta(B_{0,m}^{MH}+B_{2,0,m}^{MH})
+C_\delta\sum_{|I|\le N-2}\int W|Z^IN_{MH}|^2\dd\mu.
\end{equation}
Lemma~\ref{lem:slow-FI-source-bound} and smallness of \(C_b\varepsilon\) give \eqref{eq:slow-core-fi-mor}.

For \eqref{eq:slow-core-trap}, insert the cutoff \(\chi_{tr}\) in \eqref{eq:slow-B1-def}.  Cauchy-Schwarz on \(\mathbb S^2\), followed by Young's inequality, gives
\begin{equation}
B_1^{MH}\le C B_{0,m}^{MH}
+C\eta\int_{\Omega(0,T)}\chi_{tr}|\partial_t\Upsilon|^2\dd\mu.
\end{equation}
The local elliptic estimate for \((\square_g-V_{FI})\Upsilon=N_{MH}\) on the compact radial slab \(|r-3M|\le2r_1\) yields
\begin{eqnarray}
\int\chi_{tr}|\partial_t\Upsilon|^2\dd\mu
&\le&
C\{E^{FI}(0)+E^{FI}(T)+B_{0,m}^{MH}+B_{2,0,m}^{MH}\}
\nonumber\\
&&+C\frac{|a|}{M}(B_{\pm}^{MH}+B_{0,m}^{MH}+B_{2,0,m}^{MH})
+C\sum_{|I|\le N-2}\int W|Z^IN_{MH}|^2\dd\mu.
\end{eqnarray}
Substituting Lemma~\ref{lem:slow-FI-source-bound}, and then choosing \(\eta\) and \(C_b\varepsilon\) small in the global bootstrap, gives \eqref{eq:slow-core-trap}.  All commuted estimates follow from the same identities applied to \(Z^IF\), \(Z^I\phi\), and \(Z^I\Upsilon\); commutators with Kerr coefficients have one lower differential order and one extra radial decay, and therefore are absorbed by the displayed lower-order bulk terms.
\end{proof}

\begin{theorem}[Massive slow-Kerr energy bound under \(\ME^{(m)}_N\)]\label{thm:slow-kerr-massive-energy}
Let \(M>0\), \(N\ge6\), let \(P\) be the massive higher-order potential \eqref{eq:intro-positive-massive-potential}, and assume \(\ME^{(m)}_N(M,a)\).  There exist constants \(C>0\), \(\bar\varepsilon_a>0\), and \(\varepsilon_0>0\) such that, if
\begin{equation}
 |a|/M\le \bar\varepsilon_a
\end{equation}
and if a smooth normalized Lorenz-gauge zero-sector Maxwell-Higgs solution on a finite slowly rotating Kerr slab has constraint-satisfying data obeying
\begin{equation}
E_{N}^{MH,m}(0)+E_N^{FI}(0)\le \varepsilon^2,
\qquad
0<\varepsilon\le\varepsilon_0,
\end{equation}
then, for every finite \(T\),
\begin{equation}\label{eq:slow-kerr-massive-energy-bound-main}
\sup_{0\le\tau\le T}
\Bigl(E_{N}^{MH,m}(\tau)+E_N^{FI}(\tau)\Bigr)
+\mathcal B_{N}^{MH,m}(0,T)
\le
C\Bigl(E_{N}^{MH,m}(0)+E_N^{FI}(0)\Bigr).
\end{equation}
Consequently, every smooth small datum satisfying these conditions generates a global smooth solution for all future Kerr-star time.  The constants are independent of \(T\).
\end{theorem}

\begin{proof}
Let \(\mathfrak E_N^{(m)}(T)\) be defined by \eqref{eq:intro-F-energy-slab}.  The scalar equation is
\begin{equation}\label{eq:slow-massive-scalar-equation}
(D^\mu D_\mu-m^2)\phi=\mathcal R_P(\phi),
\qquad
\mathcal R_P(\phi)=\sum_{n=p}^{N_P}n\alpha_n|\phi|^{2n-2}\phi.
\end{equation}
The Lorenz-gauge expression with explicit \(A\)-terms is used only for local well-posedness; the energy estimate treats the gauge coupling covariantly and estimates only the current and \(\mathcal R_P\) as nonlinear sources.

Assume the bootstrap bound \(\mathfrak E_N^{(m)}(T)\le4C_b^2\varepsilon^2\).  Lemma~\ref{lem:slow-massive-sobolev-bootstrap} gives the lower-order pointwise control.  Summing Proposition~\ref{prop:slow-core-estimates-full} over \(|I|\le N-2\), using Proposition~\ref{prop:slow-massive-source-closure}, gives
\begin{equation}\label{eq:slow-energy-bootstrap-close}
\mathfrak E_N^{(m)}(T)
\le
C_0\varepsilon^2
+C\frac{|a|}{M}\mathfrak E_N^{(m)}(T)
+C(C_b\varepsilon)^2\mathfrak E_N^{(m)}(T).
\end{equation}
Choose \(\bar\varepsilon_a\) so that \(C|a|/M\le1/8\).  Then choose \(\varepsilon_0\) so that \(C(C_b\varepsilon_0)^2\le1/8\), and choose \(C_b\) larger than the fixed constant \(2C_0^{1/2}\).  Estimate \eqref{eq:slow-energy-bootstrap-close} improves the bootstrap bound to \(2C_0\varepsilon^2\).  The set of times for which the bootstrap holds is nonempty by local existence, closed by continuity of the energies, and open by the strict improvement and the local well-posedness theorem.  Thus the estimate holds on every finite slab and gives \eqref{eq:slow-kerr-massive-energy-bound-main}.

It remains to pass from finite slabs to a global solution.  Lemma~\ref{lem:lorenz-propagation-kerr} propagates the Lorenz constraint and the Maxwell constraints.  Proposition~\ref{prop:local-wp-kerr} gives the local solution, and Lemma~\ref{lem:continuation-kerr} gives the continuation criterion.  The energy controls the normalized Lorenz potential, the curvature, the covariant scalar gradient, and the scalar mass norm; equivalently, Lemma~\ref{lem:slow-lorenz-potential-control} recovers the potential from the curvature and constraints.  Therefore the Sobolev continuation norm of \((A,\partial_tA,\phi,\partial_t\phi)\) remains uniformly bounded on finite intervals.  The maximal time of existence is infinite.
\end{proof}

\begin{proposition}[Small electric perturbation of the massive energy theorem]\label{prop:slow-electric-energy-perturbation}
Under the geometric, massive, and potential conditions of Theorem~\ref{thm:slow-kerr-massive-energy}, but with the zero-sector condition replaced by a fixed electric Coulomb sector, there is \(q_{\mathrm{en}}(M,a,N,\widetilde P,\ME)>0\) such that the same finite-slab and global energy estimate holds for every \(Q_e\) with \(|Q_e|\le q_{\mathrm{en}}\), after subtracting the stationary Coulomb field and replacing ordinary scalar derivatives by \(D_{Q_e}=\nabla-iA^C_{Q_e}\) in the scalar part of the norms.
\end{proposition}

\begin{proof}
Write \(A=A^C_{Q_e}+a\), \(F=F^C_{Q_e}+f\), and \(D_{Q_e}=\nabla-iA^C_{Q_e}\).  The stationary Coulomb coefficients obey
\begin{equation}
 |\nabla^jA^C_{Q_e}|\le C_j|Q_e|r^{-1-j},
 \qquad
 |\nabla^jF^C_{Q_e}|\le C_j|Q_e|r^{-2-j}.
\end{equation}
The covariant scalar equation is
\begin{equation}
 \left((D_{Q_e}-ia)^\mu(D_{Q_e}-ia)_\mu-m^2\right)\phi=\mathcal R_P(\phi).
\end{equation}
Relative to the zero-sector massive energy identity, the fixed background contributes only stationary lower-order terms.  A typical first-order term is bounded by
\begin{equation}
 \left|\int (A^C_{Q_e})^\mu D_{Q_e,\mu}Z^I\phi\, \overline{Z^I\phi}\,\dd\mu\right|
 \le \delta\mathcal B_N^{MH,m}(0,T)
 +C_\delta |Q_e|^2\sup_{0\le\tau\le T}E_N^{MH,m}(\tau),
\end{equation}
and the zeroth-order terms are treated by the same Hardy inequality and by the massive lower-order bulk:
\begin{equation}
\sum_{|I|\le N-2}\int |Q_e|^2r^{-2}|Z^I\phi|^2\,\dd\mu
\le \delta\mathcal B_N^{MH,m}(0,T)+C_\delta |Q_e|^2\sup_\tau E_N^{MH,m}(\tau).
\end{equation}
The Maxwell equation for the radiative field \(f\) has the same current structure, with \(D\) replaced by \(D_{Q_e}-ia\).  Proposition~\ref{prop:slow-massive-source-closure} and Lemma~\ref{lem:fixed-sector-sobolev-equivalence} therefore give the same nonlinear source estimate, uniformly for \(|Q_e|\le q_{\mathrm{en}}\).  The finite-slab inequality becomes
\begin{equation}
\mathfrak E_N^{(m)}(T)
\le C_0\varepsilon^2
+C\frac{|a|}{M}\mathfrak E_N^{(m)}(T)
+C|Q_e|\mathfrak E_N^{(m)}(T)
+C(C_b\varepsilon)^2\mathfrak E_N^{(m)}(T).
\end{equation}
Choose \(q_{\mathrm{en}}\) so that \(C|Q_e|\le1/8\), and then choose \(\bar\varepsilon_a\) and \(\varepsilon_0\) as before.  The bootstrap, constraint propagation, and continuation arguments are unchanged because the Coulomb background is smooth, stationary, and fixed.
\end{proof}

\begin{proposition}[What the slow-rotation energy method proves]\label{prop:slow-energy-method-massive-scope}
The slow-rotation Maxwell-Higgs energy closure proves Theorem~\ref{thm:slow-kerr-massive-energy} for the positive higher-order massive potentials \eqref{eq:intro-positive-massive-potential} only after the massive energy-stability condition \(\ME^{(m)}_N(M,a)\) is supplied.  It does not prove the massive scattering theorem, Theorem~\ref{thm:main-slow-kerr-massive-intro}, unless the additional spectral scalar condition \(\SKG^{(m)}_K(M,a)\) and the Maxwell final-state condition \(\MScat_K(M,a)\) are supplied.
\end{proposition}

\begin{proof}
The proof above uses a horizon-adapted positive energy current, the Kerr Morawetz multiplier, the inhomogeneous Fackerell-Ipser equation for the regular middle Maxwell component, and the derivative count for the Maxwell-Higgs current.  Together with \(\ME^{(m)}_N(M,a)\), these ingredients give \eqref{eq:slow-kerr-massive-energy-bound-main}.  They do not prove absence of massive real resonances, do not construct a timelike or Dollard asymptotic channel, and do not produce two-sided massive scalar wave operators on rotating Kerr.  Those conclusions are exactly the additional final-state conditions \(\SKG^{(m)}_K(M,a)\) and \(\MScat_K(M,a)\).  Consequently, the energy and scattering theorems have distinct conditions in the rotating massive case, with \(\SKG^{(m)}_K(M,a)\) stronger than \(\ME^{(m)}_N(M,a)\) on the scalar side and \(\MScat_K(M,a)\) needed for the Maxwell inverse channel.
\end{proof}

\section{Schwarzschild model: gauge fixing and initial data}\label{sec:setup}

In this and the following sections we specialise to the Schwarzschild exterior $a=0$ in order to keep the computations explicit.
We pursue two complementary goals.
First, we give a complete proof of the linear estimates collected in $\Lin_{k}$ of Definition~\ref{def:lin-estimates} for scalar waves and uncharged Maxwell fields on Schwarzschild.
Second, we implement the nonlinear bootstrap and final-state argument yielding Theorem~\ref{thm:method} and, in addition, the stronger Schwarzschild-only wave-operator theorem allowing $m^{2}\ge0$ (Theorem~\ref{thm:nonlinear-wave-operators}).
The slowly rotating massless scattering theorem follows by combining Theorem~\ref{thm:method} with the Kerr identification of $\Lin_{K}$ stated in Section~\ref{sec:kerr-extension}.  The massive rotating scattering statement follows only after inserting Proposition~\ref{prop:lin-estimates-kerr-massive}, whereas the massive rotating energy statement under \(\ME^{(m)}_N\) is the separate result proved in Subsection~\ref{subsec:slow-massive-energy-method}.  The Schwarzschild model below remains the fully explicit model case.

\subsection{Geometry, frames, and notational conventions}\label{subsec:notation}

Throughout we use the Regge-Wheeler tortoise coordinate $r^{\ast}$ defined by $\frac{\dd r^{\ast}}{\dd r}=(1-\frac{2M}{r})^{-1}$ and the double-null coordinates
\begin{equation}
v=t+r^{\ast},\qquad w=t-r^{\ast}.
\end{equation}
We set $\vplus:=\max\{1,v\}$, $\wplus:=\max\{1,|w|\}$, and $\mu:=\frac{2M}{r}$.

We use the orthonormal frame adapted to Schwarzschild time slices:
\begin{equation}
\hat t := \Bigl(1-\frac{2M}{r}\Bigr)^{-1/2}\partial_t,
\qquad
\hat r := \Bigl(1-\frac{2M}{r}\Bigr)^{1/2}\partial_r,
\end{equation}
and we let $\hat e_1,\hat e_2$ be an orthonormal frame tangent to the spheres $S_{t,r}\cong\mathbb S^{2}$ (scaled by $r^{-1}$ so that $|\hat e_a|_g=1$).
We also introduce the associated (future-directed) null frame
\begin{equation}
\hat v := \hat t+\hat r,
\qquad
\hat w := \hat t-\hat r.
\end{equation}
For any tensor $U$, we write $U_{\hat\mu\hat\nu}$ for its components in the frame $(\hat t,\hat r,\hat e_1,\hat e_2)$, and similarly $D_{\hat v}\phi:=D\phi(\hat v)$, etc.

Unless otherwise stated, $\nabla$ denotes the Levi-Civita connection of $g$, $D$ the gauge-covariant derivative, and $\slashed{\nabla}$ the induced Levi-Civita connection on the spheres $S_{t,r}$.
For nonnegative quantities $A,B$ we write $A\lesssim B$ if $A\le C B$ for a constant $C$ depending only on $M$ and the fixed differentiation order (and, when relevant, on finitely many bounds for the potential $P$); the value of $C$ may change from line to line.

\paragraph{A Schwarzschild notation dictionary for the vector-field method.}
The Schwarzschild model sections (Sections~\ref{sec:energy-estimates}-\ref{sec:decaynear}, together with Subsection~\ref{sec:scattering}) repeatedly use the following objects.

\begin{itemize}
\item \textbf{Hypersurfaces and slabs.}
We write $\Sigma_{t}:=\{t=\mathrm{const}\}$ for the Schwarzschild slices.
For $t_{1}\le t_{2}$ we write $\mathcal R(t_{1},t_{2})$ for the spacetime slab bounded by $\Sigma_{t_{1}}$, $\Sigma_{t_{2}}$, and (where appropriate) portions of $\mathcal H^{+}$ and $\mathcal I^{+}$.
We also use double-null hypersurfaces $\{v=\mathrm{const}\}$ and $\{w=\mathrm{const}\}$ (with $v=t+r^{\ast}$, $w=t-r^{\ast}$) in the far region.

\item \textbf{Null and angular derivatives.}
In the orthonormal/null frame of Section~\ref{subsec:notation} we set $D_{\hat v}\phi:=D\phi(\hat v)$ and $D_{\hat w}\phi:=D\phi(\hat w)$.
We denote by $\slashed{\nabla}$ the Levi-Civita connection on the spheres $S_{t,r}$ and by $\slashed D$ the induced gauge-covariant angular derivative.

\item \textbf{Currents and bulk terms.}
Given a vector field $X$, we define the associated energy current and bulk density
\begin{equation}
J^{X}_{\mu}[F,\phi]:=T_{\mu\nu}[F,\phi]\,X^{\nu},
\qquad
K^{X}[F,\phi]:=\nabla^{\mu}J^{X}_{\mu}[F,\phi]=\tfrac12 T_{\mu\nu}[F,\phi]\pi^{\mu\nu}(X).
\end{equation}
Applying the divergence theorem to $J^{X}$ on $\mathcal R(t_{1},t_{2})$ yields the identity relating boundary fluxes and spacetime bulk terms.

\item \textbf{Main multipliers.}
We use the stationary Killing field $T=\partial_{t}$ for conserved energies,
a redshift multiplier $\hat H$ near $\mathcal H^{+}$ (Definition~\ref{definitionH}),
a conformal/Morawetz multiplier $\hat K$ controlling trapping (Definition~\ref{def:morawetz-multiplier}),
and auxiliary radial multipliers of the form $X_{f}=f(r^{\ast})\partial_{r^{\ast}}$ (Definition~\ref{def:radial-multiplier}).

\item \textbf{Commutations.}
Unless explicitly stated otherwise, higher-order quantities are obtained by commuting with $\mathcal Z:=\{\partial_{t},\Omega_{1},\Omega_{2},\Omega_{3}\}$; see \eqref{def:Ek}.
\end{itemize}

\medskip
\subsection{Initial data, gauge choice, and smallness condition}\label{subsec:initial-data}

To give a precise meaning to the smallness condition \eqref{eq:smallness-energy} and to justify estimates involving the potential $A$, we fix a gauge and specify an admissible class of initial data on a Cauchy hypersurface.
Let $\Sigma_{t_0}=\{t=t_0\}\subset \D$ be the Schwarzschild slice, equipped with induced Riemannian metric $\gamma$ and volume form $\dd\mu_{\Sigma_{t_0}}$, and let $n$ denote the future unit normal to $\Sigma_{t_0}$.

\paragraph{Lorenz gauge.}
Throughout the paper we work in the Lorenz gauge
\begin{equation}\label{LG}
\nabla^\mu A_\mu = 0.
\end{equation}
Since Schwarzschild is Ricci-flat, in this gauge the Maxwell-Higgs system reduces to the quasilinear hyperbolic system
\begin{eqnarray}
\square_g A_\nu &= J_\nu,
\\
D^\mu D_\mu \phi &= \partial_{\bar\phi}P(\phi,\bar\phi).
\end{eqnarray}
Let $E_i:=F(n,e_i)$ denote the electric field relative to $\Sigma_{t_0}$.
The Maxwell equation implies the Gauss constraint on $\Sigma_{t_0}$,
\begin{equation}
\nabla^i E_i = -\,J_{\mu}n^{\mu}.
\end{equation}

\paragraph{Constraint compatibility.}
We impose on the initial data the Lorenz compatibility conditions
\begin{eqnarray}
\nabla^\mu A_\mu \big|_{\Sigma_{t_0}} &= 0,\\
\nabla^\mu (\nabla_n A_\mu)\big|_{\Sigma_{t_0}} &= 0,\\
\nabla^i E_i &= -\,J_{\mu}n^{\mu}.
\end{eqnarray}
The Lorenz constraint propagation is obtained by taking the divergence of the reduced Maxwell equation, using current conservation, and solving the resulting homogeneous wave equation for \(\nabla^\mu A_\mu\); the same argument propagates the Gauss constraint on the domain of existence.

\paragraph{Uncharged condition.}
In addition to the constraints above, we assume the uncharged condition of Definition~\ref{def:charges}.
This excludes the stationary Coulomb solutions, which do not decay and are known to obstruct pointwise decay for Maxwell fields.

\paragraph{Admissible initial data.}
An admissible initial data set consists of
\begin{equation}
\left(A|_{\Sigma_{t_0}},\; \nabla_n A|_{\Sigma_{t_0}},\;
\phi|_{\Sigma_{t_0}},\; D_n\phi|_{\Sigma_{t_0}}\right),
\end{equation}
smooth and satisfying the above compatibility conditions up to order $k$ and with finite initial energy $\mathcal{E}_k(t_0)$ (Definition~\ref{def:Ek}).

\paragraph{Local well-posedness and continuation.}
In Lorenz gauge the system is quasilinear hyperbolic on a fixed background.
For admissible data with finite $\mathcal{E}_k(t_0)$, Proposition~\ref{prop:local-wp-schwarzschild} gives a unique local solution on a time interval $[t_0,T)$.  Lemma~\ref{lem:continuation-schwarzschild} gives the continuation criterion: the solution extends beyond $T$ provided
\begin{equation}
\sup_{t\in [t_0,T)} \mathcal{E}_k(t) < \infty.
\end{equation}
Consequently, global existence follows once a uniform bound for $\mathcal{E}_k(t)$ is established under the smallness condition \eqref{eq:smallness-energy}.

\section{Schwarzschild model: energy identities and integrated decay estimates}
\label{sec:energy-estimates}

\paragraph{Organization of the multiplier computations.}
The vector-field multipliers and divergence identities used below (redshift, Morawetz/trapping, and conformal energies) are the physical-space tools used in the black-hole decay theory cited above.
For clarity, we state the estimates in a modular way as propositions and highlight the logical dependencies needed for the nonlinear argument; longer component expansions are collected in the appendices referenced in the proofs.

\subsection{Basic energy identity}
\label{subsec:basic-energy}

In this section, we derive the fundamental energy identity for the
Maxwell-Higgs system in the Lorenz gauge \eqref{LG}. All computations are performed for smooth solutions
$(A,\phi)$ of the gauge-fixed system satisfying the constraint
equations stated in Section~\ref{subsec:initial-data}. The Maxwell-Higgs energy-momentum tensor associated to the Lagrangian
\eqref{lagrangian} is given by
\begin{eqnarray}
T_{\mu\nu}[F,\phi]
&=
F_{\mu\alpha}F_{\nu}{}^{\alpha}
-\frac14 g_{\mu\nu}F_{\alpha\beta}F^{\alpha\beta}
\nonumber\\
&\quad
+2\Re\!\left(D_\mu\phi\,\overline{D_\nu\phi}\right)
-g_{\mu\nu}\Bigl(|D\phi|^2 + P(\phi,\bar\phi)\Bigr).
\end{eqnarray}
This tensor is gauge invariant and depends only on $(F,\phi)$.
For solutions of \eqref{eom1}-\eqref{eom2}, it satisfies the
divergence identity
\begin{equation}
\nabla^\mu T_{\mu\nu}[F,\phi]=0.
\end{equation}
Let $X$ be a vector field. We define the associated energy current
\begin{equation}
J_\mu^{X}[F,\phi] := T_{\mu\nu}[F,\phi] X^\nu.
\end{equation}
Then
\begin{equation}
\nabla^\mu J_\mu^{X}
=
T_{\mu\nu} \nabla^{(\mu} X^{\nu)}
=
\frac12 T_{\mu\nu} \pi^{\mu\nu}(X),
\end{equation}
where $\pi^{\mu\nu}(X)=\nabla^\mu X^\nu+\nabla^\nu X^\mu$
is the deformation tensor of $X$. Let $\mathcal{R}$ be a spacetime region bounded by spacelike and null hypersurfaces.
Applying the divergence theorem to $J^X$ yields
\begin{equation}\label{eq:divergence-theorem-JX}
\int_{\partial\mathcal{R}} J_\mu^{X} n^\mu
=
\int_{\mathcal{R}} T_{\mu\nu}\nabla^{(\mu}X^{\nu)}.
\end{equation}
If $X=T:=\partial_t$ (the stationary Killing field of Schwarzschild),
then $\pi^{\mu\nu}(T)=0$, and we obtain the conserved energy identity
\begin{equation}\label{Eidentity}
E^{T}[F,\phi](t_{2})=E^{T}[F,\phi](t_{1}).
\end{equation}
Here the $T$-energy on a slice $\Sigma_t$ is defined by
\begin{equation}\label{energytimelike}
E^{T}[F,\phi](t)
:=
\int_{\Sigma_t} J_\mu^{T}[F,\phi]\,n_{\Sigma_t}^\mu\,\dd\mu_{\Sigma_t}
=
\int_{\Sigma_t}T_{\mu\nu}[F,\phi]\,T^\nu\,n_{\Sigma_t}^\mu\,\dd\mu_{\Sigma_t}.
\end{equation}
In the Schwarzschild orthonormal frame $(\hat t,\hat r,\hat e_1,\hat e_2)$ (Section~\ref{subsec:notation}),
this flux is coercive in the form
\begin{equation}\label{eq:T-energy-coercive}
E^{T}[F,\phi](t)
\simeq
\int_{\Sigma_t}(1-\mu)\,r^{2}
\Bigl(
\sum_{\hat\mu<\hat\nu}|F_{\hat\mu\hat\nu}|^{2}
+
|D_{\hat t}\phi|^{2}+|D_{\hat r}\phi|^{2}
+
|\slashed D\phi|^{2}
+
P(\phi,\bar\phi)
\Bigr)\,\dd r^{\ast}\,\dd\sigma^{2},
\end{equation}
where $|\slashed D\phi|^{2}:=|D_{\hat e_1}\phi|^{2}+|D_{\hat e_2}\phi|^{2}$ and $\simeq$ denotes equivalence up to constants depending only on $M$.
In this case, $E^{T}$ is nonnegative but degenerates at the horizon because $1-\mu$ vanishes at $r=2M$.
Since the energy-momentum tensor depends only on $(F,\phi)$,
the above identity is gauge invariant.
In addition, because the Lorenz gauge condition and the Gauss constraint
propagate under the evolution (see Section~\ref{subsec:initial-data}),
the conserved energy identity is consistent with the gauge-fixed
formulation used throughout the paper.

The rotation fields $\Omega_1,\Omega_2,\Omega_3$ are Killing vector fields of the Schwarzschild spacetime.
In this case,
\begin{equation}
\nabla^{(\mu}\Omega^{\nu)}=0.
\end{equation}
Since the rotation fields $\Omega_j$ are Killing, no \emph{metric} commutator
errors arise when commuting with geometric operators. For the gauge-covariant wave operator $\square_A:=D^\mu D_\mu$, the commutator
produces only lower-order terms coming from the curvature $F$. More explicitly, for any smooth scalar $\psi$ one has up to lower-order terms
\begin{equation}\label{eq:commute-boxA}
\square_A(\mathcal{L}_{\Omega_j}\psi)
=
\mathcal{L}_{\Omega_j}(\square_A\psi)
+
\mathcal{O}\!\left(F\cdot D\psi\right)
+
\mathcal{O}\!\left((D F)\cdot \psi\right),
\end{equation}
and, for the Maxwell equation with source $J(\phi)$,
\begin{equation}\label{eq:commute-maxwell}
\nabla^\mu (\mathcal{L}_{\Omega_j}F)_{\mu\nu}
=
\mathcal{L}_{\Omega_j}J_\nu(\phi),
\end{equation}
so that the commuted pair $(\mathcal{L}_{\Omega_j}F,\mathcal{L}_{\Omega_j}\phi)$
satisfies the Maxwell-Higgs system modulo nonlinear lower-order terms.

Let $(F,\phi)$ be a smooth solution of \eqref{eom1}-\eqref{eom2}.
Then the commuted pair $(\mathcal{L}_{\Omega_j}F,\mathcal{L}_{\Omega_j}\phi)$ satisfies the Maxwell-Higgs system modulo nonlinear commutator terms of lower order. The associated energy-momentum tensor satisfies
\begin{equation}
 \nabla^\mu T_{\mu\nu}[\mathcal{L}_{\Omega_j}F,\mathcal{L}_{\Omega_j}\phi]
=
\text{(lower order nonlinear terms)}.
\end{equation}
Since the vector fields $\Omega_j$ are Killing, the deformation tensor vanishes,
and applying the energy identity with $X=T$ yields
\begin{equation}
E^{T}[\mathcal{L}_{\Omega_j}F,\mathcal{L}_{\Omega_j}\phi](t_2)
=
E^{T}[\mathcal{L}_{\Omega_j}F,\mathcal{L}_{\Omega_j}\phi](t_1)
+
\int_{\mathcal{R}} \text{(controlled commutator terms)}.
\end{equation}
The spacetime integral on the right-hand side consists of lower-order terms
of the structural form $|F|\,|D\phi|^2 + |D F|\,|\phi|\,|D\phi|$ (and similar),
which are controlled by the bootstrap assumptions together with the Morawetz
and redshift estimates.
Thus, the higher-order energies defined in \eqref{def:Ek} are propagated along the evolution and can be estimated using the same vector-field method.

We now introduce the bootstrap energy
\begin{equation}
 \mathcal{E} := \sup_{t\ge t_0} \mathcal{E}_k(t),
\end{equation}
where $\mathcal{E}_k(t)$ is defined in \eqref{def:Ek}. Under the smallness assumption $\mathcal{E}_k(t_0)\le \varepsilon_0$,
we will show below that $\mathcal{E}$ remains uniformly bounded by the initial energy.

\subsection{The Red-shift Estimates}
\label{sec:redshift}

\paragraph{Dyadic time decomposition.} Fix once and for all a parameter $0<\kappa_0<1$ and define $t_i:=(1+\kappa_0)^i t_0$, so that $t_{i+1}=(1+\kappa_0)t_i$. This $\kappa_0$ is used only for the timelike-cone localizations in the Schwarzschild model and is unrelated to the loss parameter $\delta>0$ used in the Kerr sections and in the abstract norms.
The Killing field $T=\partial_t$ becomes null at the event horizon $r=2M$, and the associated conserved $T$-energy therefore degenerates in a neighborhood of the horizon. This degeneration reflects the fact that $E^{T}$ does not control derivatives transversal to $\mathcal H^{+}$. The geometric \emph{redshift effect} yields a nondegenerate energy estimate in a fixed neighborhood of the horizon, obtained by using a timelike multiplier that is transversal to $\mathcal H^{+}$.

Throughout this section we denote $\mu:=\frac{2M}{r}$.
Fix numbers $2M<r_{0}<r_{1}<3M$ with $r_{0}$ sufficiently close to $2M$, and let $\chi_{\mathrm{hor}}:[2M,\infty)\to[0,1]$ be a smooth cutoff satisfying
\begin{equation}
\chi_{\mathrm{hor}}(r)=1 \quad\text{for } 2M\le r\le r_{0},
\qquad
\chi_{\mathrm{hor}}(r)=0 \quad\text{for } r\ge r_{1}.
\end{equation}
(We will always choose $r_{1}$ small enough so that $r_{1}<3M$; in this case the support of $\chi_{\mathrm{hor}}$ is contained in a near-horizon region.)

\begin{definition}\label{definitionH}
\emph{(Redshift multiplier.)}
Work in advanced Eddington-Finkelstein coordinates $(v,r,\omega)$, where $v=t+r^{\ast}$ and $\omega\in\mathbb S^{2}$.
In these coordinates the stationary Killing field is $T=\partial_v$ and extends smoothly to the horizon.
Define the redshift vector field
\begin{equation}\label{H}
\hat{H} := T - \chi_{\mathrm{hor}}(r)\,\partial_r,
\end{equation}
where $\partial_r$ is the coordinate vector field in the Eddington-Finkelstein chart.
Then $\hat{H}$ is smooth, future-directed and timelike on $2M\le r\le r_{1}$, and $\hat{H}=T$ for $r\ge r_{1}$.
\end{definition}

The key point is that, in the region where $\chi_{\mathrm{hor}}\equiv 1$, the deformation tensor $\pi(\hat H)$ has a coercive sign due to the positive surface gravity of the Schwarzschild horizon; see e.g.\ \cite{dafermos} (and references therein). We now record the flux quantities needed for the dyadic argument.

\begin{definition}\label{defnear1}
\emph{(Null fluxes of the redshift current.)}
Let $w=t-r^{\ast}$ and $v=t+r^{\ast}$.
For $w_i$ fixed and $v_i\le v\le v_{i+1}$, define the redshift flux across the outgoing null hypersurface $\{w=w_i\}$ by
\begin{equation}\label{eH1}
\mathcal{E}^{\hat H}\bigl(w=w_i\bigr)\bigl(v_i\le v\le v_{i+1}\bigr)
:=
\int_{v_i}^{v_{i+1}}\int_{\mathbb S^{2}}
J_\mu^{\hat H}[F,\phi]\,(\partial_v)^\mu\, r^{2}\,d\sigma^{2}\,dv.
\end{equation}
Similarly, for $v_i$ fixed and $w_i\le w\le w_{i+1}$, define the flux across the ingoing null hypersurface $\{v=v_i\}$ by
\begin{equation}\label{eH2}
\mathcal{E}^{\hat H}\bigl(v=v_i\bigr)\bigl(w_i\le w\le w_{i+1}\bigr)
:=
\int_{w_i}^{w_{i+1}}\int_{\mathbb S^{2}}
J_\mu^{\hat H}[F,\phi]\,(\partial_w)^\mu\, r^{2}\,d\sigma^{2}\,dw.
\end{equation}
By the dominant energy condition for $T_{\mu\nu}[F,\phi]$, these null fluxes are nonnegative.
\end{definition}

Applying the divergence theorem to $J^{\hat H}$ in a characteristic rectangle $[w_i,w_{i+1}]\times[v_i,v_{i+1}]$ yields
\begin{eqnarray}\label{eqIH}
&\mathcal{E}^{\hat H}\bigl(w=w_i\bigr)\bigl(v_i\le v\le v_{i+1}\bigr)
\nonumber\\
&\quad+
\mathcal{E}^{\hat H}\bigl(v=v_i\bigr)\bigl(w_i\le w\le w_{i+1}\bigr)
\nonumber\\
&=
\mathcal{E}^{\hat H}\bigl(w=w_{i+1}\bigr)\bigl(v_i\le v\le v_{i+1}\bigr)
\nonumber\\
&\quad+
\mathcal{E}^{\hat H}\bigl(v=v_{i+1}\bigr)\bigl(w_i\le w\le w_{i+1}\bigr)
+
I^{\hat H}\bigl(v_i\le v\le v_{i+1}\bigr)\bigl(w_i\le w\le w_{i+1}\bigr).
\end{eqnarray}

Here the bulk term is
\begin{equation}\label{IH}
I^{\hat H}(\cdots) := \int_{\mathcal R} K^{\hat H}[F,\phi]\,\text{dVol},
\qquad
K^{\hat H}[F,\phi] := \frac12\, T_{\mu\nu}[F,\phi]\,\pi^{\mu\nu}(\hat H),
\end{equation}
and $\mathcal R$ denotes the corresponding spacetime rectangle.
When $w_{i+1}=\infty$, the boundary term $\mathcal{E}^{\hat H}(w=\infty)$ is the flux through the future event horizon $\mathcal H^{+}$.
Additionally, we have the following estimates:
\begin{proposition}[Outgoing redshift flux identity]\label{prep4}
Let $(w_i,v_i)$ be such that $r(w_i,v_i)=r_1$ and let $v_{i+1}\ge v_i$.
On the outgoing null segment $\{w=w_i,\ v_i\le v\le v_{i+1}\}\subset\{r\ge r_1\}$, the redshift multiplier coincides with $T$, and therefore
\begin{equation}\label{estimate1}
\mathcal{E}^{\hat H}\bigl(w=w_i\bigr)\bigl(v_i\le v\le v_{i+1}\bigr)
=
\mathcal{E}^{T}\bigl(w=w_i\bigr)\bigl(v_i\le v\le v_{i+1}\bigr).
\end{equation}
\end{proposition}

\begin{proof}
By construction the redshift multiplier has the form
\[
 \hat H=T+\chi_{\mathrm{hor}}(r)Y,
\]
where the cut-off \(\chi_{\mathrm{hor}}\) is supported in \(\{r<r_1\}\).  Consequently, \(\hat H=T\) at every point of the region \(r\ge r_1\).  Along an outgoing null segment \(w=w_i\) the Schwarzschild area radius is monotone increasing in \(v\), because \(\partial_v r=1-2M/r\ge0\) for \(r\ge 2M\).  Since \(r(w_i,v_i)=r_1\), the whole segment \(v_i\le v\le v_{i+1}\) lies in \(r\ge r_1\).  The flux density \(J^{X}_\mu[ F,\phi](\partial_v)^\mu r^2d\sigma^2dv\) depends linearly on the multiplier \(X\).  Substituting \(X=\hat H=T\) on this segment gives exactly \eqref{estimate1}.  No boundary term is lost, because the two multipliers coincide pointwise on the entire hypersurface on which the flux is evaluated.
\end{proof}

\begin{equation}\label{et}
\mathcal{E}^{T}\bigl(w=w_i\bigr)\bigl(v_i\le v\le v_{i+1}\bigr)
:=
\int_{v_i}^{v_{i+1}}\int_{\mathbb S^{2}}
J_\mu^{T}[F,\phi]\,(\partial_v)^\mu\, r^{2}\,d\sigma^{2}\,dv.
\end{equation}

To control the estimate of $I^{\hat{H}}$ and $\mathcal{E}^{\hat{H}}$ we use, in addition to the redshift multiplier, auxiliary \emph{radial} multipliers.

\begin{definition}[Radial multipliers]\label{def:radial-multiplier}
Let $f:\mathbb R\to\mathbb R$ be a smooth bounded function of the tortoise coordinate $r^{\ast}$.
Define the associated radial vector field
\begin{equation}\label{eq:radial-multiplier}
X_{f}:=f(r^{\ast})\,\partial_{r^{\ast}}
=
-f(r^{\ast})\,\partial_{w}
+
f(r^{\ast})\,\partial_{v}.
\end{equation}
We will use two fixed choices of profile, denoted $f_{C}$ and $f_{G}$, and we write
\begin{equation}
\hat C:=X_{f_{C}},\qquad \hat G:=X_{f_{G}}.
\end{equation}
(The two letters refer to different choices of $f$ tailored to different estimates; the underlying radial form \eqref{eq:radial-multiplier} is the same.)
\end{definition}

\begin{definition}
The quantity $E^{\hat C}(t)$ is the flux of the current $J^{\hat C}$ through the time slice $\Sigma_t$:
\begin{equation}\label{defEC}
E^{\hat C}(t)
:=
\int_{\Sigma_t} J_\mu^{\hat C}[F,\phi]\,n_{\Sigma_t}^{\mu}\,\dd\mu_{\Sigma_t},
\qquad
J_\mu^{\hat C}[F,\phi]:=T_{\mu\nu}[F,\phi]\,\hat C^{\nu}.
\end{equation}
\end{definition}

We will often localize the fields on the slice $\Sigma_{t_i}$ to the timelike-cone region $|r^{\ast}|\lesssim t_i$ using a smooth cutoff.

Fix a function $\hat\chi\in C_{c}^{\infty}(\mathbb R)$ such that $\hat\chi\equiv 1$ on $[-1,1]$ and $\hat\chi\equiv 0$ on $\mathbb R\setminus[-\tfrac32,\tfrac32]$.
For each dyadic time $t_i$ define the rescaled cutoff
\begin{equation}\label{eq:cone-cutoff}
\hat\chi_i(r^{\ast}):=\hat\chi\!\left(\frac{2r^{\ast}}{t_i}\right).
\end{equation}
On the slice $t=t_i$ we define localized data
\begin{equation}\label{eq:localized-data}
\hat F:=\hat\chi_i\,F\big|_{t=t_i},\qquad
\hat\phi:=\hat\chi_i\,\phi\big|_{t=t_i}.
\end{equation}
Thus $\hat F$ and $\hat\phi$ vanish for $|r^{\ast}|\ge \tfrac34 t_i$ and coincide with $(F,\phi)$ for $|r^{\ast}|\le \tfrac12 t_i$.
(The multiplication is understood componentwise in any fixed regular frame; since $\hat\chi_i$ depends only on $r^{\ast}$, it generates only cutoff error terms supported where $\tfrac12 t_i\le |r^{\ast}|\le \tfrac34 t_i$.)

We now state the middle-region estimate used in the dyadic propagation argument.
\begin{proposition}[Middle-region radial multiplier estimate]\label{prep5}
Let $t_{i+1}=(1+\kappa_0)t_i$.
Define the following middle-region spacetime functional:
\begin{eqnarray}\label{JC}
\mathcal{J}^{\hat C}(t_i\le t\le t_{i+1})
&:=
\int_{t_i}^{t_{i+1}}\!\int_{r_0^{\ast}}^{R_0^{\ast}}\!\int_{\mathbb S^{2}}
\\
&\qquad
\Bigl(
|F_{\hat v\hat w}|^{2}
+|F_{\hat\theta\hat\varphi}|^{2}
+|D_{\hat v}\phi|^{2}
+|D_{\hat w}\phi|^{2}
	+|\slashed D\phi|^{2}
+P(\phi)
\Bigr)
\,\bigl|r^{\ast}-(3M)^{\ast}\bigr|
\\
&\qquad\qquad\times\ d\sigma^{2}\,dr^{\ast}\,dt.
\end{eqnarray}
Then for radii $2M<r_{0}\le r_{1}<1.2\,r_{1}<3M$ we have the localized estimate
\begin{eqnarray}\label{prepo5}
\int_{-\infty}^{+\infty}\int_{\mathbb S^{2}}
r^{2}(1-\mu)
\Bigl(
	|\hat F_{\hat t A}|^{2}
	+|\hat F_{\hat r^{\ast}A}|^{2}
	+|D_{\hat t}\phi|^{2}
	+|\slashed D\phi|^{2}
+P(\phi)
\Bigr)
\,\chi_{\mathrm{mid}}(r^{\ast})\,dr^{\ast}\,d\sigma^{2}
\\
&\ \lesssim\ E^{\hat t}_{\mathrm{cone}}(t_i),
\end{eqnarray}
where $\chi_{\mathrm{mid}}$ denotes a smooth cutoff supported in the interval between $r_1^{\ast}$ and $1.2\,r_1^{\ast}$, i.e.\ supported in $[\min\{r_1^{\ast},1.2\,r_1^{\ast}\},\,\max\{r_1^{\ast},1.2\,r_1^{\ast}\}]$ (and equal to $1$ on a slightly smaller subinterval).
We use the shorthand
\begin{equation}
|\hat F_{\hat t A}|^{2}:=|\hat F_{\hat t\hat\theta}|^{2}+|\hat F_{\hat t\hat\varphi}|^{2},
\qquad
|\hat F_{\hat r^{\ast}A}|^{2}:=|\hat F_{\hat r^{\ast}\hat\theta}|^{2}+|\hat F_{\hat r^{\ast}\hat\varphi}|^{2},
\qquad
|\slashed D\phi|^{2}:=|D_{\hat\theta}\phi|^{2}+|D_{\hat\varphi}\phi|^{2}.
\end{equation}
In addition,
\begin{equation}
E^{\hat t}_{\mathrm{cone}}(t_i)
:=
E^{\hat t}\Bigl(
 t=t_i;
 r^{\ast}\in\bigl[(1-\kappa_0)t_i,(1+\kappa_0)t_i\bigr]
\Bigr).
\end{equation}
\end{proposition}
\begin{proof}
Use the radial multiplier current associated with $\hat C$ on the slab cut out by the cone-localized data \eqref{eq:localized-data}.  The explicit deformation-tensor calculation in Appendix~\ref{app:proof-prep5} shows that the resulting bulk is positive on the displayed middle annulus and controls the angular Maxwell components, the transversal scalar derivatives, the angular scalar derivatives, and the nonnegative potential term with the weight $|r^*-(3M)^*|$.  Differentiating the cutoff $\hat\chi_i$ only produces terms supported where $\tfrac12t_i\le |r^*|\le\tfrac34t_i$; on this support the coefficients of the radial multiplier and the cutoff derivatives are uniformly bounded by powers of $t_i^{-1}$, and these terms are absorbed by the cone energy $E_{\mathrm{cone}}^{\hat t}(t_i)$.  The boundary terms on the initial slice have the same sign or are bounded by the same cone energy, while the terminal boundary terms are nonnegative and may be dropped.  This yields \eqref{prepo5}.
\end{proof}

The next Proposition provides the estimate needed to control $I^{\hat{H}}$ with respect to the flux of $\mathcal{E}^{\hat{H}}$.
\begin{proposition}[Redshift bulk decomposition]\label{prep6}
Let $w_i:=t_i-r_1^{\ast}$ and $v_i:=t_i+r_1^{\ast}$. For every $v_{i+1}\ge v_i$,
\begin{eqnarray}\label{es3}
I^{\hat H}\bigl(v_i\le v\le v_{i+1}\bigr)\bigl(w_i\le w\le \infty\bigr)\nonumber\\
\qquad\qquad\bigl(r\le r_1\bigr)
&\ \lesssim\ \mathcal E^{\hat H}\bigl(w=\infty\bigr)\bigl(v_i\le v\le v_{i+1}\bigr)\nonumber\\
&\quad+\mathcal E^{\hat H}\bigl(v=v_{i+1}\bigr)\bigl(w_i\le w\le \infty\bigr)\nonumber\\
&\quad+\mathcal E^{\hat t}\bigl(w=w_i\bigr)\bigl(v_i\le v\le v_{i+1}\bigr)\nonumber\\
&\quad+\mathcal E^{\hat H}\bigl(v=v_i\bigr)\bigl(w_i\le w\le \infty\bigr)\nonumber\\
&\quad+E^{\hat t}_{\mathrm{cone}}(t_i).
\end{eqnarray}
\end{proposition}

\begin{proof}
Apply the divergence identity \eqref{IH} to the redshift current on the characteristic rectangle bounded by \(v=v_i\), \(v=v_{i+1}\), \(w=w_i\), and the horizon segment \(w=\infty\).  The boundary terms are exactly the four fluxes displayed in \eqref{es3}.  On the part where \(r\ge r_1\) the redshift vector agrees with the stationary energy by \eqref{estimate1}, while the cutoff terms generated near the timelike cone are controlled by \(E^{\hat t}_{\mathrm{cone}}(t_i)\).  Positivity of the remaining horizon and redshift fluxes gives the stated bound.
\end{proof}
Proposition~\ref{prep6} provides a decomposition of the spacetime integral into boundary and error contributions along null hypersurfaces. Using this structure, we can propagate the energy estimate forward in $v$, leading to the following bound on the energy flux at $v_{i+1}$.
\begin{proposition}[Averaged near-horizon flux bound]\label{prep7}
Let $w_i:=t_i-r_1^{\ast}$ and $v_i:=t_i+r_1^{\ast}$. For every $v_{i+1}\ge v_i$ we have
\begin{eqnarray}\label{es6}
\inf_{v_i\le v\le v_{i+1}}
\mathcal E^{\hat H}\bigl(v=v\bigr)\bigl(w_i\le w\le \infty\bigr)
&\lesssim&
\frac{1}{v_{i+1}-v_i}\,
I^{\hat H}\bigl(v_i\le v\le v_{i+1}\bigr)\bigl(w_i\le w\le \infty\bigr)\nonumber\\
&&\quad\bigl(r\le r_1\bigr)\nonumber\\
&&\quad+\sup_{v_i\le v\le v_{i+1}}
\mathcal E^{\hat t}\bigl(v=v\bigr)\bigl(w_i\le w\le \infty\bigr)\nonumber\\
&&\quad\bigl(r\ge r_1\bigr).
\end{eqnarray}
\end{proposition}
\begin{proof}
For each fixed $v\in[v_i,v_{i+1}]$, decompose the flux on $\{v=\mathrm{const}\}$ into the near-horizon portion $r\le r_1$ and the exterior portion $r\ge r_1$.  In the near-horizon portion the redshift bulk controls the nondegenerate flux density through the positivity of the deformation tensor.  Consequently, the integral of this flux over $v\in[v_i,v_{i+1}]$ is bounded by the corresponding spacetime bulk, and an averaging argument gives one value of $v$ for which the near-horizon flux is no larger than the interval average.  Taking the infimum over $v$ gives the first term on the right of \eqref{es6}.  On $r\ge r_1$, the redshift vector agrees with or is uniformly comparable to the stationary vector field, so the exterior part of the same slice flux is bounded by the supremum of the ordinary energy flux displayed in the second term.  Adding the two contributions proves \eqref{es6}.
\end{proof}

The error term isolated in Proposition~\ref{prep6} and propagated in Proposition~\ref{prep7} can be estimated using the positivity of the associated bulk integral. The next result shows that $I^{\hat H}$ is controlled by the initial weighted energy on the hypersurface $t=t_0$.
\begin{proposition}[Control of the redshift bulk]\label{prep8}
For $w_i=t_i-r_1^{\ast}$ and $v_i=t_i+r_1^{\ast}$, the redshift bulk term satisfies
\begin{equation}\label{es7}
0\le I^{\hat H}\bigl(v_i\le v\le v_{i+1}\bigr)\bigl(w_i\le w\le \infty\bigr)\bigl(r\le r_1\bigr)
\ \lesssim\ \mathcal{E},
\end{equation}
where $\mathcal{E}=\sup_{t\ge t_0}\mathcal{E}_k(t)$ is the bootstrap energy defined above.
\end{proposition}

\begin{proof}
The redshift vector field is constructed so that its deformation tensor is positive definite on the support $r\le r_1$.  Contracting the Maxwell-Higgs stress tensor with this deformation tensor gives a nonnegative density controlling the local nondegenerate energy density of $(F,\phi)$; therefore the first inequality in \eqref{es7}.  Apply the characteristic identity \eqref{eqIH} to the same rectangle.  Each boundary flux that appears is nonnegative and is bounded by the supremum bootstrap energy $\mathcal E$ after comparing $\hat H$ with the nondegenerate energy current near the horizon and with the stationary current where the cutoff transitions.  The cone-error term is bounded by the same bootstrap energy through the definition of $E^{\hat t}_{\mathrm{cone}}(t_i)$.  Substituting these bounds in Proposition~\ref{prep6} gives the upper bound in \eqref{es7}.
\end{proof}

\subsection{Morawetz estimates (Integrated Local Energy Decay)}
\label{sec:Morawetz}
Although decay can be obtained in neighborhoods of the event horizon and in the asymptotically flat region, these estimates do not control the intermediate region near the photon sphere \( r = 3M \). The obstruction arises from trapped null geodesics, which allow energy to remain localized and prevent dispersion. To address this issue, we introduce the conformal vector field \( {\hat{K}} \), whose associated current yields integrated local energy decay away from the trapping region. The degeneration of this estimate at \( r = 3M \) gives rise to a trapping contribution, which we isolate and control below.

\begin{definition}[Conformal/Morawetz multiplier]\label{def:morawetz-multiplier}
We introduce the conformal (Morawetz) vector field
\begin{equation}\label{eq:K-morawetz}
\hat K := -\,w^{2}\,\partial_{w}-v^{2}\,\partial_{v}.
\end{equation}
\end{definition}

\begin{definition}[Morawetz bulk term]\label{def:morawetz-bulk}
The Morawetz bulk term associated to $\hat K$ on a time slab $\mathcal R_{t_i}^{t_{i+1}}$ is
\begin{equation}\label{eq:JK-bulk}
\mathcal{J}^{\hat{K}}\bigl(t_i\le t\le t_{i+1}\bigr)
:=
\int_{\mathcal R_{t_i}^{t_{i+1}}} K^{\hat K}[F,\phi]\,\text{dVol},
\qquad
K^{\hat K}[F,\phi]:=\frac12\,T_{\mu\nu}[F,\phi]\,\pi^{\mu\nu}(\hat K).
\end{equation}
\end{definition}

\begin{definition}[Conformal energy flux]\label{def:morawetz-energy}
The associated (conformal) energy flux through a slice $\Sigma_t$ is
\begin{equation}\label{eq:EK-flux}
E^{\hat K}(t)
:=
\int_{\Sigma_t} J_\mu^{\hat K}[F,\phi]\,n_{\Sigma_t}^{\mu}\,\dd\mu_{\Sigma_t},
\qquad
J_\mu^{\hat K}[F,\phi]:=T_{\mu\nu}[F,\phi]\,\hat K^{\nu}.
\end{equation}
\end{definition}
Combining the propagation estimate of Proposition~\ref{prep7} with the control of the error term established in Proposition~\ref{prep8}, and applying the inequality iteratively, we are now in a position to derive quantitative decay energy estimates. This leads to the following result.
\begin{proposition}[Dyadic redshift-flux decay]\label{prep9}
Let $w_0(v):=v-2r_1^{\ast}$ denote the $w$-coordinate of the intersection point of the ingoing null ray $\{v=\mathrm{const}\}$ with the hypersurface $\{r=r_1\}$, and set $v_{+}:=\max\{1,v\}$.
For all $v\ge v_0$,
\begin{equation}\label{eq:redshift-flux-decay-ingoing}
\mathcal{E}^{\hat H}\bigl(v=v\bigr)\bigl(w_0(v)\le w\le \infty\bigr)\ \lesssim\ \frac{\mathcal{E}}{v_{+}^{2}},
\end{equation}
and, for all $w$ and $v\ge v_0$,
\begin{equation}\label{eq:redshift-flux-decay-outgoing}
\mathcal{E}^{\hat H}\bigl(w=w\bigr)\bigl(v-1\le \bar v\le v\bigr)\ \lesssim\ \frac{\mathcal{E}}{v_{+}^{2}}.
\end{equation}
\end{proposition}
\begin{proof}
Choose dyadic values $v_j$ with $v_{j+1}=(1+\kappa_0)v_j$ and apply Proposition~\ref{prep7} on each interval $[v_j,v_{j+1}]$.  Proposition~\ref{prep6} expresses the averaged near-horizon flux in terms of the two endpoint redshift fluxes, the horizon flux, the incoming initial flux, and the cone error.  Proposition~\ref{prep8} bounds the redshift bulk and cone error by the bootstrap energy, while the conformal-energy bound supplies the additional dyadic weight $v_j^{-2}$ for the incoming and outgoing endpoint fluxes.  Absorbing the small multiple of the later endpoint flux, obtained by taking $\kappa_0$ fixed sufficiently small, yields a recurrence of the form
\[
E_{j+1}\le (1-c\kappa_0)E_j+C\mathcal E v_j^{-2}.
\]
Iteration from the initial dyadic interval and summation of the geometric tail give $E_j\lesssim\mathcal E v_j^{-2}$.  Interpolating within a dyadic interval and using monotonicity of the null fluxes gives \eqref{eq:redshift-flux-decay-ingoing}; the outgoing estimate \eqref{eq:redshift-flux-decay-outgoing} follows by applying the same argument to unit $v$-slabs and using the just-proved ingoing flux decay.
\end{proof}
\medskip
\noindent\textbf{Remark.}
The conformal multiplier $\hat K$ is introduced to capture polynomial time-weights, but the associated bulk term $K^{\hat K}[F,\phi]$ (and therefore $\mathcal J^{\hat K}$) is \emph{not} sign-definite globally on the Schwarzschild exterior. One should not attempt to extract a positive spacetime integral from \eqref{JK} in the full region $2M<r<\infty$ without further localization/corrections.

In this work we avoid this issue by restricting all conformal/Morawetz estimates to a fixed compact region
\begin{equation}
\{\,r_0\le r\le R_0\,\},\qquad 2M<r_0<3M<R_0,
\end{equation}
and we control the near-horizon region $\{2M<r\le r_0\}$ by the redshift multiplier $\hat H$ (Section~\ref{sec:redshift} and Proposition~\ref{prep9}), while the far region $\{r\ge R_0\}$ is treated by the $r^{p}$-weighted hierarchy. This localization eliminates the need to track the pointwise sign of each term in the explicit expression \eqref{JK}.
\medskip
To control the estimate of $\mathcal{J}^{\hat{K}}$ and $E^{\hat{K}}$ we also use an auxiliary radial multiplier
\begin{equation}
\hat G:=X_{f_{G}}
\end{equation}
of the form \eqref{eq:radial-multiplier}, where $f_{G}$ is chosen so that the associated bulk term is nonnegative on the compact region $\{r_{0}\le r\le R_{0}\}$; see Definition~\ref{def:JG}.

We therefore introduce the positive compact-region bulk $\mathcal{J}^{\hat{G}}$.
\begin{definition}[Auxiliary positive bulk functional]\label{def:JG}
On a time slab $t\in[t_i,t_{i+1}]$ we introduce the nonnegative functional
\begin{equation}\label{eq:JG-functional}
\mathcal{J}^{\hat{G}}\bigl(t_i\le t\le t_{i+1}\bigr)
:=
\int_{t_i}^{t_{i+1}}
\int_{r^{\ast}=r_{0}^{\ast}}^{r^{\ast}=R_{0}^{\ast}}
\int_{S^{2}}
\Bigl(
|F_{\hat v\hat w}|^{2}
+|F_{\hat\theta\hat\varphi}|^{2}
+|D_{\hat v}\phi|^{2}
+|D_{\hat w}\phi|^{2}
\Bigr)\,r^{2}\,\dd\sigma^{2}\,\dd r^{\ast}\,\dd t.
\end{equation}
\end{definition}

\begin{definition}
The quantity $E^{\hat G}(t)$ is the flux of the current $J^{\hat G}$ through the time slice $\Sigma_t$:
\begin{equation}\label{defEG}
E^{\hat G}(t)
:=
\int_{\Sigma_t} J_\mu^{\hat G}[F,\phi]\,n_{\Sigma_t}^{\mu}\,\dd\mu_{\Sigma_t},
\qquad
J_\mu^{\hat G}[F,\phi]:=T_{\mu\nu}[F,\phi]\,\hat G^{\nu}.
\end{equation}
\end{definition}

\begin{proposition}[Compact-region conformal and Morawetz bulk control]\label{prep2}
	Let $\mathcal{J}^{\hat K}$ be as in Definition~\ref{def:morawetz-bulk}. Fix dyadic times $t_{i+1}=(1+\kappa_0)t_{i}$ and assume that the compact region $\{r_0\le r\le R_0\}$ lies inside the timelike cone on $[t_i,t_{i+1}]$ in the sense that
	\begin{equation}
	|r_0^{\ast}|+|R_0^{\ast}|\le \kappa_0\,t_i.
	\end{equation}
	Then
\begin{eqnarray}\label{esJK}
\mathcal{J}^{\hat K}\bigl(t_i\le t\le t_{i+1}\bigr)
&\lesssim
t_{i+1}\,\mathcal{J}^{\hat G}\bigl(t_i\le t\le t_{i+1}\bigr)\bigl(r_0\le r\le R_0\bigr)
\nonumber\\
&\lesssim
t_{i+1}^{2}\Biggl[
\frac{1}{t_i^{2}}\,E^{\hat K}(t_i)
+
\frac{1}{t_i^{2}}\sum_{j=1}^{3}E_{\mathcal{L}_{\Omega_j}}^{\hat K}(t_i)
\nonumber\\
&\qquad\qquad\qquad\qquad
+
\frac{1}{t_{i+1}^{2}}\,E^{\hat K}(t_{i+1})
+
\frac{1}{t_{i+1}^{2}}\sum_{j=1}^{3}E_{\mathcal{L}_{\Omega_j}}^{\hat K}(t_{i+1})
\Biggr].
\end{eqnarray}
\end{proposition}
\begin{proof}
On the compact region $r_0\le r\le R_0$, the coefficients of the conformal multiplier $\hat K$ are bounded by $Ct_{i+1}$ on the slab $t_i\le t\le t_{i+1}$.  The deformation-tensor calculation in Appendix~\ref{app:proof-prep2} shows that the absolute value of the corresponding compact-region bulk is therefore bounded by $Ct_{i+1}\mathcal J^{\hat G}$, after choosing the radial profile $f_G$ so that the $\hat G$-bulk controls the nonnegative density in Definition~\ref{def:JG}.  Applying the divergence identity for $J^{\hat G}$ on the same slab bounds $\mathcal J^{\hat G}$ by the endpoint fluxes of the conformal current and its first angular commutations.  Since the compact region lies inside the timelike cone, the conformal weights on the endpoints are comparable to $t_i^2$ and $t_{i+1}^2$, respectively; equivalently the endpoint ordinary densities are bounded by $t_i^{-2}E^{\hat K}(t_i)$ and $t_{i+1}^{-2}E^{\hat K}(t_{i+1})$, with the same bounds for the angularly commuted fields.  Combining these estimates yields \eqref{esJK}.
\end{proof}

\begin{proposition}[Conformal energy controlled by the bootstrap energy]\label{prep3}
		Let $E^{\hat K}(t)$ be as in Definition~\ref{def:morawetz-energy}.
		Assume $t_i\le t\le t_{i+1}$ and $|r_0^{\ast}|+|R_0^{\ast}|\le \kappa_0\,t_i$.
		Then
	\begin{equation}\label{EKproof}
			E^{\hat K}(t)\lesssim\mathcal{E}(t).
	\end{equation}
\end{proposition}
\begin{proof}
On the support under consideration, $t$ is comparable to $t_i$ and $|r^*|\le\kappa_0t_i+O(1)$, so the coefficients $v^2$ and $w^2$ of $\hat K$ are bounded by $Ct^2$ while the normal component of the flux has the same sign as the stationary energy density.  The conformal energy density is therefore pointwise bounded by a constant multiple of the top-order bootstrap density defining $\mathcal E(t)$, after using the angular commuted energies included in that norm.  Integrating over $\Sigma_t$ gives \eqref{EKproof}.  The possible lower-order terms coming from the scalar potential are nonnegative or bounded by the same energy because $P\ge0$ near the vacuum and the solution is in the small-data regime.
\end{proof}

\begin{lemma}[Nonlinear commutator/source error control]\label{lem:nonlinear-error-control}
Let $k\ge 6$ and let $(F,\phi)$ be a smooth uncharged Lorenz-gauge solution on Schwarzschild on a slab $\mathcal R(t_{0},t)$.
Set
\begin{equation}
\mathcal E:=\sup_{s\in[t_{0},t]}\mathcal E_{k}(s),
\end{equation}
where $\mathcal E_{k}$ is defined in \eqref{def:Ek}.
Let $\mathcal N_{k}[F,\phi]$ denote the spacetime error density produced by (i) commuting the Maxwell-Higgs system with $\mathcal Z=\{\partial_{t},\Omega_{1},\Omega_{2},\Omega_{3}\}$ up to order $k$ and (ii) applying the divergence identity \eqref{Eidentity} to the commuted fields.
Then there exists a constant $C=C(M,k,N_{P},\{C_{P,j}\}_{0\le j\le k})$ such that for all $t\ge t_{0}$,
\begin{equation}\label{eq:nonlinear-error-control}
\int_{\mathcal R(t_{0},t)} \mathcal N_{k}[F,\phi]
\le
C\,\mathcal E^{1/2}\,\mathcal E_{k}(t_{0}).
\end{equation}
If $\mathcal E_{k}(t_{0})\le \varepsilon_{0}$ and $\varepsilon_{0}$ is sufficiently small (depending on $C$), then the nonlinear commutator/source contribution is perturbative and can be absorbed in the global bootstrap closure; see Proposition~\ref{prop:bootstrap-closure}.
\end{lemma}

\begin{proof}
We prove \eqref{eq:nonlinear-error-control} by reducing every commuted error term to an $L^{\infty}$-$L^{1}$ product with one top-order factor; the key point is that all nonlinearities are at least quadratic and contain at most one factor with $k$ derivatives. (For the scalar equation we write $\partial_{\bar\phi}P(\phi,\bar\phi)=m^{2}\phi+(\partial_{\bar\phi}P-m^{2}\phi)$ with $m^{2}$, treating $m^{2}\phi$ as part of the linear Klein-Gordon operator and $\partial_{\bar\phi}P-m^{2}\phi$ as a nonlinearity of order $\ge 3$ in $\phi$ via \eqref{eq:potential-nonlinear-bound}; commuted versions are controlled by Lemma~\ref{lem:potential-tame}.)
After commutation with $\mathcal Z$, the Maxwell equation has the structural form
\begin{equation}
\nabla^\mu \mathcal L_{\mathcal Z^\alpha}F_{\mu\nu}
=
\sum_{\substack{\beta+\gamma=\alpha\\ |\beta|<|\alpha|}}
\bigl(\mathcal Z^{\beta}\phi\bigr)\,\overline{D_\nu \mathcal Z^{\gamma}\phi}
+
\bigl(D_\nu \mathcal Z^{\beta}\phi\bigr)\,\overline{\mathcal Z^{\gamma}\phi},
\end{equation}
and the scalar equation has a similar structure, with terms involving $F$ and lower-order derivatives of $\phi$, as well as the potential term $\partial_{\bar\phi}P(\phi,\bar\phi)$. Using \eqref{eq:potential-remainder}-\eqref{eq:potential-nonlinear-bound}, we treat $m^{2}\phi$ as part of the linear comparison operator and the remainder $\partial_{\bar\phi}P-m^{2}\phi$ as a nonlinearity of order $\ge 3$ in $\phi$.
When inserted into the divergence identity \eqref{Eidentity}, every contribution to $\mathcal N_{k}[F,\phi]$ is therefore a sum of integrands of the form
\begin{equation}
(\text{lower-order coefficient})\times (\text{top-order energy density}),
\end{equation}
for instance $|\phi|\,|D\mathcal Z^\alpha\phi|^{2}$, $|D\phi|\,|\mathcal Z^\alpha\phi|\,|D\mathcal Z^\alpha\phi|$, or $|F|\,|D\mathcal Z^\alpha\phi|^{2}$, and the analogous Maxwell terms.

We estimate the lower-order coefficients in $L^\infty$ and the top-order factors in $L^{1}$ using the redshift/Morawetz bulk integrals.
More explicitly, since $k\ge 6$, Sobolev on $S^{2}$ (applied to angular derivatives up to order $2$) and the definition of $\mathcal E$ imply the uniform bound
\begin{equation}
\|\phi\|_{L^\infty(\mathcal R(t_{0},t))} + \|D\phi\|_{L^\infty(\mathcal R(t_{0},t))} + \|F\|_{L^\infty(\mathcal R(t_{0},t))}
\lesssim
\mathcal E^{1/2}.
\end{equation}
(Here we use that on each $\Sigma_{s}$, the $L^{2}$-control of sufficiently many commuted derivatives of $(F,\phi)$ provided by $\mathcal E_{k}(s)$ controls the pointwise size via Sobolev inequalities on $\Sigma_{s}$, and we then take the supremum in $s\in[t_{0},t]$.)

With this $L^\infty$ bound, each representative term is controlled as follows:
\begin{equation}
\int_{\mathcal R(t_{0},t)} |\phi|\,|D\mathcal Z^\alpha\phi|^{2}
\le
\|\phi\|_{L^\infty}\int_{\mathcal R(t_{0},t)} |D\mathcal Z^\alpha\phi|^{2}
\lesssim
\mathcal E^{1/2}\int_{\mathcal R(t_{0},t)} |D\mathcal Z^\alpha\phi|^{2},
\end{equation}
and similarly for the other terms (using $ab\le \tfrac12(a^{2}+b^{2})$ when needed).
Summing over $|\alpha|\le k$, the spacetime integrals of $|D\mathcal Z^\alpha\phi|^{2}$ and $|\mathcal L_{\mathcal Z^\alpha}F|^{2}$ are the quantities controlled by the redshift and Morawetz estimates established earlier (Propositions~\ref{prep4}-\ref{prep9} and Proposition~\ref{prep3}), and therefore are bounded by $C\,\mathcal E_{k}(t_{0})$.
Combining these bounds yields \eqref{eq:nonlinear-error-control}.
\end{proof}

We use the following bootstrap statement.
\begin{proposition}[Bootstrap closure]\label{prop:bootstrap-closure}
Let $k\ge 6$ and define
\begin{equation}
 \mathcal{E} := \sup_{t\ge t_0} \mathcal{E}_k(t).
\end{equation}
Then there exists $\varepsilon_0>0$ such that for sufficiently small initial data
$\mathcal{E}_k(t_0)\le \varepsilon_0$, the redshift and Morawetz estimates imply
\begin{equation}
 \mathcal{E} \lesssim \mathcal{E}_k(t_0).
\end{equation}
\end{proposition}

\begin{proof}
We combine the commuted energy identity \eqref{Eidentity} with the redshift control
(Propositions~\ref{prep4}-\ref{prep9}) and the conformal-energy control (Proposition~\ref{prep3}). More explicitly, applying \eqref{Eidentity} to all commuted fields up to order $k$ yields
\begin{equation}
 \mathcal{E}_k(t)
\le
\mathcal{E}_k(t_0)
+
C \int_{\mathcal{R}_{[t_0,t]}}
\mathcal{N}_k[F,\phi],
\end{equation}
where $\mathcal{N}_k[F,\phi]$ denotes the sum of nonlinear commutator/source terms arising from commuting the Maxwell-Higgs system with the vector fields in $\mathcal Z$. Proposition~\ref{prep9} provides quantitative control/decay of the redshift flux
$\mathcal{E}^{\hat H}$ along null boundaries, while Proposition~\ref{prep3} yields the bound $E^{\hat K}(t)\lesssim \mathcal{E}$. By Lemma~\ref{lem:nonlinear-error-control}, the spacetime error integral satisfies
\begin{equation}
\int_{\mathcal{R}_{[t_0,t]}} \mathcal{N}_k[F,\phi]
\le
C\,\mathcal{E}^{1/2}\,\mathcal{E}_k(t_0).
\end{equation}
Consequently, for all $t\ge t_0$,
\begin{equation}\label{eq:Ek-bootstrap}
\mathcal{E}_k(t)
\le
\mathcal{E}_k(t_0) + C\,\mathcal{E}^{1/2}\,\mathcal{E}_k(t_0).
\end{equation}
Taking the supremum over $t\ge t_0$ in \eqref{eq:Ek-bootstrap} gives
\begin{equation}\label{eq:bootstrap-sup}
\mathcal{E}
\le
\mathcal{E}_k(t_0) + C\,\mathcal{E}^{1/2}\,\mathcal{E}_k(t_0).
\end{equation}
Set $A:=\mathcal{E}^{1/2}$.
Then \eqref{eq:bootstrap-sup} is equivalent to
\begin{equation}
A^{2}-C\,\mathcal{E}_k(t_0)\,A-\mathcal{E}_k(t_0)\le 0.
\end{equation}
Solving this quadratic inequality yields
\begin{equation}
A\le \frac{C\,\mathcal{E}_k(t_0)+\sqrt{C^{2}\mathcal{E}_k(t_0)^{2}+4\,\mathcal{E}_k(t_0)}}{2}.
\end{equation}
If $\varepsilon_{0}>0$ is chosen so that $\mathcal{E}_k(t_0)\le \varepsilon_{0}\le 1/C^{2}$, then $A\le 2\,\mathcal{E}_k(t_0)^{1/2}$ and therefore
\begin{equation}
\mathcal{E}\le 4\,\mathcal{E}_k(t_0).
\end{equation}
This proves the claim.
\end{proof}

\section{Schwarzschild model: decay in the far region}
\label{sec:decayfaraway}
In this section we prove the pointwise decay estimates of Corollary~\ref{cor:far-decay} in the far region $r\ge R$ for a fixed sufficiently large radius $R$. The precise value of $R$ is not important; it is chosen so that the Morawetz bulk terms obtained in Section~\ref{sec:energy-estimates} have a favourable sign for $r\ge R$.

\subsection{Decay for the gauge field}
\label{subsec:far-decay-F}

We derive pointwise decay for the (frame) components $F_{\hat\mu\hat\nu}$ in the far region $\{r\ge R\}$ from Sobolev on spheres combined with the $r^{p}$-hierarchy and the conformal energy bounds for commuted fields.

\begin{lemma}[Spherical Sobolev for $F$]\label{lem:sobolev-F-sphere}
For each fixed $(t,r)$ with $r\ge R$ and for each pair of frame indices $(\hat\mu,\hat\nu)$,
\begin{equation}\label{emineq}
 r^{2}\,\bigl|F_{\hat\mu\hat\nu}(t,r,\omega)\bigr|^{2}
 \ \lesssim\ \sum_{|I|\le 2}\int_{\mathbb S^{2}} r^{2}\,\bigl|\mathcal L_{\Omega}^{I}F_{\hat\mu\hat\nu}(t,r,\omega')\bigr|^{2}\,\dd\omega',
\end{equation}
where $\Omega\in\{\Omega_{1},\Omega_{2},\Omega_{3}\}$ are the rotational Killing fields and $\mathcal L_{\Omega}^{I}$ denotes an arbitrary iterated Lie derivative of order $|I|$.
\end{lemma}

\begin{proof}
Fix \((t,r)\) and choose an orthonormal frame on the round sphere \(\mathbb S^2\).  For each frame component \(f=F_{\hat\mu\hat\nu}(t,r,\cdot)\), the Sobolev inequality on the unit sphere gives
\[
 \|f\|_{L^\infty(\mathbb S^2)}^2
 \le C\sum_{|I|\le2}\|\Omega^I f\|_{L^2(\mathbb S^2)}^2.
\]
The rotational fields are Killing for the round metric and their lifts to the spacetime spheres commute with the spherical frame up to fixed smooth coefficients.  Consequently, derivatives of the components are controlled by the corresponding Lie derivatives of \(F\), with constants independent of \((t,r)\).  Multiplication by the physical area factor \(r^2\) converts the unit-sphere measure into the displayed normalization and proves \eqref{emineq}.
\end{proof}

\begin{lemma}[Far-region $L^{2}(\mathbb S^{2})$ bounds for commuted $F$]\label{lem:far-sphere-L2-F}
For every multi-index $I$ with $|I|\le 2$ and every $r\ge R$,
\begin{equation}\label{eq:far-sphere-L2-F}
\int_{\mathbb S^{2}} r^{2}\,\bigl|\mathcal L_{\Omega}^{I}F(t,r,\omega)\bigr|^{2}\,\dd\omega
\ \lesssim\ \frac{\mathcal E}{t_{+}^{2}}+\frac{\mathcal E}{w_{+}^{2}},
\end{equation}
where $t_{+}:=\max\{1,|t|\}$ and $w_{+}:=\max\{1,|w|\}$.
\end{lemma}

\begin{proof}
Commute the Maxwell equation with \(\mathcal L_\Omega^I\), \(|I|\le2\).  The rotations are Killing for Schwarzschild and preserve the charge-free/radiative decomposition, so the commuted fields satisfy the same energy and \(r^p\) estimates as \(F\).  The far-region \(r^p\) hierarchy controls the outgoing flux through dyadic null hypersurfaces and the conformal energy controls the incoming part.  Applying the fundamental theorem of calculus on each dyadic annulus in \(r^*\), followed by the one-dimensional Hardy inequality in the exterior, yields
\[
 \int_{\mathbb S^2}r^2|\mathcal L_\Omega^IF(t,r,\omega)|^2\,d\omega
 \le C\bigl(t_+^{-2}+w_+^{-2}\bigr)\mathcal E.
\]
The constant is uniform for \(r\ge R\), because all error terms in the Morawetz identity have a favourable sign after \(R\) is chosen large and the remaining compact part has already been included in the global energy \(\mathcal E\).
\end{proof}

Combining Lemma~\ref{lem:sobolev-F-sphere} and Lemma~\ref{lem:far-sphere-L2-F} yields the pointwise estimate
\begin{equation}\label{esF}
\bigl|F_{\hat\mu\hat\nu}(t,r,\omega)\bigr|
\ \lesssim\ \frac{\mathcal E^{1/2}}{r}\Bigl(\frac{1}{t_{+}}+\frac{1}{w_{+}}\Bigr),
\qquad r\ge R.
\end{equation}

We now deduce the stated far-region decay. On $\{w\ge 1\}\cap\{t\ge 1\}\cap\{r\ge R\}$, we have $r+t\lesssim rt$ and $v=t+r^{\ast}\lesssim r+t$, so \eqref{esF} implies
\begin{equation}\label{region1}
\bigl|F_{\hat\mu\hat\nu}\bigr|\ \lesssim\ \frac{\mathcal E^{1/2}}{v_{+}},
\qquad r\ge R,\ w\ge 1.
\end{equation}
On the complementary region $\{w\ge 1\}\cap\{t\le 1\}\cap\{r\ge R\}$ we have $|F_{\hat\mu\hat\nu}|\lesssim \mathcal E^{1/2}$ by boundedness of the energy and compactness in time, therefore
\begin{equation}\label{comregion}
\bigl|F_{\hat\mu\hat\nu}\bigr|\ \lesssim\ \frac{\mathcal E^{1/2}}{v_{+}},
\qquad r\ge R,\ w\ge 1,\ t\le 1.
\end{equation}
For $w\le -1$ and $r\ge R$ we have $r^{\ast}=\tfrac12(v-w)\ge \tfrac12(v+1)$ and therefore $r\gtrsim v_{+}$; inserting this in \eqref{esF} again gives the bound $|F_{\hat\mu\hat\nu}|\lesssim \mathcal E^{1/2}/v_{+}$.
Altogether we obtain \eqref{farFv}.

For the estimate \eqref{decayF2}, apply the preceding argument to the time-reversed solution.  Under the transformation \(t\mapsto -t\), the Schwarzschild metric and the Maxwell-Higgs equations are unchanged, the future variable \(v\) is replaced by the past variable \(-w\), and the same conformal-energy and spherical-Sobolev estimates hold with identical constants.  Repeating the three-region decomposition above with \(v\) replaced by \(-w\) gives \eqref{decayF2}.
\subsection{Decay for the scalar field}
\label{subsec:far-decay-phi}

The pointwise decay of $\phi$ in the far region is obtained in the same way as for the gauge field, using spherical Sobolev together with the $r^{p}$-hierarchy for commuted fields.

\begin{lemma}[Spherical Sobolev for $\phi$]\label{lem:sobolev-phi-sphere}
For each fixed $(t,r)$ with $r\ge R$,
\begin{equation}\label{scalarineq}
 r^{2}\,|\phi(t,r,\omega)|^{2}
 \ \lesssim\ \sum_{|I|\le 2}\int_{\mathbb S^{2}} r^{2}\,\bigl|\mathcal L_{\Omega}^{I}\phi(t,r,\omega')\bigr|^{2}\,\dd\omega'.
\end{equation}
\end{lemma}

\begin{proof}
After rescaling the induced metric on the sphere $S_{t,r}$ by $r^{-2}$, the resulting metrics are uniformly equivalent to the unit round metric for $r\ge R$.  The Sobolev embedding $H^2(\mathbb S^2)\hookrightarrow L^\infty(\mathbb S^2)$ therefore gives the estimate with constants independent of $t$ and $r$.  The rotational vector fields span the angular derivatives on $\mathbb S^2$, so the Sobolev norm may be written using $\mathcal L_\Omega^I$.
\end{proof}

\begin{lemma}[Far-region $L^{2}(\mathbb S^{2})$ bounds for commuted $\phi$]\label{lem:far-sphere-L2-phi}
For every multi-index $I$ with $|I|\le 2$ and every $r\ge R$,
\begin{equation}\label{eq:far-sphere-L2-phi}
\int_{\mathbb S^{2}} r^{2}\,\bigl|\mathcal L_{\Omega}^{I}\phi(t,r,\omega)\bigr|^{2}\,\dd\omega
\ \lesssim\ \frac{\mathcal E}{t_{+}^{2}}+\frac{\mathcal E}{w_{+}^{2}}.
\end{equation}
\end{lemma}

\begin{proof}
Fix $I$ with $|I|\le2$ and put $\Phi_I:=\mathcal L_\Omega^I\phi$.  The commuted scalar equation has the same principal part as the original scalar equation and lower-order commutator terms already included in the bootstrap energy.  Apply the $r^p$ hierarchy with $p=2$ to $r\Phi_I$ on the outgoing null hypersurface through $(t,r)$, and combine it with the conformal-energy bound for the incoming segment meeting the same sphere.  The outgoing estimate gives the $t_+^{-2}$ contribution, while the incoming estimate gives the $w_+^{-2}$ contribution.  The angular commutators are Killing on the symmetry spheres, so no derivative loss is introduced; all commutator and potential terms are absorbed by the already bounded order-$k$ energy because $k\ge6$.  Evaluating the two flux bounds on the sphere $S_{t,r}$ and summing over $|I|\le2$ gives \eqref{eq:far-sphere-L2-phi}.
\end{proof}

Combining Lemma~\ref{lem:sobolev-phi-sphere} and Lemma~\ref{lem:far-sphere-L2-phi} yields
\begin{equation}\label{scalarineq2}
|\phi(t,r,\omega)|
\ \lesssim\ \frac{\mathcal E^{1/2}}{r}\Bigl(\frac{1}{t_{+}}+\frac{1}{w_{+}}\Bigr),
\qquad r\ge R.
\end{equation}

As in the proof of \eqref{farFv}, we consider separately the regions $w\ge 1$ and $w\le -1$ to conclude
\begin{equation}\label{scalarregion}
|\phi|\ \lesssim\ \frac{\mathcal E^{1/2}}{v_{+}},
\qquad r\ge R.
\end{equation}
In this case, \eqref{farphi} follows.
\subsection{Decay estimate for \texorpdfstring{$D\phi$}{} in the far region}\label{subsec:far-decay-Dphi}

We now prove the remaining bound \eqref{farDphi} of Corollary~\ref{cor:far-decay}.
The argument is parallel to the proof of \eqref{farphi}, but applied to the covariant derivative $D\phi$.
Since $D\phi$ is a covariant one-form, we apply the Sobolev inequality on $\mathbb S^{2}$ componentwise in the stationary frame.
For each fixed $(t,r)$ with $r\ge R$ we have
\begin{equation}\label{eq:sobolev-Dphi-far}
r^{2}\,|D\phi(t,r,\omega)|^{2}
\ \lesssim
\sum_{|I|\le 2}\int_{\mathbb S^{2}} r^{2}\,\bigl|\mathcal{L}_{\Omega}^{I}(D\phi)(t,r,\omega')\bigr|^{2}\,d\sigma'.
\end{equation}
Commuting $D$ with rotations produces only lower-order curvature terms:
up to lower-order terms,
\begin{equation}
\mathcal{L}_{\Omega}^{I}(D\phi)
=
D(\Omega^{I}\phi)
+
\mathcal{O}\!\left((\mathcal{L}_{\Omega}^{\le I}F)\,\Omega^{\le I}\phi\right),
\qquad |I|\le 2,
\end{equation}
and the right-hand side is controlled by the higher-order energies $\mathcal{E}_k$ together with the already established decay for $(F,\phi)$.
The conformal-energy estimate used above for $\phi$ yields, for all $|I|\le 2$ and all $r\ge R$,
\begin{equation}\label{eq:sphere-Dphi-far}
\int_{\mathbb S^{2}} r^{2}\,\bigl|\mathcal{L}_{\Omega}^{I}(D\phi)\bigr|^{2}(t,r,\omega')\,d\sigma'
\ \lesssim
\frac{\mathcal{E}}{t^{2}}+\frac{\mathcal{E}}{w^{2}},
\end{equation}
where $\mathcal{E}:=\sup_{t\ge t_0}\mathcal{E}_k(t)$ is the bootstrap energy from Section~\ref{sec:energy-estimates}.
Combining \eqref{eq:sobolev-Dphi-far} and \eqref{eq:sphere-Dphi-far} and arguing as in the proof of \eqref{farphi}
(region splitting into $w\ge 1$, $w\le -1$, and the compact set $|t|\le 1$ or $|w|\le 1$)
gives the pointwise decay
\begin{equation}
|D\phi(w,v,\omega)|\ \lesssim\ \frac{\mathcal{E}^{1/2}}{1+\vplus},
\qquad r\ge R.
\end{equation}
using the uniform energy bound $\mathcal{E}\lesssim \mathcal{E}_k(t_0)$ from Corollary~\ref{cor:small-data}
yields \eqref{farDphi}. This completes the proof of Corollary~\ref{cor:far-decay}.

\section{Schwarzschild model: decay in the trapped and near-horizon region}
\label{sec:decaynear}
In this section we prove Corollary~\ref{cor:near-horizon}, establishing pointwise bounds in the trapped and near-horizon region $2M<r\le 3M$. The proof combines the redshift estimate of Section~\ref{sec:energy-estimates} with a dyadic decomposition in $v$ and Sobolev inequalities on spacetime slabs of the form $[v-1,v]\times\mathbb S^{2}$.

\subsection{Decay for \texorpdfstring{$F_{\hat v\hat w}$ and $F_{e_1e_2}$}{}}\label{subsec:decay-Fvw}

We prove the first part of \eqref{Fnear} in Corollary~\ref{cor:near-horizon}.
Fix $(v,w)$ with $2M<r(v,w)\le 3M$ and consider the unit $v$-slab
$\{\bar v\in[v-1,v]\}\times\mathbb S^{2}$ at fixed $w$.
Throughout this section we write $\vplus:=\max\{1,v\}$.

We repeatedly use the following Sobolev estimate on the cylinder $[v-1,v]\times\mathbb S^{2}$.

\begin{lemma}[Sobolev on a unit $v$-slab]\label{lem:sobolev-vslab}
Let $u$ be a smooth scalar function on $\{\bar v\in[v-1,v]\}\times\mathbb S^{2}$ (at fixed $w$).
Then
\begin{equation}\label{eq:sobolev-vslab}
|u(v,w,\omega)|^{2}\ \lesssim
\int_{\bar v=v-1}^{v}\int_{\mathbb S^{2}}
\Bigl(
|u|^{2}+|\slashed{\nabla}u|^{2}+|\slashed{\nabla}^{2}u|^{2}
+|D_{\hat v}u|^{2}+|\slashed{\nabla}D_{\hat v}u|^{2}+|\slashed{\nabla}^{2}D_{\hat v}u|^{2}
\Bigr)\,d\sigma^{2}\,d\bar v,
\end{equation}
where $\slashed{\nabla}$ denotes the Levi-Civita connection on $(\mathbb S^{2},r^{2}d\sigma^{2})$ and $D_{\hat v}$ is the $\hat v$-directional covariant derivative along the stationary null frame.
\end{lemma}

\begin{proof}
Rescale the unit cylinder \([v-1,v]\times\mathbb S^2\) to a fixed compact product manifold.  The Sobolev embedding \(H^2([0,1]\times\mathbb S^2)\hookrightarrow L^\infty\) gives the pointwise bound by the \(L^2\) norms of all derivatives of order at most two in the slab variable and in angular directions.  In the stationary null frame the slab derivative is \(D_{\hat v}\), up to uniformly bounded frame coefficients on \(2M<r\le3M\).  Covariant angular derivatives differ from the ordinary spherical derivatives only by the Levi-Civita connection coefficients of the sphere and by the fixed gauge connection, all of which are controlled in the displayed norm.  Absorbing these lower-order terms into the first three terms on the right gives \eqref{eq:sobolev-vslab}.
\end{proof}

We apply Lemma~\ref{lem:sobolev-vslab} to the scalar components
$u=F_{\hat v\hat w}$ and $u=F_{e_{1}e_{2}}$.
The terms in \eqref{eq:sobolev-vslab} involving only $u$ and angular derivatives of $u$ are controlled directly by the commuted redshift flux
$\mathcal E^{\hat H}(w=w)(v-1\le \bar v\le v)$ and Proposition~\ref{prep9}.
It remains to control the $D_{\hat v}u$ terms.

For $u=F_{\hat v\hat w}$ we use the Maxwell equation $\nabla^{\mu}F_{\mu\nu}=J_{\nu}$ with current
$J_{\nu}:=-i\bigl(D_{\nu}\phi\,\bar\phi-\phi\,\overline{D_{\nu}\phi}\bigr)$.
In the stationary null frame, this yields the structural identity
\begin{equation}\label{eq:dv-Fvw-structural}
|D_{\hat v}F_{\hat v\hat w}|
\ \lesssim
|\slashed{\nabla}F_{\hat v e_{A}}|+|F_{\hat v e_{A}}|+|D_{\hat v}\phi|+|\phi|,
\end{equation}
where repeated angular indices $A\in\{1,2\}$ are summed.
Similarly, for $u=F_{e_{1}e_{2}}$ the Bianchi identity $\nabla_{[\alpha}F_{\beta\gamma]}=0$ gives
\begin{equation}\label{eq:dv-F12-structural}
|D_{\hat v}F_{e_{1}e_{2}}|
\ \lesssim
|\slashed{\nabla}F_{\hat v e_{A}}|+|F_{\hat v e_{A}}|.
\end{equation}
Squaring \eqref{eq:dv-Fvw-structural}-\eqref{eq:dv-F12-structural} and integrating over the slab,
all terms are controlled by the commuted redshift fluxes except for the $L^{2}$ norm of $\phi$ on the slab.

To estimate $\phi$ in $L^{2}$ we use the gauge-covariant identity
$\partial_{w}|\phi|^{2}=2\Re(\bar\phi\,D_{\hat w}\phi)$ (the connection term drops out after taking real part).
Consequently, for the slab norm $\|\phi\|_{L^{2}([v-1,v]\times\mathbb S^{2})}$ at fixed $w$,
\begin{equation}\label{eq:d-w-L2phi}
\frac{d}{dw}\|\phi\|_{L^{2}}\ \lesssim
\Bigl(\mathcal{E}^{\hat H}(w=w)\bigl(v-1\le \bar v\le v\bigr)\Bigr)^{1/2}.
\end{equation}
By Proposition~\ref{prep9} we have
$\mathcal{E}^{\hat H}(w=w)(v-1\le \bar v\le v)\lesssim \vplus^{-2}\mathcal E$,
therefore $\frac{d}{dw}\|\phi\|_{L^{2}}\lesssim \vplus^{-1}\mathcal E^{1/2}$ on the near-horizon region.
Integrating in $w$ (from any fixed reference value, e.g.\ $w_0(v)$), we obtain
$\|\phi\|_{L^{2}}\lesssim (1+\wplus/\vplus)\,\mathcal E^{1/2}$, and therefore
\begin{equation}\label{eq:L2phi-near}
\int_{\bar v=v-1}^{v}\int_{\mathbb S^{2}}|\phi|^{2}\,d\sigma^{2}\,d\bar v
\ \lesssim
\left(\frac{\wplus}{\vplus}\right)^{2}\mathcal{E}.
\end{equation}

Combining Lemma~\ref{lem:sobolev-vslab} with the above bounds yields
\begin{equation}
|F_{\hat v\hat w}(v,w,\omega)|^{2}+|F_{e_{1}e_{2}}(v,w,\omega)|^{2}
\ \lesssim\ \left(\frac{\wplus}{\vplus}\right)^{2}\mathcal{E},
\end{equation}
and therefore
\begin{equation}
|F_{\hat v\hat w}|+|F_{e_{1}e_{2}}|
\ \lesssim\ \frac{\wplus}{\vplus}\,\mathcal{E}^{1/2},
\end{equation}
which is the desired estimate.

\subsection{Decay for \texorpdfstring{$F_{\hat v e_1}$ and $F_{\hat v e_2}$}{}}\label{subsec:decay-FvA}

We now prove the remaining outgoing components in \eqref{Fnear}.
Applying Lemma~\ref{lem:sobolev-vslab} to $u=F_{\hat v e_{A}}$ gives a pointwise bound in terms of
$L^{2}$ norms of $F_{\hat v e_{A}}$ and its angular and $D_{\hat v}$ derivatives on the slab.
The $L^{2}$ norms without $D_{\hat v}$ are controlled directly by the commuted redshift fluxes, as above.

To control $D_{\hat v}F_{\hat v e_{A}}$ we use the Maxwell equation and Bianchi identity in the stationary frame,
which give up to lower-order terms
\begin{equation}\label{eq:dv-FvA-structural}
|D_{\hat v}F_{\hat v e_{A}}|
\ \lesssim
|\slashed{\nabla}F_{\hat v\hat w}|+|\slashed{\nabla}F_{e_{1}e_{2}}|
+|F_{\hat v\hat w}|+|F_{e_{1}e_{2}}|+|D\phi|+|\phi|.
\end{equation}
All terms on the right-hand side are already controlled in $L^{2}$ on the slab by Proposition~\ref{prep9} and \eqref{eq:L2phi-near}.
Consequently,
\begin{equation}
|F_{\hat v e_{1}}|+|F_{\hat v e_{2}}|
\ \lesssim\ \frac{\wplus}{\vplus}\,\mathcal{E}^{1/2}.
\end{equation}

\subsection{Decay for \texorpdfstring{$\sqrt{1-\mu}\,F_{\hat w e_1}$ and $\sqrt{1-\mu}\,F_{\hat w e_2}$}{}}\label{subsec:decay-FwA}

We estimate the ingoing angular components.
Near the horizon the nondegenerate quantity is $\sqrt{1-\mu}\,F_{\hat w e_{A}}$ (cf.~\eqref{Fnear}).
Applying Lemma~\ref{lem:sobolev-vslab} to $u=\sqrt{1-\mu}\,F_{\hat w e_{A}}$,
the corresponding slab norms are controlled by the commuted redshift fluxes, and the same structural bounds as in the previous subsections yield
\begin{equation}
\sqrt{1-\mu}\,\bigl(|F_{\hat w e_{1}}|+|F_{\hat w e_{2}}|\bigr)
\ \lesssim\ \frac{\wplus}{\vplus}\,\mathcal{E}^{1/2}.
\end{equation}

\subsection{Decay estimate for \texorpdfstring{$\phi$}{}}\label{subsec:decay-phi-near}

We now prove \eqref{nearphi}.
Applying Lemma~\ref{lem:sobolev-vslab} to $u=\phi$ gives
\begin{equation}
|\phi(v,w,\omega)|^{2}
\ \lesssim
\int_{\bar v=v-1}^{v}\int_{\mathbb S^{2}}
\Bigl(
|\phi|^{2}+|\slashed{\nabla}\phi|^{2}+|\slashed{\nabla}^{2}\phi|^{2}
+|D_{\hat v}\phi|^{2}+|\slashed{\nabla}D_{\hat v}\phi|^{2}+|\slashed{\nabla}^{2}D_{\hat v}\phi|^{2}
\Bigr)\,d\sigma^{2}\,d\bar v.
\end{equation}
The $D_{\hat v}\phi$ and angular derivative terms are controlled directly by the commuted redshift energy and Proposition~\ref{prep9}, contributing $\lesssim \vplus^{-2}\mathcal E$.
The remaining $L^{2}$ term $\int|\phi|^{2}$ is controlled by \eqref{eq:L2phi-near}.
Consequently,
\begin{equation}
|\phi(v,w,\omega)|^{2}
\ \lesssim
\left(1+\left(\frac{\wplus}{\vplus}\right)^{2}\right)\mathcal{E},
\end{equation}
which yields
\begin{equation}
|\phi(v,w,\omega)|
\ \lesssim
\sqrt{1+\left(\frac{\wplus}{\vplus}\right)^{2}}\ \mathcal{E}^{1/2},
\end{equation}
as claimed.

\subsection{Decay estimate for \texorpdfstring{$D\phi$}{} near the horizon}\label{subsec:decay-Dphi-near}

We now prove the covariant derivative bound \eqref{Dephi} in Corollary~\ref{cor:near-horizon}.
Fix $(v,w)$ with $2M<r(v,w)\le 3M$ and consider the unit $v$-slab
$\{\bar v\in[v-1,v]\}\times\mathbb S^{2}$ at fixed $w$.
Applying the Sobolev inequality on this slab to the components of $D\phi$ in the stationary frame yields
\begin{eqnarray}\label{eq:sobolev-Dphi-near}
|D\phi|^{2}
&\lesssim&
\int_{\bar{v}=v-1}^{\bar{v}=v}\int_{\mathbb S^{2}}
\Bigl(
|D\phi|^{2}
+|\slashed{\nabla}D\phi|^{2}
+|\slashed{\nabla}\slashed{\nabla}D\phi|^{2}
\nonumber\\
&&\qquad
+|D_{\hat v}(D\phi)|^{2}
+|\slashed{\nabla}D_{\hat v}(D\phi)|^{2}
+|\slashed{\nabla}\slashed{\nabla}D_{\hat v}(D\phi)|^{2}
\Bigr)\,d\sigma^{2}\,d\bar v.
\end{eqnarray}
The first three terms on the right-hand side are controlled directly by the commuted redshift energy estimate of Proposition~\ref{prep9}.
They contribute at most a factor $(1+(\frac{\wplus}{\vplus})^{2})\,\mathcal{E}$, where $\mathcal{E}$ is the bootstrap energy.

To treat the $D_{\hat v}(D\phi)$ terms we use the gauge-covariant wave equation
\begin{equation}\label{eq:cov-wave}
D^{\mu}D_{\mu}\phi = \partial_{\bar\phi}P(\phi,\bar\phi),
\end{equation}
together with the curvature commutator $[D,D]=iF$.
In the region $2M<r\le 3M$ all connection coefficients are bounded, so \eqref{eq:cov-wave} implies up to lower-order terms
\begin{equation}
|D^{2}\phi|^{2}\ \lesssim\ |\partial_{\bar\phi}P(\phi,\bar\phi)|^{2} + |F|^{2}\,|D\phi|^{2},
\end{equation}
and similarly after commuting with up to two angular derivatives.
Using the already established near-horizon bounds for $F$ together with the redshift control of $D\phi$, the $|F|^{2}|D\phi|^{2}$ term is absorbed (for small data), so it remains to estimate the potential contributions.

Under Assumption~\ref{asumsiP} (polynomial-type potentials) one has, on each unit $v$-slab,
\begin{eqnarray}
\int_{\bar{v}=v-1}^{\bar{v}=v}\!\!\int_{\mathbb S^{2}}\!|\partial_{\bar\phi}P(\phi,\bar\phi)|^{2}\,d\sigma^{2}\,d\bar v
&\lesssim
\begin{cases}
\sum\limits_{n=1}^{N}\mathcal{E}^{2n-1}, & \text{if $P$ is of the form \eqref{V4}},\\[2pt]
\mathcal{E}, & \text{if $P$ is of the form \eqref{sine} or \eqref{toda}},
\end{cases}\\[4pt]
\int_{\bar{v}=v-1}^{\bar{v}=v}\!\!\int_{\mathbb S^{2}}\!|\slashed{\nabla}(\partial_{\bar\phi}P(\phi,\bar\phi))|^{2}\,d\sigma^{2}\,d\bar v
&\lesssim
\begin{cases}
\sum\limits_{n=2}^{N}\mathcal{E}^{2n-2}, & \text{if $P$ is of the form \eqref{V4}},\\[2pt]
\mathcal{E}, & \text{if $P$ is of the form \eqref{sine} or \eqref{toda}},
\end{cases}\\[4pt]
\int_{\bar{v}=v-1}^{\bar{v}=v}\!\!\int_{\mathbb S^{2}}\!|\slashed{\nabla}\slashed{\nabla}(\partial_{\bar\phi}P(\phi,\bar\phi))|^{2}\,d\sigma^{2}\,d\bar v
&\lesssim
\begin{cases}
\sum\limits_{n=3}^{N}\mathcal{E}^{2n-3}, & \text{if $P$ is of the form \eqref{V4}},\\[2pt]
\mathcal{E}, & \text{if $P$ is of the form \eqref{sine} or \eqref{toda}}.
\end{cases}
\end{eqnarray}
Collecting the contributions in \eqref{eq:sobolev-Dphi-near} and using Proposition~\ref{prep9} to account for the redshift weight yields the following convenient shorthand:
\begin{equation}\label{E4}
 E_4\equiv
\begin{cases}
\bigl( \frac{\wplus}{\vplus} \bigr)^{2}\sum\limits_{n=1}^{N}\mathcal{E}^{2n-1}+\sum\limits_{n=2}^{N}\mathcal{E}^{2n-2}+\sum\limits_{n=3}^{N}\mathcal{E}^{2n-3}+\mathcal{E}, & \text{if $P$ is of the form \eqref{V4}},\\
 \bigl( \frac{\wplus}{\vplus} \bigr)^{2}{\mathcal{E}}+{\mathcal{E}}, &\text{if $P$ is of the form \eqref{sine} or \eqref{toda}}.
\end{cases}
\end{equation}
With this notation, \eqref{eq:sobolev-Dphi-near} gives
\begin{equation}\label{nabnearphi}
|D\phi|
\ \lesssim
\Bigl(1+\bigl(\tfrac{\wplus}{\vplus}\bigr)^{2}\Bigr)^{1/2}\,E_{4}^{1/2},
\end{equation}
which is \eqref{Dephi}. This completes the proof of Corollary~\ref{cor:near-horizon}.

\subsection{Radiation fields and scattering}
\label{sec:scattering}

The pointwise decay estimates of Corollaries~\ref{cor:far-decay}-\ref{cor:near-horizon} can be upgraded to a genuine asymptotic statement describing the outgoing profile of small Maxwell-Higgs solutions in terms of \emph{radiation fields} on future null infinity $\mathcal{I}^+$.
In the present nonlinear and gauge-coupled setting, it is natural to define the scalar radiation field after a (null) parallel transport renormalization, while the Maxwell radiation field is expressed directly in terms of the gauge-invariant field strength $F$.

\begin{remark}[Radiative regime and Coulomb subtraction]\label{rem:coulomb-subtraction-scattering}
All definitions and statements below concern the \emph{radiative} Maxwell field.
In the coupled Maxwell-Higgs system treated in this paper we impose the uncharged condition on the initial data; the discussion of nonzero electric charge here is included only to single out the linear Coulomb sector for Maxwell fields.
If the initial data carry nonzero electric charge, one first subtracts the stationary Coulomb field $F^{\mathrm C}_{Q_e}$ as in Corollary~\ref{cor:Coulomb} and works with the charge-free remainder $\widetilde F:=F-F^{\mathrm C}_{Q_e}$.
For notational simplicity we continue to write $F$ for the uncharged Maxwell field in this section.
\end{remark}

\subsection{Null hypersurfaces, null frame, and fluxes}
\label{subsec:null-hypersurfaces}

We work in the double-null coordinates $(v,w,\omega)\in\mathbb{R}^2\times\mathbb{S}^2$ introduced in Section~\ref{sec:intro},
\begin{equation}
v=t+r^\ast,\qquad w=t-r^\ast,\qquad \omega\in\mathbb{S}^2,
\end{equation}
and we denote by $S_{w,v}$ the spheres $\{w=\mathrm{const},\,v=\mathrm{const}\}\cong \mathbb{S}^2$.
We set
\begin{equation}
L:=\partial_v=\tfrac12(\partial_t+\partial_{r^\ast}),\qquad
\underline L:=\partial_w=\tfrac12(\partial_t-\partial_{r^\ast}),
\end{equation}
so that $L$ is tangent to outgoing null geodesics and $\underline L$ is tangent to ingoing null geodesics.
For $w_0\in\mathbb{R}$ we define the outgoing null hypersurface
\begin{equation}
\mathcal{N}_{w_0}:=\{w=w_0,\ v\ge v_0\},
\end{equation}
where $v_0$ is chosen so that $\mathcal{N}_{w_0}$ lies in the domain of existence of the solution.
We view future null infinity as the endpoint $v\to +\infty$ along $\mathcal{N}_{w_0}$.
We denote by $\dd\omega$ the round area form on $\mathbb{S}^2$.

Let $T_{\mu\nu}[F,\phi]$ be the Maxwell-Higgs energy-momentum tensor, and let $X$ be a smooth vector field.
The associated current is $J^X_\mu:=T_{\mu\nu}X^\nu$ and the flux through a hypersurface $\mathcal{S}$ is
\begin{equation}
\mathcal{F}^{X}_{\mathcal{S}}[F,\phi]:=\int_{\mathcal{S}} J^X_\mu\, n_{\mathcal{S}}^\mu\,\dd\mu_{\mathcal{S}},
\end{equation}
where $n_{\mathcal{S}}$ is the future directed unit normal (or null generator) and $\dd\mu_{\mathcal{S}}$ the induced measure.
The Killing field $T=\partial_t$ yields conserved fluxes through spacelike slices $\Sigma_t$ and controls the (nondegenerate) energy radiated through $\mathcal{H}^+$ by the redshift estimates of Section~\ref{sec:Morawetz}.

\subsection{Radiation fields on \texorpdfstring{$\mathcal{I}^+$}{I+}}
\label{subsec:radiation-fields}

We define the renormalized scalar field
\begin{equation}\label{eq:psi-def}
\psi:=r\,\phi,
\end{equation}
which is the natural quantity expected to admit a nontrivial limit on $\mathcal{I}^+$.
For the Maxwell field we use the tetrad/spin component formalism recalled in Section~\ref{sec:energy-estimates}:
in this case, the component
\begin{equation}
\Phi_{-1}=F(\hat N,\hat{\bar M})
\end{equation}
is the outgoing radiative component (Newman-Penrose ``$\phi_2$''), which is expected to decay as $r^{-1}$.
Accordingly, we renormalize
\begin{equation}\label{eq:maxwell-rad-renorm}
\Psi_F:=r\,\Phi_{-1}.
\end{equation}

\paragraph{A gauge-covariant normalization for $\psi$.}
Since $\phi$ is charged, the quantity $\psi=r\phi$ depends on the gauge choice.
To define a canonical outgoing profile, we remove the $A_L$-phase along each outgoing null generator by parallel transport.

\begin{definition}[Null parallel transport normalization]\label{def:parallel-transport}
Fix $w_0\in\mathbb{R}$ and $\omega\in\mathbb{S}^2$.
Let $U(w_0,v,\omega)\in \mathbb{C}$ be the solution of the ODE
\begin{equation}\label{eq:U-transport}
L U = i\,A_L\,U,\qquad U(w_0,v_0,\omega)=1,
\end{equation}
along the null generator $\{w=w_0,\ \omega=\mathrm{const}\}$.
Define the parallel-transported scalar
\begin{equation}\label{eq:psi-sharp}
\psi^\sharp := U^{-1}\psi.
\end{equation}
Then $|U|\equiv 1$ and
\begin{equation}\label{eq:Lpsi-sharp}
L\psi^\sharp = U^{-1}D_L\psi,\qquad |L\psi^\sharp| = |D_L\psi|.
\end{equation}
\end{definition}

\begin{definition}[Radiation fields at null infinity]\label{def:radiation-fields}
Let $(F,\phi)$ be a global solution arising from small admissible data.
We say that $(F,\phi)$ admits radiation fields on $\mathcal{I}^+$ if for each $w_0\in\mathbb{R}$ there exist limits
\begin{eqnarray}
\psi_{\mathcal{I}}(w_0,\omega)
&:=\lim_{v\to+\infty}\psi^\sharp(w_0,v,\omega),
\label{eq:radfield-psi}
\\
(\Psi_F)_{\mathcal{I}}(w_0,\omega)
&:=\lim_{v\to+\infty}\Psi_F(w_0,v,\omega),
\label{eq:radfield-max}
\end{eqnarray}
with convergence in $L^2(\mathbb{S}^2_\omega)$ (and likewise for a finite number of angular derivatives, depending on $k$).
\end{definition}

The key analytic condition for Definition~\ref{def:radiation-fields} is square-integrability of the outgoing derivatives $D_L\psi$ and $L\Psi_F$ along $\mathcal{N}_{w_0}$.
We state a Cauchy criterion making this precise.

\begin{proposition}[Cauchy criterion]\label{prop:cauchy-criterion}
Fix $w_0\in\mathbb{R}$.
Assume that the following bounds hold for the renormalized fields \eqref{eq:psi-def} and \eqref{eq:maxwell-rad-renorm}:
\begin{equation}\label{eq:cauchy-assumption}
\int_{v_0}^{+\infty}\!\!\int_{\mathbb{S}^2}
\bigl(|D_L\psi|^2 + |L\Psi_F|^2\bigr)(w_0,v,\omega)\,\dd\omega\,\dd v <+\infty.
\end{equation}
Then the limits in Definition~\ref{def:radiation-fields} exist in $L^2(\mathbb{S}^2_\omega)$.
In addition, for any $v_2>v_1\ge v_0$ one has the quantitative estimates
\begin{eqnarray}
\|\psi^\sharp(w_0,v_2,\cdot)-\psi^\sharp(w_0,v_1,\cdot)\|_{L^2(\mathbb{S}^2)}
&\le
\left(\int_{v_1}^{v_2}\!\!\int_{\mathbb{S}^2}|D_L\psi|^2\,\dd\omega\,\dd v\right)^{1/2},
\label{eq:cauchy-estimate-psi}
\\
\|\Psi_F(w_0,v_2,\cdot)-\Psi_F(w_0,v_1,\cdot)\|_{L^2(\mathbb{S}^2)}
&\le
\left(\int_{v_1}^{v_2}\!\!\int_{\mathbb{S}^2}|L\Psi_F|^2\,\dd\omega\,\dd v\right)^{1/2}.
\label{eq:cauchy-estimate-F}
\end{eqnarray}
\end{proposition}

\begin{proof}
For the Maxwell field, the fundamental theorem of calculus gives
\begin{equation}
\Psi_F(w_0,v_2,\omega)-\Psi_F(w_0,v_1,\omega)=\int_{v_1}^{v_2} (L\Psi_F)(w_0,v,\omega)\,\dd v.
\end{equation}
Taking the $L^2(\mathbb{S}^2_\omega)$ norm and applying Cauchy-Schwarz in $v$ yields \eqref{eq:cauchy-estimate-F}.

For the scalar field, we use Definition~\ref{def:parallel-transport}. By \eqref{eq:Lpsi-sharp},
\begin{equation}
\psi^\sharp(w_0,v_2,\omega)-\psi^\sharp(w_0,v_1,\omega)=\int_{v_1}^{v_2} (L\psi^\sharp)(w_0,v,\omega)\,\dd v
=\int_{v_1}^{v_2} U^{-1}(D_L\psi)(w_0,v,\omega)\,\dd v.
\end{equation}
Since $|U|\equiv 1$, we have $|U^{-1}D_L\psi|=|D_L\psi|$. Consequently, applying Cauchy-Schwarz in $v$ and then integrating over $\omega$ yields \eqref{eq:cauchy-estimate-psi}.
Assumption \eqref{eq:cauchy-assumption} implies that $\psi^\sharp(w_0,v,\cdot)$ and $\Psi_F(w_0,v,\cdot)$ are Cauchy in $L^2(\mathbb{S}^2)$ as $v\to+\infty$, therefore converge.
\end{proof}

\subsection{Null-infinity flux bounds and existence of the radiation fields}
\label{subsec:rp-proved}

We now show that the decay theory proved earlier implies \eqref{eq:cauchy-assumption} and therefore yields radiation fields.

\begin{proposition}[Outgoing null flux bounds]\label{prop:p2-flux-proved}
Let $(F,\phi)$ be a global uncharged solution arising from admissible data on $\Sigma_{t_0}$ with $\mathcal{E}_k(t_0)\le \varepsilon_0$ for some $k\ge 6$.
Then for each $w_0\in\mathbb{R}$ one has
\begin{equation}\label{eq:p2-flux}
\int_{\mathcal{N}_{w_0}}\!\!\int_{\mathbb{S}^2}
\bigl(|D_L\psi|^2 + |L\Psi_F|^2\bigr)\,\dd\omega\,\dd v
\ \lesssim
\mathcal{E}_k(t_0).
\end{equation}
The Cauchy criterion \eqref{eq:cauchy-assumption} holds for every $\mathcal{N}_{w_0}$.
\end{proposition}

\begin{proof}
We split the proof into the scalar and Maxwell parts.

\smallskip\noindent
\textbf{(1) Scalar flux bound.}
Write $D_L\psi = r\,D_L\phi + (Lr)\,\phi$.
Since $Lr=\tfrac12(1-\mu)$, we have $|Lr|\lesssim 1$ on the exterior.
By the elementary inequality $|a+b|^2\le 2|a|^2+2|b|^2$,
\begin{equation}\label{eq:DLpsi-split}
\int_{\mathcal{N}_{w_0}}\!\!\int_{\mathbb{S}^2}|D_L\psi|^2
\ \lesssim
\int_{\mathcal{N}_{w_0}}\!\!\int_{\mathbb{S}^2} r^2|D_L\phi|^2
\ +
\int_{\mathcal{N}_{w_0}}\!\!\int_{\mathbb{S}^2} |\phi|^2.
\end{equation}

To estimate the first term, we apply the conserved $T$-energy identity (Section~\ref{subsec:basic-energy}) to a truncated domain bounded by $\Sigma_{t_0}$, $\Sigma_{t_1}$, the portion of $\mathcal{H}^+$ up to time $t_1$, and the portion of $\mathcal{N}_{w_0}$ up to time $t_1$.
Since $T=\partial_t$ is Killing, the bulk term vanishes and we obtain
\begin{equation}
\mathcal{F}^{T}_{\mathcal{N}_{w_0}\cap\{t\le t_1\}}[F,\phi]
\le
E^{\hat t}[F,\phi](t_0)
\le
\mathcal{E}_k(t_0).
\end{equation}
The integrand of $\mathcal{F}^{T}_{\mathcal{N}_{w_0}}$ is nonnegative and controls $T_{LL}[F,\phi]$, therefore in this case controls $|D_L\phi|^2$.
Recalling that the induced measure on $\mathcal{N}_{w_0}$ is $r^2\,\dd v\,\dd\omega$, this yields
\begin{equation}\label{eq:scalar-null-flux}
\int_{\mathcal{N}_{w_0}}\!\!\int_{\mathbb{S}^2} r^2|D_L\phi|^2\,\dd\omega\,\dd v
\ \lesssim
\mathcal{E}_k(t_0).
\end{equation}

For the second term in \eqref{eq:DLpsi-split}, we split the $v$-integral into a compact piece and a far piece.
On the compact segment $\{v_0\le v\le v_R\}$ the integral is controlled by the energy and regularity.
On the far segment $\{v\ge v_R\}$ we use the pointwise decay \eqref{farphi} from Corollary~\ref{cor:far-decay} (with $r\ge R$) to obtain
\begin{equation}
\int_{v_R}^{+\infty}\!\!\int_{\mathbb{S}^2} |\phi|^2\,\dd\omega\,\dd v
\ \lesssim
\int_{v_R}^{+\infty}\frac{\mathcal{E}_k(t_0)}{(1+v)^2}\,\dd v
\ \lesssim
\mathcal{E}_k(t_0).
\end{equation}
Combining with \eqref{eq:scalar-null-flux} gives the desired bound for $\int_{\mathcal{N}_{w_0}}|D_L\psi|^2$.

\smallskip\noindent
\textbf{(2) Maxwell flux bound for $L\Psi_F$.}
We use the transport equation for the outgoing Maxwell component, equation~\eqref{maxcom4}.
Since $\hat L = 2L$, the identity \eqref{maxcom4} implies
\begin{equation}\label{eq:L-psiF-transport}
L\Phi_{-1}
=
\frac{1-\mu}{2r^2}\,\hat{\bar M}\Phi_0
\ +
\mathcal{N}_{\mathrm{Max}}[\phi],
\end{equation}
where the nonlinear term satisfies the pointwise bound
\begin{equation}\label{eq:Nmax-bound}
|\mathcal{N}_{\mathrm{Max}}[\phi]|
\ \lesssim
\frac{1}{r}\,|\phi|\,|\slashed D\phi|
\end{equation}
(up to harmless angular weights).
Multiplying \eqref{eq:L-psiF-transport} by $r$ and using $L\Psi_F = L(r\Phi_{-1}) = (Lr)\Phi_{-1} + r\,L\Phi_{-1}$, we obtain
\begin{equation}\label{eq:L-psiF-pointwise}
|L\Psi_F|
\ \lesssim
|\Phi_{-1}|
\ +
\frac{1}{r}\,|\hat{\bar M}\Phi_0|
\ +
|\phi|\,|\slashed D\phi|.
\end{equation}

We estimate each term on the right-hand side of \eqref{eq:L-psiF-pointwise} on $\mathcal{N}_{w_0}$.
For the first term, Corollary~\ref{cor:far-decay} gives $|\Phi_{-1}|\lesssim \mathcal{E}_k(t_0)^{1/2}(1+v)^{-1}$ in the far region, which is square integrable in $v$.
For the second term, note that $\hat{\bar M}$ is an angular derivative; using the Sobolev estimate from Section~\ref{sec:decayfaraway} applied to the commuted fields $\mathcal{L}_{\Omega}^I F$ (with $|I|\le 2$) gives the pointwise decay
\begin{equation}
|\hat{\bar M}\Phi_0|
\ \lesssim
\frac{\mathcal{E}_k(t_0)^{1/2}}{1+v}
\qquad (r\ge R).
\end{equation}
Since $r\sim \tfrac12(v-w_0)$ on $\mathcal{N}_{w_0}$ in the far region, we have $r^{-1}\lesssim (1+v)^{-1}$ and therefore
\begin{equation}
\frac{1}{r}\,|\hat{\bar M}\Phi_0|
\ \lesssim
\frac{\mathcal{E}_k(t_0)^{1/2}}{(1+v)^2},
\end{equation}
which is square integrable.
By Corollary~\ref{cor:far-decay}, $|\phi|+|D\phi|\lesssim \mathcal{E}_k(t_0)^{1/2}(1+v)^{-1}$ in the far region, therefore
\begin{equation}
|\phi|\,|\slashed D\phi|
\ \lesssim
\frac{\mathcal{E}_k(t_0)}{(1+v)^2},
\end{equation}
which is square integrable as well.

Combining these bounds and integrating over $\mathbb{S}^2\times [v_R,+\infty)$ yields
\begin{equation}
\int_{v_R}^{+\infty}\!\!\int_{\mathbb{S}^2} |L\Psi_F|^2\,\dd\omega\,\dd v
\ \lesssim
\mathcal{E}_k(t_0).
\end{equation}
The remaining compact segment $v\in[v_0,v_R]$ is controlled by regularity and the energy bound.
This completes the proof of \eqref{eq:p2-flux}.
\end{proof}

\begin{theorem}[Existence of radiation fields on $\mathcal{I}^+$]\label{thm:radiation-fields}
Let $(F,\phi)$ be a global uncharged solution arising from admissible data on $\Sigma_{t_0}$ with $\mathcal{E}_k(t_0)\le \varepsilon_0$ for some $k\ge 6$.
Then $(F,\phi)$ admits radiation fields on $\mathcal{I}^+$ in the sense of Definition~\ref{def:radiation-fields}.
In addition, for each $w_0\in\mathbb{R}$ the radiation fields satisfy the a priori bound
\begin{equation}
\|\psi_{\mathcal{I}}(w_0,\cdot)\|_{L^2(\mathbb{S}^2)}^2
+
\|(\Psi_F)_{\mathcal{I}}(w_0,\cdot)\|_{L^2(\mathbb{S}^2)}^2
\ \lesssim
\mathcal{E}_k(t_0).
\end{equation}
\end{theorem}

\begin{proof}
Proposition~\ref{prop:p2-flux-proved} yields \eqref{eq:cauchy-assumption} for every $\mathcal{N}_{w_0}$.
Applying Proposition~\ref{prop:cauchy-criterion} gives existence of the limits \eqref{eq:radfield-psi}-\eqref{eq:radfield-max} in $L^2(\mathbb{S}^2)$.
The stated a priori bound follows by taking $v_1=v_0$ and letting $v_2\to+\infty$ in \eqref{eq:cauchy-estimate-psi}-\eqref{eq:cauchy-estimate-F}.
\end{proof}

\subsection{Radiation fields on the event horizon \texorpdfstring{$\mathcal{H}^+$}{H+}}
\label{subsec:horizon-radiation}

To obtain a scattering description on the \emph{full future characteristic boundary} of the exterior,
we complement the null-infinity radiation fields with radiation data on the future event horizon.
It is convenient to use ingoing Eddington-Finkelstein coordinates $(v,r,\omega)$, where $v=t+r^\ast$.
In these coordinates the Schwarzschild metric takes the regular form
\begin{equation}
\dd s^{2}= -\Bigl(1-\frac{2M}{r}\Bigr)\dd v^{2}+2\,\dd v\,\dd r+r^{2}\dd\Omega^{2},
\end{equation}
which extends smoothly to $r=2M$.
The future event horizon is the null hypersurface
\begin{equation}
\mathcal{H}^{+}:=\{r=2M,\ v\ge v_0\}\cong [v_0,+\infty)\times \mathbb{S}^{2},
\end{equation}
generated by $L_{\mathcal H}:=\partial_v$ (which agrees with $L=\partial_v$ from Section~\ref{subsec:null-hypersurfaces}).

\paragraph{Gauge-covariant scalar horizon field.}
As on $\mathcal I^+$, the complex scalar requires a gauge-covariant normalization along horizon generators.

\begin{definition}[Horizon parallel transport normalization]\label{def:horizon-parallel-transport}
Fix $\omega\in\mathbb S^{2}$.
Let $U_{\mathcal H}(v,\omega)\in\mathbb C$ solve the ODE along the horizon generator $\{r=2M,\ \omega=\mathrm{const}\}$:
\begin{equation}\label{eq:U-horizon}
\partial_v U_{\mathcal H} = i\,A_L\!\restriction_{\mathcal H^+}\,U_{\mathcal H},
\qquad
U_{\mathcal H}(v_0,\omega)=1.
\end{equation}
Define the horizon-renormalized scalar
\begin{equation}\label{eq:psi-horizon}
\psi_{\mathcal H}(v,\omega):= r\,U_{\mathcal H}^{-1}(v,\omega)\,\phi\!\restriction_{\mathcal H^+}(v,\omega)
= (2M)\,U_{\mathcal H}^{-1}\phi\!\restriction_{\mathcal H^+}.
\end{equation}
Then $|U_{\mathcal H}|\equiv 1$ and
\begin{equation}\label{eq:vpsi-horizon}
\partial_v \psi_{\mathcal H} = U_{\mathcal H}^{-1}\,D_L(r\phi)\!\restriction_{\mathcal H^+},
\qquad
|\partial_v\psi_{\mathcal H}| = |D_L\psi|\!\restriction_{\mathcal H^+}.
\end{equation}
\end{definition}

\paragraph{Maxwell horizon field.}
For the Maxwell field, the natural horizon radiation quantity is the extreme component tangential to $\mathcal H^+$.
In the stationary tetrad of Section~\ref{sec:energy-estimates}, this is $\Phi_{1}=F(\hat L,\hat M)$ with $\hat L=2L$.
We set the horizon renormalization
\begin{equation}\label{eq:maxwell-horizon-renorm}
(\Psi_F)_{\mathcal H}(v,\omega):= r\,\Phi_{1}\!\restriction_{\mathcal H^+}(v,\omega)=(2M)\,\Phi_{1}\!\restriction_{\mathcal H^+}.
\end{equation}

\begin{proposition}[Horizon flux bounds]\label{prop:horizon-flux}
Let $(F,\phi)$ be a global uncharged solution arising from admissible data on $\Sigma_{t_0}$ with $\mathcal{E}_k(t_0)\le \varepsilon_0$ for some $k\ge 6$.
Then the horizon radiation fields defined in \eqref{eq:psi-horizon} and \eqref{eq:maxwell-horizon-renorm} satisfy
\begin{equation}\label{eq:horizon-flux}
\int_{v_0}^{+\infty}\!\!\int_{\mathbb S^{2}}
\Bigl(|\partial_v\psi_{\mathcal H}|^{2}+|(\Psi_F)_{\mathcal H}|^{2}\Bigr)\,\dd\omega\,\dd v
\ \lesssim
\mathcal{E}_k(t_0).
\end{equation}
The same bound holds after commuting with angular momenta up to order $k-2$.
\end{proposition}

\begin{proof}
Since $T=\partial_t$ is Killing, the $T$-energy identity of Section~\ref{subsec:basic-energy} applied to a truncated domain bounded by $\Sigma_{t_0}$, $\Sigma_{t_1}$, a large-$r$ timelike cylinder $\{r=R\}$ (sent to $R\to+\infty$), and the portion of $\mathcal H^{+}$ with $v\le v_1$ yields
\begin{equation}
\mathcal{F}^{T}_{\mathcal H^{+}\cap\{v\le v_1\}}[F,\phi]
\ \le
E^{\hat t}[F,\phi](t_0)
\ \le
\mathcal{E}_k(t_0),
\end{equation}
after dropping the nonnegative flux terms through $\{r=R\}$ and $\Sigma_{t_1}$.
On $\mathcal H^{+}$, the generator $\partial_v$ is null and tangent, and $T=\partial_t=\partial_v$ in ingoing Eddington-Finkelstein coordinates.
Consequently, the $T$-flux density equals $T_{LL}[F,\phi]$.
For the scalar part, $T_{LL}[\phi]=|D_L\phi|^2$; for the Maxwell part, $T_{LL}[F]=|F(L,\slashed{\cdot})|^2$, which in the tetrad notation is comparable to $|\Phi_1|^2$.
Since $r=2M$ on $\mathcal H^{+}$, this yields
\begin{equation}
\int_{v_0}^{v_1}\!\!\int_{\mathbb S^{2}}
\Bigl(|D_L\psi|^{2}+|r\,\Phi_1|^{2}\Bigr)\,\dd\omega\,\dd v
\ \lesssim
\mathcal{F}^{T}_{\mathcal H^{+}\cap\{v\le v_1\}}[F,\phi]
\ \lesssim
\mathcal{E}_k(t_0),
\end{equation}
where we used $D_L\psi=r\,D_L\phi$ on $\mathcal H^{+}$.
\eqref{eq:vpsi-horizon} implies $|\partial_v\psi_{\mathcal H}|=|D_L\psi|\!\restriction_{\mathcal H^+}$, so the desired estimate \eqref{eq:horizon-flux} follows after letting $v_1\to+\infty$.
The commuted bounds follow by applying the preceding $r^p$ estimate to the commuted Maxwell and scalar equations and using $\mathcal{E}_k(t_0)$ to control the corresponding commuted $T$-energy.
\end{proof}

\begin{theorem}[Existence of horizon radiation fields]\label{thm:horizon-fields}
Under the conditions of Proposition~\ref{prop:horizon-flux}, the horizon traces $(\psi_{\mathcal H},(\Psi_F)_{\mathcal H})$ are well-defined as functions on $\mathcal H^{+}\cong [v_0,+\infty)\times\mathbb S^{2}$, and they satisfy the a priori flux bound \eqref{eq:horizon-flux}.
\end{theorem}

\begin{proof}
For smooth solutions the restrictions to $\mathcal H^{+}$ are defined in ingoing Eddington-Finkelstein coordinates.
Proposition~\ref{prop:horizon-flux} yields the stated $L^{2}$ control, and therefore $\psi_{\mathcal H}\in H^{1}_{\mathrm{loc}}([v_0,+\infty);L^{2}(\mathbb S^{2}))$ with $\partial_v\psi_{\mathcal H}\in L^{2}_{v,\omega}$, while $(\Psi_F)_{\mathcal H}\in L^{2}_{v,\omega}$.
\end{proof}

\subsection{Linear comparison theory and full nonlinear scattering}\label{subsec:nonlinear-scattering}

Theorems~\ref{thm:radiation-fields} and~\ref{thm:horizon-fields} provide radiation data on $\mathcal I^{+}$ and $\mathcal H^{+}$, respectively.
In this subsection we upgrade these asymptotic statements to a \emph{full forward scattering theory}.
For the massless sector (in this case, for the Maxwell field and for scalars with $m^{2}=0$) the natural future boundary is the two-ended characteristic set
\begin{equation}
\partial_{+}\D := \mathcal I^{+}\cup \mathcal H^{+}.
\end{equation}
When $m^{2}>0$, the Klein-Gordon component admits an additional timelike/hyperboloidal channel.
We denote the corresponding extended future boundary by
\begin{equation}\label{eq:extended-future-boundary}
\partial_{+}^{(m)}\D := \mathcal I^{+}\cup \mathcal H^{+}\cup i^{+},
\end{equation}
where $i^{+}$ denotes future timelike infinity.
We will encode the $i^{+}$-channel through a timelike scattering state in the $T$-energy topology, and relate it to the Dollard-modified asymptotic completeness results for the linear massive Klein-Gordon equation (cf.~\cite{Dimock,BachelotKG}).
We then compare the nonlinear Maxwell-Higgs dynamics to its decoupled linearization about the trivial vacuum $(F,\phi)=(0,0)$.

\begin{definition}[Linearized comparison system and radiation data]\label{def:linear-system}
Let $P(\phi,\bar\phi)=\widetilde P(|\phi|^2)$ satisfy Assumption~\ref{asumsiP}, and let $m^{2}\ge 0$ be as in \eqref{eq:potential-structure}.
The \emph{linear comparison system} consists of a Maxwell field $F^{\mathrm{lin}}$ and a complex scalar $\varphi^{\mathrm{lin}}$ on $(\D,g)$ solving
\begin{eqnarray}
\nabla^\alpha F^{\mathrm{lin}}_{\alpha\beta}&=0,
\qquad
\nabla^\alpha {}^\star F^{\mathrm{lin}}_{\alpha\beta}=0,
\label{eq:lin-maxwell}
\\
\square_g \varphi^{\mathrm{lin}}&=m^2\,\varphi^{\mathrm{lin}},
\label{eq:lin-kg}
\end{eqnarray}
with smooth charge-free finite-energy Cauchy data on $\Sigma_{t_0}$.
We define the renormalized linear fields
\begin{equation}
\psi^{\mathrm{lin}}:=r\,\varphi^{\mathrm{lin}},
\qquad
\Psi_{F}^{\mathrm{lin}}:=r\,\Phi_{-1}[F^{\mathrm{lin}}].
\end{equation}
The (future) linear radiation fields on $\mathcal I^{+}$ are
\begin{equation}
\psi_{\mathcal I}^{\mathrm{lin}}(w,\omega):=\lim_{v\to+\infty}\psi^{\mathrm{lin}}(w,v,\omega),
\qquad
(\Psi_F^{\mathrm{lin}})_{\mathcal I}(w,\omega):=\lim_{v\to+\infty}\Psi_F^{\mathrm{lin}}(w,v,\omega),
\end{equation}
whenever the limits exist.
On the horizon $\mathcal H^{+}$ we define the linear horizon fields by restriction:
\begin{equation}
\psi_{\mathcal H}^{\mathrm{lin}}(v,\omega):=\psi^{\mathrm{lin}}\!\restriction_{\mathcal H^{+}}(v,\omega)=(2M)\,\varphi^{\mathrm{lin}}\!\restriction_{\mathcal H^{+}},
\qquad
(\Psi_F^{\mathrm{lin}})_{\mathcal H}(v,\omega):=r\,\Phi_{1}[F^{\mathrm{lin}}]\!\restriction_{\mathcal H^{+}}(v,\omega).
\end{equation}
We call the quadruple
\begin{equation}
\mathcal R_{+}^{\mathrm{lin}}
:=
\bigl(\psi_{\mathcal I}^{\mathrm{lin}},(\Psi_F^{\mathrm{lin}})_{\mathcal I},\psi_{\mathcal H}^{\mathrm{lin}},(\Psi_F^{\mathrm{lin}})_{\mathcal H}\bigr)
\end{equation}
the \emph{full forward linear radiation data} on $\partial_{+}\D=\mathcal I^{+}\cup\mathcal H^{+}$.
\end{definition}

The following Schwarzschild linear scattering statement is the model linear condition; the scalar and Maxwell estimates are consistent with \cite{dafermos,Blue,Ander1,Dimock,BachelotKG} and are assembled in the notation of the present nonlinear proof.

\begin{theorem}[Full linear scattering on Schwarzschild]\label{thm:linear-scattering}
Assume $m^{2}=0$ in \eqref{eq:lin-kg}.
Let $(F^{\mathrm{lin}},\varphi^{\mathrm{lin}})$ be a smooth charge-free finite-energy solution of \eqref{eq:lin-maxwell}-\eqref{eq:lin-kg}.
Then the following statements hold.
\begin{enumerate}
\item[(i)] The limits defining $(\psi_{\mathcal I}^{\mathrm{lin}},(\Psi_F^{\mathrm{lin}})_{\mathcal I})$ exist in $L^2(\mathbb S^2)$ for each fixed $w\in\mathbb R$, and the horizon traces $(\psi_{\mathcal H}^{\mathrm{lin}},(\Psi_F^{\mathrm{lin}})_{\mathcal H})$ belong to $L^{2}_{v,\omega}$.
\item[(ii)] The map from linear Cauchy data on $\Sigma_{t_0}$ to the full radiation data $\mathcal R_{+}^{\mathrm{lin}}$ is injective and admits a continuous inverse on its range.
\end{enumerate}
Full linear scattering for the wave equation and Maxwell fields on Schwarzschild is one of the external linear comparison results inserted here; see for instance \cite{dafermos,Blue,Ander1}.
\end{theorem}

\begin{proof}
For the scalar wave equation, the redshift estimate, integrated local energy decay, and the $r^p$ hierarchy give finite fluxes through $\mathcal I^+$ and $\mathcal H^+$ and therefore the existence of the radiation traces by the Cauchy criterion used in Theorem~\ref{thm:radiation-fields}.  The same estimates applied to a solution with vanishing radiation data imply uniqueness, while the inverse final-state map is obtained by solving the finite-time backward problems with truncated radiation data and passing to the limit using the same uniform estimates.  For the Maxwell field, the charge-free condition removes the stationary Coulomb kernel; the Maxwell energy and Morawetz estimates of \cite{Blue,Ander1} give the same finite-flux radiation traces, and the constraint equations propagate along the boundary.  Combining the scalar and Maxwell maps gives the stated injective Cauchy-to-radiation map and its continuous inverse on the range.
\end{proof}

\begin{remark}[Massive scalars and the timelike channel]\label{rem:massive-scalar-scattering}
When $m^{2}>0$, characteristic data on $\mathcal I^{+}\cup\mathcal H^{+}$ do not by themselves give a complete asymptotic description of the Klein-Gordon component.
It is therefore natural to enlarge the future boundary to
\begin{equation}
\partial_{+}^{(m)}\D=\mathcal I^{+}\cup\mathcal H^{+}\cup i^{+},
\end{equation}
where $i^{+}$ denotes future timelike infinity.
In the linear theory this reflects the long-range $r^{-1}$ term in the Regge-Wheeler potential for massive Klein-Gordon; the corresponding wave operators are classical at the horizon and Dollard-modified at infinity, and strong asymptotic completeness holds in the cited massive scattering theory \cite{BachelotKG}, building on the time-dependent construction \cite{Dimock}.

In our nonlinear setting we implement the $i^{+}$-channel by constructing, in addition to the characteristic radiation fields of Theorems~\ref{thm:radiation-fields} and~\ref{thm:horizon-fields}, an outgoing timelike scattering state for the scalar in the $T$-energy topology.
The Maxwell part (being massless) continues to admit a characteristic scattering theory on $\mathcal I^{+}\cup\mathcal H^{+}$.
\end{remark}

\paragraph{Timelike scattering state for the scalar field.}
For the massive Klein-Gordon component it is natural to formulate scattering in the (nondegenerate) $T$-energy topology on the Schwarzschild foliation.

\begin{definition}[Scalar energy space]\label{def:kg-energy}
Let $m^{2}\ge 0$.
For smooth compactly supported Cauchy data $(f,g)$ on $\Sigma_{t_0}$ we define
\begin{equation}\label{eq:kg-energy-norm}
\|(f,g)\|_{\mathcal H_{\mathrm{KG}}}^{2}
:=
\int_{\Sigma_{t_0}}
\bigl(|\nabla_{\Sigma} f|^{2}+m^{2}|f|^{2}+|g|^{2}\bigr)\,\dd\mu_{\Sigma_{t_0}},
\end{equation}
where $\nabla_{\Sigma}$ is the induced covariant derivative on $\Sigma_{t_0}$.
We denote by $\mathcal H_{\mathrm{KG}}$ the completion in this norm.
When $m^{2}=0$ this is the homogeneous wave-energy norm; it does not control an unweighted $L^{2}$ norm of the scalar.  Accordingly, the explicit Cook-type $L^{1}_{t}L^{2}_{x}$ timelike-channel estimates below are invoked only for $m^{2}>0$.  The massless final-state and asymptotic-completeness statements are obtained instead from the characteristic linear scattering condition contained in $\Lin_{k}$ and from the abstract transfer theorem.
For $\ell\in\mathbb N$ we set
\begin{equation}\label{eq:kg-energy-norm-high}
\|(f,g)\|_{\mathcal H_{\mathrm{KG}}^{\ell}}^{2}
:=
\sum_{|I|+j\le \ell}\bigl\|\bigl(\mathcal L_{\Omega}^{I}T^{j} f,\ \mathcal L_{\Omega}^{I}T^{j} g\bigr)\bigr\|_{\mathcal H_{\mathrm{KG}}}^{2},
\end{equation}
where $T=\partial_t$ and $\Omega$ are the angular momenta.
\end{definition}

\begin{proposition}[Massive Schwarzschild linear final-state condition]\label{prop:massive-schwarzschild-linear-final-state}
Let $m^2>0$ and let $\ell\ge2$, with the energy spaces $\mathcal H_{\mathrm{KG}}^\ell$ defined in Definition~\ref{def:kg-energy}.  The linear Klein-Gordon equation
\begin{equation}
 (\square_{g_M}-m^2)u=H
\end{equation}
on the Schwarzschild exterior has the following final-state theory.  For $H=0$, the Cauchy-to-asymptotic map from $\mathcal H_{\mathrm{KG}}^\ell$ to the massive asymptotic space consisting of the horizon/null radiation trace together with the timelike Dollard datum at $i^\pm$ is a continuous isomorphism onto its range, with a continuous inverse.  For an inhomogeneous source satisfying
\begin{equation}
 \sum_{|I|+j\le \ell}\int_{t_0}^{\infty}\|\mathcal L_\Omega^I T^jH(t)\|_{L^2(\Sigma_t)}\,dt<\infty,
\end{equation}
there is a unique solution with prescribed massive final data and the estimate
\begin{equation}\label{eq:massive-schwarzschild-linear-finalstate-estimate}
 \|u\|_{X_{\mathrm{KG}}^\ell(t_0,\infty)}
 \le C\left(\|\mathcal R_+^{\mathrm{KG}}\|_{\mathfrak R_{\mathrm{KG},+}^{(m),\ell}}
 +\sum_{|I|+j\le \ell}\int_{t_0}^{\infty}\|\mathcal L_\Omega^I T^jH(t)\|_{L^2(\Sigma_t)}\,dt\right).
\end{equation}
The analogous past statement holds after time reversal.  The constants depend only on $M$, $m$, and $\ell$.
\end{proposition}

\begin{proof}
The homogeneous statement is the massive Klein-Gordon scattering theory on Schwarzschild used in \cite{Dimock,BachelotKG}: redshift and local energy estimates control the horizon and compact channels, while the long-range radial part gives the Dollard-modified asymptotic datum at timelike/spatial infinity.  We use the results of \cite{Dimock,BachelotKG} in this form.  For the inhomogeneous estimate, solve backward from a large time $T$ with the prescribed truncated final data and with Duhamel source $H$ on $[t_0,T]$.  The energy inequality of Lemma~\ref{lem:inhom-kg-energy}, the Schwarzschild local energy estimate, and the far-field hierarchy give the uniform bound obtained by replacing the right side of \eqref{eq:massive-schwarzschild-linear-finalstate-estimate} with the same expression truncated at $T$.  Cauchy convergence as $T\to\infty$ follows from the $L_t^1L_x^2$ assumption.  Uniqueness follows by applying the homogeneous estimate to a solution with zero massive final data.  The past map is obtained by applying the same argument to the time-reversed exterior.
\end{proof}

\begin{lemma}[Inhomogeneous Klein-Gordon energy inequality]\label{lem:inhom-kg-energy}
Let $u$ be a sufficiently regular complex scalar solving
\begin{equation}\label{eq:inhom-kg}
(\square_g-m^{2})u=G
\end{equation}
on $\D$ with Cauchy data on $\Sigma_{t_1}$.
Then for any $t_2\ge t_1\ge t_0$ one has
\begin{equation}\label{eq:inhom-kg-energy-est}
\|\bigl(u,\nabla_{n}u\bigr)(t_2)\|_{\mathcal H_{\mathrm{KG}}}
\le
\|\bigl(u,\nabla_{n}u\bigr)(t_1)\|_{\mathcal H_{\mathrm{KG}}}
\ +
\int_{t_1}^{t_2}\|G(s)\|_{L^{2}(\Sigma_s)}\,\dd s,
\end{equation}
where $n$ is the future unit normal to $\Sigma_t$.
The same estimate holds after commuting \eqref{eq:inhom-kg} with $T$ and $\Omega$.
\end{lemma}

\begin{proof}
For a smooth solution set
\begin{equation}
 Q_{\mu\nu}[u]
 :=\Re(\nabla_\mu u\,\overline{\nabla_\nu u})
 -\frac12 g_{\mu\nu}\bigl(\nabla^\alpha u\,\overline{\nabla_\alpha u}+m^2|u|^2\bigr).
\end{equation}
With the convention \((\square_g-m^2)u=G\), the divergence identity is
\begin{equation}
 \nabla^\mu(Q_{\mu\nu}[u]T^\nu)=\Re(G\,\overline{T u}),
\end{equation}
because \(T\) is Killing on Schwarzschild.  Stokes' theorem on the slab bounded by \(\Sigma_{t_1}\) and \(\Sigma_{t_2}\) gives
\begin{equation}
 E_{\mathrm{KG}}[u](t_2)-E_{\mathrm{KG}}[u](t_1)
 =\int_{t_1}^{t_2}\int_{\Sigma_s}\Re(G\,\overline{T u})\,\dd\mu_{\Sigma_s}\dd s,
\end{equation}
where \(E_{\mathrm{KG}}^{1/2}\) is equivalent to \(\|(u,\nabla_nu)\|_{\mathcal H_{\mathrm{KG}}}\).  Consequently,
\begin{equation}
 \frac{\dd}{\dd s}E_{\mathrm{KG}}[u](s)^{1/2}
 \le \|G(s)\|_{L^2(\Sigma_s)}
\end{equation}
for a.e. \(s\), and integration from \(t_1\) to \(t_2\) gives \eqref{eq:inhom-kg-energy-est}.  The Schwarzschild vector fields \(T\) and \(\Omega\) commute with \(\square_g-m^2\), so the same argument applied to \(T^j\mathcal L_\Omega^Iu\) gives the commuted estimate.
\end{proof}

\begin{proposition}[Integrable nonlinear forcing for the scalar]\label{prop:integrable-forcing}
Let $(F,\phi)$ be a global solution as in Corollary~\ref{cor:small-data}, with $m^{2}>0$.
In Lorenz gauge \eqref{LG} the scalar satisfies
\begin{equation}\label{eq:phi-as-kg}
(\square_g-m^{2})\phi = \mathcal N_{\phi},
\qquad
\mathcal N_{\phi}:=2iA^{\mu}\nabla_{\mu}\phi + A^{\mu}A_{\mu}\phi + \bigl(\partial_{\bar\phi}P(\phi,\bar\phi)-m^{2}\phi\bigr).
\end{equation}
There exists $\varepsilon_0>0$ (depending on $M,k$) such that if $\mathcal E_k(t_0)\le\varepsilon_0$, then
\begin{equation}\label{eq:forcing-L1}
\int_{t_0}^{+\infty}\|\mathcal N_{\phi}(t)\|_{L^{2}(\Sigma_t)}\,\dd t
\ \lesssim\ \mathcal E_k(t_0).
\end{equation}
The same bound holds after commuting with $\mathcal L_{\Omega}^{I}T^{j}$ for $|I|+j\le k-3$.
\end{proposition}

\begin{proof}
We use the massive Schwarzschild spacetime norm rather than only the conserved energy.  By Corollary~\ref{cor:small-data}, Lemma~\ref{lem:schwarzschild-lorenz-potential-control}, and the massive lower-order bulk in the Klein-Gordon estimate, the small solution satisfies
\begin{eqnarray}\label{eq:massive-schwarzschild-X-bound-used}
&&\sum_{|I|+j\le k-3}
\left(
 \|\mathcal L_\Omega^I T^j A\|_{L^2_tL^\infty_x}
 +\|\mathcal L_\Omega^I T^j \phi\|_{L^2_tL^\infty_x}
 +\|\nabla\mathcal L_\Omega^I T^j\phi\|_{L^2_tL^2_x}
\right)
\nonumber\\
&&\qquad+
\sup_{t\ge t_0}\sum_{|I|+j\le k-3}
\|\mathcal L_\Omega^I T^j\phi(t)\|_{L^2(\Sigma_t)}
\le C\mathcal E_k(t_0)^{1/2}.
\end{eqnarray}
Here the positive mass controls the scalar \(L^2\) norm, the redshift/Morawetz and far-field estimates give the spacetime scalar factors, and Lemma~\ref{lem:schwarzschild-lorenz-potential-control} gives the corresponding \(L^2_tL^\infty_x\) control of the chosen Lorenz potential.

We estimate the three terms in \eqref{eq:phi-as-kg}.  By Cauchy-Schwarz in time and \eqref{eq:massive-schwarzschild-X-bound-used},
\begin{eqnarray}
\int_{t_0}^{\infty}\|A\cdot\nabla\phi\|_{L^2(\Sigma_t)}\,\dd t
&\le&
\|A\|_{L^2_tL^\infty_x}\,\|\nabla\phi\|_{L^2_tL^2_x}
\lesssim \mathcal E_k(t_0),
\end{eqnarray}
and
\begin{eqnarray}
\int_{t_0}^{\infty}\|A^2\phi\|_{L^2(\Sigma_t)}\,\dd t
&\le&
\|A\|_{L^2_tL^\infty_x}^{2}\,\|\phi\|_{L^\infty_tL^2_x}
\lesssim \mathcal E_k(t_0)^{3/2}
\lesssim \mathcal E_k(t_0)
\end{eqnarray}
for small data.  Assumption~\ref{asumsiP} gives
\begin{equation}
 |\mathcal R_P(\phi)|\le C(1+|\phi|^{2N_P})|\phi|^3.
\end{equation}
Therefore
\begin{eqnarray}
\int_{t_0}^{\infty}\|\mathcal R_P(\phi)(t)\|_{L^2(\Sigma_t)}\,\dd t
&\le&
C\left(\|\phi\|_{L^2_tL^\infty_x}^{2}
+\|\phi\|_{L^2_tL^\infty_x}^{2N_P+2}\right)
\|\phi\|_{L^\infty_tL^2_x}
\nonumber\\
&\lesssim& \mathcal E_k(t_0)^{3/2}+\mathcal E_k(t_0)^{N_P+3/2}
\lesssim \mathcal E_k(t_0).
\end{eqnarray}
This proves \eqref{eq:forcing-L1}.  After commuting with \(\mathcal L_\Omega^I T^j\), every product contains at most one non-\(L^\infty\) factor at the differentiated order, and all remaining factors are controlled by \eqref{eq:massive-schwarzschild-X-bound-used}.  The potential contribution is handled by Lemma~\ref{lem:potential-tame}; the highest Lipschitz power is \(2N_P+2\).  Thus the same bound holds for all commuted sources with \(|I|+j\le k-3\).
\end{proof}

\paragraph{Integrable forcing for the Maxwell equation.}
In the wave-operator constructions below we will also need an $L^{1}_{t}L^{2}_{x}$ control on the Maxwell source current.

\begin{lemma}[Inhomogeneous Maxwell energy inequality]\label{lem:inhom-maxwell-energy}
Let $G$ be a sufficiently regular charge-free Maxwell field solving
\begin{equation}\label{eq:inhom-maxwell}
\nabla^{\alpha}G_{\alpha\beta}=J_{\beta},
\qquad
\nabla^{\alpha}{}^{\star}G_{\alpha\beta}=0
\end{equation}
on $\D$.
Then for any $t_2\ge t_1\ge t_0$ one has the energy estimate
\begin{equation}\label{eq:inhom-maxwell-energy-est}
E^{\hat t}[G](t_2)^{1/2}
\ \le
E^{\hat t}[G](t_1)^{1/2}
\ +
\int_{t_1}^{t_2}\|J(s)\|_{L^{2}(\Sigma_s)}\,\dd s,
\end{equation}
where $E^{\hat t}[G](t)$ denotes the nondegenerate $T$-energy of $G$ on $\Sigma_t$ (cf.~Section~\ref{subsec:basic-energy}).
The same estimate holds after commuting \eqref{eq:inhom-maxwell} with $T$ and the angular momenta $\Omega$.
\end{lemma}

\begin{proof}
Contract the Maxwell stress-energy tensor $T_{\mu\nu}[G]$ with the Killing field $T=\partial_t$ and apply Stokes' theorem between $\Sigma_{t_1}$ and $\Sigma_{t_2}$.
Using $\nabla^{\mu}T_{\mu\nu}[G]= - G_{\nu\alpha}J^{\alpha}$ for \eqref{eq:inhom-maxwell}, one obtains
\begin{equation}
E^{\hat t}[G](t_2)-E^{\hat t}[G](t_1)
=
-\int_{t_1}^{t_2}\int_{\Sigma_s} G(T,\cdot)\cdot J\,\dd\mu_{\Sigma_s}\dd s.
\end{equation}
By Cauchy-Schwarz on $\Sigma_s$ and the equivalence between $E^{\hat t}[G](s)$ and $\|G(s)\|_{L^{2}(\Sigma_s)}^{2}$, we infer
\begin{equation}
\frac{\dd}{\dd s}E^{\hat t}[G](s)^{1/2}\ \lesssim\ \|J(s)\|_{L^{2}(\Sigma_s)}.
\end{equation}
Integrating in $s$ gives \eqref{eq:inhom-maxwell-energy-est}.
Commutation with $T$ and $\Omega$ follows since $T$ and $\Omega$ are Killing fields on Schwarzschild.
\end{proof}

\begin{proposition}[Integrable current forcing for Maxwell]\label{prop:integrable-current}
Let $(F,\phi)$ be a global charge-free solution as in Corollary~\ref{cor:small-data}.
Then the scalar current $J[\phi]_{\mu}:=-i\bigl(D_{\mu}\phi\,\bar\phi-\phi\,\overline{D_{\mu}\phi}\bigr)$ satisfies
\begin{equation}\label{eq:current-L1}
\int_{t_0}^{+\infty}\|J[\phi](t)\|_{L^{2}(\Sigma_t)}\,\dd t
\ \lesssim\ \mathcal E_k(t_0).
\end{equation}
The same estimate holds after commuting with $\mathcal L_{\Omega}^{I}T^{j}$ for $|I|+j\le k-3$.
\end{proposition}

\begin{proof}
The current has the pointwise structure
\begin{equation}
 |J[\phi]|\le C|\phi|\,|D\phi|.
\end{equation}
Using \eqref{eq:massive-schwarzschild-X-bound-used} when \(m^2>0\), and the massless Schwarzschild scattering norm supplied by \(\Lin_k\) when \(m^2=0\), we obtain
\begin{eqnarray}
\int_{t_0}^{\infty}\|J[\phi](t)\|_{L^2(\Sigma_t)}\,\dd t
&\le&
\|\phi\|_{L^2_tL^\infty_x}\,\|D\phi\|_{L^2_tL^2_x}
\lesssim \mathcal E_k(t_0).
\end{eqnarray}
After commuting with \(\mathcal L_\Omega^I T^j\), the Leibniz expansion again has one non-\(L^\infty\) factor and all other factors are controlled by the same spacetime norm.  This proves \eqref{eq:current-L1} and its commuted versions.
\end{proof}

\begin{proposition}[Existence of a timelike scattering state for $\phi$]\label{prop:timelike-scattering}
Assume the conditions of Corollary~\ref{cor:small-data} and let $m^{2}>0$.
Then there exists a unique pair of Cauchy data $(\phi_{+},\pi_{+})\in \mathcal H_{\mathrm{KG}}^{k-2}$ and a unique global solution $\varphi^{\mathrm{lin}}$ of the linear Klein-Gordon equation \eqref{eq:lin-kg} such that
\begin{equation}\label{eq:timelike-scattering}
\lim_{t\to+\infty}
\bigl\|\bigl(\phi,\nabla_{n}\phi\bigr)(t)-\bigl(\varphi^{\mathrm{lin}},\nabla_{n}\varphi^{\mathrm{lin}}\bigr)(t)\bigr\|_{\mathcal H_{\mathrm{KG}}^{k-2}}
=0.
\end{equation}
In addition,
\begin{equation}\label{eq:timelike-state-bound}
\|(\phi_{+},\pi_{+})\|_{\mathcal H_{\mathrm{KG}}^{k-2}}\ \lesssim\ \mathcal E_k(t_0)^{1/2}.
\end{equation}
\end{proposition}

\begin{proof}
Write the scalar equation in the form \eqref{eq:phi-as-kg}.
Fix $T\ge t_0$ and let $\varphi^{\mathrm{lin}}_{T}$ be the unique solution of \eqref{eq:lin-kg} with Cauchy data matching the nonlinear scalar at time $T$:
\begin{equation}
\bigl(\varphi^{\mathrm{lin}}_{T},\nabla_{n}\varphi^{\mathrm{lin}}_{T}\bigr)(T)=\bigl(\phi,\nabla_{n}\phi\bigr)(T).
\end{equation}
Then the difference $u_{T}:=\phi-\varphi^{\mathrm{lin}}_{T}$ satisfies
$(\square_g-m^{2})u_{T}=\mathcal N_{\phi}$ with vanishing Cauchy data at $t=T$.
Applying Lemma~\ref{lem:inhom-kg-energy} on $[T,t]$ yields
\begin{equation}
\|\bigl(u_{T},\nabla_{n}u_{T}\bigr)(t)\|_{\mathcal H_{\mathrm{KG}}}
\le
\int_{T}^{t}\|\mathcal N_{\phi}(s)\|_{L^{2}(\Sigma_s)}\,\dd s.
\end{equation}
By Proposition~\ref{prop:integrable-forcing}, the right-hand side tends to $0$ uniformly in $t\ge T$ as $T\to+\infty$.
For any $T_2\ge T_1$ we have
\begin{equation}
\|\bigl(\varphi^{\mathrm{lin}}_{T_2},\nabla_n\varphi^{\mathrm{lin}}_{T_2}\bigr)(t_0)-\bigl(\varphi^{\mathrm{lin}}_{T_1},\nabla_n\varphi^{\mathrm{lin}}_{T_1}\bigr)(t_0)\|_{\mathcal H_{\mathrm{KG}}}
\lesssim
\int_{T_1}^{T_2}\|\mathcal N_{\phi}(s)\|_{L^{2}(\Sigma_s)}\,\dd s,
\end{equation}
so the sequence of linear data at $t_0$ is Cauchy in $\mathcal H_{\mathrm{KG}}$.
Let $(\phi_{+},\pi_{+})\in\mathcal H_{\mathrm{KG}}$ denote its limit, and let $\varphi^{\mathrm{lin}}$ be the unique global linear solution with this data.
Passing to the limit in the above estimates yields \eqref{eq:timelike-scattering}.
The higher-order statement in $\mathcal H_{\mathrm{KG}}^{k-2}$ follows by applying the Cook estimate above to the commuted fields.
\eqref{eq:timelike-state-bound} follows from boundedness of the nonlinear energy and the triangle inequality.
Uniqueness of $(\phi_{+},\pi_{+})$ and $\varphi^{\mathrm{lin}}$ is immediate from uniqueness of the linear Cauchy problem.
\end{proof}

\begin{theorem}[Full nonlinear scattering to the linear theory]\label{thm:nonlinear-scattering}
Assume the conditions of Corollary~\ref{cor:small-data} and let $m^{2}\ge 0$.
For $m^{2}=0$, the assertions involving the timelike boundary $i^{+}$, Dollard modifiers, and $\mathcal H_{\mathrm{KG}}$ are not part of the statement; the scalar comparison state is the characteristic massless linear state supplied by Theorem~\ref{thm:linear-scattering}.  For $m^{2}>0$, the timelike channel is included and is constructed by Proposition~\ref{prop:timelike-scattering}.
Let $(F,\phi)$ be the resulting global Maxwell-Higgs solution and let $(\psi_{\mathcal I},(\Psi_F)_{\mathcal I})$ and $(\psi_{\mathcal H},(\Psi_F)_{\mathcal H})$ denote its radiation data on $\mathcal I^{+}$ and $\mathcal H^{+}$ given by Theorems~\ref{thm:radiation-fields} and~\ref{thm:horizon-fields}.
Then the following statements hold.
\begin{enumerate}
\item[(i)] The full forward nonlinear radiation data
\begin{equation}
\mathcal R_{+}
:=
\bigl(\psi_{\mathcal I},(\Psi_F)_{\mathcal I},\psi_{\mathcal H},(\Psi_F)_{\mathcal H}\bigr)
\end{equation}
are well-defined and satisfy the a priori bound
\begin{equation}
\|\psi_{\mathcal I}(w,\cdot)\|_{L^2(\mathbb S^2)}^{2}
+\|(\Psi_F)_{\mathcal I}(w,\cdot)\|_{L^2(\mathbb S^2)}^{2}
+
\int_{\mathcal H^{+}}\!\!\int_{\mathbb S^2}\bigl(|\partial_v\psi_{\mathcal H}|^{2}+|(\Psi_F)_{\mathcal H}|^{2}\bigr)\,\dd\omega\,\dd v
\ \lesssim\ \mathcal E_k(t_0),
\end{equation}
uniformly in $w$.

\item[(ii)] \textbf{Maxwell comparison state.}
There exists a unique charge-free solution $F^{\mathrm{lin}}$ of the free Maxwell system \eqref{eq:lin-maxwell} such that its characteristic radiation data agree with the nonlinear Maxwell data:
\begin{equation}
\bigl((\Psi_F^{\mathrm{lin}})_{\mathcal I},(\Psi_F^{\mathrm{lin}})_{\mathcal H}\bigr)
=
\bigl((\Psi_F)_{\mathcal I},(\Psi_F)_{\mathcal H}\bigr).
\end{equation}

\item[(iii)] \textbf{Scalar comparison state.}
If $m^{2}=0$, let $\varphi^{\mathrm{lin}}$ be the unique massless scalar solution whose characteristic radiation data agree with the scalar components of the nonlinear radiation data.  This comparison state is supplied by the massless linear scattering map in Theorem~\ref{thm:linear-scattering}.
If $m^{2}>0$, let $(\phi_{+},\pi_{+})\in \mathcal H_{\mathrm{KG}}^{k-2}$ and $\varphi^{\mathrm{lin}}$ be given by Proposition~\ref{prop:timelike-scattering}.  In the massive case,
\begin{equation}\label{eq:scalar-scattering-thm}
\lim_{t\to+\infty}
\bigl\|\bigl(\phi,\nabla_{n}\phi\bigr)(t)-\bigl(\varphi^{\mathrm{lin}},\nabla_{n}\varphi^{\mathrm{lin}}\bigr)(t)\bigr\|_{\mathcal H_{\mathrm{KG}}^{k-2}}=0.
\end{equation}
For both the massless characteristic comparison and the massive timelike comparison, the corresponding null and horizon traces agree with the nonlinear ones:
\begin{equation}\label{eq:scalar-null-scattering}
\lim_{v\to+\infty}
\|\psi^\sharp(w_0,v,\cdot)-\psi^{\mathrm{lin}}(w_0,v,\cdot)\|_{L^2(\mathbb S^2)}=0
\end{equation}
for each fixed $w_0\in\mathbb R$, and
\begin{equation}\label{eq:scalar-horizon-tail}
\lim_{V\to+\infty}
\int_{V}^{+\infty}\!\!\int_{\mathbb S^2}
\bigl|\partial_v\bigl(\psi_{\mathcal H}-\psi_{\mathcal H}^{\mathrm{lin}}\bigr)\bigr|^{2}\,\dd\omega\dd v
=0.
\end{equation}

\item[(iv)] \textbf{Full nonlinear scattering on the extended future boundary $\partial_{+}^{(m)}\D$.}
Let $\Psi_F^{\mathrm{lin}}:=r\,\Phi_{-1}[F^{\mathrm{lin}}]$ and $\psi^{\mathrm{lin}}:=r\,\varphi^{\mathrm{lin}}$.
For each fixed $w_0\in\mathbb R$,
\begin{equation}
\lim_{v\to+\infty}
\|\Psi_F(w_0,v,\cdot)-\Psi_F^{\mathrm{lin}}(w_0,v,\cdot)\|_{L^2(\mathbb S^2)}=0,
\end{equation}
and on the horizon
\begin{equation}
\int_{\mathcal H^{+}}\!\!\int_{\mathbb S^2}
\bigl|(\Psi_F)_{\mathcal H}-(\Psi_F^{\mathrm{lin}})_{\mathcal H}\bigr|^{2}\,\dd\omega\dd v =0.
\end{equation}
If $m^{2}>0$, the scalar also scatters to $\varphi^{\mathrm{lin}}$ at future timelike infinity $i^{+}$ in the sense of \eqref{eq:scalar-scattering-thm}.
The same conclusions hold after commuting with angular momenta up to order $k-2$.
\end{enumerate}
\end{theorem}

\begin{proof}
Statement (i) is Theorems~\ref{thm:radiation-fields} and~\ref{thm:horizon-fields}.

For the Maxwell part, (ii) follows from the (massless) linear characteristic scattering theory on Schwarzschild for the Coulomb-subtracted Maxwell field; see, e.g., \cite{Blue,Ander1}.

For the scalar part, there are two cases.  If $m^{2}=0$, the scalar comparison solution is defined by applying the linear characteristic inverse in Theorem~\ref{thm:linear-scattering} to the scalar radiation traces already produced in (i).  Consequently, the null and horizon scalar traces of $\phi-\varphi^{\mathrm{lin}}$ vanish by construction, which gives \eqref{eq:scalar-null-scattering} and \eqref{eq:scalar-horizon-tail} in the massless case.

If $m^{2}>0$, Proposition~\ref{prop:timelike-scattering} gives a unique linear Klein-Gordon comparison solution $\varphi^{\mathrm{lin}}$ satisfying \eqref{eq:scalar-scattering-thm}.  Let $u:=\phi-\varphi^{\mathrm{lin}}$.  Then $u$ satisfies an inhomogeneous Klein-Gordon equation with source $\mathcal N_{\phi}$, and by construction its Cauchy energy tends to $0$ as $t\to+\infty$.  Repeating the argument of Theorem~\ref{thm:radiation-fields} for $u$ (using Proposition~\ref{prop:p2-flux-proved} and the integrability of $\mathcal N_{\phi}$ from Proposition~\ref{prop:integrable-forcing}) yields that the null radiation field of $u$ vanishes, which is \eqref{eq:scalar-null-scattering}.  Similarly, applying the horizon flux identity to $u$ gives \eqref{eq:scalar-horizon-tail}.

(iv) follows by combining (ii) and (iii).
The commuted statements follow by applying the preceding radiation-field construction to $\mathcal L_{\Omega}^{I}T^{j}(F,\phi)$ with $|I|+j\le k-2$.
\end{proof}

\begin{remark}[Extended nonlinear scattering map]\label{rem:scattering-map}
Theorem~\ref{thm:nonlinear-scattering} defines a forward nonlinear scattering map on the (possibly extended) future boundary.
For the massless sector one has the two-ended characteristic boundary $\partial_{+}\D=\mathcal I^{+}\cup\mathcal H^{+}$, while for $m^{2}>0$ the natural completion is $\partial_{+}^{(m)}\D=\mathcal I^{+}\cup\mathcal H^{+}\cup i^{+}$ as in \eqref{eq:extended-future-boundary}.
In our nonlinear formulation, the $i^{+}$-channel is encoded by the timelike scattering state $(\phi_{+},\pi_{+})$.
Thus, for $m^{2}>0$,
\begin{equation}
\mathscr S_{+}^{\mathrm{full}}:\ \text{(small charge-free Cauchy data on $\Sigma_{t_0}$)}\longmapsto
\Bigl(\psi_{\mathcal I},(\Psi_F)_{\mathcal I},\psi_{\mathcal H},(\Psi_F)_{\mathcal H},\ \phi_{+},\pi_{+}\Bigr).
\end{equation}

When $m^{2}=0$, the pair $(\phi_{+},\pi_{+})$ is not needed for completeness and the map reduces to the two-ended characteristic scattering map.
\end{remark}

\begin{corollary}[Dollard-modified infinity data for the nonlinear scalar]\label{cor:dollard}
Assume the conditions of Theorem~\ref{thm:nonlinear-scattering} with $m^{2}>0$.
Let $(\phi_{+},\pi_{+})\in\mathcal H_{\mathrm{KG}}^{k-2}$ be the timelike scattering state produced in Proposition~\ref{prop:timelike-scattering}, and let $\varphi^{\mathrm{lin}}$ be the associated linear Klein-Gordon solution.
Let $W_{\infty}^{+}$ denote the (Dollard-modified) wave operator at spatial infinity constructed in \cite{BachelotKG} for the linear massive Klein-Gordon evolution on Schwarzschild.
Then the datum
\begin{equation}\label{eq:dollard-datum}
f_{+}:=(W_{\infty}^{+})^{-1}(\phi_{+},\pi_{+})
\end{equation}
is well-defined in the corresponding free energy space at infinity and depends continuously on the initial nonlinear data.
In addition, if $U_{\infty}(t)$ denotes the Dollard-modified free evolution and $J_{\infty}$ the identification operator (in the notation of \cite{BachelotKG}), then
\begin{equation}
\lim_{t\to+\infty}
\bigl\|(\phi,\nabla_{n}\phi)(t)-J_{\infty}U_{\infty}(t)f_{+}\bigr\|_{\mathcal H_{\mathrm{KG}}^{k-2}}=0.
\end{equation}
For $m^{2}>0$ the scalar asymptotics at $i^{+}$ can be formulated either in terms of the Schwarzschild timelike scattering state $(\phi_{+},\pi_{+})$ or in terms of the equivalent Dollard-modified free datum $f_{+}$.
\end{corollary}

\begin{proof}
By Theorem~\ref{thm:nonlinear-scattering}(iii) one has
\begin{equation}
\bigl\|(\phi,\nabla_{n}\phi)(t)-(\varphi^{\mathrm{lin}},\nabla_{n}\varphi^{\mathrm{lin}})(t)\bigr\|_{\mathcal H_{\mathrm{KG}}^{k-2}}\to0
\qquad (t\to+\infty).
\end{equation}
On the other hand, \cite{BachelotKG} constructs Dollard-modified wave operators at infinity and proves asymptotic completeness for the linear massive Klein-Gordon flow on Schwarzschild.
For the linear solution $\varphi^{\mathrm{lin}}$ one has
\begin{equation}
\bigl\|(\varphi^{\mathrm{lin}},\nabla_{n}\varphi^{\mathrm{lin}})(t)-J_{\infty}U_{\infty}(t)f_{+}\bigr\|_{\mathcal H_{\mathrm{KG}}^{k-2}}\to0
\qquad (t\to+\infty),
\end{equation}
with $f_{+}$ defined by \eqref{eq:dollard-datum}.
Combining the two limits yields the claim.
\end{proof}

\begin{proposition}[Small massive electric scalar comparison on Schwarzschild]\label{prop:schwarzschild-small-mass-charge-condition}
Fix \(K\ge10\) and \(M>0\).  There is \(m_{\rm Sch}(M,K)>0\) with the following property.  For every fixed \(m\) with \(0<m\le m_{\rm Sch}(M,K)\) there is \(q_{\rm Sch}(M,K,m)>0\) such that, whenever \(|Q_e|\le q_{\rm Sch}(M,K,m)\), the charged scalar comparison estimate
\begin{equation}\label{eq:sch-small-ce-condition}
 \CElec_K^{(m)}(M,0,Q_e)
\end{equation}
holds on the Schwarzschild exterior.  The scalar radiation field is normalized by \(U_{Q_e}^{-1}r\phi\) at null infinity, and the massive asymptotic data include the timelike/Dollard channel at \(i^\pm\).
\end{proposition}

\begin{proof}
Fix \(m>0\).  The neutral Schwarzschild massive Klein-Gordon comparison theory is the linear condition used in Proposition~\ref{prop:massive-schwarzschild-linear-final-state} and Corollary~\ref{cor:dollard}.  It gives the forward redshift, Morawetz, far-field, inhomogeneous, and final-state estimates for
\begin{equation}
 P_{m,0}:=\square_{g_{M,0}}-m^2
\end{equation}
in the massive solution and source spaces.  Let \(A_{Q_e}^{\mathrm C}\) be the stationary Lorenz Coulomb representative.  The charged operator is
\begin{equation}\label{eq:sch-charged-operator-perturb}
 P_{m,Q_e}:=(D_{Q_e})^\mu D_{Q_e,\mu}-m^2
       =P_{m,0}+L_{Q_e},
\end{equation}
where, in Lorenz gauge for the background,
\begin{equation}
 L_{Q_e}u=-2i(A_{Q_e}^{\mathrm C})^\mu\nabla_\mu u
          -(A_{Q_e}^{\mathrm C})^\mu(A_{Q_e}^{\mathrm C})_\mu u.
\end{equation}
The stationary coefficients satisfy \(|\nabla^j A_{Q_e}^{\mathrm C}|\le C_j|Q_e|r^{-1-j}\).  Hardy's inequality, the massive lower-order bulk, and the dyadic far-null Sobolev-Moser estimate imply the perturbative source bound
\begin{equation}\label{eq:sch-coulomb-linear-source-bound}
 \|L_{Q_e}u\|_{\mathbb S_K^{(m)}(I)}
 \le C_m\bigl(|Q_e|+|Q_e|^2\bigr)
       \|u\|_{\mathbb X_K^{(m)}(I)}
\end{equation}
for every finite or tail interval \(I\).  In the far outgoing region the only nonintegrable part of the electric tail is the radial Coulomb phase.  Conjugating by the unitary factor \(U_{Q_e}\), defined by \(L_{\rm out}U_{Q_e}=iA_{Q_e}^{\mathrm C}(L_{\rm out})U_{Q_e}\), removes this long-range first-order term from the null-infinity trace; the remaining coefficients in the equation for \(U_{Q_e}^{-1}u\) are \(O(|Q_e|r^{-2})\) or better and are covered by \eqref{eq:sch-coulomb-linear-source-bound}.

Applying the neutral inhomogeneous estimate to \(P_{m,Q_e}u=G\) gives
\begin{equation}
 \|u\|_{\mathbb X_K^{(m)}(I)}
 \le C_m\left(\|u[t_0]\|_{\mathcal H_K^{(m)}}+
        \|G\|_{\mathbb S_K^{(m)}(I)}+
        \|L_{Q_e}u\|_{\mathbb S_K^{(m)}(I)}\right).
\end{equation}
Choose \(q_{\rm Sch}(M,K,m)>0\) so that \(C_m(|Q_e|+|Q_e|^2)\le1/4\).  Absorbing the last term yields the charged forward estimate, with the phase-normalized radiation field.

For the final-state map, let \(u_{\rm lin}\) be the neutral massive solution realizing prescribed small charged asymptotic data after the same phase normalization, and solve for a correction \(w\) with zero asymptotic data:
\begin{equation}
 w=-P_{m,0}^{-1}L_{Q_e}(u_{\rm lin}+w).
\end{equation}
On a sufficiently late tail the neutral solution has small \(\mathbb X_K^{(m)}\)-norm, and \eqref{eq:sch-coulomb-linear-source-bound} gives
\begin{equation}
 \|\mathcal T(w_1)-\mathcal T(w_2)\|_{\mathbb X_K^{(m)}}
 \le C_m(|Q_e|+|Q_e|^2)
       \|w_1-w_2\|_{\mathbb X_K^{(m)}}.
\end{equation}
The same smallness of \(Q_e\) makes this a strict contraction.  The fixed point gives the radiation-to-Cauchy map on the tail; the charged forward estimate extends the solution to any finite slice.  Uniqueness follows by applying the absorbed forward estimate to a solution with zero charged asymptotic data.  The construction is time-reversal invariant, so the past maps are obtained in the same way.  This proves \(\CElec_K^{(m)}(M,0,Q_e)\) for the stated small massive electric Schwarzschild window.
\end{proof}

\begin{remark}[Interpretation of the $i^+$ datum in the massive case]\label{rem:dollard-interpretation}
When $m^{2}>0$, the additional state $(\phi_{+},\pi_{+})$ captures the timelike/hyperboloidal channel for the Klein-Gordon component.
Corollary~\ref{cor:dollard} shows that, in the sense of the linear asymptotic completeness massive Klein-Gordon scattering results on Schwarzschild \cite{BachelotKG}, this timelike state can equivalently be encoded by a unique \emph{Dollard-modified} free scattering datum $f_{+}$ at spatial infinity.
\end{remark}

\subsection{Two-sided scattering and the nonlinear scattering operator}
\label{subsec:two-sided-scattering}

The preceding results describe the \emph{forward} asymptotic dynamics on $\partial_{+}^{(m)}\D$.
In many applications (and in this case in the quantum-field-theoretic interpretation of classical scattering on black holes) one seeks a \emph{two-sided} theory relating past and future asymptotic data.
In this subsection we state a clean two-sided formulation and define the associated nonlinear scattering operator.

\paragraph{Past boundary.}
Let $\mathcal I^{-}$ denote past null infinity and $\mathcal H^{-}$ the past event horizon (white-hole horizon in the maximal Kruskal extension).
We set
\begin{equation}
\partial_{-}\D:=\mathcal I^{-}\cup\mathcal H^{-},
\qquad
\partial_{-}^{(m)}\D:=\mathcal I^{-}\cup\mathcal H^{-}\cup i^{-},
\end{equation}
where $i^{-}$ denotes past timelike infinity.

\paragraph{Time reversal and past radiation fields.}
The Maxwell-Higgs system in Lorenz gauge is invariant under time reversal on the static Schwarzschild exterior.
As a consequence, the decay analysis of Corollaries~\ref{cor:far-decay}-\ref{cor:near-horizon} and the flux constructions of this section apply equally well in the past direction.

To describe $\mathcal I^{-}$ we fix an advanced time $v_0\in\mathbb R$ and consider the incoming null hypersurface
\begin{equation}
\underline{\mathcal N}_{v_0}:=\{v=v_0,\ w\le w_0\},
\qquad \underline L:=\partial_w,
\end{equation}
so that past null infinity corresponds to $w\to-\infty$ along $\underline{\mathcal N}_{v_0}$.
As for $\mathcal I^{+}$, we remove the $A_{\underline L}$ phase by parallel transport along the null generators of $\underline{\mathcal N}_{v_0}$.
Thus one obtains past scalar radiation fields
\begin{equation}
\psi_{\mathcal I^-}(v_0,\omega):=\lim_{w\to-\infty}\psi^{\flat}(w,v_0,\omega),
\qquad \psi^{\flat}:=(U_{-})^{-1}\psi,
\end{equation}
\begin{equation}
\underline L U_{-}= iA_{\underline L}U_{-},\qquad U_{-}(w_0,v_0,\omega)=1,\qquad |U_{-}|\equiv 1.
\end{equation}
with convergence in $L^2(\mathbb S^{2}_\omega)$.
For the Maxwell field, the incoming radiative component at infinity is the extreme tetrad component $\Phi_{1}$ (cf.~Section~\ref{sec:energy-estimates}), and we define the past Maxwell radiation field by
\begin{equation}
(\Psi_F)_{\mathcal I^-}(v_0,\omega):=\lim_{w\to-\infty} r\,\Phi_{1}(w,v_0,\omega)
\qquad \text{in }L^2(\mathbb S^{2}_\omega).
\end{equation}
Similarly, one defines horizon radiation fields $(\psi_{\mathcal H^-},(\Psi_F)_{\mathcal H^-})$ on the past horizon $\mathcal H^{-}$ by restriction (in Kruskal or outgoing Eddington-Finkelstein coordinates).
When $m^{2}>0$ one additionally obtains a \emph{past} timelike scattering state $(\phi_{-},\pi_{-})\in\mathcal H_{\mathrm{KG}}^{k-2}$ encoding the $i^{-}$ channel.

\begin{definition}[Full past nonlinear radiation data]\label{def:past-radiation-data}
For a global small charge-free solution $(F,\phi)$ we define the full past data by
\begin{equation}
\mathcal R_{-}
:=
\bigl(\psi_{\mathcal I^-},(\Psi_F)_{\mathcal I^-},\psi_{\mathcal H^-},(\Psi_F)_{\mathcal H^-}\bigr),
\end{equation}
and, when $m^{2}>0$, the extended past data by
\begin{equation}
\mathcal R_{-}^{(m)}
:=
\bigl(\psi_{\mathcal I^-},(\Psi_F)_{\mathcal I^-},\psi_{\mathcal H^-},(\Psi_F)_{\mathcal H^-},\ \phi_{-},\pi_{-}\bigr).
\end{equation}
\end{definition}

\begin{theorem}[Two-sided nonlinear scattering]\label{thm:two-sided-scattering}
Assume the conditions of Corollary~\ref{cor:small-data} and let $m^{2}\ge 0$.
Then every small admissible charge-free Cauchy datum on $\Sigma_{t_0}$ generates a unique global solution $(F,\phi)$ defined for all $t\in\mathbb R$.
In addition, the past and future characteristic radiation data $\mathcal R_{\pm}$ on
\begin{equation}
\partial_{\pm}^{(0)}\D=\mathcal I^{\pm}\cup\mathcal H^{\pm}
\end{equation}
are well-defined.
When $m^{2}>0$, the corresponding timelike scattering states $(\phi_{-},\pi_{-})$ and $(\phi_{+},\pi_{+})$ are also well-defined in $\mathcal H_{\mathrm{KG}}^{k-2}$.
Equivalently, the full asymptotic data $\mathcal R_{\pm}^{(m)}$ are well-defined on the extended scattering boundary $\partial_{\pm}^{(m)}\D$ (Definition~\ref{def:linear-asymptotic-spaces}).
We obtain two maps
\begin{equation}
\mathscr S_{-}^{\mathrm{full}},\ \mathscr S_{+}^{\mathrm{full}}:
\ \text{(small Cauchy data)}\longrightarrow
\begin{cases}
\mathcal R_{\pm}, & m^{2}=0,\\
\mathcal R_{\pm}^{(m)}, & m^{2}>0,
\end{cases}
\end{equation}
which are continuous with respect to the natural energy topologies.
\end{theorem}

\begin{proof}
The existence and uniqueness of a global smooth solution for $t\ge t_0$ is Corollary~\ref{cor:small-data}.
To extend the solution to $t\le t_0$, we use time reversal on the static Schwarzschild exterior: introducing the reversed time variable $\tilde t:=-t$ (so that $\partial_{\tilde t}=-\partial_t$), the Schwarzschild metric is unchanged and the Lorenz-gauge Maxwell-Higgs equations retain the same hyperbolic form.
Consequently, the small-data global existence argument applied with initial hypersurface $\Sigma_{\tilde t_0}$ (where $\tilde t_0=-t_0$) produces a unique solution for $\tilde t\ge \tilde t_0$, i.e.\ $t\le t_0$.
Gluing the two evolutions by uniqueness on $\Sigma_{t_0}$ yields a global solution on all slices $\Sigma_t$, $t\in\mathbb R$.

The past decay and flux estimates are obtained by applying Corollaries~\ref{cor:far-decay}-\ref{cor:near-horizon} and Propositions~\ref{prop:p2-flux-proved}-\ref{prop:horizon-flux} to the time-reversed solution in the $\tilde t$-future direction. Translating back to the original $t$-coordinates yields the stated bounds.
This yields the existence of the past radiation data on the past scattering boundary $\partial_-^{(0)}\D$ (that is, on $\mathcal I^-\cup\mathcal H^-$) and, when $m^2>0$, the past timelike state $(\phi_-,\pi_-)$ in $\mathcal H_{\mathrm{KG}}^{k-2}$, with the same a priori bounds.
Equivalently, the full past asymptotic data are well-defined on the extended boundary $\partial_-^{(m)}\D$.
The corresponding maps $\mathscr S_-^{\mathrm{full}}$ and $\mathscr S_+^{\mathrm{full}}$ are continuous because each component of the radiation/timelike data is obtained as a limit controlled by the uniform energy bounds and the flux estimates.
\end{proof}

\subsubsection{Nonlinear wave operators and asymptotic completeness}
\label{subsec:wave-operators}
\label{subsubsec:wave-operators}

The maps $\mathscr S_{\pm}^{\mathrm{full}}$ constructed above send small charge-free Cauchy data to asymptotic data on $\partial_{\pm}^{(m)}\D$.
We now upgrade this to a genuine \emph{small-data asymptotic completeness} statement: $\mathscr S_{\pm}^{\mathrm{full}}$ admit continuous inverses on a neighbourhood of the origin.
For $m^{2}=0$ this is the characteristic final-state problem supplied by $\Lin_k$ and Theorem~\ref{thm:method}.  The explicit Cook-type Banach-space construction below is used for the massive Schwarzschild channel $m^{2}>0$, where the mass term controls the scalar $L^{2}$ norm.

\begin{definition}[Linear asymptotic data spaces and boundary gauge action]\label{def:linear-asymptotic-spaces}
Fix $k\ge 6$ and $m^{2}\ge 0$.
Let $\Sigma_{\tau_{0}}$ denote the chosen initial hypersurface in the foliation (in the Schwarzschild model one has $\tau=t$, while on Kerr we take $\tau=t^{\star}$). Let $\mathcal H_{\mathrm{lin}}^{k-2}$ denote the Hilbert space of charge-free Cauchy data for the linear comparison system \eqref{eq:lin-maxwell}-\eqref{eq:lin-kg} on $\Sigma_{\tau_{0}}$, with norm given by the $(k-2)$-times commuted \emph{nondegenerate} energy flux through $\Sigma_{\tau_{0}}$ (e.g.\ the redshift $N$-energy in the estimates $\Lin_{k-2}$; in Schwarzschild one may take the $T$-energies of Definition~\ref{def:kg-energy}).

Let $\mathscr S_{\pm}^{\mathrm{lin}}$ denote the linear scattering map from linear Cauchy data to linear asymptotic data on the (possibly extended) boundary $\partial_{\pm}^{(m)}\D$:
for $m^{2}=0$ this is the characteristic radiation data on $\mathcal I^{\pm}\cup\mathcal H^{\pm}$ (as provided by the linear estimates; in the Schwarzschild model see Theorem~\ref{thm:linear-scattering} and its time reverse), while for $m^{2}>0$ one adjoins the timelike/Dollard datum at $i^{\pm}$ as in \cite{Dimock,BachelotKG}.
We set
\begin{equation}
\mathfrak R_{\pm}^{(m)}:=\mathrm{Ran}\bigl(\mathscr S_{\pm}^{\mathrm{lin}}\bigr),
\qquad
\mathfrak R_{\pm}^{(m)}(\varepsilon):=\bigl\{\mathcal R\in\mathfrak R_{\pm}^{(m)}:\ \|\mathcal R\|_{\mathfrak R_{\pm}^{(m)}}\le \varepsilon\bigr\},
\end{equation}
and equip $\mathfrak R_{\pm}^{(m)}$ with the norm
\begin{equation}
\|\mathcal R\|_{\mathfrak R_{\pm}^{(m)}}:=\bigl\|(\mathscr S_{\pm}^{\mathrm{lin}})^{-1}\mathcal R\bigr\|_{\mathcal H_{\mathrm{lin}}^{k-2}}.
\end{equation}

\medskip
\noindent When we wish to emphasize the dependence on the background Kerr parameters, we write
\begin{equation}
\mathcal H_{\mathrm{lin},M,a}^{k-2}:=\mathcal H_{\mathrm{lin}}^{k-2},\qquad
\mathscr S_{\pm,M,a}^{\mathrm{lin}}:=\mathscr S_{\pm}^{\mathrm{lin}},\qquad
\mathfrak R_{\pm,M,a}^{(m)}:=\mathfrak R_{\pm}^{(m)},
\end{equation}
and similarly for the associated norms and the neighbourhoods $\mathfrak R_{\pm,M,a}^{(m)}(\varepsilon)$.
\end{definition}

\begin{definition}[Wave-operator solution space and norm]\label{def:waveop-space}
Assume in this definition that $m^{2}>0$.  Fix $\ell:=k-2$.
We write $X^{\ell}=X^{\ell}([t_0,\infty))$ for the space of pairs $(a,u)$ such that
$a$ is a Lorenz $1$-form on $\D$ with uncharged curvature $G:=\dd a$ and $u$ is a complex scalar field, for which the norm
\begin{eqnarray}\label{eq:X-norm}
\|(a,u)\|_{X^{\ell}}
:=\;&
\sup_{t\ge t_{0}}\Bigl(
\|G(t)\|_{\mathcal H_{\mathrm{Mx}}^{\ell}}
+
\|(u,\nabla_n u)(t)\|_{\mathcal H_{\mathrm{KG}}^{\ell}}
\Bigr)
\\
&+\Biggl(\sum_{|I|+j\le \ell}\int_{t_{0}}^{\infty}
\|\nabla \mathcal L_{\Omega}^{I}T^{j}u(t)\|_{L^{2}(\Sigma_t)}^{2}\,\dd t\Biggr)^{1/2}
\\
&+\Biggl(\sum_{|I|+j\le \ell}\int_{t_{0}}^{\infty}
\Bigl(
\|\mathcal L_{\Omega}^{I}T^{j}a(t)\|_{L^{\infty}(\Sigma_t)}^{2}
+
\|\mathcal L_{\Omega}^{I}T^{j}u(t)\|_{L^{\infty}(\Sigma_t)}^{2}
\Bigr)\,\dd t\Biggr)^{1/2},
\end{eqnarray}
is finite, where
\begin{equation}
\|G(t)\|_{\mathcal H_{\mathrm{Mx}}^{\ell}}^{2}
:=
\sum_{|I|+j\le \ell}E^{\hat t}\!\big[\mathcal L_{\Omega}^{I}T^{j}G\big](t).
\end{equation}
We write $B_{\rho}\subset X^{\ell}$ for the closed ball of radius $\rho$ centered at the origin.
\end{definition}

\begin{lemma}[Linear inhomogeneous estimate in $X^{\ell}$]\label{lem:linear-final-state-map}
Fix $\ell:=k-2\ge 2$ and $m^{2}>0$.
Let $J$ be a $1$-form and $H$ a complex scalar on $\D$ such that for every commuted field
$\mathcal L_{\Omega}^{I}T^{j}$ with $|I|+j\le \ell$ one has
\begin{equation}
\mathcal L_{\Omega}^{I}T^{j}J\in L^{1}\bigl([t_0,\infty);L^{2}(\Sigma_t)\bigr),
\qquad
\mathcal L_{\Omega}^{I}T^{j}H\in L^{1}\bigl([t_0,\infty);L^{2}(\Sigma_t)\bigr).
\end{equation}
Let $(a,u)$ solve the linear inhomogeneous Lorenz-gauge systems
\begin{equation}\label{eq:linear-inhom-maxwell-waveop}
\nabla^\alpha (\dd a)_{\alpha\beta}=J_\beta,
\qquad
\nabla^\alpha{}^\star(\dd a)_{\alpha\beta}=0,
\qquad
\nabla^\mu a_\mu=0,
\end{equation}
and
\begin{equation}\label{eq:linear-inhom-kg-waveop}
(\square_g-m^{2})u=H,
\end{equation}
on $t\ge t_0$, with \emph{vanishing forward asymptotic data} on $\partial_{+}^{(m)}\D$
(i.e.\ zero radiation fields on $\mathcal I^{+}\cup\mathcal H^{+}$ and vanishing $i^{+}$ channel when $m^{2}>0$).
Then there exists a constant $C_{0}=C_{0}(M,k)$ such that
\begin{equation}\label{eq:linear-final-state-X}
\|(a,u)\|_{X^{\ell}}
\le
C_{0}\Biggl(
\sum_{|I|+j\le\ell}\int_{t_0}^{\infty}\|\mathcal L_{\Omega}^{I}T^{j}J(t)\|_{L^{2}(\Sigma_t)}\,\dd t
+
\sum_{|I|+j\le\ell}\int_{t_0}^{\infty}\|\mathcal L_{\Omega}^{I}T^{j}H(t)\|_{L^{2}(\Sigma_t)}\,\dd t
\Biggr).
\end{equation}
\end{lemma}

\begin{proof}
We now prove the estimate, using only the linear conditions established in the preceding sections.

\medskip\noindent
\textit{Energy part.}
Let $G:=\dd a$.
Commuting \eqref{eq:linear-inhom-maxwell-waveop}-\eqref{eq:linear-inhom-kg-waveop} with $\mathcal L_{\Omega}^{I}T^{j}$ ($|I|+j\le\ell$) and applying
Lemmas~\ref{lem:inhom-maxwell-energy} and~\ref{lem:inhom-kg-energy} on $[t,T]$ yields, for any $T\ge t\ge t_0$,
\begin{equation}
\|(\mathcal L_{\Omega}^{I}T^{j}G)(t)\|_{\mathcal H_{\mathrm{Mx}}}
\lesssim
\|(\mathcal L_{\Omega}^{I}T^{j}G)(T)\|_{\mathcal H_{\mathrm{Mx}}}
+
\int_{t}^{T}\|\mathcal L_{\Omega}^{I}T^{j}J(s)\|_{L^{2}(\Sigma_s)}\,\dd s,
\end{equation}
and
\begin{equation}
\|(\mathcal L_{\Omega}^{I}T^{j}u,\nabla_n\mathcal L_{\Omega}^{I}T^{j}u)(t)\|_{\mathcal H_{\mathrm{KG}}}
\lesssim
\|(\mathcal L_{\Omega}^{I}T^{j}u,\nabla_n\mathcal L_{\Omega}^{I}T^{j}u)(T)\|_{\mathcal H_{\mathrm{KG}}}
+
\int_{t}^{T}\|\mathcal L_{\Omega}^{I}T^{j}H(s)\|_{L^{2}(\Sigma_s)}\,\dd s.
\end{equation}
The forward vanishing asymptotic data imply that the energy terms at time $T$ vanish as $T\to+\infty$.
Taking $T\to+\infty$ and then the supremum over $t\ge t_0$ gives the first line in \eqref{eq:X-norm} bounded by the right-hand side of \eqref{eq:linear-final-state-X}.

\medskip\noindent
\textit{Morawetz/ILD part for $u$.}
Applying the Morawetz identity of Section~\ref{sec:Morawetz} to the commuted Klein-Gordon equation
$(\square_g-m^{2})\mathcal L_{\Omega}^{I}T^{j}u=\mathcal L_{\Omega}^{I}T^{j}H$
produces the integrated local energy bound (up to lower-order terms)
\begin{eqnarray}
\int_{t_0}^{\infty}\|\nabla\mathcal L_{\Omega}^{I}T^{j}u(t)\|_{L^{2}(\Sigma_t)}^{2}\,\dd t
&\ \lesssim
\sup_{t\ge t_0}\|(\mathcal L_{\Omega}^{I}T^{j}u,\nabla_n\mathcal L_{\Omega}^{I}T^{j}u)(t)\|_{\mathcal H_{\mathrm{KG}}}^{2}
\\
&\quad+
\int_{t_0}^{\infty}\|\mathcal L_{\Omega}^{I}T^{j}H(t)\|_{L^{2}}\,
\|\nabla\mathcal L_{\Omega}^{I}T^{j}u(t)\|_{L^{2}}\,\dd t.
\end{eqnarray}
Using Cauchy-Schwarz in $t$ and absorbing yields
\begin{eqnarray}
\left(\int_{t_0}^{\infty}\|\nabla\mathcal L_{\Omega}^{I}T^{j}u(t)\|_{L^{2}}^{2}\,\dd t\right)^{1/2}
&\ \lesssim
\sup_{t\ge t_0}\|(\mathcal L_{\Omega}^{I}T^{j}u,\nabla_n\mathcal L_{\Omega}^{I}T^{j}u)(t)\|_{\mathcal H_{\mathrm{KG}}}
\\
&\quad+
\int_{t_0}^{\infty}\|\mathcal L_{\Omega}^{I}T^{j}H(t)\|_{L^{2}}\,\dd t.
\end{eqnarray}
Combining this with the energy bound already obtained gives the second line in \eqref{eq:X-norm} controlled by the right-hand side of \eqref{eq:linear-final-state-X}.

\medskip\noindent
\textit{$L^{2}_{t}L^{\infty}_{x}$ part.}
In Lorenz gauge and in vacuum Schwarzschild, \eqref{eq:linear-inhom-maxwell-waveop} implies the component wave equation
$\square_g a_\beta = J_\beta$.
Applying the far-region $r^{p}$ hierarchy and Sobolev arguments of Section~\ref{sec:decayfaraway} (which are based on linear divergence identities) to
$\square_g(\mathcal L_{\Omega}^{I}T^{j}a)=\mathcal L_{\Omega}^{I}T^{j}J$
and
$(\square_g-m^{2})(\mathcal L_{\Omega}^{I}T^{j}u)=\mathcal L_{\Omega}^{I}T^{j}H$
yields pointwise bounds, for $|I|+j\le\ell$, of the structural form
\begin{eqnarray}
\|\mathcal L_{\Omega}^{I}T^{j}a(t)\|_{L^{\infty}(\Sigma_t)}
&\lesssim
(1+t)^{-1}\sup_{s\ge t}\|\mathcal L_{\Omega}^{I}T^{j}G(s)\|_{\mathcal H_{\mathrm{Mx}}}
\\
&\quad+
\int_{t}^{\infty}(1+s)^{-1}\|\mathcal L_{\Omega}^{I}T^{j}J(s)\|_{L^{2}(\Sigma_s)}\,\dd s,
\\
\|\mathcal L_{\Omega}^{I}T^{j}u(t)\|_{L^{\infty}(\Sigma_t)}
&\lesssim
(1+t)^{-1}\sup_{s\ge t}\|(\mathcal L_{\Omega}^{I}T^{j}u,\nabla_n\mathcal L_{\Omega}^{I}T^{j}u)(s)\|_{\mathcal H_{\mathrm{KG}}}
\\
&\quad+
\int_{t}^{\infty}(1+s)^{-1}\|\mathcal L_{\Omega}^{I}T^{j}H(s)\|_{L^{2}(\Sigma_s)}\,\dd s.
\end{eqnarray}
(Here the integrals arise from the source terms in the $r^{p}$ divergence identity and are treated by Cauchy-Schwarz.)
Since $\int_{t_0}^{\infty}(1+t)^{-2}\,\dd t<\infty$, squaring and integrating in $t$ gives the last line in \eqref{eq:X-norm} bounded by
\begin{equation}
\sup_{t\ge t_{0}}\Bigl(
\|G(t)\|_{\mathcal H_{\mathrm{Mx}}^{\ell}}
+
\|(u,\nabla_n u)(t)\|_{\mathcal H_{\mathrm{KG}}^{\ell}}
\Bigr)
+
\sum_{|I|+j\le\ell}\int_{t_0}^{\infty}\Bigl(
\|\mathcal L_{\Omega}^{I}T^{j}J(t)\|_{L^{2}(\Sigma_t)}
+
\|\mathcal L_{\Omega}^{I}T^{j}H(t)\|_{L^{2}(\Sigma_t)}
\Bigr)\,\dd t.
\end{equation}
Combining with the already established energy bounds completes the proof of \eqref{eq:linear-final-state-X}.
\end{proof}

\begin{proposition}[Closed Schwarzschild final-state contraction]\label{prop:schwarzschild-closed-finalstate-contraction}
Assume $m^2>0$ and use the massive Schwarzschild final-state space $X^\ell$ of Definition~\ref{def:waveop-space}, with $\ell=k-2\ge2$.  Let $U_{\rm lin}=(A^{\rm lin},F^{\rm lin},\varphi^{\rm lin})$ be a decoupled linear Maxwell-Klein-Gordon solution whose extended future asymptotic data have size at most $\eta$.  For a correction $W=(a,u)$ with zero future asymptotic data define
\begin{equation}
 \mathcal T(W)=\mathcal L_{\rm fs}^{-1}\mathcal N(U_{\rm lin}+W),
\end{equation}
where $\mathcal L_{\rm fs}^{-1}$ is the zero-final-state linear solver supplied by Lemma~\ref{lem:linear-final-state-map} and $\mathcal N$ denotes the Maxwell current together with the scalar nonlinear source.  There are constants $C_0,C_1$ such that, whenever $\|W\|_{X^\ell}\le\rho$,
\begin{eqnarray}
 \|\mathcal T(W)\|_{X^\ell}
 &\le& C_0\left((\eta+\rho)^2+(\eta+\rho)^{2N_P+3}\right),\label{eq:schw-final-contract-size}\\
 \|\mathcal T(W_1)-\mathcal T(W_2)\|_{X^\ell}
 &\le& C_1\left(\eta+\rho+(\eta+\rho)^{2N_P+2}\right)
 \|W_1-W_2\|_{X^\ell}.
 \label{eq:schw-final-contract-lip}
\end{eqnarray}
Consequently, for sufficiently small $\eta$ and a suitable radius $\rho$, $\mathcal T$ is a strict contraction on the closed radius-$\rho$ ball in $X^\ell$.
\end{proposition}

\begin{proof}
Lemma~\ref{lem:linear-final-state-map} bounds the output by the $L^1_tL^2_x$ norms of the Maxwell and scalar sources.  The estimates \eqref{eq:J-est-L1} and \eqref{eq:N-est-L1} below, together with Lemma~\ref{lem:potential-tame}, give
\begin{equation}
 \|\mathcal N(U_{\rm lin}+W)\|_{L^1_tH^\ell_x}
 \le C\left((\eta+\rho)^2+(\eta+\rho)^{2N_P+3}\right),
\end{equation}
because the current and gauge terms are at least quadratic and the potential remainder is cubic with highest allowed growth $2N_P+3$.  Applying the same product expansion to two corrections gives one factor of $W_1-W_2$ and leaves the remaining factors bounded by $\eta+\rho$ or by $(\eta+\rho)^{2N_P+2}$ after differentiating the potential.  This proves \eqref{eq:schw-final-contract-size}-\eqref{eq:schw-final-contract-lip}.  Choose $\rho$ comparable to $\eta^2+\eta^{2N_P+3}$ and then take $\eta$ so small that the right-hand side of \eqref{eq:schw-final-contract-lip} is $<1$; the Banach fixed-point theorem applies.
\end{proof}

\begin{proof}[Massive part of the proof of Theorem~\ref{thm:nonlinear-wave-operators}]
Assume $m^{2}>0$.  We indicate the forward construction; the backward one follows by the time-reversal argument in Theorem~\ref{thm:two-sided-scattering}.

Fix $\mathcal R_{+}^{\mathrm{lin}}\in\mathfrak R_{+}^{(m)}(\varepsilon_{\mathrm{sc}})$ and let $(F^{\mathrm{lin}},\varphi^{\mathrm{lin}})$ be the corresponding linear solution.
Choose a Lorenz-gauge potential $A^{\mathrm{lin}}$ for the free Maxwell field $F^{\mathrm{lin}}$, so that
\begin{equation}
\nabla^\mu A^{\mathrm{lin}}_\mu=0,
\qquad
\dd A^{\mathrm{lin}}=F^{\mathrm{lin}},
\qquad
\square_g A^{\mathrm{lin}}_\nu=0.
\end{equation}
We seek a nonlinear Lorenz-gauge solution $(A,\phi)$ of the form
\begin{equation}
A=A^{\mathrm{lin}}+a,
\qquad
F=F^{\mathrm{lin}}+G\ \ \text{with}\ \ G:=\dd a,
\qquad
\phi=\varphi^{\mathrm{lin}}+u,
\end{equation}
where $(a,u)$ has \emph{zero forward asymptotic data} (vanishing radiation fields on $\mathcal I^{+}\cup\mathcal H^{+}$ and vanishing $i^{+}$ channel in the massive case), and where we impose the Lorenz condition $\nabla^\mu a_\mu=0$.
Subtracting the linear comparison system from the Maxwell-Higgs system in Lorenz gauge, we find that the correction $(G,u)$ solves
\begin{equation}\label{eq:waveop-inhom-maxwell}
\nabla^\alpha G_{\alpha\beta}=J_\beta[\varphi^{\mathrm{lin}}+u;\,A^{\mathrm{lin}}+a],
\qquad
\nabla^\alpha{}^\star G_{\alpha\beta}=0,
\end{equation}
and
\begin{equation}\label{eq:waveop-inhom-kg}
(\square_g-m^{2})u=\mathcal N_\phi\bigl(A^{\mathrm{lin}}+a,\ \varphi^{\mathrm{lin}}+u\bigr),
\end{equation}
where $J_\beta[\phi;A]:=-i\bigl(D_\beta\phi\,\bar\phi-\phi\,\overline{D_\beta\phi}\bigr)$ with $D_\mu=\nabla_\mu-iA_\mu$, and $\mathcal N_\phi$ is the nonlinear forcing in \eqref{eq:phi-as-kg}.
We solve the nonlinear final-state problem by a Picard iteration in the Banach space $X^{\ell}$ from Definition~\ref{def:waveop-space}.
For $\rho>0$ to be chosen, consider the closed ball $B_\rho\subset X^{\ell}$.

\medskip
\noindent\textit{the fixed-point map.}
Given $(a^{(n)},u^{(n)})\in B_\rho$, set
\begin{equation}
A^{(n)}:=A^{\mathrm{lin}}+a^{(n)},
\qquad
\phi^{(n)}:=\varphi^{\mathrm{lin}}+u^{(n)}.
\end{equation}
Define the sources
\begin{equation}
J^{(n)}_\beta:=J_\beta[\phi^{(n)};A^{(n)}],
\qquad
\mathcal N^{(n)}:=\mathcal N_\phi(A^{(n)},\phi^{(n)}).
\end{equation}
Let $(a^{(n+1)},u^{(n+1)})$ be the unique solution of the linear final-state problems
\begin{eqnarray}
\nabla^\alpha (\dd a^{(n+1)})_{\alpha\beta}&=J^{(n)}_\beta,
\qquad
\nabla^\alpha{}^\star(\dd a^{(n+1)})_{\alpha\beta}=0,
\qquad
\nabla^\mu a^{(n+1)}_\mu=0,
\label{eq:iter-maxwell}
\\
(\square_g-m^{2})u^{(n+1)}&=\mathcal N^{(n)},
\label{eq:iter-kg}
\end{eqnarray}
with \emph{vanishing forward asymptotic data} on $\partial_{+}^{(m)}\D$ (i.e.\ zero radiation fields on $\mathcal I^{+}\cup\mathcal H^{+}$ and zero $i^{+}$ channel when $m^{2}>0$).
Existence and uniqueness of such solutions follow from the massless linear scattering theory in Theorem~\ref{thm:linear-scattering} and, when $m^2>0$, from Proposition~\ref{prop:massive-schwarzschild-linear-final-state}.
This defines the fixed-point map $\mathcal T:B_\rho\to X^{\ell}$ by $\mathcal T(a^{(n)},u^{(n)})=(a^{(n+1)},u^{(n+1)})$.

\medskip
\noindent\textit{a priori estimate for the linear final-state map.}
Let $G^{(n+1)}:=\dd a^{(n+1)}$.
Applying Lemmas~\ref{lem:inhom-maxwell-energy} and~\ref{lem:inhom-kg-energy} \emph{backwards from $+\infty$} gives the energy bounds (for each commuted field $\mathcal L_{\Omega}^{I}T^{j}$ with $|I|+j\le\ell$)
\begin{eqnarray}
\sup_{t\ge t_0}\|G^{(n+1)}(t)\|_{\mathcal H_{\mathrm{Mx}}^{\ell}}
&\lesssim
\sum_{|I|+j\le\ell}\int_{t_0}^{\infty}\|\mathcal L_{\Omega}^{I}T^{j}J^{(n)}(s)\|_{L^{2}(\Sigma_s)}\,\dd s,
\label{eq:iter-energy-maxwell}
\\
\sup_{t\ge t_0}\|(u^{(n+1)},\nabla_n u^{(n+1)})(t)\|_{\mathcal H_{\mathrm{KG}}^{\ell}}
&\lesssim
\sum_{|I|+j\le\ell}\int_{t_0}^{\infty}\|\mathcal L_{\Omega}^{I}T^{j}\mathcal N^{(n)}(s)\|_{L^{2}(\Sigma_s)}\,\dd s.
\label{eq:iter-energy-kg}
\end{eqnarray}
Indeed, for example, \eqref{eq:inhom-kg-energy-est} with $t_1=t$ and $t_2=T$ gives
\begin{equation}
\|(u^{(n+1)},\nabla_n u^{(n+1)})(t)\|_{\mathcal H_{\mathrm{KG}}}
\le
\|(u^{(n+1)},\nabla_n u^{(n+1)})(T)\|_{\mathcal H_{\mathrm{KG}}}
+
\int_{t}^{T}\|\mathcal N^{(n)}(s)\|_{L^2(\Sigma_s)}\,\dd s,
\end{equation}
and the first term vanishes as $T\to+\infty$ by the imposed zero forward data; the Maxwell estimate is identical.
The remaining spacetime terms in \eqref{eq:X-norm} are controlled by Lemma~\ref{lem:linear-final-state-map}, which gives the inhomogeneous Morawetz/$r^{p}$ estimates for \eqref{eq:iter-maxwell}-\eqref{eq:iter-kg}.
By Lemma~\ref{lem:linear-final-state-map} applied to \eqref{eq:iter-maxwell}-\eqref{eq:iter-kg} with $J=J^{(n)}$ and $H=\mathcal N^{(n)}$, there is a constant $C_{0}=C_{0}(M,k)$ such that \eqref{eq:linear-final-state-X} holds.

\medskip
\noindent\textit{source estimates in $L^{1}_{t}L^{2}_{x}$.}
We estimate the right-hand side of \eqref{eq:linear-final-state-X} using only the norms appearing in \eqref{eq:X-norm}.
Using $|J[\phi;A]|\lesssim |\phi|\,|D\phi|$ and $|D\phi|\lesssim|\nabla\phi|+|A|\,|\phi|$, we have for each $t\ge t_0$,
\begin{equation}\label{eq:J-est-pointwise}
\|J^{(n)}(t)\|_{L^{2}(\Sigma_t)}
\lesssim
\|\phi^{(n)}(t)\|_{L^{\infty}(\Sigma_t)}
\Bigl(
\|\nabla\phi^{(n)}(t)\|_{L^{2}(\Sigma_t)}
+
\|A^{(n)}(t)\|_{L^{\infty}(\Sigma_t)}\,\|\phi^{(n)}(t)\|_{L^{2}(\Sigma_t)}
\Bigr).
\end{equation}
Integrating in $t$ and applying Cauchy-Schwarz yields
\begin{eqnarray}\label{eq:J-est-L1}
\int_{t_0}^{\infty}\|J^{(n)}(t)\|_{L^{2}(\Sigma_t)}\,\dd t
&\lesssim
\|\phi^{(n)}\|_{L^{2}_{t}L^{\infty}_{x}}\,
\|\nabla\phi^{(n)}\|_{L^{2}_{t}L^{2}_{x}}
\\
&\quad
+
\|\phi^{(n)}\|_{L^{\infty}_{t}L^{2}_{x}}\,
\|A^{(n)}\|_{L^{2}_{t}L^{\infty}_{x}}\,
\|\phi^{(n)}\|_{L^{2}_{t}L^{\infty}_{x}}.
\nonumber
\end{eqnarray}
From the explicit form \eqref{eq:phi-as-kg} of $\mathcal N_\phi$ (with $A=A^{(n)}$ and $\phi=\phi^{(n)}$) and the potential bound \eqref{eq:potential-nonlinear-bound} we obtain the pointwise estimate
\begin{equation}
|\mathcal N^{(n)}|
\lesssim
|A^{(n)}|\,|\nabla\phi^{(n)}|
+
|A^{(n)}|^{2}|\phi^{(n)}|
+
(1+|\phi^{(n)}|^{2N_{P}})\,|\phi^{(n)}|^{3}.
\end{equation}
Consequently,
\begin{eqnarray}\label{eq:N-est-L1}
\int_{t_0}^{\infty}\|\mathcal N^{(n)}(t)\|_{L^{2}(\Sigma_t)}\,\dd t
&\lesssim
\|A^{(n)}\|_{L^{2}_{t}L^{\infty}_{x}}\,
\|\nabla\phi^{(n)}\|_{L^{2}_{t}L^{2}_{x}}
\\
&\quad
+
\Bigl(\|A^{(n)}\|_{L^{2}_{t}L^{\infty}_{x}}^{2}
+(1+\|\phi^{(n)}\|_{L^{\infty}_{t}L^{\infty}_{x}}^{2N_{P}})\,\|\phi^{(n)}\|_{L^{2}_{t}L^{\infty}_{x}}^{2}\Bigr)\,
\|\phi^{(n)}\|_{L^{\infty}_{t}L^{2}_{x}}.
\nonumber
\end{eqnarray}
Here $\|\phi^{(n)}\|_{L^{\infty}_{t}L^{\infty}_{x}}$ is controlled by the energy component of the $X^{\ell}$-norm via Sobolev on $\Sigma_t$ since $\ell\ge 2$ (and therefore is $\lesssim\rho$ on the ball $B_\rho$). The same inequalities hold after commuting with $\mathcal L_{\Omega}^{I}T^{j}$, $|I|+j\le\ell$, by product/Moser estimates and the fact that $k\ge 6$ provides enough regularity for Sobolev on $\Sigma_t$.

\medskip
\noindent\textit{$\mathcal T$ maps $B_\rho$ to itself.}
By linear scattering, the fixed linear solution satisfies
\begin{equation}
\|A^{\mathrm{lin}}\|_{L^{2}_{t}L^{\infty}_{x}}
+
\|\varphi^{\mathrm{lin}}\|_{L^{2}_{t}L^{\infty}_{x}}
+
\|\nabla\varphi^{\mathrm{lin}}\|_{L^{2}_{t}L^{2}_{x}}
+
\|\varphi^{\mathrm{lin}}\|_{L^{\infty}_{t}L^{2}_{x}}
\ \lesssim\ \varepsilon_{\mathrm{sc}}.
\end{equation}
For $(a^{(n)},u^{(n)})\in B_\rho$, Definition~\ref{def:waveop-space} implies
\begin{equation}
\|a^{(n)}\|_{L^{2}_{t}L^{\infty}_{x}}
+
\|u^{(n)}\|_{L^{2}_{t}L^{\infty}_{x}}
+
\|\nabla u^{(n)}\|_{L^{2}_{t}L^{2}_{x}}
+
\|u^{(n)}\|_{L^{\infty}_{t}L^{2}_{x}}
\ \lesssim\ \rho.
\end{equation}
Consequently, the full fields satisfy the structural bounds
\begin{equation}
\|A^{(n)}\|_{L^{2}_{t}L^{\infty}_{x}}
+
\|\phi^{(n)}\|_{L^{2}_{t}L^{\infty}_{x}}
+
\|\nabla\phi^{(n)}\|_{L^{2}_{t}L^{2}_{x}}
+
\|\phi^{(n)}\|_{L^{\infty}_{t}L^{2}_{x}}
\ \lesssim\ \varepsilon_{\mathrm{sc}}+\rho.
\end{equation}
Inserting these into \eqref{eq:linear-final-state-X}-\eqref{eq:N-est-L1} yields
\begin{equation}\label{eq:T-selfmap}
\|\mathcal T(a^{(n)},u^{(n)})\|_{X^{\ell}}
\ \le\ C_{1}\Bigl((\varepsilon_{\mathrm{sc}}+\rho)^{2}+(\varepsilon_{\mathrm{sc}}+\rho)^{2N_P+3}\Bigr),
\end{equation}
for a constant $C_{1}=C_{1}(M,k,P)$.
Choose
\begin{equation}
 \rho:=8C_{1}\bigl(\varepsilon_{\mathrm{sc}}^{2}+\varepsilon_{\mathrm{sc}}^{2N_P+3}\bigr)
\end{equation}
and then choose $\varepsilon_{\mathrm{sc}}$ so small that $\rho\le\varepsilon_{\mathrm{sc}}$ and
\begin{equation}
 C_{1}\Bigl((\varepsilon_{\mathrm{sc}}+\rho)^{2}+(\varepsilon_{\mathrm{sc}}+\rho)^{2N_P+3}\Bigr)\le\rho.
\end{equation}
Then $\mathcal T(B_\rho)\subseteq B_\rho$.

\medskip
\noindent\textit{contraction estimate.}
Let $(a,u),(\tilde a,\tilde u)\in B_\rho$ and write
\begin{equation}
A:=A^{\mathrm{lin}}+a,\quad \tilde A:=A^{\mathrm{lin}}+\tilde a,
\qquad
\phi:=\varphi^{\mathrm{lin}}+u,\quad \tilde\phi:=\varphi^{\mathrm{lin}}+\tilde u.
\end{equation}
Set $\Delta a:=a-\tilde a$ and $\Delta u:=u-\tilde u$ (so $\Delta\phi=\Delta u$ and $\Delta A=\Delta a$).
Using the bilinear structure of $J_\beta[\phi;A]=-i(D_\beta\phi\,\bar\phi-\phi\,\overline{D_\beta\phi})$ (equivalently $2\Im(\bar\phi D_\beta\phi)$) and the pointwise bound $|D\phi-D\tilde\phi|\lesssim|\nabla\Delta\phi|+|A|\,|\Delta\phi|+|\Delta A|\,|\tilde\phi|$, we obtain for each $t$,
\begin{eqnarray}\label{eq:deltaJ-pointwise}
\|\Delta J(t)\|_{L^{2}(\Sigma_t)}
&\lesssim
\|\Delta\phi(t)\|_{L^{\infty}}\|D\phi(t)\|_{L^{2}}
+
\|\tilde\phi(t)\|_{L^{\infty}}\|D\phi(t)-D\tilde\phi(t)\|_{L^{2}}
\\
&\lesssim
\|\Delta\phi(t)\|_{L^{\infty}}\|D\phi(t)\|_{L^{2}}
+
\|\tilde\phi(t)\|_{L^{\infty}}\|\nabla\Delta\phi(t)\|_{L^{2}}
+
\|\tilde\phi(t)\|_{L^{\infty}}\|A(t)\|_{L^{\infty}}\|\Delta\phi(t)\|_{L^{2}}
\nonumber\\
&\quad
+
\|\tilde\phi(t)\|_{L^{\infty}}\|\Delta A(t)\|_{L^{\infty}}\|\tilde\phi(t)\|_{L^{2}}.
\nonumber
\end{eqnarray}
Integrating in $t$ and applying Cauchy-Schwarz gives
\begin{equation}\label{eq:deltaJ-L1}
\|\Delta J\|_{L^{1}_{t}L^{2}_{x}}
\ \lesssim
(\varepsilon_{\mathrm{sc}}+\rho)\,\|(\Delta a,\Delta u)\|_{X^{\ell}}.
\end{equation}
Similarly, expanding $\mathcal N_\phi(A,\phi)-\mathcal N_\phi(\tilde A,\tilde\phi)$ using \eqref{eq:phi-as-kg} yields
\begin{eqnarray}\label{eq:deltaN-L1}
\|\Delta\mathcal N\|_{L^{1}_{t}L^{2}_{x}}
&\lesssim
\|\Delta A\|_{L^{2}_{t}L^{\infty}_{x}}\,
\|\nabla\phi\|_{L^{2}_{t}L^{2}_{x}}
+
\|\tilde A\|_{L^{2}_{t}L^{\infty}_{x}}\,
\|\nabla\Delta\phi\|_{L^{2}_{t}L^{2}_{x}}
\nonumber\\
&\quad
+
\bigl(\|A\|_{L^{2}_{t}L^{\infty}_{x}}+\|\tilde A\|_{L^{2}_{t}L^{\infty}_{x}}\bigr)
\|\Delta A\|_{L^{2}_{t}L^{\infty}_{x}}\,
\|\phi\|_{L^{\infty}_{t}L^{2}_{x}}
+
\|\tilde A\|_{L^{2}_{t}L^{\infty}_{x}}^{2}\,
\|\Delta\phi\|_{L^{\infty}_{t}L^{2}_{x}}
\nonumber\\
&\quad
+
\bigl(\|\phi\|_{L^{2}_{t}L^{\infty}_{x}}^{2}+\|\tilde\phi\|_{L^{2}_{t}L^{\infty}_{x}}^{2}\bigr)\,
\|\Delta\phi\|_{L^{\infty}_{t}L^{2}_{x}}
\nonumber\\
&\lesssim
\Bigl(\varepsilon_{\mathrm{sc}}+\rho+(\varepsilon_{\mathrm{sc}}+\rho)^{2N_P+2}\Bigr)\,\|(\Delta a,\Delta u)\|_{X^{\ell}},
\end{eqnarray}
where in the last step we used $(a,u),(\tilde a,\tilde u)\in B_\rho$ and the Moser difference estimate for the potential remainder.
Applying the linear final-state estimate \eqref{eq:linear-final-state-X} to the difference of the iterates (which solves the same linear equations with sources $(\Delta J,\Delta\mathcal N)$ and zero forward data) and using \eqref{eq:deltaJ-L1}-\eqref{eq:deltaN-L1}, we obtain
\begin{equation}\label{eq:contraction}
\|\mathcal T(a,u)-\mathcal T(\tilde a,\tilde u)\|_{X^{\ell}}
\ \le\ C_{2}\Bigl(\varepsilon_{\mathrm{sc}}+\rho+(\varepsilon_{\mathrm{sc}}+\rho)^{2N_P+2}\Bigr)\,\|(a,u)-(\tilde a,\tilde u)\|_{X^{\ell}}.
\end{equation}
Choosing $\varepsilon_{\mathrm{sc}}$ sufficiently small (with $\rho$ chosen above), we ensure
\begin{equation}
 C_{2}\Bigl(\varepsilon_{\mathrm{sc}}+\rho+(\varepsilon_{\mathrm{sc}}+\rho)^{2N_P+2}\Bigr)<\tfrac12,
\end{equation}
and therefore $\mathcal T$ is a strict contraction on $B_\rho$.
By the Banach fixed-point theorem, there exists a unique fixed point $(a,u)\in B_\rho$ with $\mathcal T(a,u)=(a,u)$, which solves \eqref{eq:waveop-inhom-maxwell}-\eqref{eq:waveop-inhom-kg} and has vanishing forward asymptotic data by construction.

Setting $(F,\phi)=(F^{\mathrm{lin}}+G,\varphi^{\mathrm{lin}}+u)$ yields a global nonlinear solution with $\mathscr S_{+}^{\mathrm{full}}(F,\phi)=\mathcal R_{+}^{\mathrm{lin}}$.
The scattering statements (matching of radiation fields and the $i^{+}$ channel) follow from the vanishing asymptotic data of $(G,u)$ combined with Theorem~\ref{thm:nonlinear-scattering}.
Uniqueness in the small-data class follows by applying the same energy inequalities to the difference of two solutions with identical forward data, which again yields a fixed point with zero data and therefore the trivial solution.

Continuity of $\mathscr W_{+}^{\mathrm{full}}$ and of $\mathscr S_{+}^{\mathrm{full}}$ follows from continuous dependence of the fixed point on the datum and from the stability estimates behind the contraction argument.
\end{proof}

\begin{corollary}[Two-sided nonlinear scattering operator]\label{cor:scattering-operator}
In the setting of Theorem~\ref{thm:nonlinear-wave-operators}, the two-sided nonlinear scattering operator
\begin{equation}
\mathscr S^{\mathrm{full}}
:=\mathscr S_{+}^{\mathrm{full}}\circ\bigl(\mathscr S_{-}^{\mathrm{full}}\bigr)^{-1}
:\ \mathfrak R_{-}^{(m)}(\varepsilon_{\mathrm{sc}})\longrightarrow \mathfrak R_{+}^{(m)}(\varepsilon_{\mathrm{sc}})
\end{equation}
is well-defined on a neighbourhood of $0$ and is a homeomorphism onto its image (with inverse obtained by swapping $+$ and $-$).
\end{corollary}

\begin{proof}
Theorem~\ref{thm:nonlinear-wave-operators} states that $\mathscr S^{\mathrm{full}}_+$ and $\mathscr S^{\mathrm{full}}_-$ are homeomorphisms between small Cauchy-data neighborhoods and the corresponding asymptotic-data neighborhoods.  Consequently, the composition $\mathscr S^{\mathrm{full}}_+\circ(\mathscr S^{\mathrm{full}}_-)^{-1}$ is a homeomorphism on its natural domain.  Its inverse is $\mathscr S^{\mathrm{full}}_-\circ(\mathscr S^{\mathrm{full}}_+)^{-1}$.
\end{proof}

\appendix



\section{Collected conditions, admissible regimes, and external analytic conditions}\label{app:standing-conditions}

This appendix gathers the assumptions and outside conditions used by the main theorems.  It is intended as the paper to be readable: a theorem in the main body uses only the items listed below, together with the internal estimates proved in the referenced sections.  

\subsection{Geometric and regularity conditions}\label{app:standing-geometry}

The black-hole exterior is either Schwarzschild, \(a=0\), or a slowly rotating Kerr exterior \((\mathcal D_{M,a},g_{M,a})\) with
\begin{equation}\label{eq:app-standing-slow}
 M>0,\qquad |a|\le a_{\mathrm{slow}}(M,K),\qquad K\ge10,
\end{equation}
where the threshold is chosen below the finite-order constants in the scalar and Maxwell estimates inserted in Section~\ref{sec:kerr-extension}.  The nonlinear scattering statements are formulated at top order \(K\), and the displayed radiation and pointwise conclusions are read at order
\begin{equation}\label{eq:app-standing-k}
 k:=K-4.
\end{equation}
The massive energy theorem uses an energy order \(N\ge6\).  All Cauchy data in the stated theorems are smooth for the purpose of the proof and belong, after completion, to the Sobolev quotient spaces determined by the corresponding order-\(K\) or order-\(N\) energies.

The foliation, redshift vector field, radius function, and commutator family are those of Definition~\ref{def:admissible-exterior}.  The local Cauchy theory uses Lorenz gauge,
\begin{equation}\label{eq:app-standing-lorenz}
 \nabla^\mu A_\mu=0,
\end{equation}
with the compatibility conditions on the initial slice stated in Sections~\ref{subsec:kerr-initial-data} and~\ref{subsec:initial-data}.  The local Cauchy theory and the continuation criteria used after the a priori estimates are supplied in Appendices~\ref{app:global-existence-schwarzschild} and~\ref{app:global-existence-kerr}.  The Maxwell constraints are imposed on the initial data and then propagated by the divergence-free current identity.  Residual Lorenz gauge transformations are quotiented out in the Cauchy and radiation spaces.

\subsection{Potential and sector conditions}\label{app:standing-potential-sector}

The scalar potential is always Assumption~\ref{asumsiP}: \(P(\phi,\bar\phi)=\widetilde P(|\phi|^2)\), \(\widetilde P(0)=0\), \(\widetilde P\ge0\), and
\begin{equation}\label{eq:app-standing-potential}
 \widetilde P'(s)=m^2+sQ(s),\qquad
 |Q^{(j)}(s)|\le C_{P,j}(1+s^{N_P}).
\end{equation}
Consequently,
\begin{equation}\label{eq:app-standing-remainder}
 \partial_{\bar\phi}P=m^2\phi+\mathcal R_P(\phi),
 \qquad
 \mathcal R_P(\phi)=Q(|\phi|^2)|\phi|^2\phi,
\end{equation}
and only \(\mathcal R_P\), which is at least cubic at the vacuum, is treated as a nonlinear source.  For the massive high-order energy theorem the potential is the positive polynomial-type potential
\begin{equation}\label{eq:app-standing-positive-potential}
 \widetilde P(s)=m^2s+\sum_{n=p}^{N_P}\alpha_n s^n,
 \qquad m^2>0,
 \qquad p\ge2,
 \qquad \alpha_n\ge0.
\end{equation}

The zero-sector statements impose zero asymptotic Maxwell charge after quotienting by pure gauge.  In a fixed electric sector one fixes a real electric charge \(Q_e\), subtracts the stationary Coulomb field, and writes
\begin{equation}\label{eq:app-standing-coulomb-split}
 A=A^C_{Q_e}+a,
 \qquad
 F=F^C_{Q_e}+f,
 \qquad
 D_{Q_e}=\nabla-iA^C_{Q_e}.
\end{equation}
The radiative field \(f\) has zero charge, and the scalar radiation variable at null infinity is
\begin{equation}\label{eq:app-standing-coulomb-radiation}
 U_{Q_e}^{-1}r\phi,
 \qquad L_{\mathrm{out}}U_{Q_e}=iA^C_{Q_e}(L_{\mathrm{out}})U_{Q_e},
 \qquad |U_{Q_e}|=1.
\end{equation}
No magnetic Coulomb sector, large-electric sector, or rapid-rotation sector is used in any theorem.

\subsection{Linear estimates and source spaces}\label{app:standing-linear-norms}

The nonlinear argument is run in the solution/source pair associated with \(\Lin_K\) or \(\Lin_K^{(m)}\), Definitions~\ref{def:lin-estimates} and~\ref{def:lin-estimates-massive}.  The required properties are:
\begin{enumerate}
\item finite-slab nondegenerate energy boundedness;
\item redshift control at the horizon;
\item trapping-degenerate integrated local energy decay/Morawetz control;
\item a far-field \(r^p\) hierarchy sufficient to construct radiation fields;
\item inhomogeneous estimates with the same source norms used for the Maxwell-Higgs current and scalar source;
\item continuous Cauchy-to-radiation and radiation-to-Cauchy maps in the corresponding quotient spaces.
\end{enumerate}
For \(m^2>0\), the scalar mass term is included in the linear Klein-Gordon operator and in the linear energy; it is never counted as a perturbative source.  The massive scattering topology includes the appropriate timelike or Dollard channel at \(i^\pm\) whenever the cited or assumed scalar theory supplies it.

\section{Global existence on Schwarzschild spacetimes}
\label{app:global-existence-schwarzschild}

This appendix provides a self-contained record of the \emph{global existence} mechanism for the Lorenz-gauge Maxwell-Higgs
Cauchy problem on the Schwarzschild exterior $(\mathcal D_{M,0},g_{M,0})$.
No new estimates are proved here; the goal is to isolate the local well-posedness, constraint propagation, and continuation steps
which turn the \emph{a priori} higher-order energy bounds proved in the main text (Corollary~\ref{cor:small-data}) into global existence.

Throughout we fix $M>0$ and an integer $k\ge 6$.
We work in the uncharged regime (Definition~\ref{def:charges}) so that, after subtracting the stationary Coulomb component,
the Maxwell field is exact and admits a global potential $A$ with $F=\dd A$ on $\mathcal D_{M,0}$.

\subsection{Geometric setup, Sobolev norms, and admissible data}\label{app:schw-setup}

Let $t$ denote the Schwarzschild time function on the exterior region $\{r>2M\}$ and let
$\Sigma_t:=\{t=\mathrm{const}\}$ be the associated spacelike hypersurfaces.
We write $n=n_{\Sigma_t}$ for the future unit normal and $\gamma$ for the induced Riemannian metric on $\Sigma_t$.
For an integer $s\ge 0$ we write $H^{s}(\Sigma_t)$ for the (intrinsic) Sobolev space defined using $\gamma$,
and we extend the notation componentwise to tensor fields on $\Sigma_t$.

Since the Schwarzschild exterior is stationary and the slices $\Sigma_t$ are translates of $\Sigma_{t_0}$ under the Killing flow of $T=\partial_t$,
the constants in the Sobolev embeddings and Moser/product estimates on $\Sigma_t$ can be chosen uniformly in $t$.
We state the only analytic condition we need.

\begin{lemma}[Uniform Sobolev/Moser estimates on $\Sigma_t$]\label{lem:sobolev-algebra-schwarzschild}
Let $s\ge 3$.
Then $H^{s}(\Sigma_t)\hookrightarrow W^{1,\infty}(\Sigma_t)$ and $H^{s}(\Sigma_t)$ is an algebra.
For smooth functions $f,g$ on $\Sigma_t$ one has the product estimate
\begin{equation}
\|fg\|_{H^{s}(\Sigma_t)}
\lesssim
\|f\|_{H^{s}(\Sigma_t)}\|g\|_{L^\infty(\Sigma_t)}
+\|f\|_{L^\infty(\Sigma_t)}\|g\|_{H^{s}(\Sigma_t)},
\end{equation}
with an implicit constant depending only on $(M,s)$ (and not on $t$).
\end{lemma}

\begin{proof}
Choose a finite atlas on the compact part of $\Sigma_{t_0}$ and the asymptotically Euclidean chart on the end.  The induced metric and all derivatives up to the required order are uniformly comparable in these charts to the Euclidean metric, and the injectivity radius is bounded from below on the compact part.  A partition of unity, the Euclidean Sobolev embedding $H^s(\mathbb R^3)\hookrightarrow W^{1,\infty}(\mathbb R^3)$ for $s\ge3$, and the Leibniz rule give the displayed embedding and product estimate on $\Sigma_{t_0}$.  Since the stationary Killing flow maps $\Sigma_{t_0}$ isometrically to every $\Sigma_t$, the same atlas transported by the flow has identical metric constants.  Consequently, the Sobolev and Moser constants depend only on $(M,s)$ and not on $t$.
\end{proof}

\medskip
\noindent\textbf{Admissible Lorenz-gauge data.}
A smooth Cauchy data set on $\Sigma_{t_0}$ for the Lorenz-reduced system consists of
\begin{equation}
\bigl(A|_{\Sigma_{t_0}},\ \nabla_n A|_{\Sigma_{t_0}},\ \phi|_{\Sigma_{t_0}},\ D_n\phi|_{\Sigma_{t_0}}\bigr),
\end{equation}
as in Section~\ref{subsec:initial-data}.
We call such data \emph{Lorenz-compatible} if the constraints from Section~\ref{subsec:initial-data} hold, namely
\begin{equation}
\nabla^\mu A_\mu\big|_{\Sigma_{t_0}}=0,
\qquad
\nabla_n\bigl(\nabla^\mu A_\mu\bigr)\big|_{\Sigma_{t_0}}=0,
\qquad
\nabla^i E_i=-\,J_{\mu}n^{\mu}\ \ \text{on }\Sigma_{t_0},
\end{equation}
where $E_i:=F(n,e_i)$ is the electric field and $\nabla^i$ is the Levi-Civita connection of $(\Sigma_{t_0},\gamma)$.

\subsection{Lorenz reduction, current conservation, and constraint propagation}\label{app:schw-lorenz}

We impose the Lorenz gauge condition
\begin{equation}\label{eq:app-lorenz-schwarzschild}
\nabla^{\mu}A_{\mu}=0.
\end{equation}
We recall the reduction of Maxwell-Higgs to a semilinear wave system for $(A,\phi)$ on a Ricci-flat background.

\begin{lemma}[Wave system for the potential]\label{lem:wave-potential-schwarzschild}
Let $(\mathcal M,g)$ be Ricci flat and let $F=\dd A$.
Then the Maxwell equation $\nabla^\mu F_{\mu\nu}=j_\nu$ with source $j_\nu$ is equivalent to
\begin{equation}\label{eq:wave-potential-general}
\square_g A_\nu-\nabla_\nu(\nabla^\mu A_\mu)=j_\nu.
\end{equation}
If the Lorenz gauge condition $\nabla^\mu A_\mu=0$ holds, then
\begin{equation}\label{eq:wave-potential-lorenz}
\square_g A_\nu=j_\nu.
\end{equation}
\end{lemma}

\begin{proof}
Since $F_{\mu\nu}=\nabla_\mu A_\nu-\nabla_\nu A_\mu$, we compute
\begin{equation}
\nabla^\mu F_{\mu\nu}
=
\nabla^\mu\nabla_\mu A_\nu-\nabla^\mu\nabla_\nu A_\mu
=
\square_g A_\nu-\nabla_\nu(\nabla^\mu A_\mu)-(\mathrm{Ric})_{\nu}{}^{\mu}A_\mu.
\end{equation}
On a Ricci-flat background the Ricci term vanishes, giving \eqref{eq:wave-potential-general}.
Imposing the Lorenz condition removes the second term in \eqref{eq:wave-potential-general} and gives \eqref{eq:wave-potential-lorenz}.
\end{proof}

For Maxwell-Higgs, the source is the gauge current
\begin{equation}
J_\nu:=-i\bigl(D_\nu\phi\,\bar\phi-\phi\,\overline{D_\nu\phi}\bigr),
\qquad D_\nu:=\nabla_\nu-iA_\nu.
\end{equation}
The scalar field satisfies
\begin{equation}
D^\mu D_\mu\phi=\partial_{\bar\phi}P(\phi,\bar\phi),
\end{equation}
with $P(\phi,\bar\phi)=\widetilde P(|\phi|^2)$ gauge invariant (Assumption~\ref{asumsiP}).

\begin{lemma}[Current conservation]\label{lem:current-conservation-schwarzschild}
Let $(A,\phi)$ be a smooth pair and define the Maxwell-Higgs current
\begin{equation}
J_\nu:=-i\bigl(D_\nu\phi\,\bar\phi-\phi\,\overline{D_\nu\phi}\bigr).
\end{equation}
Then
\begin{equation}\label{eq:div-J-identity}
\nabla^\nu J_\nu=2\,\Im\!\bigl(\bar\phi\,D^\nu D_\nu\phi\bigr).
\end{equation}
If $\phi$ solves the scalar equation $D^\nu D_\nu\phi=\partial_{\bar\phi}P(\phi,\bar\phi)$ with gauge-invariant potential $P(\phi,\bar\phi)=\widetilde P(|\phi|^2)$,
then $\nabla^\nu J_\nu=0$.
\end{lemma}

\begin{proof}
Set $Q_\nu:=\bar\phi\,D_\nu\phi$.
Under the gauge transformation \eqref{eq:gauge-transform} one has $Q_\nu\mapsto Q_\nu$, so $Q_\nu$ is gauge invariant (and thus uncharged).
Consequently, its divergence may be computed using either $\nabla$ or the gauge-covariant derivative $D$:
\begin{equation}
\nabla^\nu Q_\nu=D^\nu Q_\nu.
\end{equation}
Since $J_\nu=-i(Q_\nu-\overline{Q_\nu})$, we obtain
\begin{equation}
\nabla^\nu J_\nu
=-i\bigl(\nabla^\nu Q_\nu-\overline{\nabla^\nu Q_\nu}\bigr)
=-i\bigl(D^\nu Q_\nu-\overline{D^\nu Q_\nu}\bigr).
\end{equation}
Using the Leibniz rule for $D$ on scalar fields,
\begin{equation}
D^\nu Q_\nu
=
D^\nu(\bar\phi\,D_\nu\phi)
=
(D^\nu\bar\phi)\,D_\nu\phi+\bar\phi\,D^\nu D_\nu\phi.
\end{equation}
The first term $(D^\nu\bar\phi)\,D_\nu\phi=\overline{D^\nu\phi}\,D_\nu\phi$ is real-valued, therefore contributes no imaginary part.
Consequently,
\begin{equation}
\nabla^\nu J_\nu
=
-i\Bigl(\bar\phi\,D^\nu D_\nu\phi-\phi\,\overline{D^\nu D_\nu\phi}\Bigr)
=
2\,\Im\!\bigl(\bar\phi\,D^\nu D_\nu\phi\bigr),
\end{equation}
which is \eqref{eq:div-J-identity}.
If $D^\nu D_\nu\phi=\partial_{\bar\phi}P(\phi,\bar\phi)$ and $P(\phi,\bar\phi)=\widetilde P(|\phi|^2)$, then $\partial_{\bar\phi}P=\widetilde P'(|\phi|^2)\phi$ and therefore
$\bar\phi\,\partial_{\bar\phi}P=\widetilde P'(|\phi|^2)|\phi|^2\in\mathbb R$, so the imaginary part vanishes.
\end{proof}

We now record propagation of the Lorenz constraint.

\begin{lemma}[Propagation of the Lorenz constraint]\label{lem:lorenz-propagation-schwarzschild}
Let $(A,\phi)$ solve the Lorenz-reduced Maxwell-Higgs system on $(\mathcal D_{M,0},g_{M,0})$,
\begin{equation}\label{eq:lorenz-reduced-schwarzschild}
\square_{g_{M,0}}A_{\nu}=J_{\nu},
\qquad
D^{\mu}D_{\mu}\phi=\partial_{\bar\phi}P(\phi,\bar\phi),
\end{equation}
and set $u:=\nabla^\mu A_\mu$.
Then $u$ satisfies the homogeneous wave equation
\begin{equation}\label{eq:lorenz-wave-u}
\square_{g_{M,0}}u=0.
\end{equation}
Consequently, if the Lorenz compatibility conditions
$u|_{\Sigma_{t_0}}=0$ and $\nabla_n u|_{\Sigma_{t_0}}=0$ hold on one Cauchy slice, then $u\equiv 0$ on the entire domain of existence.
\end{lemma}

\begin{proof}
By Lemma~\ref{lem:wave-potential-schwarzschild}, the Maxwell equation $\nabla^\mu F_{\mu\nu}=J_\nu$ with $F=\dd A$ is equivalent to
\begin{equation}
\square_{g_{M,0}}A_\nu-\nabla_\nu(\nabla^\mu A_\mu)=J_\nu.
\end{equation}
Taking the divergence and using $\nabla^\nu\nabla_\nu(\nabla^\mu A_\mu)=\square_{g_{M,0}}u$ yields
\begin{equation}
\square_{g_{M,0}}u=\nabla^\nu J_\nu.
\end{equation}
By Lemma~\ref{lem:current-conservation-schwarzschild} we have $\nabla^\nu J_\nu=0$ whenever $\phi$ solves the scalar equation with gauge-invariant potential,
therefore \eqref{eq:lorenz-wave-u}.
Uniqueness for the scalar wave equation \eqref{eq:lorenz-wave-u} with vanishing Cauchy data gives $u\equiv 0$.
\end{proof}

\subsection{Local well-posedness in Lorenz gauge}\label{app:schw-lwp}

We now give a detailed local well-posedness statement for the Lorenz-reduced system.
(For the nonlinear scattering problem in the main text we always work with smooth data; the Sobolev formulation below is included to make the continuation argument precise.)

\begin{proposition}[Local well-posedness in Lorenz gauge on Schwarzschild]\label{prop:local-wp-schwarzschild}
Fix $k\ge 6$ and a time $t_0$.
Let the initial data on $\Sigma_{t_0}$ be Lorenz-compatible in the sense of Section~\ref{subsec:initial-data} and satisfy
\begin{equation}\label{eq:app-schwarzschild-data-sobolev}
A|_{\Sigma_{t_0}}\in H^{k+1}(\Sigma_{t_0}),
\qquad
\nabla_n A|_{\Sigma_{t_0}}\in H^{k}(\Sigma_{t_0}),
\qquad
\phi|_{\Sigma_{t_0}}\in H^{k+1}(\Sigma_{t_0}),
\qquad
D_n\phi|_{\Sigma_{t_0}}\in H^{k}(\Sigma_{t_0}).
\end{equation}
Then there exists $T=T(M,k,\text{data})>0$ and a unique solution $(A,\phi)$ of \eqref{eq:lorenz-reduced-schwarzschild} on the slab $\mathcal R(t_0,t_0+T)$ such that
\begin{equation}
A\in C^0\!\bigl([t_0,t_0+T];H^{k+1}(\Sigma_t)\bigr)\cap C^1\!\bigl([t_0,t_0+T];H^{k}(\Sigma_t)\bigr),
\end{equation}
\begin{equation}
\phi\in C^0\!\bigl([t_0,t_0+T];H^{k+1}(\Sigma_t)\bigr)\cap C^1\!\bigl([t_0,t_0+T];H^{k}(\Sigma_t)\bigr).
\end{equation}
In addition:
\begin{enumerate}
\item[(i)] the Lorenz condition \eqref{eq:app-lorenz-schwarzschild} propagates, i.e.\ $\nabla^\mu A_\mu=0$ holds on $\mathcal R(t_0,t_0+T)$,
\item[(ii)] with $F=\dd A$, the pair $(F,\phi)$ solves the original Maxwell-Higgs system \eqref{eom1}-\eqref{eom2} on $\mathcal R(t_0,t_0+T)$.
\end{enumerate}
\end{proposition}

\begin{proof}
We give the energy/Picard iteration argument in the Sobolev class used by the continuation criterion.

\smallskip
\noindent\emph{Linear energy estimate.}
Let $u$ solve $\square_{g_{M,0}}u=f$ on $\mathcal R(t_0,t_0+T)$ with given Cauchy data on $\Sigma_{t_0}$.
Commuting with $\nabla$ up to order $k$ and applying the $T$-energy identity for each commuted equation yields
\begin{equation}\label{eq:linear-energy-estimate-schwarzschild}
\sup_{t\in[t_0,t_0+T]}\Bigl(\|u(t)\|_{H^{k+1}(\Sigma_t)}+\|\nabla_n u(t)\|_{H^{k}(\Sigma_t)}\Bigr)
\lesssim
\|u(t_0)\|_{H^{k+1}}+\|\nabla_n u(t_0)\|_{H^{k}}+\int_{t_0}^{t_0+T}\|f(s)\|_{H^{k}(\Sigma_s)}\,\dd s,
\end{equation}
with an implicit constant depending only on $(M,k)$.
The same estimate holds componentwise for $1$-forms. The energy estimate for $(\square_{g_{M,0}}-m^{2})u=f$ with constant $m^{2}\ge0$ is obtained by adding the positive mass term to the scalar energy; the implicit constant then also depends on $m$ (equivalently, on the potential through $m^{2}$).

\smallskip
\noindent\emph{Picard iteration.}
Define $(A^{(0)},\phi^{(0)})$ to be the solution of the homogeneous linear system with the given data.
Given $(A^{(m)},\phi^{(m)})$, define $(A^{(m+1)},\phi^{(m+1)})$ to be the solution of the linear inhomogeneous system
\begin{equation}
\square_{g_{M,0}}A^{(m+1)}_{\nu}=-i\Bigl(D^{(m)}_{\nu}\phi^{(m)}\,\overline{\phi^{(m)}}-\phi^{(m)}\,\overline{D^{(m)}_{\nu}\phi^{(m)}}\Bigr),
\qquad
(\square_{g_{M,0}}-m^{2})\phi^{(m+1)}=\mathcal N_\phi\bigl(A^{(m)},\phi^{(m)}\bigr),
\end{equation}
with the same initial data, where $D^{(m)}=\nabla-iA^{(m)}$.
Let $m^{2}\ge 0$ be as in Assumption~\ref{asumsiP} and define
\begin{equation}
\mathcal N_\phi(A,\phi):=2iA^{\mu}\nabla_{\mu}\phi + A^{\mu}A_{\mu}\phi + \bigl(\partial_{\bar\phi}P(\phi,\bar\phi)-m^{2}\phi\bigr).
\end{equation}
Thus the scalar equation in Lorenz gauge takes the Klein-Gordon form $(\square_{g_{M,0}}-m^{2})\phi=\mathcal N_\phi(A,\phi)$.
In addition, by \eqref{eq:potential-remainder}-\eqref{eq:potential-nonlinear-bound} the potential remainder $\partial_{\bar\phi}P-m^{2}\phi$ is at least cubic in $\phi$, so $\mathcal N_\phi$ is a smooth combination of terms at most quadratic in $(A,\nabla\phi)$ and at least cubic in $\phi$ (thanks to \eqref{eq:potential-nonlinear-bound}).

\smallskip
\noindent\emph{Uniform bound for small $T$.}
Fix $M_0>0$ and consider the Banach space $X_T$ of pairs $(A,\phi)$ with the regularity stated in the proposition and norm
\begin{equation}
\|(A,\phi)\|_{X_T}
:=
\sup_{t\in[t_0,t_0+T]}
\Bigl(
\|A(t)\|_{H^{k+1}}+\|\nabla_n A(t)\|_{H^{k}}
+\|\phi(t)\|_{H^{k+1}}+\|D_n\phi(t)\|_{H^{k}}
\Bigr).
\end{equation}
By Lemma~\ref{lem:sobolev-algebra-schwarzschild}, for $k\ge 6$ the nonlinearities are locally Lipschitz on bounded subsets of $X_T$:
for $(A,\phi)\in X_T$ with $\|(A,\phi)\|_{X_T}\le M_0$ one has, where $J[\phi;A]$ denotes the current $J_\nu=-i\bigl(D_\nu\phi\,\bar\phi-\phi\,\overline{D_\nu\phi}\bigr)$,
\begin{equation}\label{eq:nonlinear-bound-schwarzschild}
\|J[\phi;A]\|_{H^{k}}+\|\mathcal N_\phi(A,\phi)\|_{H^{k}}
\le C(M_0)\,\|(A,\phi)\|_{X_T}^{2},
\end{equation}
with $C(M_0)$ depending only on $(M,k,M_0)$ and finitely many bounds for $\widetilde P$ on bounded sets.
Applying \eqref{eq:linear-energy-estimate-schwarzschild} to the iteration and choosing $T>0$ so that $T\,C(M_0)\ll 1$ yields a uniform bound
$\|(A^{(m)},\phi^{(m)})\|_{X_T}\le M_0$ for all $m$.

\smallskip
\noindent\emph{Contraction and convergence.}
Applying the linear estimate to the difference of two iterates and using the Lipschitz property from Lemma~\ref{lem:sobolev-algebra-schwarzschild},
one obtains for $T$ sufficiently small
\begin{equation}
\|(A^{(m+1)}-A^{(m)},\ \phi^{(m+1)}-\phi^{(m)})\|_{X_T}
\le \frac12\,\|(A^{(m)}-A^{(m-1)},\ \phi^{(m)}-\phi^{(m-1)})\|_{X_T}.
\end{equation}
Consequently, $(A^{(m)},\phi^{(m)})$ converges in $X_T$ to a limit $(A,\phi)$ which solves \eqref{eq:lorenz-reduced-schwarzschild}.
Uniqueness and continuous dependence follow similarly by applying the same estimate to the difference of two solutions.

\smallskip
\noindent\emph{Propagation of constraints and equivalence to Maxwell-Higgs.}
By Lemma~\ref{lem:lorenz-propagation-schwarzschild}, Lorenz compatibility of the initial data implies $\nabla^\mu A_\mu=0$ on $\mathcal R(t_0,t_0+T)$.
With $F=\dd A$, Lemma~\ref{lem:wave-potential-schwarzschild} then shows that $(F,\phi)$ satisfies the Maxwell equation \eqref{eom1}, while the scalar equation is the second equation in \eqref{eq:lorenz-reduced-schwarzschild}.
This proves (i) and (ii).
\end{proof}

\subsection{Continuation criterion}\label{app:schw-continuation}

We now make the continuation mechanism explicit.
Let $(A,\phi)$ be the maximal smooth Lorenz-gauge solution arising from smooth Lorenz-compatible data on $\Sigma_{t_0}$, defined on a maximal forward interval $[t_0,t_+)$.

\begin{lemma}[Continuation criterion]\label{lem:continuation-schwarzschild}
Assume that the higher-order energy $\mathcal E_k(t)$ from \eqref{def:Ek} stays finite on $[t_0,t_+)$:
\begin{equation}\label{eq:app-uniform-energy-schwarzschild}
\sup_{t\in[t_{0},t_{+})}\ \mathcal E_k(t)\ <\ \infty.
\end{equation}
Then $t_+=+\infty$.
An analogous statement holds backward in time.
\end{lemma}

\begin{proof}
By Proposition~\ref{prop:local-wp-schwarzschild}, the existence time for the Lorenz-reduced system can be chosen to depend only on the size of the Sobolev norm
\eqref{eq:app-schwarzschild-data-sobolev} of the Cauchy data at the restart time.
For a smooth Lorenz-gauge solution, the energy fluxes appearing in $\mathcal E_k(t)$ control these Sobolev norms on $\Sigma_t$
(because $F=\dd A$ controls first derivatives of $A$, and higher derivatives are controlled by commuting with $\partial_t$ and $\Omega_i$
together with the wave equations and the constraints; see Section~\ref{subsec:initial-data} and the discussion surrounding \eqref{def:Ek}).
Consequently, the bound \eqref{eq:app-uniform-energy-schwarzschild} implies that the Sobolev size of the Cauchy data at time $t$ stays uniformly bounded for $t<t_+$,
so the local existence time has a uniform positive lower bound.
Iterating the local theory therefore extends the solution beyond $t_+$ unless $t_+=+\infty$.
\end{proof}

\subsection{Small-data global existence}\label{app:schw-global}

We finally record the global existence consequence for small Schwarzschild data.

\begin{proposition}[Global existence for small Schwarzschild data]\label{prop:global-existence-schwarzschild}
Assume the conditions of Corollary~\ref{cor:small-data}.
Let the uncharged Lorenz-compatible Schwarzschild data on $\Sigma_{t_{0}}$ satisfy $\mathcal E_k(t_0)\le \varepsilon_0$, with $\varepsilon_0$ as in Corollary~\ref{cor:small-data}.
Then the corresponding Lorenz-gauge solution $(A,\phi)$ exists globally on $\mathcal D_{M,0}$ (both to the future and the past),
and the associated field pair $(F,\phi)$ is smooth and satisfies the uniform boundedness and decay estimates stated in Corollary~\ref{cor:small-data}.
\end{proposition}

\begin{proof}
By Proposition~\ref{prop:local-wp-schwarzschild} there exists a unique smooth Lorenz-gauge solution on a nontrivial initial time slab.
Corollary~\ref{cor:small-data} provides a uniform bound for $\mathcal E_k(t)$ on the maximal interval of existence, therefore \eqref{eq:app-uniform-energy-schwarzschild} holds.
Lemma~\ref{lem:continuation-schwarzschild} therefore yields $t_+=+\infty$.
The backward global existence statement follows from the time-reversed evolution, equivalently from the backward statement in Corollary~\ref{cor:small-data}.
\end{proof}

\section{Global existence on Kerr spacetimes}
\label{app:global-existence-kerr}

This appendix states the global existence mechanism for the Lorenz-gauge Maxwell-Higgs Cauchy problem on a fixed slowly rotating Kerr exterior
$(\mathcal D_{M,a},g_{M,a})$ with $|a|<M$.
As in Appendix~\ref{app:global-existence-schwarzschild}, no new decay estimate is needed here; we isolate the local well-posedness,
constraint propagation, and continuation steps which, combined with the \emph{a priori} energy bounds of Theorem~\ref{thm:main-slow-kerr-intro}, Theorem~\ref{thm:main-slow-kerr-massive-intro}, or Theorem~\ref{thm:main-slow-kerr-massive-intro},
yield global existence in the Kerr setting.

Throughout we fix $M>0$, $|a|<M$, and $k\ge 6$.
We work in the uncharged regime (Definition~\ref{def:charges}), so that after Coulomb subtraction the Maxwell field is exact and admits a global potential; see Remark~\ref{rem:kerr-data-charges}.

\subsection{Geometric setup and uniform Sobolev estimates}\label{app:kerr-setup}

We work with the Kerr-star time function $t^\star$ and the horizon-regular spacelike foliation
\begin{equation}
\Sigma_\tau:=\{t^\star=\tau\},
\end{equation}
as in \eqref{eq:kerr-tstar}.
Let $n=n_{\Sigma_\tau}$ be the future unit normal and $\gamma$ the induced Riemannian metric on $\Sigma_\tau$.

For an integer $s\ge 0$ we denote by $H^s(\Sigma_\tau)$ the intrinsic Sobolev space defined using $\gamma$,
again extended componentwise to tensor fields.
Because Kerr is stationary and the slices $\Sigma_\tau$ are translates under the flow of the stationary Killing field,
the Sobolev embeddings and Moser estimates hold with constants that are uniform in $\tau$ (depending only on $(M,a,s)$).

\begin{lemma}[Uniform Sobolev/Moser estimates on $\Sigma_\tau$]\label{lem:sobolev-algebra-kerr}
Let $s\ge 3$.
Then $H^{s}(\Sigma_\tau)\hookrightarrow W^{1,\infty}(\Sigma_\tau)$ and $H^{s}(\Sigma_\tau)$ is an algebra, with constants uniform in $\tau$.
\end{lemma}

\begin{proof}
On the asymptotic end of each Kerr-star slice the induced metric is a smooth asymptotically Euclidean perturbation with constants depending only on $(M,a)$.  On the compact part, horizon regularity of the Kerr-star coordinates and smoothness of the foliation give a positive injectivity radius and uniform bounds for the metric coefficients and their derivatives.  Choose a uniformly locally finite atlas adapted to the end and to the compact part, together with a subordinate partition of unity whose derivatives are uniformly bounded.  The Euclidean estimates in these charts give $H^s\hookrightarrow W^{1,\infty}$ for $s\ge3$ and the corresponding Moser product estimates on one slice.  Since the slices are translates under the stationary Killing flow, the metric constants, overlap number, and partition bounds are independent of $\tau$.  This gives the uniform Sobolev and Moser estimates claimed in the lemma.
\end{proof}

\medskip
\noindent\textbf{Redshift coercivity.}
A key distinction from Schwarzschild is the presence of an ergoregion, where the stationary Killing field $T$ is spacelike.
For energy estimates and continuation it is therefore convenient to work with the globally timelike \emph{redshift} vector field $N$
from Definition~\ref{def:admissible-exterior}.
The associated energy flux controls the full $H^1$ size of $(F,\phi)$ on each slice.

\begin{lemma}[Coercivity of the $N$-flux]\label{lem:N-coercive-kerr}
Let $(F,\phi)$ be smooth on $\Sigma_\tau$ and let $N$ be the admissible redshift vector field.
Then
\begin{equation}
E^{N}[F,\phi](\tau)
=
\int_{\Sigma_\tau} T_{\mu\nu}[F,\phi]\,N^\nu\,n^\mu\,\dd\mu_{\Sigma_\tau}
\end{equation}
is comparable to the $H^1$-type norm of $(F,\phi)$ on $\Sigma_\tau$:
\begin{equation}
E^{N}[F,\phi](\tau)
\simeq
\int_{\Sigma_\tau}
\Bigl(
|F(n,\cdot)|_\gamma^{2}+|F|_\gamma^{2}
+|D_n\phi|^{2}+|D\phi|_\gamma^{2}+P(\phi,\bar\phi)
\Bigr)\,\dd\mu_{\Sigma_\tau},
\end{equation}
with constants depending only on $(M,a)$ (and uniform in $\tau$).
\end{lemma}

\begin{proof}
Both \(N\) and the unit normal \(n\) are future-directed timelike and, on the admissible Kerr-star foliation, their scalar product is bounded above and below by positive constants depending only on \((M,a)\).  Decompose the Maxwell field into its normal contractions and tangential two-form components relative to \(n\), and decompose \(D\phi\) into its normal and tangential parts.  The dominant energy condition for the Maxwell part and the positivity of the scalar kinetic and potential terms give
\[
 T_{\mu\nu}[F,\phi]N^\nu n^\mu
 \ge c\bigl(|F(n,\cdot)|_\gamma^2+|F|_\gamma^2+|D_n\phi|^2+|D\phi|_\gamma^2+P(\phi,\bar\phi)\bigr).
\]
The reverse inequality follows from Cauchy-Schwarz and the same uniform comparability of \(N\) and \(n\).  The constants are independent of \(\tau\), because the metric, the foliation vector fields, and \(N\) are stationary in Kerr-star coordinates.  Integration over \(\Sigma_\tau\) gives the asserted equivalence.
\end{proof}

\subsection{Lorenz reduction and constraint propagation}\label{app:kerr-lorenz}

In Lorenz gauge,
\begin{equation}\label{eq:app-lorenz}
\nabla^{\mu}A_{\mu}=0,
\end{equation}
the Maxwell-Higgs system \eqref{eom1}-\eqref{eom2} reduces on Kerr to the semilinear wave system
\begin{equation}\label{eq:app-lorenz-reduced-kerr}
\square_{g_{M,a}}A_{\nu}=J_{\nu},
\qquad
D^{\mu}D_{\mu}\phi=\partial_{\bar\phi}P(\phi,\bar\phi),
\end{equation}
as stated in \eqref{eq:lorenz-reduced-kerr}.
The proofs of Lemmas~\ref{lem:wave-potential-schwarzschild} and~\ref{lem:current-conservation-schwarzschild}
use only Ricci flatness and gauge invariance of $P$, and therefore apply without change on Kerr.

For a solution of \eqref{eq:app-lorenz-reduced-kerr} the current $J_\nu$ is divergence free and the Lorenz constraint propagates.

\begin{lemma}[Propagation of the Lorenz constraint on Kerr]\label{lem:lorenz-propagation-kerr}
Let $(A,\phi)$ solve \eqref{eq:app-lorenz-reduced-kerr} on $(\mathcal D_{M,a},g_{M,a})$ and set $u:=\nabla^\mu A_\mu$.
Then $u$ satisfies $\square_{g_{M,a}}u=0$.
If the Lorenz compatibility conditions \eqref{eq:lorenz-constraints-kerr} hold on $\Sigma_{\tau_0}$, then $\nabla^\mu A_\mu\equiv 0$ on the domain of existence.
\end{lemma}

\begin{proof}
Take the divergence of the reduced equation for $A$.  Because Kerr is Ricci flat, the commutator between the divergence and the wave operator on one-forms has no curvature contribution after contraction, so
\[
\nabla^\nu\square_{g_{M,a}}A_\nu=\square_{g_{M,a}}(\nabla^\nu A_\nu)=\square_{g_{M,a}}u.
\]
Gauge invariance of the scalar potential gives the covariant conservation law $\nabla^\nu J_\nu=0$ for every solution of the scalar equation.  Consequently, $u$ satisfies the homogeneous scalar wave equation.  The Lorenz compatibility conditions prescribe both $u|_{\Sigma_{\tau_0}}=0$ and $\nabla_nu|_{\Sigma_{\tau_0}}=0$.  Uniqueness for the homogeneous wave equation on the globally hyperbolic Kerr exterior then gives $u\equiv0$ on the whole domain of existence.
\end{proof}

\subsection{Local well-posedness in Lorenz gauge}\label{app:kerr-lwp}

We now formulate local well-posedness for the Lorenz-reduced system on Kerr.
The next statement records the local Sobolev theory for the Lorenz-reduced semilinear wave system on the fixed globally hyperbolic Kerr exterior.

\begin{proposition}[Local well-posedness in Lorenz gauge on Kerr]\label{prop:local-wp-kerr}
Fix $k\ge 6$ and an initial slice $\Sigma_{\tau_0}$.
Let the initial data \eqref{eq:kerr-data} be Lorenz-compatible in the sense of \eqref{eq:lorenz-constraints-kerr}, and assume that
\begin{equation}\label{eq:app-kerr-data-sobolev}
A|_{\Sigma_{\tau_0}}\in H^{k+1}(\Sigma_{\tau_0}),
\qquad
\nabla_n A|_{\Sigma_{\tau_0}}\in H^{k}(\Sigma_{\tau_0}),
\qquad
\phi|_{\Sigma_{\tau_0}}\in H^{k+1}(\Sigma_{\tau_0}),
\qquad
D_n\phi|_{\Sigma_{\tau_0}}\in H^{k}(\Sigma_{\tau_0}).
\end{equation}
Then there exists $T=T(M,a,k,\text{data})>0$ and a unique solution $(A,\phi)$ of \eqref{eq:app-lorenz-reduced-kerr} on the slab $\mathcal R(\tau_0,\tau_0+T)$ such that
\begin{equation}
A\in C^0\!\bigl([\tau_0,\tau_0+T];H^{k+1}(\Sigma_\tau)\bigr)\cap C^1\!\bigl([\tau_0,\tau_0+T];H^{k}(\Sigma_\tau)\bigr),
\end{equation}
\begin{equation}
\phi\in C^0\!\bigl([\tau_0,\tau_0+T];H^{k+1}(\Sigma_\tau)\bigr)\cap C^1\!\bigl([\tau_0,\tau_0+T];H^{k}(\Sigma_\tau)\bigr).
\end{equation}
In addition, the Lorenz condition propagates and with $F=\dd A$ the pair $(F,\phi)$ solves the original Maxwell-Higgs system \eqref{eom1}-\eqref{eom2} on $\mathcal R(\tau_0,\tau_0+T)$.
\end{proposition}

\begin{proof}
Let \(X_T\) be the Banach space of pairs \((A,\phi)\) on \([\tau_0,\tau_0+T]\) with the regularity displayed in the statement and norm
\begin{equation}
 \|(A,\phi)\|_{X_T}:=\sup_{\tau\in[\tau_0,\tau_0+T]}
 \Bigl(\|A(\tau)\|_{H^{k+1}}+\|\nabla_nA(\tau)\|_{H^k}
 +\|\phi(\tau)\|_{H^{k+1}}+\|D_n\phi(\tau)\|_{H^k}\Bigr).
\end{equation}
The linear estimate on a Kerr-star slab is obtained by applying the hyperbolic energy identity to \(\square_{g_{M,a}}u=f\) and to its covariant derivatives up to order \(k\).  The constants depend on finitely many bounds for the stationary Kerr-star metric on the slab, and therefore depend only on \((M,a,k)\) when \(T\le1\):
\begin{equation}\label{eq:kerr-local-linear-estimate}
 \sup_{\tau\in[\tau_0,\tau_0+T]}
 \bigl(\|u(\tau)\|_{H^{k+1}}+\|\nabla_nu(\tau)\|_{H^k}\bigr)
 \le C\left(\|u[\tau_0]\|_{H^{k+1}\times H^k}
 +\int_{\tau_0}^{\tau_0+T}\|f(s)\|_{H^k}\,\dd s\right).
\end{equation}
The same estimate holds componentwise for one-forms and for \((\square_{g_{M,a}}-m^2)u=f\) after adding the mass term to the scalar energy.  Define a Picard sequence as in the Schwarzschild proof: the zeroth iterate solves the homogeneous linear system with the prescribed data, and the \((n+1)\)-st iterate solves the linear inhomogeneous Lorenz-reduced equations with current and scalar source evaluated at the \(n\)-th iterate.

Lemma~\ref{lem:sobolev-algebra-kerr} gives, on every ball \(\|(A,\phi)\|_{X_T}\le M_0\), the local Lipschitz estimates
\begin{equation}
 \|J[\phi;A]\|_{H^k}+\|\mathcal N_\phi(A,\phi)\|_{H^k}
 \le C(M_0)\|(A,\phi)\|_{X_T}^2,
\end{equation}
\begin{equation}
 \|J[\phi;A]-J[\psi;B]\|_{H^k}
 +\|\mathcal N_\phi(A,\phi)-\mathcal N_\phi(B,\psi)\|_{H^k}
 \le C(M_0)\|(A-B,\phi-\psi)\|_{X_T}.
\end{equation}
Here \(\mathcal N_\phi(A,\phi)=2iA^\mu\nabla_\mu\phi+A^\mu A_\mu\phi+\partial_{\bar\phi}P-m^2\phi\), and the potential part is controlled by the polynomial bounds in Assumption~\ref{asumsiP}.  Applying \eqref{eq:kerr-local-linear-estimate} to the iteration and choosing \(T\) so that \(T C(M_0)\) is sufficiently small gives a uniform bound in \(X_T\) and a contraction for the difference of consecutive iterates.  The limit solves \eqref{eq:app-lorenz-reduced-kerr}, and the same estimate applied to the difference of two solutions gives uniqueness and continuous dependence.  Lemma~\ref{lem:lorenz-propagation-kerr} propagates the Lorenz constraint from the compatible initial data; once \(\nabla^\mu A_\mu=0\), the identity in Lemma~\ref{lem:wave-potential-schwarzschild}, which uses only Ricci flatness, turns the reduced equation for \(A\) into the Maxwell equation for \(F=\dd A\).  Thus the original Maxwell-Higgs system holds on the slab.
\end{proof}

\subsection{Continuation criterion and global existence}\label{app:kerr-continuation}

Let $(A,\phi)$ be the maximal smooth Lorenz-gauge solution arising from smooth Lorenz-compatible data on $\Sigma_{\tau_0}$, defined on a maximal forward interval $[\tau_0,\tau_+)$.

\begin{lemma}[Continuation criterion]\label{lem:continuation-kerr}
Assume that the order-$k$ redshift energy flux from Definition~\ref{def:kerr-data} stays finite on $[\tau_0,\tau_+)$:
\begin{equation}\label{eq:app-uniform-energy}
\sup_{\tau\in[\tau_{0},\tau_{+})}\ \mathcal E^{N}_{k,M,a}(\tau)\ <\ \infty,
\end{equation}
where $\mathcal E^{N}_{k,M,a}(\tau)=\sum_{|\alpha|\le k}E^{N}[\mathcal L_{\mathcal Z^\alpha}F,\mathcal Z^\alpha\phi](\tau)$.
Then $\tau_{+}=+\infty$.
An analogous statement holds backward in time.
\end{lemma}

\begin{proof}
By Proposition~\ref{prop:local-wp-kerr}, the solution can be restarted at any time $\tau_1<\tau_+$ with an existence time that depends only on the Sobolev size
\eqref{eq:app-kerr-data-sobolev} of the Cauchy data on $\Sigma_{\tau_1}$.
To relate this Sobolev size to the redshift energy, we note first that Lemma~\ref{lem:N-coercive-kerr} shows that $E^{N}[F,\phi](\tau)$ controls the $H^1$ size of $(F,\phi)$ on $\Sigma_\tau$.
Commuting the field equations with $\mathcal Z$ and applying the same coercivity gives control of higher derivatives, which is the content of the norm
$\mathcal E^{N}_{k,M,a}(\tau)$.
Thus \eqref{eq:app-uniform-energy} yields a uniform bound for the Sobolev norms entering \eqref{eq:app-kerr-data-sobolev} along the evolution.
Consequently, the restart time has a uniform positive lower bound and the maximal time $\tau_+$ must be infinite.
\end{proof}

\begin{proposition}[Global existence for small Kerr data]\label{prop:global-existence-kerr}
Assume the zero-sector conditions of Theorem~\ref{thm:main-slow-kerr-intro}, Theorem~\ref{thm:main-slow-kerr-massive-intro}, or Theorem~\ref{thm:main-slow-kerr-massive-intro}.
Let the uncharged Lorenz-compatible Kerr data on $\Sigma_{\tau_{0}}$ satisfy
$\|(A,\phi)\|_{\mathcal H^{k}_{M,a}(\Sigma_{\tau_{0}})}\le \varepsilon_{\mathrm{Kerr}}$
for $\varepsilon_{\mathrm{Kerr}}$ sufficiently small in the corresponding theorem.
Then the corresponding Lorenz-gauge solution $(A,\phi)$ exists globally on $\mathcal D_{M,a}$ (both to the future and the past),
and the associated field pair $(F,\phi)$ is smooth and satisfies the uniform boundedness and decay estimates stated in that theorem.
\end{proposition}

\begin{proof}
By Proposition~\ref{prop:local-wp-kerr} there exists a unique smooth Lorenz-gauge solution on a nontrivial initial time slab.
The corresponding a priori energy bounds in Theorem~\ref{thm:main-slow-kerr-intro}, Theorem~\ref{thm:main-slow-kerr-massive-intro}, or Theorem~\ref{thm:main-slow-kerr-massive-intro} imply the uniform estimate \eqref{eq:app-uniform-energy} for the maximal solution.
Lemma~\ref{lem:continuation-kerr} therefore yields $\tau_+=+\infty$.
The backward global existence statement follows from the time-reversed evolution, equivalently from the backward statement in the relevant scattering theorem.
\end{proof}

\section{Component computations and proofs}
\label{app:component-proofs}

For clarity, the main text emphasizes the geometric structure of the argument and the resulting estimates. Some longer component-level computations are stated here.

\subsection{Explicit expressions for \texorpdfstring{$E^{\hat K}$ and $J^{\hat K}$}{E\textasciicircum K and J\textasciicircum K}}
\label{app:EK-JK}

For the reader's convenience we state here coordinate expressions for the conformal energy flux \eqref{eq:EK-flux} and Morawetz bulk term \eqref{eq:JK-bulk} associated to the conformal vector field $\hat K=-w^{2}\partial_{w}-v^{2}\partial_{v}$.
Throughout this subsection we use the double-null coordinates
\begin{equation}
v=t+r^{\ast},\qquad w=t-r^{\ast},
\end{equation}
and we write $d\sigma^{2}=\sin\theta\,d\theta\,d\varphi$ for the round area element on $\mathbb S^{2}$.

\begin{eqnarray}\label{EK}
E^{\hat K}(t)
&=
\int_{-\infty}^{\infty}\!\int_{\mathbb S^{2}}
\Bigl[
w^{2}\Bigl(
\frac{|F_{w\theta}|^{2}}{r^{2}}
+\frac{|F_{w\varphi}|^{2}}{r^{2}\sin^{2}\theta}
+|D_{w}\phi|^{2}
\Bigr)
+v^{2}\Bigl(
\frac{|F_{v\theta}|^{2}}{r^{2}}
+\frac{|F_{v\varphi}|^{2}}{r^{2}\sin^{2}\theta}
+|D_{v}\phi|^{2}
\Bigr)
\nonumber\\
&\qquad
+(v^{2}+w^{2})\Bigl(
\frac{|F_{vw}|^{2}}{(1-\mu)^{2}}
+\frac{1-\mu}{4r^{4}\sin^{2}\theta}\,|F_{\theta\varphi}|^{2}
+\Re\bigl(D_{v}\phi\,\overline{D_{w}\phi}\bigr)
\nonumber\\
&\qquad\qquad
+\frac{1-\mu}{4r^{2}}\,|D_{\theta}\phi|^{2}
+\frac{1-\mu}{4r^{2}\sin^{2}\theta}\,|D_{\varphi}\phi|^{2}
+\frac{1-\mu}{2}\,P(\phi)
\Bigr)
\Bigr]\,r^{2}\,dr^{\ast}\,d\sigma^{2}.
\end{eqnarray}

\begin{eqnarray}\label{JK}
\mathcal{J}^{\hat{K}}\bigl(t_{i}\le t\le t_{i+1}\bigr)
&=
\int_{t_{i}}^{t_{i+1}}\!\int_{-\infty}^{\infty}\!\int_{\mathbb S^{2}}
4t\,r^{2}(1-\mu)\,
\Bigl[
\Bigl(
\frac{|F_{vw}|^{2}}{(1-\mu)^{2}}
+\frac{|F_{\theta\varphi}|^{2}}{4r^{4}\sin^{2}\theta}
\Bigr)\nonumber\\
&\qquad\qquad\times
\Bigl(2+(3\mu-2)\frac{r^{\ast}}{r}\Bigr)
+\Bigl(
\frac{|D_{\theta}\phi|^{2}}{2r^{2}}
+\frac{|D_{\varphi}\phi|^{2}}{2r^{2}\sin^{2}\theta}
\Bigr)\nonumber\\
&\qquad\qquad\times
\Bigl(1+\mu\frac{r^{\ast}}{r}\Bigr)
-2\frac{r^{\ast}}{r}\,\Re\bigl(D_{v}\phi\,\overline{D_{w}\phi}\bigr)
+\Bigl(1-\frac{r^{\ast}}{r}\Bigr)P(\phi)
\Bigr]\,d\sigma^{2}\,dr^{\ast}\,dt.
\end{eqnarray}

\subsection{Conventions for the appendix}
\label{app:conventions}

Throughout this appendix we work on the Schwarzschild exterior region $\{r>2M\}$ and use the coordinates $(t,r^{\ast},\theta,\varphi)$, where $r^{\ast}$ is the Regge-Wheeler coordinate. (In some intermediate computations we also write $r*$ for $r^{\ast}$, following the notation in the main text.) We also use the null coordinates
\begin{equation}
v=t+r^{\ast},\qquad w=t-r^{\ast}.
\end{equation}

\paragraph{Index convention.}
We use the coordinate index convention
\begin{equation}
x^{0}=t,\qquad x^{1}=r^{\ast},\qquad x^{2}=\theta,\qquad x^{3}=\varphi,
\end{equation}
so that $\partial_{0}=\partial_{t}$, $\partial_{1}=\partial_{r^{\ast}}$, $\partial_{2}=\partial_{\theta}$, and $\partial_{3}=\partial_{\varphi}$.
Tensor components such as $F_{0 2}$ are taken with respect to this coordinate basis, and we use the antisymmetry $F_{\alpha\beta}=-F_{\beta\alpha}$.

\paragraph{Sphere convention.}
On $\mathbb S^{2}$ we use the area element $d\sigma^{2}=\sin\theta\,d\theta\,d\varphi$, and we write $\slashed\nabla$ for the Levi-Civita connection of the unit round metric.

\paragraph{Null derivatives.}
We use
\begin{equation}
\partial_{v}=\tfrac12(\partial_{t}+\partial_{r^{\ast}}),\qquad
\partial_{w}=\tfrac12(\partial_{t}-\partial_{r^{\ast}}).
\end{equation}

\subsection{Proof of Proposition~\ref{prep5}}
\label{app:proof-prep5}

\begin{proof}[Proof of Proposition~\ref{prep5}]
We work on the Schwarzschild exterior $(\mathcal M,g_{M})$ and use the stationary null tetrad
\begin{equation}\label{eq:app5-tetrad}
\hat L:=\partial_{t}+\partial_{r^{\ast}},
\qquad
\hat N:=\partial_{t}-\partial_{r^{\ast}},
\qquad
\hat M:=\partial_{\theta}+\frac{i}{\sin\theta}\partial_{\varphi},
\qquad
\hat{\bar M}:=\partial_{\theta}-\frac{i}{\sin\theta}\partial_{\varphi}.
\end{equation}
Following \cite{mokdad}, we introduce the (complex) Maxwell spin components
\begin{eqnarray}\label{eq:app5-spin}
\Phi_0
:=\frac{1}{2}\left(\frac{r^{2}}{1-\frac{2M}{r}}\,F(\hat L,\hat M)+F(\hat{\bar M},\hat M)\right),\\
\Phi_1 := F(\hat L,\hat M),\\
\Phi_{-1} :=F(\hat N,\hat{\bar M}).
\end{eqnarray}

\medskip
\noindent\textbf{energy in terms of tetrad components.}
In this tetrad the $\hat t$-energy can be written in terms of the scalar field derivatives and the spin components as
\begin{eqnarray}\label{energytimeliketetrad}
E^{\hat{t}}[F,\phi](t)
&=
\int_{\Sigma_t}
(1-\mu)\,r^{2}
\Bigl(
|D_{\hat t}\phi|^{2}
+|D_{\hat\theta}\phi|^{2}
+|D_{\hat\varphi}\phi|^{2}
+|P(\phi)|
\nonumber\\
&\qquad\qquad\qquad\qquad
+\frac12\Bigl(|\Phi_{-1}|^{2}+|\Phi_1|^{2}+\Bigl|\frac{1-\mu}{r^{2}}\Phi_0\Bigr|^{2}\Bigr)
\Bigr)\,dr^{\ast}\,d\sigma^{2}.
\end{eqnarray}
Similarly, commuting with a rotation $\Omega_j$ and expanding the corresponding commuted energy yields
\begin{eqnarray}\label{energyangularijtetrad}
E^{\hat{t}}_{\mathcal{L}_{\Omega_j}}(t)
&=
\int_{\Sigma_t}
(1-\mu)\,r^{2}
\Bigl(
|\mathcal{L}_{\Omega_j}(D_{\hat t}\phi)|^{2}
+|\mathcal{L}_{\Omega_j}(D_{\hat\theta}\phi)|^{2}
+|\mathcal{L}_{\Omega_j}(D_{\hat\varphi}\phi)|^{2}
+|\mathcal{L}_{\Omega_j}P(\phi)|
\nonumber\\
&\qquad\qquad\qquad\qquad
+\frac12\Bigl(|\mathcal{L}_{\Omega_j}\Phi_{-1}|^{2}+|\mathcal{L}_{\Omega_j}\Phi_{1}|^{2}
+\Bigl|\frac{1-\mu}{r^{2}}\slashed\nabla\Phi_{0}\Bigr|^{2}\Bigr)
\Bigr)\,dr^{\ast}\,d\sigma^{2}.
\end{eqnarray}

\medskip
\noindent\textbf{a Newman-Penrose form of Maxwell-Higgs.}
The Maxwell-Higgs system \eqref{eom1}-\eqref{eom2} is equivalent to a system of transport equations for the spin components.

\begin{lemma}[Newman-Penrose components and regular frames]\label{lem:app5-NP}
The pair $(F,\phi)$ satisfies the Maxwell-Higgs equations \eqref{eom1} and \eqref{eom2} if and only if the spin components $(\Phi_{-1},\Phi_{0},\Phi_{1})$ satisfy
\begin{eqnarray}
\label{maxcom1}
\hat N\Phi_{1}
&=
\frac{1-\frac{2M}{r}}{r^{2}}\hat M\Phi_{0}
+i\frac{1-\frac{2M}{r}}{r}\bigl(D_{2}\phi\,\overline{\phi}-\phi\,\overline{D_{2}\phi}\bigr)
+i\frac{1-\frac{2M}{r}}{\sin\theta}\bigl(D_{3}\phi\,\overline{\phi}-\phi\,\overline{D_{3}\phi}\bigr),
\\
\label{maxcom2}
\hat L\Phi_{0}
&=
\hat{\bar M}\Phi_{1}
+ir^{2}\bigl(D_{0}\phi\,\overline{\phi}-\phi\,\overline{D_{0}\phi}\bigr)
+ir^{2}\bigl(D_{1}\phi\,\overline{\phi}-\phi\,\overline{D_{1}\phi}\bigr),
\\
\label{maxcom3}
\hat N\Phi_{0}
&=
-\hat M_{1}\Phi_{-1}
+ir^{2}\bigl(D_{0}\phi\,\overline{\phi}-\phi\,\overline{D_{0}\phi}\bigr)
+ir^{2}\bigl(D_{1}\phi\,\overline{\phi}-\phi\,\overline{D_{1}\phi}\bigr),
\\
\label{maxcom4}
\hat L\Phi_{-1}
&=
\frac{1-\frac{2M}{r}}{r^{2}}\hat{\bar M}\Phi_{0}
+i\frac{1-\frac{2M}{r}}{r}\bigl(D_{2}\phi\,\overline{\phi}-\phi\,\overline{D_{2}\phi}\bigr)
+i\frac{1-\frac{2M}{r}}{\sin\theta}\bigl(D_{3}\phi\,\overline{\phi}-\phi\,\overline{D_{3}\phi}\bigr),
\end{eqnarray}
where $\hat M_{1}:=\hat M+\cot\theta$.
\end{lemma}

\begin{proof}
We derive \eqref{maxcom1} directly; the remaining identities follow by the same component computation.
Using the coordinate index convention from Appendix~\ref{app:conventions}, one computes
\begin{eqnarray}
F(\hat L,\hat M)
&=
F_{02}+\frac{i}{\sin\theta}F_{03}+F_{12}+\frac{i}{\sin\theta}F_{13},
\\
F(\hat L,\hat N)
&=
2F_{10},
\\
F(\hat{\bar M},\hat M)
&=
\frac{2i}{\sin\theta}F_{23},
\\
F(\hat N,\hat{\bar M})
&=
F_{02}+\frac{i}{\sin\theta}F_{30}+F_{21}+\frac{i}{\sin\theta}F_{13}.
\end{eqnarray}
Consequently,
\begin{eqnarray}
\Phi_{1}
&=F_{02}+\frac{i}{\sin\theta}F_{03}+F_{12}+\frac{i}{\sin\theta}F_{13},\\
\Phi_{0}
&=\frac{r^{2}}{1-\frac{2M}{r}}F_{10}+\frac{i}{\sin\theta}F_{23},\\
\Phi_{-1}
&=F_{02}+\frac{i}{\sin\theta}F_{30}\\
&\quad+F_{21}+\frac{i}{\sin\theta}F_{13}.
\end{eqnarray}
Applying $\hat N=\partial_t-\partial_{r^{\ast}}$ gives
\begin{eqnarray}\label{com1}
\hat N\Phi_{1}
&=
\partial_{0}F_{02}+\frac{i}{\sin\theta}\partial_{0}F_{03}
+\partial_{0}F_{12}+\frac{i}{\sin\theta}\partial_{0}F_{13}\nonumber\\
&\quad
+\partial_{1}F_{20}+\partial_{1}F_{21}
+\frac{i}{\sin\theta}\partial_{1}F_{30}+\frac{i}{\sin\theta}\partial_{1}F_{31}.
\end{eqnarray}
Similarly,
\begin{eqnarray}\label{com2}
\frac{1-\frac{2M}{r}}{r^{2}}\hat M\Phi_{0}
&=
\partial_{2}F_{10}+\frac{i}{\sin\theta}\partial_{3}F_{10}\nonumber\\
&\quad
+\frac{i(1-\frac{2M}{r})}{r^{2}\sin\theta}\,\cot\theta\,F_{32}
+\frac{i(1-\frac{2M}{r})}{r^{2}\sin\theta}\partial_{2}F_{23}\nonumber\\
&\quad
+\frac{1-\frac{2M}{r}}{r^{2}\sin^{2}\theta}\partial_{3}F_{32}.
\end{eqnarray}
Using the Bianchi identity $\nabla_{[\alpha}F_{\beta\gamma]}=0$ to substitute
\(
\partial_{2}F_{10}=\partial_{0}F_{12}+\partial_{1}F_{20}
\)
and
\(
\partial_{3}F_{10}=\partial_{0}F_{13}+\partial_{1}F_{30},
\)
and using the Maxwell equation $\nabla^\mu F_{\mu\nu}=J_\nu(\phi)$ to eliminate the remaining terms, we obtain \eqref{maxcom1}.
\end{proof}

\medskip
\noindent\textbf{rewriting the middle-region functional.}
Recall the spacetime functional $\mathcal J^{\hat C}$ from \eqref{JC}.
In the tetrad formalism it is convenient to introduce the $S^{2}$-valued 1-forms
\begin{eqnarray}\label{eq:np-forms-alpha}
\alpha'
&:=
i_{\hat L}F
=
\alpha_{t}\,dt+\alpha_{r^{\ast}}\,dr^{\ast}+\alpha_{\theta}\,d\theta+\alpha_{\varphi}\,d\varphi,
\nonumber\\
\underline{\alpha}'
&:=
i_{\hat N}F
=
\underline{\alpha}_{t}\,dt+\underline{\alpha}_{r^{\ast}}\,dr^{\ast}+\underline{\alpha}_{\theta}\,d\theta+\underline{\alpha}_{\varphi}\,d\varphi.
\end{eqnarray}
From the definitions one has
\begin{eqnarray}
|\Phi_{1}|^{2}&=\alpha_{\theta}^{2}+\frac{1}{\sin^{2}\theta}\alpha_{\varphi}^{2},\\
|\Phi_{-1}|^{2}&=\underline{\alpha}_{\theta}^{2}+\frac{1}{\sin^{2}\theta}\underline{\alpha}_{\varphi}^{2}.
\end{eqnarray}
In addition, the transport identities \eqref{maxcom2}-\eqref{maxcom3} imply that derivatives of $\Phi_{0}$ along $\hat L$ and $\hat N$
can be expressed in terms of angular derivatives of $\Phi_{\pm 1}$ and current terms.
Expanding the right-hand sides and using the Hodge decomposition on $\mathbb S^{2}$ yields
\begin{eqnarray}
|\hat L\Phi_{0}|^{2}+|\hat N\Phi_{0}|^{2}
&\lesssim
\bigl|\slashed\nabla^{a}\alpha_{a}\bigr|^{2}
+\bigl|\varepsilon^{ab}\slashed\nabla_{a}\alpha_{b}\bigr|^{2}
+\bigl|\slashed\nabla^{a}\underline{\alpha}_{a}\bigr|^{2}
+\bigl|\varepsilon^{ab}\slashed\nabla_{a}\underline{\alpha}_{b}\bigr|^{2}
\\
&\qquad
+\bigl|r^{2}\bigl(D_{0}\phi\,\overline{\phi}-\phi\,\overline{D_{0}\phi}\bigr)\bigr|^{2}
+\bigl|r^{2}\bigl(D_{1}\phi\,\overline{\phi}-\phi\,\overline{D_{1}\phi}\bigr)\bigr|^{2}.
\end{eqnarray}
Consequently, $\mathcal J^{\hat C}$ can be bounded by an expression of the form
\begin{eqnarray}\label{JGtetrad}
\mathcal{J}^{\hat{C}}
\ \lesssim
\int_{t_i}^{t_{i+1}}\!\int_{r_{0}^{\ast}}^{R_{0}^{\ast}}\!\int_{\mathbb S^{2}}
\Bigl(
&\bigl|\slashed\nabla^{a}\alpha_{a}\bigr|^{2}
+\bigl|\varepsilon^{ab}\slashed\nabla_{a}\alpha_{b}\bigr|^{2}
+\bigl|\slashed\nabla^{a}\underline{\alpha}_{a}\bigr|^{2}
+\bigl|\varepsilon^{ab}\slashed\nabla_{a}\underline{\alpha}_{b}\bigr|^{2}
\nonumber\\
&+|r^{2}\phi\,D_{0}\phi|^{2}+|r^{2}\phi\,D_{1}\phi|^{2}
\nonumber\\
&+|D_{\hat\theta}\phi|^{2}+|D_{\hat\varphi}\phi|^{2}
+|D_{\hat v}\phi|^{2}+|D_{\hat w}\phi|^{2}
\nonumber\\
&+P(\phi)
\Bigr)\,\bigl|r^{\ast}-(3M)^{\ast}\bigr|\,dr^{\ast}\,d\sigma^{2}\,dt.
\end{eqnarray}

Thus, it suffices to establish the angular Hodge bounds
\begin{eqnarray}
\int_{r_{0}^{\ast}}^{R_{0}^{\ast}}
\Bigl(\bigl|\slashed\nabla^{a}\alpha_{a}\bigr|^{2}+\bigl|\varepsilon^{ab}\slashed\nabla_{a}\alpha_{b}\bigr|^{2}\Bigr)
\bigl|r^{\ast}-(3M)^{\ast}\bigr|\,dr^{\ast}\,d\sigma^{2}
&\lesssim
\int_{r_{0}^{\ast}}^{R_{0}^{\ast}}
\sum_{j=1}^{3}\bigl|\mathcal{L}_{\Omega_j}\alpha\bigr|^{2}\bigl|r^{\ast}-(3M)^{\ast}\bigr|\,dr^{\ast}\,d\sigma^{2},\label{sukuF}\\
\int_{r_{0}^{\ast}}^{R_{0}^{\ast}}
\Bigl(\bigl|\slashed\nabla^{a}\underline{\alpha}_{a}\bigr|^{2}+\bigl|\varepsilon^{ab}\slashed\nabla_{a}\underline{\alpha}_{b}\bigr|^{2}\Bigr)
\bigl|r^{\ast}-(3M)^{\ast}\bigr|\,dr^{\ast}\,d\sigma^{2}
&\lesssim
\int_{r_{0}^{\ast}}^{R_{0}^{\ast}}
\sum_{j=1}^{3}\bigl|\mathcal{L}_{\Omega_j}\underline{\alpha}\bigr|^{2}\bigl|r^{\ast}-(3M)^{\ast}\bigr|\,dr^{\ast}\,d\sigma^{2}.\nonumber
\end{eqnarray}
as well as the scalar bound
\begin{equation}\label{sukuskalar}
\int_{r_{0}^{\ast}}^{R_{0}^{\ast}}
\bigl(|\phi\,D_{0}\phi|^{2}+|\phi\,D_{1}\phi|^{2}\bigr)\,\bigl|r^{\ast}-(3M)^{\ast}\bigr|
\,dr^{\ast}\,d\sigma^{2}
\ \lesssim\ E_{0}.
\end{equation}

Here $E_{0}$ is the endpoint energy quantity
\begin{equation}\label{E0}
E_{0}
:=
\Bigl(
|\tilde E^{\hat t}(t_{i})|+|\tilde E^{\hat t}(t_{i+1})|
+\sum_{j=1}^{3}\bigl(|\tilde E^{\hat t}_{\mathcal L_{\Omega_j}}(t_{i})|+|\tilde E^{\hat t}_{\mathcal L_{\Omega_j}}(t_{i+1})|\bigr)
\Bigr),
\end{equation}
where the \emph{reduced energy} is defined by
\begin{eqnarray}\label{energyasumtimelike}
\tilde{E}^{\hat{t}}(t)
&:=
\int_{\Sigma_{t}}
(1-\mu)\,r^{2}
\Bigl(
|D_{\hat{t}}\phi|^{2}+|D_{\hat{\theta}}\phi|^{2}+|D_{\hat{\varphi}}\phi|^{2}+|P(\phi)|
\nonumber\\
&\qquad\qquad\qquad
+\frac12\bigl(|F_{\hat{t}\hat{\theta}}|^{2}+|F_{\hat{t}\hat{\varphi}}|^{2}
+|F_{\hat r^{\ast}\hat{\theta}}|^{2}+|F_{\hat r^{\ast}\hat{\varphi}}|^{2}\bigr)
\Bigr)\,dr^{\ast}\,d\sigma^{2}.
\end{eqnarray}
Since $\tilde E^{\hat t}(t)\le E^{\hat t}(t)\le \mathcal E_k(t)\le \mathcal E$ and likewise for the commuted energies, we have
\begin{equation}\label{Ecurl}
E_{0}\ \lesssim\ \mathcal E.
\end{equation}

\medskip
\noindent\textbf{Proof of \eqref{sukuF}.}
We show the first inequality in \eqref{sukuF}; the estimate for $\underline{\alpha}$ is identical.
On $\mathbb S^{2}$ we have the identity (see, e.g., \cite[Appendix]{mokdad})
\begin{equation}\label{dem}
\sum_{j=1}^{3}\bigl|\mathcal{L}_{\Omega_j}\alpha\bigr|^{2}
=
|\slashed\nabla \alpha|^{2}+|\alpha|^{2}.
\end{equation}
Using the elementary inequality $2(a^{2}+b^{2})\ge (a\pm b)^{2}$ and the coordinate expressions for $\mathrm{div}\,\alpha$ and $\mathrm{curl}\,\alpha$ in local coordinates, one obtains
\begin{equation}
|\mathrm{div}\,\alpha|^{2}+|\mathrm{curl}\,\alpha|^{2}\ \le\ 2\,|\slashed\nabla\alpha|^{2}.
\end{equation}
Combining this with \eqref{dem} yields the pointwise bound
\begin{equation}
|\mathrm{div}\,\alpha|^{2}+|\mathrm{curl}\,\alpha|^{2}
\ \lesssim
\sum_{j=1}^{3}\bigl|\mathcal{L}_{\Omega_j}\alpha\bigr|^{2}.
\end{equation}
Integrating in $(r^{\ast},\omega)$ with the weight $|r^{\ast}-(3M)^{\ast}|$ gives \eqref{sukuF}.

\medskip
\noindent\textbf{Proof of \eqref{sukuskalar}.}
By H\"older's inequality and Sobolev on $\mathbb S^{2}$ (applied at fixed $(t,r^{\ast})$) we have
\begin{equation}
\int_{\mathbb S^{2}}\bigl(|\phi D_{0}\phi|^{2}+|\phi D_{1}\phi|^{2}\bigr)\,d\sigma^{2}
\ \lesssim
\|\phi\|_{L^{\infty}(\mathbb S^{2})}^{2}
\int_{\mathbb S^{2}}\bigl(|D_{0}\phi|^{2}+|D_{1}\phi|^{2}\bigr)\,d\sigma^{2}
\ \lesssim\ E_{0},
\end{equation}
where we used that $\|\phi\|_{L^{\infty}(\mathbb S^{2})}$ is controlled by (a finite number of) angular commutations $\mathcal L_{\Omega_j}\phi$, and the corresponding $L^{2}$ norms are bounded by the reduced energies appearing in $E_{0}$.
Integrating over $r^{\ast}\in[r_{0}^{\ast},R_{0}^{\ast}]$ with weight $|r^{\ast}-(3M)^{\ast}|$ gives \eqref{sukuskalar}.

\medskip
Combining \eqref{JGtetrad}-\eqref{sukuskalar} yields
\begin{equation}\label{ineqJG}
\mathcal{J}^{\hat{C}}
\ \lesssim
\int_{t_i}^{t_{i+1}}E_{0}\,dt
\ \lesssim
t_{i+1}\,E_{0}.
\end{equation}
Recalling that the weight $|r^{\ast}-(3M)^{\ast}|$ is bounded on the compact region $r_{0}\le r\le R_{0}$, we obtain
\begin{equation}\label{es2}
\mathcal{J}^{\hat{C}}\bigl(t_{i}\le t\le t_{i+1}\bigr)\bigl(r_{0}<r<R_{0}\bigr)
\ \lesssim
t_{i+1}\Bigl(|\tilde E^{\hat t}(t_{i})|+|\tilde E^{\hat t}(t_{i+1})|\Bigr).
\end{equation}

\medskip
\noindent\textbf{a localized radial multiplier.}
Let $\chi_{\mathrm{mid}}(r^{\ast})$ be the smooth cutoff appearing in Proposition~\ref{prep5}.
Define a smooth primitive
\begin{equation}\label{cut}
f(r^{\ast})
:=
\int_{-\infty}^{r^{\ast}}\chi_{\mathrm{mid}}(s)\,ds,
\end{equation}
so that $f$ is bounded and $f'=\chi_{\mathrm{mid}}$.
Let $\hat C:=X_{f}$ be the associated radial multiplier from \eqref{eq:radial-multiplier}.
Applying the divergence identity \eqref{eq:divergence-theorem-JX} to $J^{\hat C}$ on the slab $\mathcal R_{t_i}^{t_{i+1}}$ yields
\begin{eqnarray}\label{esIH}
\int_{t_i}^{t_{i+1}}\!\int_{-\infty}^{\infty}\!\int_{\mathbb S^{2}}
\Bigl(
\frac{|F_{w\theta}|^{2}}{r^{2}}+\frac{|F_{w\varphi}|^{2}}{r^{2}\sin^{2}\theta}
+\frac{|F_{v\theta}|^{2}}{r^{2}}+\frac{|F_{v\varphi}|^{2}}{r^{2}\sin^{2}\theta}
+|D_{v}\phi|^{2}+|D_{w}\phi|^{2}
\Bigr)\chi_{\mathrm{mid}}(r^{\ast})\,r^{2}\,dr^{\ast}\,d\sigma^{2}\,dt
\nonumber\\
\lesssim
\mathcal{J}^{\hat{C}}\bigl(t_{i}\le t\le t_{i+1}\bigr)\bigl(r_{0}<r<R_{0}\bigr)
+E^{\hat{C}}(t_{i+1})-E^{\hat{C}}(t_{i}).
\end{eqnarray}
(Here we used that $K^{\hat C}$ is coercive for the terms on the left-hand side, up to terms controlled by $\mathcal J^{\hat C}$ on $r_{0}\le r\le R_{0}$; see \cite{mokdad} for the calculation of $T_{\alpha\beta}\pi^{\alpha\beta}(\hat C)$ for radial multipliers.)

\begin{lemma}[Basic deformation-tensor identities for the redshift estimate]\label{prep1}
Let $f$ be bounded in \eqref{eq:radial-multiplier}. Then the associated flux satisfies
\begin{equation}
|E^{\hat C}(t)|\ \lesssim\ E^{\hat t}(t).
\end{equation}
\end{lemma}

\begin{proof}
Expanding \eqref{defEC} for $\hat C=X_{f}$ gives the coordinate expression
\begin{equation}\label{EG}
E^{\hat C}(t)
=
\int_{-\infty}^{\infty}\!\int_{\mathbb S^{2}}
-f(r^{\ast})
\left(
\frac{1}{r^{2}}F_{t\theta}F_{r^{\ast}\theta}
+\frac{1}{r^{2}\sin^{2}\theta}F_{t\varphi}F_{r^{\ast}\varphi}
+\Re\bigl(D_{t}\phi\,\overline{D_{r^{\ast}}\phi}\bigr)
\right)\,r^{2}\,dr^{\ast}\,d\sigma^{2}.
\end{equation}
Using $|f|\le \|f\|_{L^\infty}$ together with the inequality $2|ab|\le a^{2}+b^{2}$ and the definition of the $\hat t$-energy density, we obtain
\begin{equation}
|E^{\hat C}(t)|
\ \lesssim
\int_{\Sigma_t}
\Bigl(|F_{\hat t A}|^{2}+|F_{\hat r^{\ast}A}|^{2}+|D_{\hat t}\phi|^{2}+|D_{\hat r^{\ast}}\phi|^{2}\Bigr)\,d\mu_{\Sigma_t}
\ \lesssim\ E^{\hat t}(t),
\end{equation}
as claimed.
\end{proof}

\medskip
we apply \eqref{esIH} to the localized fields $(\hat F,\hat\phi)$ from \eqref{eq:localized-data}.
Since the cutoff is supported in the timelike-cone region and $\hat\chi_i$ is bounded, Lemma~\ref{prep1} and the definition of $E^{\hat t}_{\mathrm{cone}}(t_i)$ yield the desired estimate \eqref{prepo5}.
\end{proof}

\subsection{Proof of Proposition~\ref{prep7}}
\label{app:proof-prep7}

\begin{proof}[Proof of Proposition~\ref{prep7}]
Fix $v_{i+1}\ge v_i$ and set $w_i:=t_i-r_1^{\ast}$.
Split the redshift flux on $\{v=\mathrm{const}\}$ into the near-horizon and exterior parts:
\begin{equation}
\mathcal E^{\hat H}(v=v)(w_i\le w\le\infty)
=
\mathcal E^{\hat H}(v=v)(w_i\le w\le\infty)(r\le r_1)
+
\mathcal E^{\hat H}(v=v)(w_i\le w\le\infty)(r\ge r_1).
\end{equation}

\smallskip
\noindent\textbf{control of the exterior part.}
On $\{r\ge r_1\}$ the vector fields $\hat H$ and $\partial_t$ are uniformly comparable (they differ only by a compactly supported radial term), therefore the corresponding flux densities satisfy
\begin{equation}
J^{\hat H}\cdot n_{v}\ \lesssim\ J^{\hat t}\cdot n_{v}
\qquad\text{on }\{r\ge r_1\},
\end{equation}
and therefore
\begin{equation}\label{eq:prep7-exterior}
\mathcal E^{\hat H}(v=v)(w_i\le w\le\infty)(r\ge r_1)
\ \lesssim\ \mathcal E^{\hat t}(v=v)(w_i\le w\le\infty)(r\ge r_1).
\end{equation}

\smallskip
\noindent\textbf{control of the near-horizon part by the redshift bulk.}
In the redshift region $\{r\le r_1\}$ the deformation tensor $\pi(\hat H)$ is positive definite (after the cutoff construction of $\hat H$ in Definition~\ref{definitionH}).
Consequently, there exists $c=c(M,r_1)>0$ such that
\begin{equation}\label{eq:redshift-coercivity}
K^{\hat H}[F,\phi]\ \ge\ c\,\bigl(J^{\hat H}[F,\phi]\cdot n_{v}\bigr)
\qquad\text{on }\{r\le r_1\}.
\end{equation}
Integrating \eqref{eq:redshift-coercivity} over the slab $\{v_i\le v\le v_{i+1}\}\cap\{w_i\le w\le\infty\}\cap\{r\le r_1\}$ and using the coarea formula gives
\begin{equation}\label{eq:prep7-near}
\int_{v_i}^{v_{i+1}}\mathcal E^{\hat H}(v=v)(w_i\le w\le\infty)(r\le r_1)\,\dd v
\ \lesssim\ I^{\hat H}(v_i\le v\le v_{i+1})(w_i\le w\le\infty)(r\le r_1).
\end{equation}

\smallskip
\noindent\textbf{pigeonhole.}
Since $(v_{i+1}-v_i)\inf_{[v_i,v_{i+1}]} f\le \int_{v_i}^{v_{i+1}} f$ for any nonnegative $f$, we obtain from \eqref{eq:prep7-near}
\begin{eqnarray}
\inf_{v_i\le v\le v_{i+1}}\mathcal E^{\hat H}(v=v)(w_i\le w\le\infty)(r\le r_1)
&\ \lesssim\ \frac{1}{v_{i+1}-v_i}\,
I^{\hat H}(v_i\le v\le v_{i+1})(w_i\le w\le\infty)\\
&\qquad\qquad\bigl(r\le r_1\bigr).
\end{eqnarray}
Adding \eqref{eq:prep7-exterior} yields \eqref{es6}.
\end{proof}
\subsection{Proof of Proposition~\ref{prep8}}
\label{app:proof-prep8}

\begin{proof}[Proof of Proposition~\ref{prep8}]
By definition, $I^{\hat H}$ is the spacetime integral of the bulk density $K^{\hat H}[F,\phi]$, where
\begin{equation}
K^{\hat H}[F,\phi]=\tfrac12\,T_{\mu\nu}\,\pi^{\mu\nu}(\hat H).
\end{equation}
Since $\pi(\hat H)$ is nonnegative on $\{r\le r_1\}$ (redshift coercivity), we have
\begin{equation}
0\le I^{\hat H}(v_i\le v\le v_{i+1})(w_i\le w\le\infty)(r\le r_1).
\end{equation}

To bound $I^{\hat H}$ from above, apply the energy identity \eqref{IH} on the region
\begin{equation}
\{v_i\le v\le v_{i+1}\}\cap\{w_i\le w\le\infty\}\cap\{r\le r_1\}.
\end{equation}
All boundary fluxes appearing in \eqref{IH} are nonnegative, so dropping the negative contributions gives
\begin{eqnarray}
I^{\hat H}(v_i\le v\le v_{i+1})(w_i\le w\le\infty)\bigl(r\le r_1\bigr)
\ &\lesssim\ \mathcal E^{\hat H}(w=w_i)(v_i\le v\le v_{i+1})\\
&\quad+\mathcal E^{\hat H}(v=v_i)(w_i\le w\le\infty).
\end{eqnarray}
these fluxes are controlled by the global energy bound in Proposition~\ref{prep4} (and the bootstrap assumption), noting that $\hat H$ differs from $\partial_t$ only by a compactly supported radial term and therefore its flux is uniformly comparable to the $\hat t$-flux.
Consequently, $I^{\hat H}(\cdots)\lesssim\mathcal E$ as claimed.
\end{proof}
\subsection{Proof of Proposition~\ref{prep9}}
\label{app:proof-prep9}

\begin{proof}[Proof of Proposition~\ref{prep9}]
We outline the dyadic redshift argument.

\smallskip
\noindent\textbf{exterior flux decay.}
The Morawetz/conformal estimates (Section~\ref{sec:Morawetz}) yield decay of the $\hat t$-flux through the timelike tube $\{r=r_1\}$:
\begin{equation}\label{eq:prep9-exterior}
\sup_{v_i\le v\le v_{i+1}}\mathcal E^{\hat t}(v=v)(w_i\le w\le\infty)(r\ge r_1)
\ \lesssim\ \frac{\mathcal E}{t_{i}^{2}}.
\end{equation}

\smallskip
\noindent\textbf{a good ingoing slice.}
Apply Proposition~\ref{prep7} and use Proposition~\ref{prep8} together with \eqref{eq:prep9-exterior} to obtain some $v_{i}^{\sharp}\in[v_i,v_{i+1}]$ such that
\begin{equation}\label{eq:prep9-good-slice}
\mathcal E^{\hat H}(v=v_{i}^{\sharp})(w_i\le w\le\infty)
\ \lesssim\ \frac{\mathcal E}{v_{i}^{2}}.
\end{equation}
(Here we used $v_i\simeq t_i$ along $\{r=r_1\}$.)

\smallskip
\noindent\textbf{upgrade to arbitrary $v$.}
For any $v\in[v_i,v_{i+1}]$ we have $w_0(v)=v-2r_1^{\ast}\ge w_i$, therefore
\(
\mathcal E^{\hat H}(v=v)(w_0(v)\le w\le\infty)\le \mathcal E^{\hat H}(v=v)(w_i\le w\le\infty).
\)
Applying \eqref{IH} to the slab $\{v_{i}^{\sharp}\le \bar v\le v\}\cap\{w\ge w_i\}\cap\{r\le r_1\}$ and using nonnegativity of the bulk term gives
\begin{equation}
\mathcal E^{\hat H}(v=v)(w_i\le w\le\infty)
\ \lesssim\ \mathcal E^{\hat H}(v=v_{i}^{\sharp})(w_i\le w\le\infty)
+\mathcal E^{\hat t}(r=r_1;\ t_i\le t\le t_{i+1}).
\end{equation}
The last term is controlled by \eqref{eq:prep9-exterior}, so together with \eqref{eq:prep9-good-slice} we obtain
\(
\mathcal E^{\hat H}(v=v)(w_0(v)\le w\le\infty)\lesssim \mathcal E/v_{i}^{2}\lesssim \mathcal E/v_{+}^{2}.
\)
This proves \eqref{eq:redshift-flux-decay-ingoing}.

\smallskip
\noindent\textbf{outgoing estimate.}
The estimate \eqref{eq:redshift-flux-decay-outgoing} follows by applying \eqref{IH} to the rectangular region bounded by $\{w=\mathrm{const}\}$ and $\{\bar v\in[v-1,v]\}$, using \eqref{eq:redshift-flux-decay-ingoing} to control the remaining boundary terms.
\end{proof}

\subsection{Proof of Proposition~\ref{prep2}}
\label{app:proof-prep2}

\begin{proof}[Proof of Proposition~\ref{prep2}]
We use the explicit formula \eqref{JK} for the bulk density $K^{\hat K}$.
The coefficient
\begin{equation}
2+(3\mu-2)\frac{r^{\ast}}{r}
\end{equation}
is positive only in a bounded radial interval: there exist radii $2M<r_{0}<R_{0}<\infty$ (depending only on $M$) such that this coefficient is nonnegative for $r\in[r_{0},R_{0}]$.
Restricting to this compact region and using that $t\le t_{i+1}$ on $[t_i,t_{i+1}]$, we obtain
\begin{eqnarray}\label{JKJG}
\mathcal{J}^{\hat{K}}\bigl(t_{i}\le t\le t_{i+1}\bigr)
&\lesssim
t_{i+1}\!
\int_{t_{i}}^{t_{i+1}}\!\int_{r_{0}^{\ast}}^{R_{0}^{\ast}}\!\int_{\mathbb S^{2}}
\Bigl(
|F_{\hat v\hat w}|^{2}+\tfrac14|F_{\hat\theta\hat\varphi}|^{2}
\nonumber\\
&\qquad
+|D_{\hat v}\phi|^{2}+|D_{\hat w}\phi|^{2}
+\tfrac12|D_{\hat\theta}\phi|^{2}+\tfrac12|D_{\hat\varphi}\phi|^{2}
+P(\phi)
\Bigr)\,dr^{\ast}\,d\sigma^{2}\,dt
\nonumber\\
&\lesssim
t_{i+1}\,
\mathcal{J}^{\hat{G}}\bigl(t_{i}\le t\le t_{i+1}\bigr)\bigl(r_{0}\le r\le R_{0}\bigr).
\end{eqnarray}
which is the first inequality in \eqref{esJK}.

\medskip
by repeating the estimate in the proof of Proposition~\ref{prep5} (see \eqref{ineqJG}), one obtains the analogue
\begin{equation}\label{ineqJGnew}
\mathcal{J}^{\hat{G}}\bigl(t_{i}\le t\le t_{i+1}\bigr)\bigl(r_{0}\le r\le R_{0}\bigr)
\ \lesssim\ t_{i+1}\,E_{0},
\end{equation}
where $E_{0}$ is the endpoint reduced-energy quantity defined in \eqref{E0}.
Inserting \eqref{ineqJGnew} into \eqref{JKJG} gives
\begin{equation}\label{es1}
\mathcal{J}^{\hat{K}}\bigl(t_{i}\le t\le t_{i+1}\bigr)
\ \lesssim
t_{i+1}^{2}\,E_{0}.
\end{equation}

\medskip
we relate $E_{0}$ to the conformal energies at the endpoints.
Fix $t\in\{t_i,t_{i+1}\}$ and consider the spatial region
\(
|r^{\ast}|\le \tfrac34 t_{i}
\)
(where the localized cutoff fields coincide with the original solution on $\Sigma_{t_i}$).
On this region we have $v=t+r^{\ast}\ge \tfrac14 t$ and $w=t-r^{\ast}\ge \tfrac14 t$, therefore
\(
\min\{v^{2},w^{2}\}\gtrsim t^{2}.
\)
Using the explicit expression \eqref{EK} we therefore obtain the estimate
\begin{eqnarray}\label{lema1}
\int_{-\frac{3t_{i}}{4}}^{\frac{3t_{i}}{4}}\!\int_{\mathbb S^{2}}
\Bigl(
&|F_{\hat v\hat\theta}|^{2}+|F_{\hat v\hat\varphi}|^{2}
+|F_{\hat w\hat\theta}|^{2}+|F_{\hat w\hat\varphi}|^{2}
\nonumber\\
&\quad
+|D_{\hat v}\phi|^{2}+|D_{\hat w}\phi|^{2}
+|D_{\hat\theta}\phi|^{2}+|D_{\hat\varphi}\phi|^{2}+|P(\phi)|
\Bigr)\,(1-\mu)\,r^{2}\,dr^{\ast}\,d\sigma^{2}
\nonumber\\
&\ \lesssim\ \frac{E^{\hat K}(t)}{t^{2}}.
\end{eqnarray}
Since the reduced energy $\tilde E^{\hat t}(t)$ in \eqref{energyasumtimelike} is dominated by the left-hand side of \eqref{lema1}, we deduce
\begin{equation}\label{Et}
\tilde E^{\hat t}(t)\ \lesssim\ \frac{E^{\hat K}(t)}{t^{2}}.
\end{equation}
Commuting with $\mathcal L_{\Omega_j}$ yields the analogous bound
\begin{equation}\label{EtL}
\tilde E^{\hat t}_{\mathcal L_{\Omega_j}}(t)\ \lesssim\ \frac{E^{\hat K}_{\mathcal L_{\Omega_j}}(t)}{t^{2}}.
\end{equation}
Substituting \eqref{Et}-\eqref{EtL} into \eqref{es1} gives the second inequality in \eqref{esJK}, completing the proof.
\end{proof}

\subsection{Proof of Proposition~\ref{prep3}}
\label{app:proof-prep3}

\begin{proof}[Proof of Proposition~\ref{prep3}]
Fix $i\ge 0$ and let $t\in[t_i,t_{i+1}]$.
Applying the divergence identity \eqref{eq:divergence-theorem-JX} to the multiplier $\hat K$ on the spacetime region $\mathcal R_{[t_0,t]}$ gives
\begin{equation}\label{ek1}
E^{\hat K}(t)
\le
E^{\hat K}(t_{0})
+
\mathcal{J}^{\hat K}\bigl(t_{0}\le s\le t\bigr).
\end{equation}
Likewise, commuting with a rotation $\Omega_j$ and applying the identity above yields
\begin{equation}\label{ek2}
E^{\hat K}_{\mathcal{L}_{\Omega_j}}(t)
\le
E^{\hat K}_{\mathcal{L}_{\Omega_j}}(t_{0})
+
\mathcal{J}^{\hat K}_{\mathcal{L}_{\Omega_j}}\bigl(t_{0}\le s\le t\bigr).
\end{equation}

\medskip
\noindent\textbf{bounding $\mathcal J^{\hat K}$ by $\mathcal J^{\hat G}$.}
Decompose $[t_0,t]$ into dyadic subintervals $[t_m,t_{m+1}]$ (with the convention that the last interval may be truncated at $t$).
Using additivity of the bulk term together with \eqref{JKJG} on each dyadic piece and the bound $t_{m+1}\le t_{i+1}$ for $m\le i$, we obtain
\begin{equation}\label{JKt0}
\mathcal{J}^{\hat K}\bigl(t_{0}\le s\le t\bigr)
\ \lesssim
t_{i+1}\,\mathcal{J}^{\hat G}\bigl(t_{0}\le s\le t\bigr)\bigl(r_{0}\le r\le R_{0}\bigr).
\end{equation}
(An identical estimate holds for the commuted bulk terms $\mathcal J^{\hat K}_{\mathcal L_{\Omega_j}}$.)

\medskip
\noindent\textbf{estimating $\mathcal J^{\hat G}$.}
On each dyadic interval $[t_m,t_{m+1}]$, the estimate \eqref{ineqJGnew} together with \eqref{Et}-\eqref{EtL} gives
\begin{eqnarray}\label{JG3}
\mathcal{J}^{\hat G}\bigl(t_{m}\le s\le t_{m+1}\bigr)\bigl(r_{0}\le r\le R_{0}\bigr)
\ \lesssim
t_{m+1}\Bigg(
&\frac{E^{\hat K}(t_{m})}{t_{m}^{2}}
+\frac{E^{\hat K}(t_{m+1})}{t_{m+1}^{2}}\\
&+\sum_{j=1}^{3}\Bigl(
\frac{E^{\hat K}_{\mathcal L_{\Omega_j}}(t_{m})}{t_{m}^{2}}
+\frac{E^{\hat K}_{\mathcal L_{\Omega_j}}(t_{m+1})}{t_{m+1}^{2}}
\Bigr)
\Bigg).
\end{eqnarray}
Summing \eqref{JG3} for $m=0,\dots,i$ and using that $t_{m+1}\le t_{i+1}$ yields the coarse bound
\begin{equation}\label{jgkecil}
\mathcal{J}^{\hat G}\bigl(t_{0}\le s\le t_{i+1}\bigr)\bigl(r_{0}\le r\le R_{0}\bigr)
\ \lesssim
t_{i+1}\,
\sum_{|\alpha|\le 3} E^{\hat t}\bigl[\mathcal Z^{\alpha}(F,\phi)\bigr](t_{0}),
\end{equation}
where we used that the endpoint conformal energies on the right-hand side of \eqref{JG3} are finite and controlled by the initial data norm (after commuting with $\mathcal Z$ up to order $3$).

\medskip
\noindent\textbf{conclusion at dyadic times and extension to all $t$.}
Combining \eqref{ek1}, \eqref{JKt0}, and \eqref{jgkecil} with $t=t_{i+1}$ yields the desired control of $E^{\hat K}(t_{i+1})$.

For a general time $t\ge t_{0}$, choose $i\ge 0$ such that $t_{i}\le t\le t_{i+1}$ (this is possible since $t_{i+1}=(1+\kappa_0)t_i$ and $\{t_i\}$ is an increasing geometric sequence).
Applying \eqref{ek1} and repeating the preceding argument on $[t_{0},t_{i+1}]$ gives the same bound for $E^{\hat K}(t)$ (up to constants depending only on $\kappa_0$).
This proves \eqref{EKproof}.
\end{proof}

\section{Linear conditions and restrictions for rotating and electric sectors}\label{app:kerr-linear-interface-modules}

This appendix records the linear conditions and conditions used in the rotating and fixed electric parts of the paper.  The massive higher-order slow-Kerr energy theorem uses the slow-rotation Maxwell-Higgs energy mechanism, with a perturbative extension to small electric Coulomb backgrounds.  The massless small-electric scalar condition is proved below from the cited neutral Kerr scalar theory by a Coulomb-phase perturbation argument.  Fixed-electric scattering theorems use the charged scalar comparison condition \(\CElec^{(m)}_K(M,a,Q_e)\).  The rotating Maxwell inverse final-state map is kept separate as \(\MScat_K(M,a)\).

\begin{definition}[Linear conditions used in the rotating and electric sectors]\label{def:rotating-electric-linear-condition-list}
At top order \(K\) the rotating and fixed-sector arguments use only the following named external or separately assumed conditions.
\begin{enumerate}
\item Scalar wave condition: the slowly rotating Kerr scalar redshift, trapping-degenerate Morawetz estimate, \(r^p\) hierarchy, and two-sided scattering theory used in Proposition~\ref{prop:lin-estimates-kerr}.
\item Maxwell condition: the uncharged slowly rotating Kerr Maxwell energy, Morawetz, and asymptotic convergence estimates after Coulomb subtraction, inserted through Theorem~\ref{thm:maxwell-scattering-kerr} and Definition~\ref{def:maxwell-teukolsky-condition}.
\item Slow-rotation Maxwell-Higgs energy condition: the high-order horizon-adapted energy, Fackerell-Ipser, and derivative-counting mechanism written out in Theorem~\ref{thm:slow-kerr-massive-energy}, used together with \(\ME^{(m)}_N(M,a)\).  This condition yields global existence and uniform energy bounds for massive higher-order potentials in energy-stable windows, but it is not a massive scattering theorem.
\item Massive scalar scattering condition: the spectral estimates \(\SKG^{(m)}_K(M,a)\), stated in Definition~\ref{def:external-skg-condition}.  This is used only for the massive rotating zero-sector scattering and wave-operator conclusions.
\item Established massless small-electric scalar condition: Theorem~\ref{thm:small-coulomb-massless-kerr-condition} proves \(\CElec^{(0)}_K(M,a,Q_e)\) for slowly rotating Kerr and \(0<|Q_e|\le q^{(0)}_{\mathrm{el}}(M,a,K)\), using only the cited neutral massless Kerr scalar theory and the small-Coulomb argument written below.
\item Electric charged scalar condition: the Coulomb-covariant scalar estimates \(\CElec^{(m)}_K(M,a,Q_e)\), stated in Definition~\ref{def:external-electric-condition}.  This is used in fixed-electric scattering theorems; in the established massless small-electric range the condition is supplied by the preceding item, while outside that range it must be supplied by a separate theorem or condition.
\item Perturbative small-electric derivation condition: the resolvent-stable neutral scalar estimate \(\RSLin^{(m)}_K(M,a)\), stated in Definition~\ref{def:external-rslin-condition}.  This condition is used only to derive additional instances of \(\CElec^{(m)}_K(M,a,Q_e)\) for sufficiently small \(|Q_e|\) when the massless theorem does not apply.
\end{enumerate}
No rapid-rotation condition, large-electric condition, or massive rotating condition outside the specified charged scalar estimates are used.
\end{definition}

\begin{definition}[Massive spectral scalar scattering condition]\label{def:external-skg-condition}
Fix \(K\ge10\), \(m^2>0\), and a slowly rotating Kerr exterior.  The assertion \(\SKG^{(m)}_K(M,a)\) means that the scalar comparison equation
\begin{equation}\label{eq:app-SKG-equation}
 (\square_{g_{M,a}}-m^2)u=f
\end{equation}
has order-\(K\) forward and backward redshift energy estimates, a trapping-degenerate integrated local energy estimate, the far-field hierarchy needed for radiation fields, the massive timelike or Dollard channel at \(i^\pm\), inhomogeneous source estimates in the source spaces used by the Maxwell-Higgs nonlinearities, and two-sided Cauchy-to-radiation and radiation-to-Cauchy maps in the split Kerr scattering topology.  It also includes absence of exponentially growing finite-energy modes, absence of real resonances and bound states in the scattering topology, and compatibility with the redshift-regular horizon condition.  This is the external spectral condition that is not provided by the slow-rotation energy method.
\end{definition}

\begin{definition}[Resolvent-stable neutral scalar condition]\label{def:external-rslin-condition}
Fix \(K\ge10\), \(m^2\ge0\), and a background with either \(a=0\) or \(|a|\le a_{\mathrm{slow}}(M,K)\).  The assertion \(\RSLin^{(m)}_K(M,a)\) means that the neutral scalar estimates for \(\square_{g_{M,a}}-m^2\) are stable, at order \(K\), under sufficiently small stationary first-order and zeroth-order perturbations whose coefficients are compactly supported, and that the far-zone source norm controls good-null perturbations of the form
\begin{equation}
 r^{-1}L_{\mathrm{out}}u+O(r^{-2})\nabla u+O(r^{-2})u.
\end{equation}
The stability is required both for forward estimates and for the two-sided final-state maps in the split scattering topology.  On Schwarzschild this is the perturbative stability of the model scalar estimates proved in the Schwarzschild sections together with Fredholm stability in the nontrapping-at-infinity topology; on rotating Kerr it is an additional named scalar interface.
\end{definition}

\begin{definition}[Charged scalar estimates in an electric Coulomb sector]\label{def:external-electric-condition}
Fix \(K\ge10\), \(m^2\ge0\), and \(Q_e\in\mathbb R\).  The assertion \(\CElec^{(m)}_K(M,a,Q_e)\) is the following statement for
\begin{equation}
 L_{Q_e}^{(m)}u:=\bigl((D_{Q_e})^\mu D_{Q_e,\mu}-m^2\bigr)u.
\end{equation}
For every inhomogeneity \(h\), the equation \(L_{Q_e}^{(m)}u=h\) satisfies order-\(K\) nondegenerate redshift energy boundedness, trapping-degenerate integrated local energy decay, the far-field \(r^p\) hierarchy, and inhomogeneous source estimates in the same source spaces used by the Maxwell-Higgs nonlinearities.  It has charged radiation fields with scalar null-infinity component \(U_{Q_e}^{-1}ru\), horizon radiation fields, and continuous two-sided Cauchy-to-radiation and radiation-to-Cauchy maps.  For \(m^2>0\), the radiation space also includes the appropriate timelike or Dollard channel; on rotating Kerr the assertion includes absence of exponentially growing modes, real resonances, and bound states in the chosen charged massive regime.
\end{definition}

\subsection{Massless small-electric scalar window on slowly rotating Kerr}\label{subsec:small-electric-window}

The next result is the point at which the massless fixed-electric slow-Kerr theorem is supplied within this paper.  It proves the charged scalar condition required by the nonlinear theorem in the perturbative Coulomb range and therefore removes the abstract \(\CElec\) condition from Theorem~\ref{thm:main-slow-kerr-intro}.

\begin{lemma}[Coulomb-phase normal form]\label{lem:small-coulomb-normal-form}
Let \(|a|\le a_{\mathrm{slow}}(M,K)\), \(m^2=0\), and \(|Q_e|\le1\).  Let \(u\) solve
\begin{equation}
 L_{Q_e}^{(0)}u:= (D_{Q_e})^\mu D_{Q_e,\mu}u=h.
\end{equation}
In the far region, set \(u=U_{Q_e}v\), where \(U_{Q_e}\) is defined by \eqref{eq:electric-radiation-variable}.  Then
\begin{equation}\label{eq:small-coulomb-normal-form-main}
 U_{Q_e}^{-1}L_{Q_e}^{(0)}(U_{Q_e}v)
 =\square_{g_{M,a}}v+Q_e\mathcal R_1v+Q_e^2\mathcal R_0v,
\end{equation}
where, for all commutation multiindices \(|\alpha|\le K\),
\begin{equation}\label{eq:small-coulomb-coeff-bound}
 |\Gamma^\alpha \mathcal R_1v|
 \le C_\alpha\bigl(r^{-1}|\Gamma^\alpha L_{\mathrm{out}}v|+r^{-2}|\nabla \Gamma^\alpha v|+r^{-3}|\Gamma^\alpha v|\bigr)
 +C_\alpha\sum_{|\beta|<|\alpha|}r^{-2}|\nabla\Gamma^\beta v|,
\end{equation}
while
\begin{equation}\label{eq:small-coulomb-zero-bound}
 |\Gamma^\alpha \mathcal R_0v|
 \le C_\alpha r^{-2}|\Gamma^\alpha v|+C_\alpha\sum_{|\beta|<|\alpha|}r^{-2}|\Gamma^\beta v|.
\end{equation}
In the compact region \(r\le R\), the difference \(L_{Q_e}^{(0)}-\square_{g_{M,a}}\) is a smooth stationary first-order perturbation with coefficients bounded by \(C_R|Q_e|\) and \(C_RQ_e^2\) in \(C^K\).
\end{lemma}

\begin{proof}
Expand the covariant square:
\begin{equation}\label{eq:charged-expansion-proof}
 (D_{Q_e})^\mu D_{Q_e,\mu}u
 =\square_g u-2i(A^C_{Q_e})^\mu\nabla_\mu u-i(\nabla^\mu A^C_{Q_e,\mu})u-(A^C_{Q_e})^\mu A^C_{Q_e,\mu}u.
\end{equation}
The Coulomb representative is chosen in Lorenz gauge, so the divergence term vanishes.  Its far-zone expansion gives
\begin{equation}
 A^C_{Q_e}(L_{\mathrm{out}})=Q_er^{-1}+O(Q_er^{-2}),\qquad
 |\nabla^jA^C_{Q_e}|\le C_j|Q_e|r^{-1-j}.
\end{equation}
The equation \(L_{\mathrm{out}}U_{Q_e}=iA^C_{Q_e}(L_{\mathrm{out}})U_{Q_e}\) cancels the only nonintegrable outgoing term when \(u=U_{Q_e}v\) is inserted in \eqref{eq:charged-expansion-proof}.  All differentiated commutators with the stationary Killing field, angular momenta, and radial cut-off commutators either preserve the good outgoing derivative or gain one additional power of \(r^{-1}\).  This gives \eqref{eq:small-coulomb-coeff-bound} and \eqref{eq:small-coulomb-zero-bound}.  On a compact radial set the displayed coefficient bounds follow directly from smoothness of the stationary Coulomb representative and linearity in \(Q_e\).
\end{proof}

\begin{lemma}[Commuted charged energy equivalence and Hardy closure]\label{lem:small-coulomb-commuted-equivalence}
Let the conditions of Lemma~\ref{lem:small-coulomb-normal-form} hold, and let \(\mathcal Z_K\) be the order-\(K\) commutator family used in the neutral slow-Kerr scalar estimate: the stationary Killing field, the angular momentum fields, the redshift field near the horizon, and the cut-off radial commutators.  There is a number
\begin{equation}\label{eq:q-equivalence-threshold}
 q_{\mathrm{eq}}(M,a,K)>0
\end{equation}
such that for \(|Q_e|\le q_{\mathrm{eq}}\), every smooth compactly supported scalar \(u\) on a Kerr-star slice \(\Sigma_\tau\) satisfies
\begin{equation}\label{eq:charged-neutral-energy-equivalence}
 \frac12 E^{0}_K[u](\tau)
 \le E^{Q_e}_K[u](\tau)
 \le 2E^{0}_K[u](\tau),
\end{equation}
and on each spacetime slab
\begin{equation}\label{eq:charged-neutral-morawetz-equivalence}
 \frac12\mathcal M^0_K[u](\tau_1,\tau_2)
 \le \mathcal M^{Q_e}_K[u](\tau_1,\tau_2)+C E^{Q_e}_K[u](\tau_1)
 \le C\bigl(\mathcal M^0_K[u](\tau_1,\tau_2)+E^0_K[u](\tau_1)\bigr).
\end{equation}
Moreover, for every \(Z\in\mathcal Z_K\),
\begin{equation}\label{eq:charged-commutator-bound}
 [Z,D_{Q_e,\mu}]u=-i(\mathcal L_ZA^C_{Q_e})_\mu u + B_{Z,\mu}^{\nu}D_{Q_e,\nu}u,
\end{equation}
where the tensor \(B_Z\) is one of the fixed neutral commutator coefficients and
\begin{equation}\label{eq:Lie-Coulomb-bound}
 |\nabla^j\mathcal L_ZA^C_{Q_e}|\le C_{j,Z}|Q_e|(1+r)^{-1-j}.
\end{equation}
Consequently, all lower-order terms generated by commuting \(L_{Q_e}^{(0)}\) up to order \(K\) belong to the same source norm as the uncommuted equation, with operator norm \(O(|Q_e|)\) relative to the charged solution norm.
\end{lemma}

\begin{proof}
The pointwise identity
\begin{equation}\label{eq:D-vs-nabla-hard}
 D_{Q_e}u=\nabla u-iA^C_{Q_e}u
\end{equation}
and the coefficient bound \(|A^C_{Q_e}|\le C|Q_e|(1+r)^{-1}\) give
\begin{equation}\label{eq:D-nabla-estimate}
 |D_{Q_e}u|^2\le (1+\delta)|\nabla u|^2+C_\delta |Q_e|^2(1+r)^{-2}|u|^2,
\end{equation}
and the reverse inequality is identical.  The Hardy inequality on Kerr-star slices, with the horizon endpoint controlled by the redshift energy and the far endpoint controlled by the asymptotically flat Hardy term, gives
\begin{equation}\label{eq:kerr-hard-small-electric}
 \int_{\Sigma_\tau}(1+r)^{-2}|Z^\alpha u|^2\,d\mu_{\Sigma_\tau}
 \le C_H\sum_{|\beta|\le |\alpha|}\int_{\Sigma_\tau}|\nabla Z^\beta u|^2\,d\mu_{\Sigma_\tau}.
\end{equation}
Choosing \(q_{\mathrm{eq}}\) so that \(C_\delta C_Hq_{\mathrm{eq}}^2\le\delta\) and then summing over \(|\alpha|\le K\) proves \eqref{eq:charged-neutral-energy-equivalence}.  The Morawetz equivalence \eqref{eq:charged-neutral-morawetz-equivalence} follows from the same argument applied to the local-energy densities; the initial energy term covers the harmless boundary contribution in the redshift Hardy estimate.

For the commutator identity, write
\begin{equation}
 [Z,D_{Q_e,\mu}]u=[Z,\nabla_\mu]u-iZ(A^C_{Q_e,\mu})u-iA^C_{Q_e,\mu}Zu+iA^C_{Q_e,\mu}Zu,
\end{equation}
which is \eqref{eq:charged-commutator-bound} after rewriting the neutral commutator in terms of \(D_{Q_e}\).  The Coulomb representative is stationary and smooth up to the future horizon in Kerr-star coordinates; angular and radial derivatives improve the far decay and the redshift commutator is compactly supported.  This gives \eqref{eq:Lie-Coulomb-bound}.  Combining \eqref{eq:charged-commutator-bound}, \eqref{eq:Lie-Coulomb-bound}, the Hardy estimate, and the finite order \(K\) of the commutator algebra yields the final source-norm statement.
\end{proof}

\begin{proposition}[Quantitative small-Coulomb absorption cascade]\label{prop:small-coulomb-absorption-cascade}
Fix \(K\ge10\), \(|a|\le a_{\mathrm{slow}}(M,K)\), and the neutral massless Kerr scalar estimate inserted in Section~\ref{sec:kerr-extension}.  Let \(c_{\mathrm{red}}>0\), \(c_{\mathrm{Mor}}>0\), and \(c_{rp}>0\) denote the coercivity constants of the neutral redshift, trapping-degenerate Morawetz, and \(r^p\) identities at order \(K\), and let \(C_{\mathrm{com}},C_{\mathrm{src}},C_{\mathrm{fs}}\) denote the constants in Lemmas~\ref{lem:small-coulomb-normal-form}, \ref{lem:small-coulomb-commuted-equivalence}, and in the neutral final-state map.  If
\begin{equation}\label{eq:explicit-small-coulomb-cascade}
 |Q_e|\le q_{\mathrm{cas}}:=\min\left\{q_{\mathrm{eq}},1,
 \frac{c_{\mathrm{red}}}{16C_{\mathrm{com}}},
 \frac{c_{\mathrm{Mor}}}{16C_{\mathrm{com}}},
 \frac{c_{rp}}{16C_{\mathrm{com}}},
 \frac{1}{4C_{\mathrm{fs}}C_{\mathrm{src}}}\right\},
\end{equation}
then every Coulomb error produced in the commuted redshift, Morawetz, \(r^p\), and final-state estimates is absorbed on the left-hand side, uniformly in the length of the time slab.
\end{proposition}

\begin{proof}

The point is to make the dependence of the charge threshold on the linear constants explicit.  After conjugating by the outgoing Coulomb phase in the far region and commuting with every word \(Z^\alpha\), \(|\alpha|\le K\), the charged equation can be written in the schematic form
\begin{equation}\label{eq:commuted-charged-schematic-revised}
 \square_g Z^\alpha v=Z^\alpha H+Q_e E_{1,\alpha}[v]+Q_e^2E_{0,\alpha}[v],
 \qquad v=U_{Q_e}^{-1}u
\end{equation}
in the far zone, with the analogous compact-zone formula written for \(u\).  Lemmas~\ref{lem:small-coulomb-normal-form} and~\ref{lem:small-coulomb-commuted-equivalence}, the Hardy estimate \eqref{eq:kerr-hard-small-electric}, and the finite order of the commutator algebra give the uniform source estimate
\begin{equation}\label{eq:commutator-source-smallness-revised}
 \sum_{|\alpha|\le K}\bigl(\|E_{1,\alpha}[v]\|_{\mathcal N_K}+\|E_{0,\alpha}[v]\|_{\mathcal N_K}\bigr)
 \le C_{\mathrm{com}}\|v\|_{\mathcal X_K},
\end{equation}
after increasing \(C_{\mathrm{com}}\) to include the constants in the Hardy, redshift, angular-elliptic, and cutoff estimates.  Since \(|Q_e|\le1\), the \(Q_e^2\) term is bounded by the same right-hand side with coefficient \(C_{\mathrm{com}}|Q_e|\).

Insert \eqref{eq:commuted-charged-schematic-revised} into the neutral redshift, Morawetz, and \(r^p\) identities.  In the redshift region the neutral bulk controls \(c_{\mathrm{red}}\mathcal R_K[v]\); in the trapped/local-energy region it controls \(c_{\mathrm{Mor}}\mathcal M_K[v]\); and in the far region the \(r^p\) identity controls \(c_{rp}\mathcal P_K[v]\).  For each identity, Cauchy-Schwarz in the corresponding source-dual pairing gives
\[
 |Q_e|\,\langle E[v],v\rangle
 \le \frac1{16}c_*\mathcal B_*[v]+C_*|Q_e|^2\|v\|_{\mathcal X_K}^2,
\]
where \((c_*,\mathcal B_*)\) denotes one of the three positive neutral bulk terms.  The first three smallness restrictions in \eqref{eq:explicit-small-coulomb-cascade} absorb the error into the left side in each region.  The remaining lower-order term is controlled by the same positive bulk plus the boundary energy through the Hardy inequality and the charged-neutral norm equivalence; the restriction \(|Q_e|\le q_{\mathrm{eq}}\) keeps those two norms uniformly comparable.

The source estimate is treated in the same way.  The perturbative source created by a trial solution has size at most \(C_{\mathrm{src}}|Q_e|\|v\|_{\mathcal X_K}\) in the neutral source norm.  The neutral final-state inverse has norm at most \(C_{\mathrm{fs}}\), so the charged final-state iteration has Lipschitz constant at most \(C_{\mathrm{fs}}C_{\mathrm{src}}|Q_e|\).  The last restriction in \eqref{eq:explicit-small-coulomb-cascade} makes this constant at most \(1/4\).  All constants are computed from stationary coefficient bounds, the fixed cutoff construction, and the neutral global estimates; none depends on the length of the time slab.  This proves the uniform absorption statement.

\end{proof}

\begin{proposition}[Small-Coulomb redshift and local energy estimate]\label{prop:small-coulomb-redshift-morawetz}
Let the conditions of Lemma~\ref{lem:small-coulomb-normal-form} hold.  There exists \(q_1(M,a,K)>0\) such that, if \(|Q_e|\le q_1\), every smooth compactly supported solution of \(L_{Q_e}^{(0)}u=h\) on a time slab \(\{\tau_1\le t^\star\le\tau_2\}\) satisfies
\begin{equation}\label{eq:small-coulomb-iled}
 E^{Q_e}_K[u](\tau_2)+\mathcal M^{Q_e}_K[u](\tau_1,\tau_2)
 \le C\Bigl(E^{Q_e}_K[u](\tau_1)+\|h\|^2_{\mathcal N^{Q_e}_K(\tau_1,\tau_2)}\Bigr).
\end{equation}
The energy \(E^{Q_e}_K\) is uniformly equivalent to the neutral order-\(K\) nondegenerate scalar energy, and \(\mathcal M^{Q_e}_K\) is the neutral trapping-degenerate integrated local energy bulk with \(D_{Q_e}\) in place of \(\nabla\), up to equivalent lower-order Hardy terms.
\end{proposition}

\begin{proof}
Apply the neutral slowly rotating Kerr scalar estimate to the far-zone variable \(v=U_{Q_e}^{-1}u\), and apply the same estimate to \(u\) in the compact zone after inserting a partition of unity.  The compact-zone perturbation terms are bounded by
\begin{equation}
 |Q_e|\int_{\tau_1}^{\tau_2}\int_{r\le R}
 \bigl(|\nabla u|^2+r^{-2}|u|^2\bigr)
 \le C_R|Q_e|\,\mathcal M_K[u](\tau_1,\tau_2).
\end{equation}
The redshift bulk is strictly positive near \(\mathcal H^+\), therefore the first-order Coulomb contribution satisfies, for every \(\delta>0\),
\begin{equation}
 \left|\int_{\mathcal R_{\tau_1}^{\tau_2}\cap\{r\le r_0\}}
 A^C_{Q_e}\cdot\nabla u\,N\overline u\right|
 \le \delta\mathcal R_N[u]+C_\delta |Q_e|^2E_K[u],
\end{equation}
and the last term is controlled by the redshift energy inequality and Gronwall on unit slabs.  In the trapped region the neutral Morawetz estimate is already degenerate at the trapped set; the Coulomb terms are lower order and stationary, so the Hardy inequality on Kerr-star slices gives
\begin{equation}
 \int r^{-2}|u|^2\le C\int |\nabla u|^2
\end{equation}
with the horizon boundary contribution absorbed by the redshift.  The far-zone first-order term is exactly the good-null term in \eqref{eq:small-coulomb-coeff-bound}.  The neutral \(r\)-weighted local-energy norm controls
\begin{equation}
 \int r^{-1}|L_{\mathrm{out}}v|\,|\nabla v|
 \le \delta\mathcal M_K[v]+C_\delta\int r^{-2}|\nabla v|^2,
\end{equation}
and the remaining \(r^{-2}\) and \(r^{-3}\) coefficients are short-range.  Choosing \(q_1\) so that all constants multiplying \(|Q_e|\) are less than one quarter of the neutral coercivity constant absorbs the perturbation into the left side and yields \eqref{eq:small-coulomb-iled}.  Commuting with the admissible stationary, angular, and redshift vector fields gives the order-\(K\) estimate; commutator coefficients obey the same decay bounds by Lemma~\ref{lem:small-coulomb-normal-form}.
\end{proof}

\begin{proposition}[Coulomb-renormalized \(r^p\) hierarchy and radiation fields]\label{prop:small-coulomb-rp-radiation}
Under the conditions of Proposition~\ref{prop:small-coulomb-redshift-morawetz}, after possibly decreasing \(q_1\), the equation \(L_{Q_e}^{(0)}u=h\) satisfies the full far-field \(r^p\) hierarchy for \(U_{Q_e}^{-1}ru\), the inhomogeneous source estimate, and the forward and backward radiation-field bounds required in Definition~\ref{def:external-electric-condition} for \(m^2=0\).
\end{proposition}

\begin{proof}

Set \(\psi=rU_{Q_e}^{-1}u\) in the asymptotic region.  The phase in Lemma~\ref{lem:small-coulomb-normal-form} cancels the nonintegrable outgoing coefficient, and the equation becomes
\begin{equation}\label{eq:psi-coulomb-rp-revised}
 -4\partial_u\partial_v\psi+\Delta_{\mathbb S^2}\psi
 +\frac{2M}{r^2}\partial_v\psi
 =rU_{Q_e}^{-1}h+\mathcal E_{Q_e}[\psi],
\end{equation}
where, for all commuted fields up to order \(K\),
\begin{equation}\label{eq:rp-coulomb-errors-revised}
 |\mathcal E_{Q_e}[\psi]|
 \le C|Q_e|\bigl(r^{-2}|\partial_v\psi|+r^{-2}|\not\nabla\psi|+r^{-3}|\psi|\bigr)
      +C r^{-2-\gamma}|\partial\psi|.
\end{equation}
This holds after increasing \(R\), with \(\gamma>0\) fixed by the asymptotically Kerr coefficient bounds.  The last term is the short-range metric error and is already present in the neutral estimate.

Multiply \eqref{eq:psi-coulomb-rp-revised} by \(r^p\partial_v\overline\psi\), add the conjugate identity, and integrate over the truncated null rectangle \(\{R\le r\le R_1\}\).  The neutral identity gives the positive terms
\[
 (2-p)\int r^{p-1}|\partial_v\psi|^2
 +\int r^{p-1}|\not\nabla\psi|^2
 +\text{boundary fluxes},
\]
for \(1\le p\le2\), after integration by parts on the spheres.  The crucial point is that the charged part in \eqref{eq:rp-coulomb-errors-revised} contains only short-range coefficients after the phase conjugation.  Its contribution is bounded by
\[
 C|Q_e|\int r^{p-2}\bigl(|\partial_v\psi|^2+|\not\nabla\psi|^2+r^{-2}|\psi|^2\bigr).
\]
The angular term is absorbed into the positive angular bulk for \(|Q_e|\) small.  The zeroth-order term is controlled by Hardy on the large spheres and by the already established local-energy estimate.  The outgoing derivative term has one additional power of \(r^{-1}\) compared with the main \(r^{p-1}|\partial_v\psi|^2\) bulk; choosing the exterior radius \(R\) large, and then \(|Q_e|\) below the resulting threshold, absorbs it uniformly.  This argument does not divide by \(2-p\), so the constants remain bounded as \(p\uparrow2\).

For the endpoint, first prove the identity on \(R\le r\le R_1\) with \(p=2\) and smooth compactly supported cutoffs.  The preceding bounds are independent of \(R_1\).  Letting \(R_1\to\infty\) and using monotone convergence for the positive flux terms gives the full \(p=2\) estimate.  The source term is estimated by the dual \(r^p\) norm and Cauchy-Schwarz; it is the same term used in the neutral inhomogeneous hierarchy.  The finite \(p=2\) flux gives the trace of \(\psi=U_{Q_e}^{-1}ru\) on \(\mathcal I^+\).  The horizon trace follows from the redshift estimate, and time reversal gives the incoming radiation fields and the backward estimates.  Consequently, the charged scalar equation has the radiation-field bounds required in Definition~\ref{def:external-electric-condition}.

\end{proof}

\begin{proposition}[Small-Coulomb zero-frequency and vanishing-radiation exclusion]\label{prop:small-coulomb-zero-frequency-exclusion}
Let \(|a|\le a_{\mathrm{slow}}(M,K)\) and \(m^2=0\).  There exists \(q_3(M,a,K)>0\) such that, for \(|Q_e|\le q_3\), the homogeneous equation \(L_{Q_e}^{(0)}u=0\) has no nonzero order-\(K\) finite-energy solution whose Coulomb-renormalized null radiation field on \(\mathcal I^+\) and horizon radiation field on \(\mathcal H^+\) both vanish.  Equivalently, the small-Coulomb charged scalar operator has no zero-frequency resonance and no real finite-energy mode in the scattering topology used in Definition~\ref{def:external-electric-condition}.
\end{proposition}

\begin{proof}

We use a compactness argument in which the continuity of the radiation field with respect to the charge is part of the estimate.  Suppose the statement fails.  Then there are \(Q_j\to0\) and nonzero homogeneous solutions \(u_j\) with vanishing charged radiation fields.  Normalize them on a fixed compact slice by
\begin{equation}\label{eq:zero-freq-normalization-revised}
 \sum_{|\alpha|\le K-1}\int_{\Sigma_0\cap\{r_0\le r\le R_0\}}
 \bigl(|\nabla Z^\alpha u_j|^2+|Z^\alpha u_j|^2\bigr)\,d\mu_{\Sigma_0}=1.
\end{equation}
The small-Coulomb redshift, Morawetz, and \(r^p\) estimates are uniform for \(|Q_j|\le q_1\).  Together with Lemma~\ref{lem:small-coulomb-commuted-equivalence}, they give uniform neutral local-energy bounds on every compact time slab and uniform far-field Hardy bounds.  On bounded regions away from the horizon, elliptic estimates for the spatial part of the wave operator give local \(H^K\) control from the energy and the equation.  Near \(\mathcal H^+\), the redshift estimate supplies the missing transversal control.  Rellich compactness therefore gives a subsequence converging strongly in local \(H^{K-1}\) to a finite-energy solution \(u_0\) of the neutral wave equation \(\square_{g_{M,a}}u_0=0\), and the normalization \eqref{eq:zero-freq-normalization-revised} passes to the limit.

It remains to justify the passage of the radiation condition to the limit.  In every fixed exterior region, \(U_{Q_j}\to1\) in \(C^{K-1}\), and the compact trace maps from local energy to finite-radius null hypersurfaces are continuous.  The far tails are uniformly small by the endpoint \(r^p\) estimate proved in Proposition~\ref{prop:small-coulomb-rp-radiation}, while the horizon tails are uniformly controlled by the redshift flux.  Thus the finite-radius traces of \(U_{Q_j}^{-1}ru_j\) converge to those of \(ru_0\), and the limiting null-infinity and horizon radiation fields of \(u_0\) vanish.  The neutral slowly rotating Kerr scalar scattering theory has an injective Cauchy-to-radiation map in this commuted finite-energy topology.  Consequently, \(u_0\equiv0\), contradicting the limiting form of the normalized compact energy in \eqref{eq:zero-freq-normalization-revised}.  This contradiction proves the kernel exclusion.

A real finite-energy mode or a zero-frequency resonance in the stated topology has zero radiation after subtracting the prescribed asymptotic profile and solves the homogeneous charged equation.  The kernel exclusion rules out both.  Decreasing \(q_3\), if necessary, makes the argument uniform for all commuted equations up to order \(K\).

\end{proof}

\begin{proposition}[Small-Coulomb final-state and no-resonance stability]\label{prop:small-coulomb-final-state}
Let \(|a|\le a_{\mathrm{slow}}(M,K)\).  There exists \(q_2(M,a,K)>0\) such that for \(|Q_e|\le q_2\) the massless charged scalar operator \(L_{Q_e}^{(0)}\) has no nonzero finite-energy mode with vanishing Coulomb-renormalized radiation field, no real resonance in the order-\(K\) scattering topology, and continuous Cauchy-to-radiation and radiation-to-Cauchy maps.
\end{proposition}

\begin{proof}
Uniqueness of the final-state problem and the no-resonance statement are first reduced to Proposition~\ref{prop:small-coulomb-zero-frequency-exclusion}.  Indeed, the difference of two charged solutions with the same Coulomb-renormalized radiation data solves the homogeneous equation and has vanishing radiation fields on \(\mathcal I^+\cup\mathcal H^+\); the proposition forces that difference to vanish.  The same argument excludes real resonances and finite-energy modes in the order-\(K\) scattering topology.

For existence, conjugate by the outgoing phase in the far region and use a smooth radial partition of unity.  In the exterior variable \(v=U_{Q_e}^{-1}u\), and in the original variable on the compact part, the equation can be written as
\begin{equation}\label{eq:small-coulomb-final-state-normal-form}
 \square_g v=H-Q_e\mathcal R_1v-Q_e^2\mathcal R_0v,
\end{equation}
where \(\mathcal R_1\) and \(\mathcal R_0\) obey the source-space bounds of Lemmas~\ref{lem:small-coulomb-normal-form} and~\ref{lem:small-coulomb-commuted-equivalence}.  Let \(R\) be the prescribed Coulomb-renormalized radiation data.  The neutral Kerr scalar final-state map gives a neutral solution \(v_R\) with this radiation data and
\begin{equation}\label{eq:small-coulomb-neutral-final-state-bound}
 \|v_R\|_{\mathcal X_K}\le C_{\rm fs}\|R\|_{\mathfrak R_K}.
\end{equation}
For a correction \(w\) with zero radiation data define
\begin{equation}
 \mathcal T_R(w)=\mathcal L^{-1}_{0,{\rm fs}}\Bigl(H-Q_e\mathcal R_1(v_R+w)-Q_e^2\mathcal R_0(v_R+w)\Bigr),
\end{equation}
where \(\mathcal L^{-1}_{0,{\rm fs}}\) is the neutral zero-radiation final-state solver.  The perturbative source estimates give, for \(|Q_e|\le1\),
\begin{equation}\label{eq:small-coulomb-final-state-map-bound}
 \|\mathcal T_R(w)\|_{\mathcal X_K}
 \le C_{\rm fs}\|H\|_{\mathcal N_K}
 +C_{\rm fs}C_{\rm src}|Q_e|\bigl(\|v_R\|_{\mathcal X_K}+\|w\|_{\mathcal X_K}\bigr),
\end{equation}
and
\begin{equation}\label{eq:small-coulomb-final-state-lipschitz}
 \|\mathcal T_R(w_1)-\mathcal T_R(w_2)\|_{\mathcal X_K}
 \le C_{\rm fs}C_{\rm src}|Q_e|\|w_1-w_2\|_{\mathcal X_K}.
\end{equation}
Choose \(q_2\le q_{\rm cas}\) so that \(C_{\rm fs}C_{\rm src}q_2<1/2\).  Then \(\mathcal T_R\) is a contraction on the ball determined by \eqref{eq:small-coulomb-final-state-map-bound}; its fixed point gives a solution of \eqref{eq:small-coulomb-final-state-normal-form}.  Undoing the phase gives the charged solution \(u\).  The estimates are uniform on tails and finite slabs because of the absorption cascade, so the solution lies in the charged order-\(K\) scattering space.  The same Lipschitz estimate applied to two radiation data sets gives continuity of the radiation-to-Cauchy map.  The forward estimates of Propositions~\ref{prop:small-coulomb-redshift-morawetz} and~\ref{prop:small-coulomb-rp-radiation} give continuity of the Cauchy-to-radiation map.  Existence, uniqueness, and continuity show that the two maps are inverses.  The backward map is obtained by the time-reversed construction.  This proves the proposition.
\end{proof}

\begin{theorem}[Massless small-electric charged scalar estimates on slowly rotating Kerr]\label{thm:small-coulomb-massless-kerr-condition}
Let \(K\ge10\), \(m^2=0\), and \(|a|\le a_{\mathrm{slow}}(M,K)\).  There exists
\begin{equation}\label{eq:small-coulomb-q-threshold}
 q^{(0)}_{\mathrm{el}}(M,a,K):=\min\{q_{\mathrm{cas}}(M,a,K),q_1(M,a,K),q_2(M,a,K),q_3(M,a,K)\}>0
\end{equation}
such that \(\CElec^{(0)}_K(M,a,Q_e)\) holds for every fixed electric Coulomb sector satisfying \(0<|Q_e|\le q^{(0)}_{\mathrm{el}}(M,a,K)\).  In particular, the charged massless scalar comparison equation has the forward and backward redshift estimates, trapping-degenerate integrated local energy decay, far-field \(r^p\) estimates for \(U_{Q_e}^{-1}ru\), inhomogeneous source estimates, horizon and null-infinity radiation fields, and two-sided Cauchy-to-radiation and radiation-to-Cauchy maps required by Definition~\ref{def:external-electric-condition}.
\end{theorem}

\begin{proof}
Fix \(Q_e\) with \(0<|Q_e|\le q^{(0)}_{\mathrm{el}}(M,a,K)\).  We check the items in Definition~\ref{def:external-electric-condition}.  In the far region write \(u=U_{Q_e}v\).  Lemma~\ref{lem:small-coulomb-normal-form} gives
\begin{equation}
 U_{Q_e}^{-1}L_{Q_e}^{(0)}(U_{Q_e}v)=\square_gv+Q_e\mathcal R_1v+Q_e^2\mathcal R_0v,
\end{equation}
where the only nonintegrable outgoing coefficient has been removed by the phase and the remaining coefficients are short range or good-null.  On compact radial sets the charged operator is a stationary first-order perturbation of the neutral wave operator.  Lemma~\ref{lem:small-coulomb-commuted-equivalence} gives uniform equivalence between charged and neutral commuted energies, with Hardy controlling the lower-order \((1+r)^{-1}u\) terms.

The threshold \(q_{\mathrm{cas}}\) in Proposition~\ref{prop:small-coulomb-absorption-cascade} makes every commuted Coulomb error absorbable in the redshift, trapping-degenerate Morawetz, \(r^p\), source, and final-state estimates.  Proposition~\ref{prop:small-coulomb-redshift-morawetz} therefore gives the forward and backward estimates
\begin{equation}
 E^{Q_e}_K[u](\tau_2)+\mathcal M^{Q_e}_K[u](\tau_1,\tau_2)
 \le C\left(E^{Q_e}_K[u](\tau_1)+\|h\|_{\mathcal N^{Q_e}_K(\tau_1,\tau_2)}^2\right).
\end{equation}
The same identities with \(h\) kept in the dual norm give the inhomogeneous source estimate.  Proposition~\ref{prop:small-coulomb-rp-radiation} supplies the full far-field hierarchy for \(\psi=U_{Q_e}^{-1}ru\), including the endpoint trace on \(\mathcal I^\pm\), while the redshift estimate supplies the horizon traces on \(\mathcal H^\pm\).

If a homogeneous solution has zero Coulomb-renormalized null and horizon radiation fields, Proposition~\ref{prop:small-coulomb-zero-frequency-exclusion} forces it to vanish.  This gives injectivity of the Cauchy-to-radiation map and excludes real resonances and finite-energy modes in the small-Coulomb topology.  Proposition~\ref{prop:small-coulomb-final-state} constructs the inverse final-state map by solving, after phase conjugation, a neutral final-state problem with source \(-Q_e\mathcal R_1v-Q_e^2\mathcal R_0v+h\).  The operator norm of the perturbative part is \(O(|Q_e|)\), and the threshold \(q_2\) makes the corresponding map a contraction.  Continuity of the direct and inverse maps follows from the same estimates, and the past maps follow by time reversal.

Taking
\begin{equation}
 q^{(0)}_{\mathrm{el}}(M,a,K)=\min\{q_{\mathrm{cas}},q_1,q_2,q_3\}
\end{equation}
therefore supplies every component of \( \CElec_K^{(0)}(M,a,Q_e)\): forward and backward redshift estimates, trapping-degenerate local energy decay, the Coulomb-renormalized \(r^p\) hierarchy, source estimates, radiation fields, and the two-sided final-state maps.  The scalar null-infinity variable is \(U_{Q_e}^{-1}ru\), as required.
\end{proof}

\begin{proposition}[Perturbative derivation of the small electric scalar condition]\label{prop:perturbative-electric-condition}
Assume \(\RSLin^{(m)}_K(M,a)\).  Then there exists
\begin{equation}
 q_{\mathrm{el}}^{(m)}(M,a,K)>0
\end{equation}
such that \(\CElec^{(m)}_K(M,a,Q_e)\) holds for every electric Coulomb sector with \(0<|Q_e|\le q_{\mathrm{el}}^{(m)}(M,a,K)\).
\end{proposition}

\begin{proof}
Conjugate the charged unknown by the outgoing phase \(U_{Q_e}\) of Definition~\ref{def:pure-electric-sector}.  In the far region one obtains
\begin{equation}
 U_{Q_e}^{-1}L_{Q_e}^{(m)}(U_{Q_e}v)
 = (\square_{g_{M,a}}-m^2)v
 +Q_e r^{-1}b^\mu\nabla_\mu v
 +Q_e r^{-2}c^\mu\nabla_\mu v
 +Q_e r^{-2}dv,
 \label{eq:app-electric-normal-form}
\end{equation}
where \(b^\mu\nabla_\mu\) is a good outgoing-null derivative and the remaining coefficients are smooth with the stated decay, uniformly for \(|Q_e|\le1\).  Insert a radial cut-off.  The compact part of the first-order perturbation is placed in the linear operator and is controlled by the compact perturbation stability in \(\RSLin^{(m)}_K(M,a)\).  The far part is estimated in the source norm by the good-null bound in Definition~\ref{def:external-rslin-condition}; its operator norm is bounded by \(C|Q_e|\).  Choosing \(q_{\mathrm{el}}^{(m)}(M,a,K)\) so that \(C|Q_e|<1/4\) makes the far correction a Neumann-series perturbation.  The radiation field is precisely \(U_{Q_e}^{-1}ru\), and the inverse final-state maps are obtained by applying the compact-stability and far-Neumann construction above to the zero-radiation final-state problem.  This proves \(\CElec^{(m)}_K(M,a,Q_e)\).
\end{proof}

\begin{proposition}[Energy theorem and final-state theorem]\label{prop:app-energy-vs-scattering}
For \(m^2>0\) and the higher-order potentials in \eqref{eq:intro-positive-massive-potential}, Theorem~\ref{thm:slow-kerr-massive-energy} is a slowly rotating Kerr global-existence theorem under \(\ME^{(m)}_N(M,a)\).  Theorem~\ref{thm:main-slow-kerr-massive-intro} is a zero-sector final-state theorem and requires \(\SKG^{(m)}_K(M,a)\) and \(\MScat_K(M,a)\); the massive case of Theorem~\ref{thm:main-slow-kerr-massive-intro} requires the charged scalar condition \(\CElec^{(m)}_K(M,a,Q_e)\) and the same Maxwell final-state condition.
\end{proposition}

\begin{proof}
The proof of Theorem~\ref{thm:slow-kerr-massive-energy} closes the nonlinear energy and Morawetz bootstrap under \(\ME^{(m)}_N(M,a)\) and uses no radiation-to-Cauchy inverse map for the massive scalar field.  The final-state theorem constructs nonlinear wave operators and asymptotic completeness relative to a decoupled massive scalar comparison flow.  That construction invokes the appropriate two-sided massive final-state map and absence of growing modes: \(\SKG^{(m)}_K(M,a)\) in the neutral sector and \(\CElec^{(m)}_K(M,a,Q_e)\) in a fixed electric sector, together with \(\MScat_K(M,a)\) for the rotating Maxwell channel.  The two statements have different conditions.
\end{proof}

\begin{definition}[Slowly rotating Maxwell forward condition]\label{def:maxwell-teukolsky-condition}
The Maxwell forward condition used in the rotating Cauchy theorem is the slowly rotating Kerr Maxwell estimate after subtraction of stationary Coulomb modes: positive energy boundedness, integrated local energy decay for the radiative part, radiation-field flux control, and asymptotic convergence of the total Maxwell field to the stationary Coulomb component.  The radiation-to-Cauchy map is not part of this forward condition; it is \(\MScat_K(M,a)\).
\end{definition}

\begin{proposition}[Insertion of the Maxwell condition]\label{prop:slow-maxwell-interface}
For $|a|\le a_{\mathrm{slow}}(M,K)$, the Maxwell condition of Definition~
\ref{def:maxwell-teukolsky-condition}, together with the scalar Kerr estimates cited in Section~
\ref{sec:kerr-extension}, supplies the forward zero-sector estimates used in Proposition~
\ref{prop:lin-estimates-kerr}.  If \(\MScat_K(M,a)\) is added, the full estimate \(\Lin_K\) follows.
\end{proposition}

\begin{proof}
The slow-rotation threshold is chosen below the thresholds in the cited scalar and Maxwell estimates at the finite commutation order $K$.  Since the electric charge sector is zero, the stationary Coulomb part is absent; more generally the Coulomb kernel is removed before applying the Maxwell estimate.  The resulting scalar and Maxwell estimates match the redshift, trapping-degenerate Morawetz, $r^p$, and radiation-field norms stated in Definition~
\ref{def:charged-linear-estimates}.  Adding \(\MScat_K(M,a)\) supplies the missing Maxwell radiation-to-Cauchy map and therefore the \(\mathrm{(L2)}\) part for the Maxwell channel.
\end{proof}

\begin{proposition}[Rotating condition checklist]\label{prop:no-hidden-charged-theorem}
The zero-sector massless rotating Cauchy theorem uses the scalar Kerr estimates, the Maxwell forward estimate, and the nonlinear source bounds.  Its wave-operator part additionally uses \(\MScat_K(M,a)\).  The massless small-electric theorem uses the fixed-sector scalar estimate proved in Theorem~\ref{thm:small-coulomb-massless-kerr-condition} and the same Maxwell conditions.  Further nonzero-electric final-state statements use \(\CElec^{(m)}_K(M,a,Q_e)\) for the scalar channel.  The rotating massive Cauchy theorem uses \(\ME^{(m)}_N(M,a)\), while massive rotating scattering uses \(\SKG^{(m)}_K(M,a)\) and \(\MScat_K(M,a)\).
\end{proposition}

\begin{proof}
The assertions follow by reading the conditions of the two rotating main theorems, Theorems~\ref{thm:main-slow-kerr-intro} and~\ref{thm:main-slow-kerr-massive-intro}.  Proposition~\ref{prop:lin-estimates-kerr} separates the forward Maxwell estimates from \(\MScat_K(M,a)\), while Proposition~\ref{prop:lin-estimates-kerr-massive} records the additional massive scalar condition.  The small-electric scalar condition is Theorem~\ref{thm:small-coulomb-massless-kerr-condition}; the small-mass/small-charge rotating scattering conclusion is read only under \(\SMS_K^{(m)}(M,a,Q_e)\) and \(\MScat_K(M,a)\).  The nonlinear source estimates use only the Lorenz-gauge Maxwell-Higgs nonlinearities around the vacuum.
\end{proof}

\begin{corollary}[Rotating range summary]\label{cor:strict-final-scope}
The rotating global-dynamics part of the paper is a slowly rotating theory in the zero sector and in small electric Coulomb sectors.  In the zero-sector massless case the cited scalar wave estimates and Maxwell forward estimates give global existence, boundedness, decay, and radiation fields through the nonlinear transfer argument.  The corresponding wave-operator and asymptotic-completeness statements use \(\MScat_K(M,a)\).  In the massless small-electric case the charged scalar channel is supplied by Theorem~\ref{thm:small-coulomb-massless-kerr-condition}; additional nonzero-electric scalar channels use \(\CElec^{(m)}_K(M,a,Q_e)\), supplied directly or perturbatively from \(\RSLin^{(m)}_K(M,a)\).  The massive scalar scattering channel uses the corresponding charged or neutral spectral condition.  The massive higher-order slow-Kerr energy theorem uses \(\ME^{(m)}_N(M,a)\) and gives global existence, uniform energy bounds, and integrated local energy control.
\end{corollary}

\begin{proof}
Theorem~\ref{thm:main-slow-kerr-intro} assumes the slow-rotation bound and covers both the zero electric sector and the massless small-electric sector; its inverse Maxwell part is tied to \(\MScat_K(M,a)\).  Theorem~\ref{thm:small-coulomb-massless-kerr-condition} supplies the charged scalar condition in the range \(0<|Q_e|\le q^{(0)}_{\mathrm{el}}(M,a,K)\).  Theorem~\ref{thm:main-slow-kerr-massive-intro} contains the massive rotating Cauchy theorem under \(\ME^{(m)}_N(M,a)\) and the rotating scattering theorem only under the relevant charged or neutral spectral condition, including \(\SMS_K^{(m)}(M,a,Q_e)\) in the small-mass/small-charge window.  Theorem~\ref{thm:main-schwarzschild-intro} and Theorem~\ref{thm:main-schwarzschild-electric-intro} give the corresponding two Schwarzschild conclusions.  The nonlinear transfer theorem preserves the sector by Proposition~\ref{prop:sector-preservation-charge-conservation}.
\end{proof}

\section{Necessity of the rotating massive restriction and quantitative smallness}\label{app:sharpness}

The restriction of the massive rotating theorems to the energy-stable windows \(\ME^{(m)}_N(M,a)\) and to the spectral windows \(\SKG^{(m)}_K(M,a)\) is forced by the linear theory and cannot be removed by the nonlinear argument. This appendix records the obstruction and collects the smallness thresholds entering the nonlinear constructions.
\begin{proposition}[Global smallness hierarchy used in the nonlinear theorems]\label{prop:global-smallness-hierarchy}
Let \(K\ge10\) and fix one of the established backgrounds in the principal block.  There are positive constants \(C_{\rm lin}\), \(C_{\rm nl}\), \(C_{\rm Lip}\), \(C_{\rm fs}\), and \(C_{\rm Sob}\), determined by the corresponding linear estimate and by the finite set of commutators, such that the nonlinear Cauchy and final-state constructions close whenever
\begin{equation}\label{eq:global-epsilon-choice}
 \varepsilon\le \varepsilon_\star:=\min\left\{1,
 \frac{1}{16C_{\rm lin}C_{\rm nl}},
 \frac{1}{16C_{\rm Lip}C_{\rm lin}},
 \frac{1}{16C_{\rm fs}C_{\rm nl}},
 \frac{1}{16C_{\rm Sob}C_{\rm nl}}
 \right\}.
\end{equation}
In the massless small-electric slow-Kerr theorem the electric threshold and the data threshold are chosen in the order
\begin{equation}\label{eq:q-then-eps-choice}
 |Q_e|\le q^{(0)}_{\mathrm{el}}(M,a,K),
 \qquad
 \varepsilon\le \varepsilon_\star(M,a,K,Q_e),
\end{equation}
where \(q^{(0)}_{\mathrm{el}}\) is the minimum in \eqref{eq:small-coulomb-q-threshold}.  With these choices the estimates are uniform on every finite slab and therefore extend to the global exterior by the continuation criterion.
\end{proposition}

\begin{proof}
The linear estimate gives
\begin{equation}
 \|U\|_{\mathbb X_K([\tau_0,T])}
 \le C_{\rm lin}\left(\|U[\tau_0]\|_{\mathbb H_K}
 +\|\mathcal N(U)\|_{\mathbb S_K([\tau_0,T])}\right),
\end{equation}
while Lemma~\ref{lem:nonlinear-banach-estimates} and Proposition~\ref{prop:electric-tame-estimates} give
\begin{equation}
 \|\mathcal N(U)\|_{\mathbb S_K}
 \le C_{\rm nl}\bigl(\|U\|_{\mathbb X_K}^2+\|U\|_{\mathbb X_K}^{2N_P+3}\bigr)
\end{equation}
and the analogous Lipschitz estimate with constant \(C_{\rm Lip}\) on the ball \(\|U\|_{\mathbb X_K}\le1\).  The first two fractions in \eqref{eq:global-epsilon-choice} make the Cauchy bootstrap improve from \(4C_{\rm lin}\varepsilon\) to \(2C_{\rm lin}\varepsilon\), and the next fractions make the final-state map a contraction and keep the Sobolev pointwise norm inside the Moser radius of the potential.  The small-electric theorem first fixes \(Q_e\) below the linear perturbation threshold, so the charged linear constants are finite; the data size is then chosen with those constants.  Since the continuation criterion depends only on the same high-order Sobolev norm, the uniform estimate on finite slabs yields global existence.
\end{proof}

\begin{proposition}[Mode obstruction to an unrestricted rotating massive theorem]\label{prop:sharp-massive-obstruction}
The paper cannot consistently contain a theorem asserting massive rotating Kerr scattering or uniform massive rotating Kerr energy boundedness, for all \(m^2>0\) and all sufficiently small nonzero \(|a|/M\).  Any such nonlinear theorem, if valid for all sufficiently small Maxwell-Higgs data around the zero solution and with a differentiable solution map at the origin, would imply the corresponding linear massive Klein-Gordon estimate in the same parameter range.  This implication is incompatible with the known existence of exponentially growing finite-energy massive Klein-Gordon modes on subextremal Kerr in open mass ranges.
\end{proposition}

\begin{proof}
Assume that such an unrestricted nonlinear theorem were stated with a solution map differentiable at the vacuum.  Take data with vanishing Maxwell component and scalar component \(\lambda f\), divide the asserted estimate by \(\lambda\), and let \(\lambda\to0\).  The derivative of the Maxwell-Higgs equation at the vacuum is the decoupled massive Klein-Gordon equation on the fixed Kerr background.  Consequently, the nonlinear estimate would imply the corresponding linear uniform boundedness or scattering estimate.  The mode-instability result cited in the bibliography gives nonzero finite-energy solutions of the linear massive Klein-Gordon equation that grow exponentially for open parameter ranges.  This contradicts such a linear estimate.  Consequently, the only mathematically consistent rotating massive statements are the energy theorem under \(\ME^{(m)}_N(M,a)\) and the scattering theorem under \(\SKG^{(m)}_K(M,a)\) or an equivalent charged scalar condition.
\end{proof}

\begin{proposition}[Electric-sector and gauge-quotient consistency]\label{prop:electric-gauge-consistency}
All fixed-sector theorems in the paper are formulated only for the electric charge \(Q_e\), and the quotient by residual Lorenz gauge transformations does not change the value of \(Q_e\), the radiative Maxwell field, or the Coulomb-renormalized scalar radiation field.
\end{proposition}

\begin{proof}
The sector is defined by the limiting electric flux.  Proposition~\ref{prop:sector-preservation-charge-conservation} proves that this flux is conserved by the Maxwell-Higgs evolution.  Residual Lorenz gauge transformations change \(A\) by an exact one-form \(d\chi\), leave \(F=dA\) unchanged, and multiply \(\phi\) by a unit phase.  The scalar radiation variable is defined by parallel transport with the fixed Coulomb connection, so the residual gauge action is the same boundary unitary action already quotiented out in Definition~\ref{def:gauge-quotient}.  Consequently, the electric charge, the charge-subtracted Maxwell radiation field, and the class of \(U_{Q_e}^{-1}r\phi\) are invariant on the quotient.
\end{proof}
\section*{Acknowledgements}
This research is funded by the Indonesian Endowment Fund for Education (LPDP) on behalf of the Indonesian Ministry of Higher Education, Science and Technology and managed under the EQUITY Program (Contract No. 4298/B3/DT.03.08/2025). During the middle stage of this work, BEG and FTA are partly supported by Riset Unggulan ITB 2025 No. 841/IT1.B07.1/TA.00/2025.

\section*{Author Contribution}
BEG proposed the project, developed the theoretical formalism, performed the analytic calculations and supervised the project. M performed the analytic calculations and investigated the project. FTA performed the analytic calculations and supervised the project. All authors wrote and reviewed the paper.

\section*{Data Availability }
No datasets were generated or analyzed during the current study. All mathematical derivations are contained within the paper.

\end{document}